%% file: thesis.tex
\documentclass{ucetd}
\usepackage{subfigure,epsfig,amsfonts}
\usepackage{amsmath}
\usepackage{amssymb}
\usepackage{amsthm}
\usepackage{algpseudocode}
\usepackage{caption}
\usepackage{epstopdf}
\usepackage{bigints}
\usepackage{color}
\usepackage{fontenc}

\def\etal{{\it et al. }}

\DeclareCaptionType{algorithm}
\DeclareMathOperator*{\argmin}{arg\,min}
\newcommand{\new}[1]{\textcolor{black}{#1}}

\graphicspath{{./Chapter2/Figs/}{./Chapter3/Figs/}{./Chapter4/Figs/}{./Chapter5/Figs/}{./Chapter6/Figs/}}

\title{Task-Based Optimization of Computed Tomography Imaging Systems}
\author{Adrian A. S\'anchez}
\committee{Medical Physics}
\division{Biological Sciences}
\divisiontwo{Pritzker School of Medicine}
\degree{Doctor of Philosophy}
\date{June 2015}

\dedication{To My Wife Jesse}

\epigraph{J  \hspace{0.08cm} + M \hspace{0.08cm} + J

\vspace{2.0cm}

\it Ad Majorem Dei Gloriam
}

\begin{document}
\maketitle

\makecopyright
\makededication
\makeepigraph

\tableofcontents
\listoffigures
\listoftables

\acknowledgments
I would first like to gratefully acknowledge the guidance and support of Dr.\ Emil Sidky and Dr.\ Xiaochuan Pan, who 
through their example have given me a scientific standard to which I aspire. 
I have been privileged to watch several more senior students mature under their guidance, and it was evident 
at an early stage of this project that they were dedicated mentors and advisors. As time progressed, I learned more and more
that each of them is deeply invested in the scientific and professional success of each of their students, and I 
consider myself fortunate to have benefited from their guidance. I sincerely thank them for countless hours of discussion, 
support, consultation, and advice, the rewards of which have impacted me profoundly.

I wish also to thank Dr.\ Ingrid Reiser for her support and encouragement at every stage of my doctoral studies. Likewise, 
I appreciate the help I have received from my committee members, Dr.\ Patrick La Rivi\`{e}re and Dr.\ Charles Pelizzari. 
I have been exceedingly fortunate in my many friends and lab members who have provided tremendous scientific and personal support. 
Lab members I especially wish to acknowledge are Zheng Zhang, Andrew Davis, Buxin Chen, Sean Rose, Dr.\ Xiao Han, Dr.\ Erik Pearson, and Dr.\ Junguo Bian.
My classmates Jonathon Rosenfield, David Rigie, and William Weiss were a tremendous help throughout my studies, and particularly in the early years of my Ph.D. 
In particular, I owe a special debt of gratitude to David Rigie for his unwavering friendship and support.

Ultimately, none of this work would have been possible without my wife Jessica. Her love and encouragement have guided me through these past several years, and
will undoubtedly be my most important source of strength in years to come. Lastly, I want to thank my sister and parents for their love and support, as well as for the examples they provided
through their own professional achievements. 

Grant support from the Ruth L. Kirchstein NRSA F-31 EB018742 and the Lawrence H. Lanzl Award helped to make this work possible, along with NIH R01 Nos. CA158446, 
CA182264, EB018102, and EB000225.

\abstract
The goal of this thesis is to provide a framework for the use of task-based metrics of image quality to aid in the design, implementation, and evaluation of CT image reconstruction algorithms
and CT systems in general.
We support the view that task-based metrics of image quality can be useful in guiding the algorithm design and implementation process 
in order to yield images of objectively superior quality and higher utility for a given task. Further, we believe that metrics such as the Hotelling observer (HO) SNR 
can be used as summary scalar metrics of image quality for the evaluation of images produced by novel reconstruction algorithms. In this work, we aim 
to construct a concise and versatile formalism for image reconstruction algorithm design, implementation, and assessment. The bulk 
of the work focuses on linear analytical algorithms, specifically the ubiquitous filtered back-projection (FBP) algorithm.
However, due to the demonstrated importance of optimization-based algorithms in a wide variety of 
CT applications, we devote one chapter to the characterization of noise properties in TV-based iterative reconstruction, as the understanding of image statistics 
in optimization-based reconstruction is the limiting factor in applying HO metrics.

\mainmatter

\include{./Chapter1-Introduction/Introduction}

\include{./Chapter2/Chapter2}

\include{./Chapter3/Chapter3}

\include{./Chapter4/Chapter4}

\include{./Chapter5/Chapter5}

\include{./Chapter6/Chapter6}

\include{./Chapter7-Conclusion/Conclusion}



\end{document}

%% file: Chapter1-Introduction/Introduction.tex
\chapter{Introduction}

From its inception in 1972, x-ray computed tomography (CT) quickly developed a role as an indispensable tool in the diagnosis of disease. 
By illuminating a patient with x-rays at varying angles, three-dimensional information could be obtained and, along with the invention of MRI, 
CT enabled doctors to non-invasively visualize volumetric anatomical information for the first time in history. In the decades that followed,
diagnostic CT has become a staple of clinical radiology, while other more specialized applications of CT have rapidly developed as well. While 
standard diagnostic CT consists of a circular gantry through which a patient couch is translated, CT in modern clinical practice takes on many forms, 
from interventional cone-beam CT which guides surgical procedures to dedicated-purpose CT systems for dentistry or orthopedics. 

A number of technological advancements have enabled this blossoming of new CT applications. One vital aspect has been the development and refinement of 
CT image reconstruction algorithms, or the mathematical processes that transform the many x-ray projections into a single volumetric image. 
Work on CT reconstruction has been as diverse as CT itself, and in the research community reconstruction is nearly always listed as a potential source of improvement for 
existing CT systems or as an enabling force for novel, previously infeasible CT system designs. 

Unfortunately, many of the advancements which reconstruction methods could enable are difficult to realize in practice, since even the simplest reconstruction algorithms 
involve a wide array of implementation decisions. 
For well established CT applications, contributing marginal image quality improvements through better reconstruction is therefore challenging, since a delicate balance 
of optimal implementation options must be achieved in order to observe any benefit from the improved reconstruction method.
For more novel CT applications, the situation is even worse, since we lack the benefit 
of several decades of experience to guide our search for the ``best'' reconstruction methods, and we often find ourselves paralyzed by the incredible variety of 
methods and algorithms which now exist. Even after the field is narrowed by considering only a single class of algorithms, there is still a wide array of parameters 
for any algorithm, and the selection of appropriate values for these parameters is often difficult. This causes many problems in CT research and development, not the 
least of which is that it is then surprisingly difficult to objectively compare two reconstruction methods to determine which is better.

In general terms, this thesis aims to address this issue by defining image quality metrics which facilitate the objective assessment and design of CT reconstruction algorithms. 
Further, the metrics which we will apply here can be used not only to guide reconstruction development, but also in the assessment of a completed CT system, 
including the impact of system design, hardware, and even the patient being imaged. For the remainder of this introductory chapter, we will provide necessary background 
in the fundamentals of CT image reconstruction, as well as introduction to basic concepts of what is known as task-based image quality. Specifically, we will describe
the concept of the Hotelling observer, which forms the basis for the image quality metrics we apply throughout the thesis. 

In Chapter \ref{ch:humans}, we attempt to provide some insight into what the image quality metrics we develop actually mean, with regard to the usefulness of an 
image for a human with a given task to perform. This will serve as motivation for designing CT systems in a way that is optimal with respect to the metrics we develop.
In Chapter \ref{ch:sampling}, we investigate the dependency of our metrics on the number of image pixels used. Our primary motivation for this chapter is to attempt
to design a metric which requires only a subset of image pixels, as the large number of pixels in a full image currently presents a computational burden for the type of image quality 
metrics we advocate. Next, in Chapter \ref{ch:roi}, guided by the preceding chapter, we validate the use of metrics which rely only on a small region-of-interest in the 
reconstructed image. Finally, in Chapter \ref{ch:breastct}, we apply the developed metrics to an extensive optimization of dedicated breast CT system, including the reconstruction 
algorithm. 

Chapters \ref{ch:humans} through \ref{ch:breastct} all pertain to the most widely used class of reconstruction algorithms, namely direct analytic algorithms. More recently, 
another class of algorithms known as optimization-based or iterative image reconstruction (IIR) algorithms has become popular. These IIR algorithms have the potential 
to enable a wide variety of previously implausible CT systems and methods, but present unique challenges for image quality assessment. In Chapter \ref{ch:tv}, we 
lay the groundwork for the extension of the methods developed in previous chapters to a specific type of IIR algorithm which has seen increasing popularity in recent years.
Chapter \ref{ch:conclusion} gives conclusions and some final considerations for the thesis.  

\section{Background: Computed Tomography}

\subsection{An Imaging Model for CT}
\label{data_model_sec}

\begin{figure}
 \centering
 \includegraphics[width = 0.6\columnwidth]{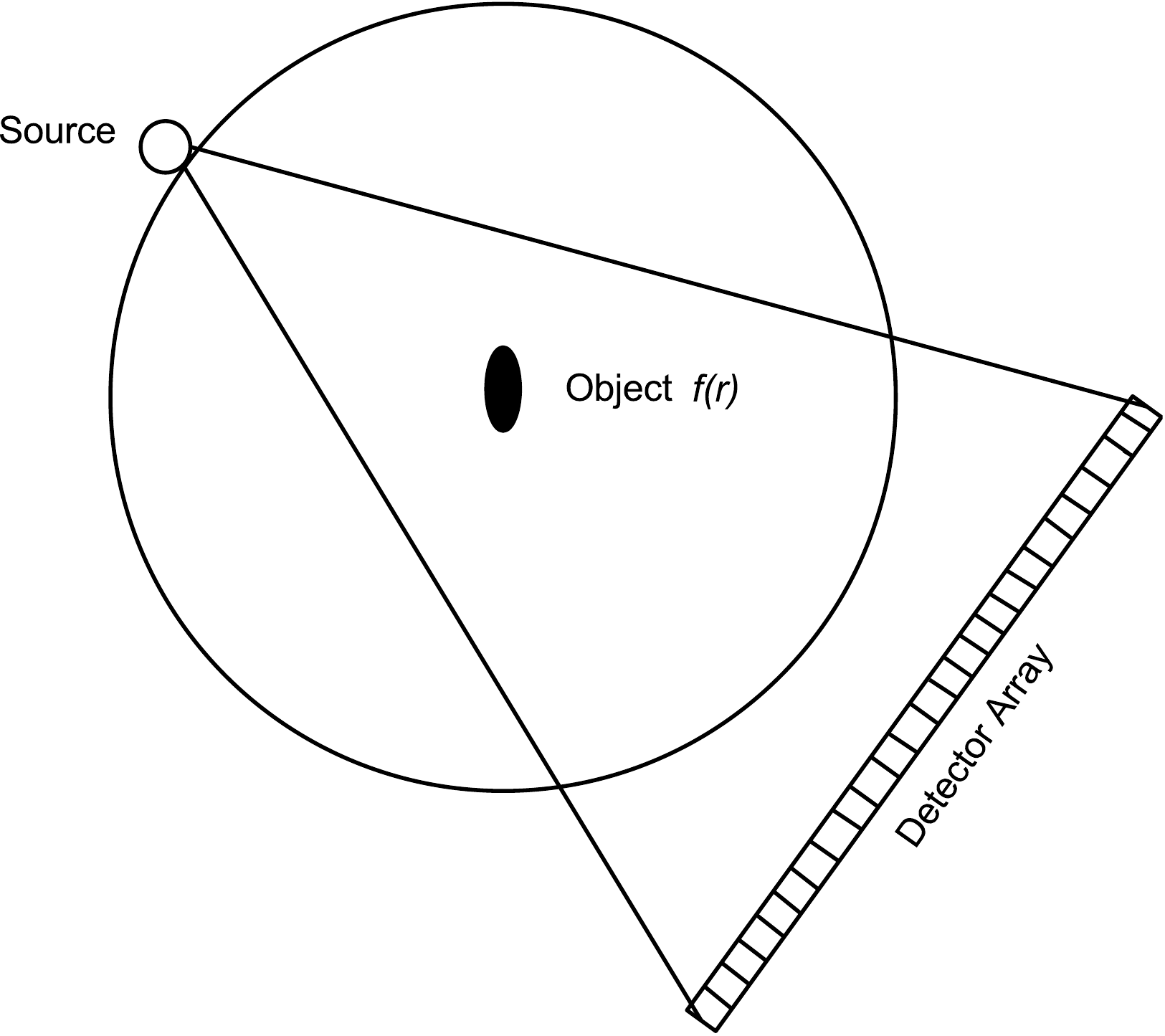}
 \caption{Fan-beam geometry with a flat detector and uniform detector element spacing. \label{fanbeam}}
 \end{figure}

The CT configuration considered throughout this work is fan-beam, in which a source and detector pair rotate about a central field-of-view, acquiring 
discrete projection data at a fixed number of views. In particular, we consider a flat detector with uniform detector bin spacing as in Fig. \ref{fanbeam}. 
A fan-beam CT data model incorporating a statistical description of the data relates the continuous object function $f(\vec{r})$ to the transmitted x-ray intensity $I_i$ via a line 
integral over $f(\vec{r})$ along the $i$th ray, defined by the source position $s_i$ and the ray direction $\hat{\theta}_i$:
\begin{equation}
\begin{split}
\mathbf{I}_i = \Omega_i \int_E \int_u \text{Poisson}\left\{I_0 \left(u,E\right)e^{\left(-\int_{-\infty}^\infty \mu\left( s_i + l \hat{\theta}_i, E\right) dl \right)}\text{d}u \text{d}E \right\} + 
\text{Gaussian}\left\{\mu_i,\sigma_i^2\right\}, \\
\mbox{ where } i\in [1,N]
\end{split}
\end{equation}
where $\Omega_i$ is the gain in the $i^\text{th}$ detector, $I_0$ is the mean incident intensity along the ray defined by the detector, $u$ denotes the position along the detector bin, 
$E$ denotes the incident x-ray energy, $\mu$ is the energy-dependent object attenuation along path $l$ from source position $s_i$ in ray direction $\hat{\theta}_i$, and $\mu_i$ and 
$\sigma_i^2$ are the mean and variance of additive electronic noise, respectively. We will typically reserve the use of bold letters for random variables. The exception in Chapter \ref{ch:tv}, 
where because of a shift in the formulation of our metrics, we adopt a different notational system, denoting all vector quantities with bold. 
 
Note that, unless stated elsewhere we choose to neglect the effects of scattered radiation, a finite focal spot, and stochastic gain in the detectors. Further, we will introduce several other 
simplifying assumptions to this data model. Firstly, we typically choose to assume a monochromatic x-ray beam, so that we drop explicit dependence on $E$, and operate in the log data domain. We also assume that the 
detector response is invariant across the width of a single detector bin, resulting in the model:
\begin{equation}
\label{model}
\mathbf{g}_i = -\mathrm{log}\left(\frac{\mathbf{I}_i}{u\Omega_i I_0}\right) = \int_{L} f\left(s_i + l\hat{\theta_i}\right)\,dl + \mathbf{n}_i
\end{equation}
where $u$ is the width of a single detector bin. The domain of integration $L$ is defined as the intersection of the $i$th ray with the compact support of the continuous object function $f$. 
Finally, we assume that the electronic noise and compound Poisson noise in each detector reading can be modeled as 
additive Gaussian noise, via the stochastic term $\mathbf{n}_i$.

Note that the additive noise term $\mathbf{n}_i$ 
is added in the post-log projection data domain as opposed to being considered in the incident photon fluence, hence the term $\mathbf{n}_i$ is not contained in 
the integral of Eqn. \ref{model}. Also observe that since this data model produces a discretized projection data vector $\mathbf{g}$, all subsequent 
image reconstruction steps map an intermediate discrete data vector to the final discretized image and are, therefore, discrete-to-discrete transformations. 
Further, since the filtered back-projection (FBP) algorithm which we employ for most of this thesis
is a linear algorithm, the reconstruction operator constitutes a linear, discrete-to-discrete operator which can be represented by a matrix $A$. 

When dealing with ellipse-based phantoms, we model the CT data generation by the action of a continuous-to-discrete projection operator $\mathcal{P}$, 
whose action is defined by Eqn. \ref{model}.
We employ a concise notation for the action of $\mathcal{P}$ as follows:
\begin{equation}
\label{data_model}
\mathbf{g} = \mathcal{P}f + \mathbf{n}.
\end{equation}
For projection of discretized phantoms, we consider the equivalent ray-driven projection model, however we then compute the path integral of 
each ray through each pixel or in the phantom. Whenever discrete phantoms are used, we restrict the pixel size of the phantom 
to be roughly an order of magnitude smaller than the detector bins, when projected onto the detector face. 
Chapter \ref{ch:tv} is again an exception, since for certain tests in IIR, the discrete phantom is identical to the ``true'' image which we hope to recover.

\subsubsection{CT Noise Model}
The fullest CT noise model would include the physical effects discussed in Sec. \ref{data_model_sec}, namely a compound Poisson model for incident x-rays, an additive 
Gaussian electronic noise model, and detector correlations resulting from detector cross-talk. For simplicity, we will adopt a noise model which is strictly Gaussian and 
assumes uncorrelated noise in the 
detector elements. Under this assumption, the data covariance vector 
$K_g = \mathrm{E}\left\{ \left(\mathbf{g}-\bar{g}\right)\left(\mathbf{g}-\bar{g}\right)^T\right\}$ is diagonal, consisting only of variances with all covariances between detector elements equal to 0.
Two models will be investigated. The first assumes uniform variance in each of the detector bins, so that the data covariance matrix is given by
\begin{equation}
\left(K_g\right)_{i,j} = \alpha \delta_{ij},
\end{equation}
where $\left(K_g\right)_{i,j}=\text{Cov}\left(g_i,g_j\right)$, $\delta_{ij}$ is the Kronecker delta and $\alpha$ is a constant.

In order to determine the detector variances for an object-dependent noise model, we consider the CT data noise model put forward by Barrett and Swindell \cite{barrett_radiological_1996}. 
In particular, the noise model begins with the assumption that the variance in the
data (after applying the negative logarithm) goes as:
\begin{equation}
\text{Var}\left\{g_{i}\right\} = \frac{1}{\bar{N}_{i}}+\frac{1}{\bar{N}_0}
\end{equation}
where $\bar{N}_0$ is a constant denoting the average number of photons at the $i$th ray in a blank scan, and $\bar{N}_{i}$ is the mean number of photons transmitted through the object and incident on the 
$i$th detector element. The final covariance matrix $K_g$ is then given by 
\begin{equation}
\left(K_g\right)_{i,j}=
	\begin{cases}
	\frac{e^{\bar{g}_{i}}+1}{\bar{N}_0} &: i=j \\
	0 &: \text{else}.
	\end{cases}
\label{cov}
\end{equation}
This covariance is object-dependent, which will affect the computation of subsequent model observer metrics, as discussed below.

\subsubsection{Linear Reconstruction Algorithm Approach}
The class of linear reconstruction algorithms can be expressed in terms of linear systems. 
In general terms, we consider a linear image reconstruction algorithm $A$ that takes a discrete set of
data $\mathbf{g}$ and produces $\mathbf{y}$, a discrete representation of an image in the form
of pixel coefficients: $\mathbf{y} = A \mathbf{g}$.
As many such algorithms consist of several linear processing steps, it is useful
to consider $A$ as the product of matrices representing each processing step: $A = \prod_i A_i$.
The filtered back-projection (FBP) algorithm is one such algorithm, and the various operations involved such as discrete 
Fourier transforms and back-projections can be represented in the form of matrix operations. The 
resulting algorithm is then also interpreted as the action of a matrix.

\subsection{Fan-Beam CT Reconstruction - The FBP Algorithm}
\label{recon}

Here we briefly summarize the fan-beam FBP algorithm. Our description is based on that contained in Ref. \cite{kak_principles_1988}, and we 
refer the reader to that work for a more thorough introduction. Fan-beam FBP reconstruction follows a three-step process: 
First, each $\mathbf{g}_i$ is weighted by the factor $w(n)=D/\sqrt{D^2 + n^2du^2}$,
where $D$ is the source-to-detector distance, $n$ is an integer denoting the detector element where $n=0$ corresponds to the central ray, 
and $du$ is the detector element spacing. Each value of $w(n)$ can be combined into a vector $\mathbf{w}$. Numerically, the implementation of this weighting is equivalent to operating on the discrete projection data with a 
diagonal weighting matrix such that
\begin{equation}
\mathbf{g}^\prime = \text{diag}\left(\mathbf{w}\right)\mathbf{g}
\end{equation}
where $\text{diag}\left(\mathbf{w}\right)$ is the diagonal weighting matrix. Second, the modified projection data vector $\mathbf{g}^\prime$ is transformed to the frequency domain using 
a discrete Fourier transform (DFT) algorithm, also representable by a matrix which we shall label $F$. 
In this domain, the conventional ramp filtering can be performed via multiplication. Alternatively, one may apply an apodization 
window such as a Hanning window to perform regularization in the reconstruction. This is also achieved through a multiplication in the DFT domain. 
Finally, one applies the matrix $F^{-1}$ and performs a weighted back-projection of each of the 
filtered projections onto a discrete image array via a back-projection matrix operator $B$, resulting in the reconstructed image pixels $\mathbf{y}_{j}$.
The full reconstruction operation can then be written as 
\begin{equation}
\mathbf{y} = A\mathbf{g} = BF^{-1}\text{diag}(\mathbf{a})\text{diag}(\mathbf{r})F\text{diag}\left(\mathbf{w}\right)\mathbf{g}
\end{equation}
where $\text{diag}(\mathbf{a})$ and $\text{diag}(\mathbf{r})$ are diagonal matrices with an apodization window and the ramp kernel along their diagonals, respectively, and $B$, $F$, and 
$\text{diag}\left(\mathbf{w}\right)$ are again matrices representing back-projection, the discrete FFT operation and data weighting. 
 In general, this reconstruction matrix $A$ can be singular, meaning that no left-inverse of $A$ exists. The corresponding  
null-space can then play a direct role in the loss of data information which could be useful for a given task using the reconstructed image. 
By \emph{information}, we specifically refer to components of the the data vector $\mathbf{g}$ which can directly contribute to 
the improved performance of some task. We consider this information to be lost when this component of $\mathbf{g}$ lies in the 
null-space of the matrix $A$.
The theme of information loss through $A$ will play a recurring role in our image quality analysis in subsequent chapters. 
Finally, the representation of a linear algorithm as a matrix $A$ allows the 
computation of the image covariance matrix given by $K_y = AK_gA^T$, where $K_g$ is the data covariance matrix.

\subsection{Iterative Image Reconstruction}

While analytic reconstruction algorithms like FBP are still the most ubiquitous reconstruction algorithms, 
optimization-based image reconstruction in CT has shown great promise for improving the utility of images, particularly in cases where the assumptions 
underlying analytic methods are violated. Specifically, these algorithms are more robust than their analytical counterparts, facilitating a wide variety of dose-reduction 
strategies \cite{sidky_constrained_2011, tian_low-dose_2011}, novel imaging techniques \cite{bian_evaluation_2010,han_algorithm-enabled_2011,sidky_enhanced_2009}, and specialized CT systems 
\cite{sidky_enhanced_2009}. 
In many cases, striking examples can be shown wherein optimization-based algorithms lead to images that are vastly different in appearance from 
images employing conventional algorithms. 
However, these drastic changes in subjective quality need to be objectively characterized 
in order to ensure that these algorithms yield images which retain as much diagnostic utility as possible. 

In order to ensure that optimization-based reconstruction algorithms meet necessary clinical standards of quality, we will investigate the extension of
the task-based formalism presented in Chapters \ref{ch:humans} through \ref{ch:breastct} to these novel algorithms. The mathematical framework involved will necessarily be 
more involved than that which is relevant to analytic 
algorithms, since these algorithms are generally nonlinear. Further, development of optimization-based reconstruction is still in a rapid state of flux 
due to the novelty of these techniques in CT. 

Investigation of the feasibility of objective assessment of iterative algorithms is necessary 
in part because these algorithms are already commonly included in clinical systems as optional costly upgrades, typically with promises of preservation of image quality 
with reduced radiation dose, or of improvement of image quality at equivalent patient radiation dose. An investigation into summary metrics 
of image quality in iterative reconstruction is therefore prudent. A task-based approach to determination of such metrics is currently the most promising 
option, as alternative metrics such as contrast-to-noise ratio (CNR) may have little meaning in the context of optimization programs which include prior information, such as total 
variation (TV), which incorporates prior gradient magnitude image sparsity. As an important preliminary step in this direction, we will investigate the approximation of the image covariance matrix $K_y$ for 
images reconstructed with TV penalties.

\section{Background: Objective Image Quality Assessment}
Ultimately, the quality of a CT image depends on the task for which it is intended. 
Therefore, the image quality metrics which we will 
investigate in this thesis are task-specific. While the endpoint utility of an imaging system can be 
measured based on the tasks it enables a radiologist to perform, this endpoint can only be evaluated for 
isolated, often well refined, imaging systems. 
For example, in the case of lesion detection, human observer detectability is arguably 
the most meaningful metric of image quality. However, 
human observer studies needed to evaluate detectability are too expensive and burdensome to be useful 
for exhaustive systems optimization. Therefore, mathematical model observers are often employed in order to either 
estimate human observer performance (e.g. channelized Hotelling observers \cite{barrett_foundations_2004}) or provide a 
theoretical upper bound on human observer performance (the Bayesian ideal observer \cite{trees_detection_1968}). 

In this section, we will motivate the use of task-based image quality metrics based on model observers as a reasonable 
direction for the assessment of CT algorithms and systems. We will describe, in particular, the Hotelling observer (HO) and 
its associated metrics in order to provide a foundation for the chapters which follow.


\subsection{Task-Based Assessment}

In the context of medical imaging, early development of image quality metrics such as the modulation 
transfer function (MTF) was 
focused on applications such as planar radiography \cite{wagner_assortment_1972}. These metrics of quality were based on 
several simplifying assumptions such as stationarity and shift-invariance \cite{barrett_radiological_1996}. 
In CT, however, the assumptions upon which conventional 
metrics are based are not always justifiable \cite{pineda_analysis_2008}. Therefore, the overall image properties of CT such as noise are difficult to characterize due to the lack of 
applicable simplifying assumptions. As a result, the quality of CT images is usually either described 
with metrics that have very limited meaning in the context of tomography, or is not described quantitatively at all. 
The result is that claims of image quality from researchers and companies alike are often unsubstantiated, either because 
the metrics employed rely on assumptions which are violated, or (particularly in academic research), because subjective 
evaluation of single images is used.

The outcome of patient treatment depends crucially on accurate diagnosis, and CT images of high utility are a crucial component in obtaining such diagnoses. Further, 
many of the parameters used in CT image reconstruction algorithms have demonstrated impacts on image utility. Therefore, 
since conventional metrics of image quality often do not apply to CT, it is imperative that meaningful measures related to image utility be developed. 
Currently, task-based metrics such as mathematical model observer performance show great promise in addressing this missing link to ensuring optimal patient diagnosis, 
namely enabling the design and assessment of reconstruction algorithms with objectively high levels of image quality \cite{abbey_observer_1996, abbey_linear_1995, soares_noise_1995}.

Task-based assessment has a long history in the field of nuclear medicine, where it has been used to study the effects on image quality of various system parameters, 
such as the aperture opening, the collimator, and even the reconstruction algorithm \cite{barrett_foundations_2004, barrett_model_1993, fiete_psychophysical_1987, tsui_comparison_1978}. 
However, its application to CT is relatively novel. This is partially 
due to the fact 
that images in CT are of a decidedly higher quality than those in nuclear medicine, generally speaking, so that advanced assessment strategies were not typically 
seen as necessary for CT. Recently, however, various considerations such as dose reduction strategies have prompted progress toward CT systems or acquisitions that push the limits 
of the reconstruction algorithms used. Specifically, it is common that the preservation of fine structures in noise-dominated data is of importance, thereby mimicking the situation 
of task-performance in nuclear medicine.

It is necessary to objectively evaluate the quality of images produced by novel systems 
and to 
ensure that the associated reconstruction algorithms are adequate for enabling the specific tasks for which such systems are designed. Taking dedicated breast CT as an 
example, it is necessary that if breast CT should be used as digital mammography is used today, that the utility of the images produced can be evaluated objectively 
as being comparable to those in mammography. Clearly, there exists a pressing need to 
ensure that rapid technical progress seen in CT development does not connote a sacrifice in terms of image utility, and in light of the lack of versatile metrics applicable to tomography,  
in this thesis we attempt to develop a widely applicable formalism for task-based assessment of image reconstruction algorithms in CT.     

\subsection{Model Observer Approach}
As we have already suggested, quantification of image quality in CT is useful not only in evaluating a completed imaging system, but also in 
optimizing decisions in the system design and implementation. A wide array of parameters exist in the CT imaging chain, including various physical parameters such as 
patient size, acquisition parameters such as the number of projection views, 
as well as reconstruction algorithm parameters such as image pixel size. Each of these parameters has an effect 
on the noise properties of the CT images, as well as the resolution in the image, however the impact of these factors on the utility of the images produced 
is often unclear. Further, for parameters which are controllable, such as acquisition and reconstruction parameters, the optimal selection of implementation choices 
is rarely obvious. We propose the use of a mathematical model observer for efficient and objectively meaningful optimization of CT systems and reconstruction algorithms. 

In general terms, there are two approaches to constructing mathematical model observers: (1) estimate as accurately as possible the performance of a human observer for a given task
 \cite{barrett_foundations_2004,myers_visual_1985, burgess_human_2001} or (2) establish an upper-bound on human observer performance by determining the performance of the Bayesian ideal observer \cite{trees_detection_1968}. 
We generally prefer the latter approach because our goal is to optimize the reconstruction algorithm relative to the absolute upper-bound on human observer performance, thereby 
ensuring that there is minimal information loss in the reconstruction algorithm. This leaves room for post-reconstruction image processing methods to transform the information in the reconstructed image into a form that is amenable to 
human task performance. The human visual system is complex and difficult to model adequately, meaning that the former approach can also
introduce errors from improper human modeling. The latter approach avoids these errors and will provide non-stochastic and stable upper bounds on human performance. 
We will therefore avoid the use of so-called perceptual channel mechanisms in general.

The specific observer we adopt is the Hotelling observer (HO) \cite{myers_effect_1985, 
fiete_hotelling_1987, barrett_model_1993, rolland_effect_1992}.
The particular HO implementation
we propose is exact in that it does not rely on assumptions such as shift-invariance or noise stationarity. These assumptions are commonly 
invoked, at least locally, because they enable the construction of image quality metrics based on discrete Fourier transform domain image properties, resulting in familiar metrics such 
as the noise power spectrum (NPS) \cite{ barrett_radiological_1996,
jenkins_spectral_1968,dainty_image_1974, blackman_measurement_1959} and modulation transfer function (MTF) 
\cite{rossmann_measurement_1964, rossmann_spatial_1968, barrett_radiological_1996, dainty_image_1974} or local NPSs \cite{baek_local_2011, pineda_beyond_2012} and MTFs
\cite{defrise_performance_1994}.
Instead, the approach we propose is not limited by assumptions regarding stationarity.
Meanwhile, the propagation of the signal and noise into the image domain is fully and exactly 
accounted for through a linear systems theory approach.
While the assessment performed here makes an additional assumption of a fixed (non-random) background, the framework described is 
sufficiently general that additional terms could be added to account for anatomical noise covariance when the task of interest is limited by anatomical noise. 
Likewise, performance-degrading psychophysical factors could be modeled if direct estimation of human performance were of interest; however, in this work we are concerned instead with determining ideal observer performance.

An additional benefit of our proposed method for HO assessment is that we do not rely on statistical realizations (samples of noisy reconstructed images) in order to 
construct the relevant metrics. Performing noise realizations can be incredibly time-consuming, particularly for large image volumes, since many realizations 
are necessary in order to compute accurate sample statistics. Further, regardless of the number of realizations performed, there is an inherent statistical uncertainty 
in the estimated quantities. Throughout this work, however, the metrics that we construct are based on analytic formulas which are non-stochastic and do not rely on 
noisy images or statistical estimates from samples. 

\subsubsection{Binary Detection}
\label{bi_detect}
To this point, we have spoken in general terms regarding task performance. In order to clarify our meaning, we take a specific example of a classification task - 
the task of binary detection. 
In order to demonstrate the two hypotheses which make up the binary detection task, recall the data model of Eqn. \ref{data_model}, defining a
continuous to discrete operator $\mathcal{P}$, which maps the continuous object function $f(\vec{r})$ to the discrete $M\times 1$ data vector $\mathbf{g}$:
\begin{equation}
\mathbf{g} = \mathcal{P}f + \mathbf{n}
\end{equation}
where $\mathbf{n}$ is an additive measurement noise term in the projection data domain, and the sole source of noise considered. 

The two hypotheses relevant to a signal detection task are the signal-absent and signal-present hypotheses, which we shall label $H_0$ and $H_1$, 
respectively. Since we consider detection in the reconstructed image domain, we must apply the reconstruction operator $A$ discussed in Section 
\ref{recon} to the data vectors representing each hypothesis. The resulting hypotheses may then be expressed mathematically as
\begin{equation}
\label{hypotheses}
\begin{split}
&H_0 \hspace{0.3cm}:\hspace{0.3cm} \mathbf{y} = A\left( \mathcal{P}f_b + \mathbf{n}\right) \\
&H_1 \hspace{0.3cm}:\hspace{0.3cm} \mathbf{y} = A\left( \mathcal{P}\left(f_b+f_s\right) +\mathbf{n}\right)
\end{split}
\end{equation}
where $f_b$ and $f_s$ represent the background object and signal object, respectively. It should be noted that the 
statistics of the additive noise term in each hypothesis are generally assumed to be the same in this thesis, since the signals of interest are usually small.
As discussed in later chapters, it is trivial to extend our analysis to the case of differing statistics between hypotheses, but this impacts the interpretation of 
the resulting metrics \cite{barrett_foundations_2004}. 
The task we consider is a signal-known-exactly, 
background-known-exactly (SKE/BKE) task, meaning that the Hotelling observer has full knowledge of the reconstructed signal and 
the image noise statistics. A generalization of this paradigm is also occasionally investigated in this work, and is known as a signal-known-exactly-but-variable (SKEV) paradigm. 
Essentially, this paradigm accounts for signal variability by averaging observer performance over a set of possible signal locations. 
Further, the human observer studies in Chapter \ref{ch:humans} are also in the SKE or SKEV paradigm since the reconstructed signal is presented to the 
observer study participants.

\subsection{Mathematics of the Hotelling Observer}

The Hotelling Observer (HO) is the optimal linear model observer in the limit of Gaussian noise, and it has been used for prediction of human observer task performance \cite{myers_visual_1985, 
fiete_psychophysical_1987}. Even in 
cases where the HO does not accurately predict human performance, it often still constitutes a useful upper bound on human task performance \cite{yao_predicting_1992,
barrett_model_1993, rolland_effect_1992}. 
In a classification task, the HO operates by computing a scalar value which is a linear combination of the 
image pixel values. It then uses this value to classify an image as corresponding to one of two hypotheses, and it
performs this classification in a way which is optimal with respect to 
all other linear observers. The scalar value used for classification is termed the Hotelling test statistic and is computed as
\begin{equation}
t = w_y^T\textbf{y},
\end{equation}
where $w_y$ is the optimal set of weights for the image pixels, known as the Hotelling template. The Hotelling template is in turn defined as
\begin{equation}
\label{kwd}
\bar{K}_yw_y = \Delta\bar{y},
\end{equation}
where $\bar{K}_y = A\bar{K}_gA^T$ is the image covariance matrix under the two hypotheses (an average is taken if the covariance differs between hypotheses), 
and $\Delta\bar{y}$ is the mean difference between images from the two hypotheses. 
The quantity $\Delta\bar{y}$ can be obtained by generating noiseless reconstructed images of each of the two 
hypotheses. Since the detector noise is additive and zero-mean and the reconstruction operator is linear, these noiseless images correspond to the mean image
under each hypothesis.
Eq. (\ref{kwd}) is solved via the Moore-Penrose pseudo-inverse for cases where $K_y$ is rank-deficient. Otherwise,
for large system models Eq. (\ref{kwd}) can be solved by LU decomposition followed by backsubstitution, and for very large
system models iterative methods such as conjugate gradients can be applied.

\begin{figure}[h]
\centering
\includegraphics[bb = 92 169 657 426, width=0.9\columnwidth,clip=True]{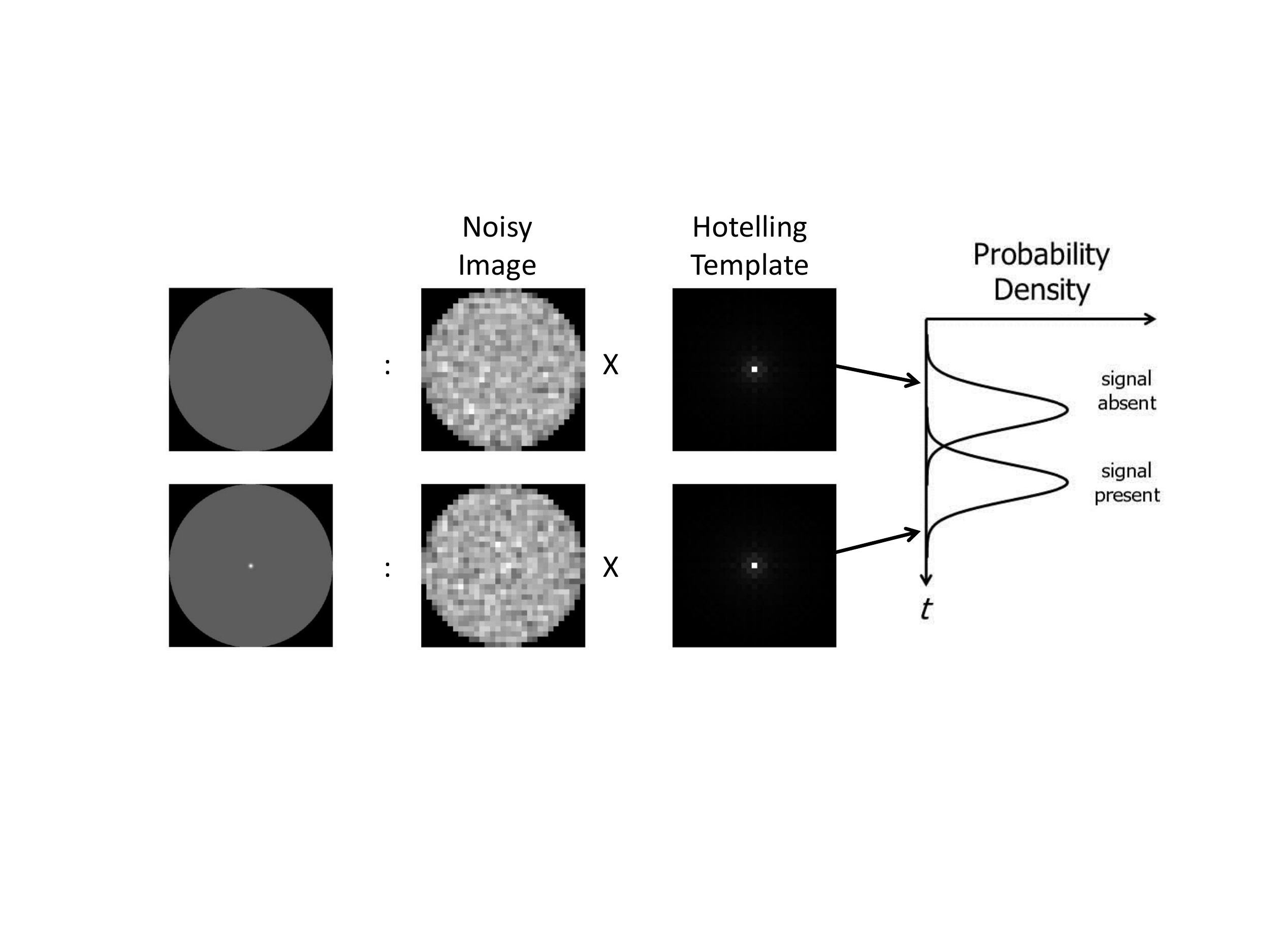}
\caption{In order to classify an image as corresponding to a single hypothesis, the HO computes the inner product of a noisy image with 
the Hotelling template. The resulting scalar test statistic $t$ is then compared to a threshold value and assigned to a class accordingly. This is the approach 
of any linear observer performing a classification task, and the HO is the optimal linear observer.
\label{fig:flowchart}}
\end{figure}

The resulting image covariance 
matrix can be stored directly in computer memory and inverted via a Moore-Penrose pseudo-inversion
\begin{equation}
\label{wydirect}
w_y = \bar{K}_y^\dagger \Delta\bar{y}.
\end{equation}
The process of HO classification is outlined schematically in Fig. \ref{fig:flowchart}. 
Note that noise in the images introduces statistical variability in the outcome of the test statistic for each hypothesis. The task of the HO can then be interpreted as the construction of a linear 
test statistic which has maximally separated statistical distributions under the two hypotheses. 

One HO figure of merit is the HO SNR$_y$, defined as
\begin{equation}
\label{HOSNR}
\text{SNR}_y^2 = w_y^T \Delta\bar{y}.
\end{equation}
Likewise, the HO can operate in the data domain, and all of the foregoing equations hold, but with $y$ replaced by $g$. 
Readers unfamiliar with the SNR of an oberver as defined above can consider a related metric, the percentage of correct decisions, $P_C$, which can be illustrated
through a common psychophysical study called a 2-alternative forced choice or 2AFC experiment.
If we consider a task in which the HO is presented with one image from each hypothesis and made to assign the images to their respective hypotheses, we can define 
the percentage of correct decisions $P_C$. Since the data we consider are Gaussian distributed, the HO SNR can be related to $P_C$ as
\begin{equation}
\label{eqn:pc}
P_C = \frac{1}{2} + \frac{1}{2}\text{erf}\left(\frac{\text{SNR}_y}{2}\right).
\end{equation}
$P_C$ defined in this way is the optimal percent of correct decisions for any linear observer, and will be interpreted in this work as an upper bound on a human's percent of correct decisions. 
In other words, when $P_C=100\%$, we will assume that the upper bound on a human's percentage of correct decisions is
100\%, while when $P_C=50\%$, we will 
assume that a human is incapable of performing the task, with correct decisions resulting from chance alone. In this work, $P_C$ is equivalent to the area under the ROC curve (AUC).

Typically, we use $P_C$ as the figure of merit when studying the impact of any parameter which affects the inherent aspects of the projection data, such as the 
size of the object being imaged. However, for 
optimization of the reconstruction algorithm or dose allocation between views, which do not significantly affect the HO's SNR$_g$ (HO performance in the projection 
data domain), 
we define an additional metric, the HO efficiency:
\begin{equation}
\varepsilon = \frac{\text{SNR}^2_y}{\text{SNR}^2_g} = \frac{w_y\Delta\bar{y}}{w_g \Delta\bar{g}},
\end{equation}
where $\Delta \bar{g}$ is the mean difference in projection data under each hypothesis, and $w_g$ is the Hotelling template in the data domain, 
defined as
\begin{equation}
K_g w_g = \Delta\bar{g}.
\end{equation}
The efficiency metric $\varepsilon$ is 1 whenever the reconstruction algorithm perfectly preserves the information in the projection data used by the HO to 
classify an image, and will be less than 1 whenever information is lost in the reconstruction operation or in the restriction of the image to an ROI. 
Note that each of the metrics we have presented have no statistically variable quantities. 

We consider $\varepsilon$ a more useful parameter than SNR for algorithm optimization since for many CT applications, one would not operate at the threshold for observer 
performance within which SNR is a meaningful metric. Instead, $\varepsilon$ is a useful metric for algorithm and system evaluation that remains independent 
of the difficulty of a given task, while still retaining the exactness of computation inherent in HO performance calculation through the exact covariance matrix
$K_y$. 
A further motivation for selection of the efficiency metric over more conventional metrics such as SNR or area under the ROC curve (AUC\cite{hanley_meaning_1982}) is that we are interested in performing system 
component optimization rather than full system evaluation. A single value of SNR or AUC carries little information regarding how much one can hope to improve
performance by optimizing a single component. Rather, we would prefer a metric like efficiency, which relates the quality of the input to the system component to the quality of the component's output.
Here, the component which we optimize is a parameter of the reconstruction algorithm, so that the efficiency values are a reflection of how well the reconstruction preserves information relevant to the classification task.

One source of difficulty in computing the HO SNR or HO efficiency is the inversion of the covariance matrix $K_y$. 
Although the form of Eqn. \ref{kwd} is compact, obtaining a direct solution for $w_y$ is computationally nontrivial, since 
the matrix $K_y$ can be too large to be stored in computer memory and is non-diagonal, as discussed in Section \ref{noise}. 
In fact, the computational burden of solving Eqn. \ref{kwd} is the basis 
for much active research. For instance, approximate inversions of the image covariance in 
data space either by efficient channels \cite{park_singular_2009, barrett_foundations_2004, beutel_handbook_2000,witten_partial_2010} or assumptions of stationarity in
 the image covariance are sometimes made to ease 
the computational burden. In Chapter \ref{ch:humans}, for example, 1024 view angles and 
256 detector bins are used, implying that $\mathbf{g}$ and $K_g$ have dimensions $262,144 \times 1$ and 
$262,144 \times 262,144$ respectively. Similarly, $\mathbf{y}$ and $K_y$ have dimensions 
$147,456 \times 1$ and $147,456 \times 147,456$ respectively.  These covariance matrices are too large to be directly stored in memory on many systems, implying the 
need for methods that require only a function which outputs matrix-vector products, without direct access to the matrices. 
In the work presented in this thesis, whenever the dimensionality of Eqn. \ref{kwd} is too large to store in computer memory, we solve for the Hotelling template iteratively via the method of
 conjugate gradients, as described in section 5.1 of \cite{nocedal_numerical_2006}. We find this method preferable to the use of channels (either efficient or 
psychophysical) because we wish to accurately determine the performance of the HO unhindered by information loss through channels. Further, although 
efficient channels might exist which can be found using a method like that in Witten et al.\cite{witten_partial_2010}, most work involving efficient channels imposes circular 
symmetry on the signal and template, an assumption violated by the templates in our case.
The calculated HO performance can then provide an absolute upper bound on measured human observer performance for several cases of 
information loss (see discussion of nullspace in Section \ref{recon}).

%% file: Chapter2/Chapter2.tex
\chapter{The Hotelling Observer and Human Observers}
\label{ch:humans}
\section{Introduction}
In this chapter, we demonstrate the relationship between model observer metrics presented in the previous chapter and human task performance in CT.
Our purpose is two-fold: First, comparison of model observers with human observers provides intuition for the meaning of model observer metrics. Second, 
quantitative comparison of model observers with humans lends greater credibility to task-based system optimization. The specific task which we investigate here 
is lesion detection. 

Lesion detection is one prominent clinical task for which CT is employed. 
Therefore, lesion detectability has been frequently investigated as a useful metric for systems 
optimization and evaluation in CT \cite{laroque_evaluation_2007, sidky_accurate_2008, sidky_-depth_2008, park_singular_2009, wunderlich_image_2008, abbey_human-_2001}. 
In order to provide an introduction to detectability as a metric for CT, this chapter demonstrates the computation of detectability of the Hotelling observer (HO) 
and provides a comparison between HO performance and human performance for a simple detection task. We investigate the 
behavior of human and HO performance as a function of regularization in the image (either smoothing or increased pixel size), and 
provide some preliminary findings which could explain discrepancies between humans and the HO and ensure that system optimization with the HO is 
beneficial to human observers as well.  

For the present case, the HO 
is equivalent to the ideal observer. Seeking to provide a link between existing
 studies that have extensively investigated human and model observer performance for systems that
 mimic CT \cite{abbey_human-_2001} and those which simplify the computation of observer performance via 
channels \cite{wunderlich_image_2008, park_singular_2009}, this chapter investigates exact Hotelling observer performance for a binary 
lesion detection task and compares this result to measured human observer performance for the 
case of detecting a small, high-contrast lesion in noise correlated via the filtered 
back-projection (FBP) algorithm. 

Three separate implementations of fan-beam FBP are investigated. The first serves as a reference case 
and utilizes the conventional ramp filter in reconstruction. The remaining two reconstruction 
implementations are intended to degrade HO SNR and illustrate the corresponding effect on human 
observer performance. One implementation accomplishes this through regularization via a Hanning 
filter, and the other increases the reconstruction image pixel size. Each of these implementations of the FBP 
algorithm takes the form of a matrix operator whose null-space determines the specific loss of inherent signal 
detectability from the sinogram domain to the reconstructed image domain. In the case of regularization, some components 
of the filtered data in the discrete Fourier transform domain are set to zero, thereby introducing a null-space corresponding 
to these components. In the case of increased pixel size, the $M\times N$ reconstruction matrix acquires a null-space 
by virtue of the fact that $M < N$. 

The chapter is organized as follows: Section \ref{sec:background} reviews necessary background regarding 2AFC studies. 
Section \ref{sec:methods} outlines the methods for generating the 
images used in the 2AFC studies and performing the psychophysical and model observer studies. The 
results of these studies are presented in Section \ref{results}.  We then describe the incorporation of internal noise into the HO model in 
Section \ref{sec:int_noise}, and present results of this incorporation in Section \ref{sec:int_noise_res}. Finally, a brief discussion of these results and conclusion 
are given in Section \ref{sec:conc}.

\label{}

\section{The 2AFC Study}
\label{sec:background}
In a two-alternative forced choice (2AFC) experiment, an observer is presented with two images $\textbf{y}$ and 
$\textbf{y}^\prime$, where $\textbf{y}$ is drawn from $\text{pr}(\textbf{y}|H_0)$ and $\textbf{y}^\prime$ is drawn from $\text{pr}(\textbf{y}|H_1)$ \cite{barrett_foundations_2004}. In 
the case of a signal detection task, the hypotheses $H_0$ and $H_1$ will represent a signal-absent and signal-present case (see Section \ref{bi_detect}). The observer computes a test statistic 
$t(y)$, commonly referred to as the decision variable, for each of the two images. The observer then assigns the signal-present decision to the image that produces the higher value of the decision variable. For any 
decision variable, the probability of a correct decision in a 2AFC process is then equal to the area under the ROC curve (AUC) for that observer and 
detection task. The mean number of correct decisions in a human observer study can then be taken as an estimator for the AUC. AUC can then 
be related to an effective SNR by the equation
\begin{equation}
\label{AUC2SNR}
\text{SNR}= 2\text{erf}^{-1}\left[2(\text{AUC})-1\right],
\end{equation}
which is simply the inversion of Eqn. \ref{eqn:pc}. By defining observer SNR in this way, in relation to the 2AFC study, the comparison of human observer performance measured through 2AFC trials and the HO SNR is
straightforward.

\section{Methods}
\label{sec:methods}
\subsection{Generation of Images}
\subsubsection{The Reconstructed Signal}
The signal of interest was chosen to be an ellipse with major and minor axes of lengths 6 detector bin
 widths and 3 
detector bin widths, respectively, and its location was fixed for each of the studies to the center of 
the field of view for reconstruction. We investigate a small high-contrast signal, which can be seen as a 
model of micro-calcification detection in breast CT. The signal was defined in the continuous object domain and discretized by a continuous-to-discrete forward projection operator, as in Eqn. \ref{data_model}. 

\begin{figure}[h!]
\begin{center} $
\begin{array}{cc}
 \includegraphics[width=123pt, bb = 168 120  377 333, clip]{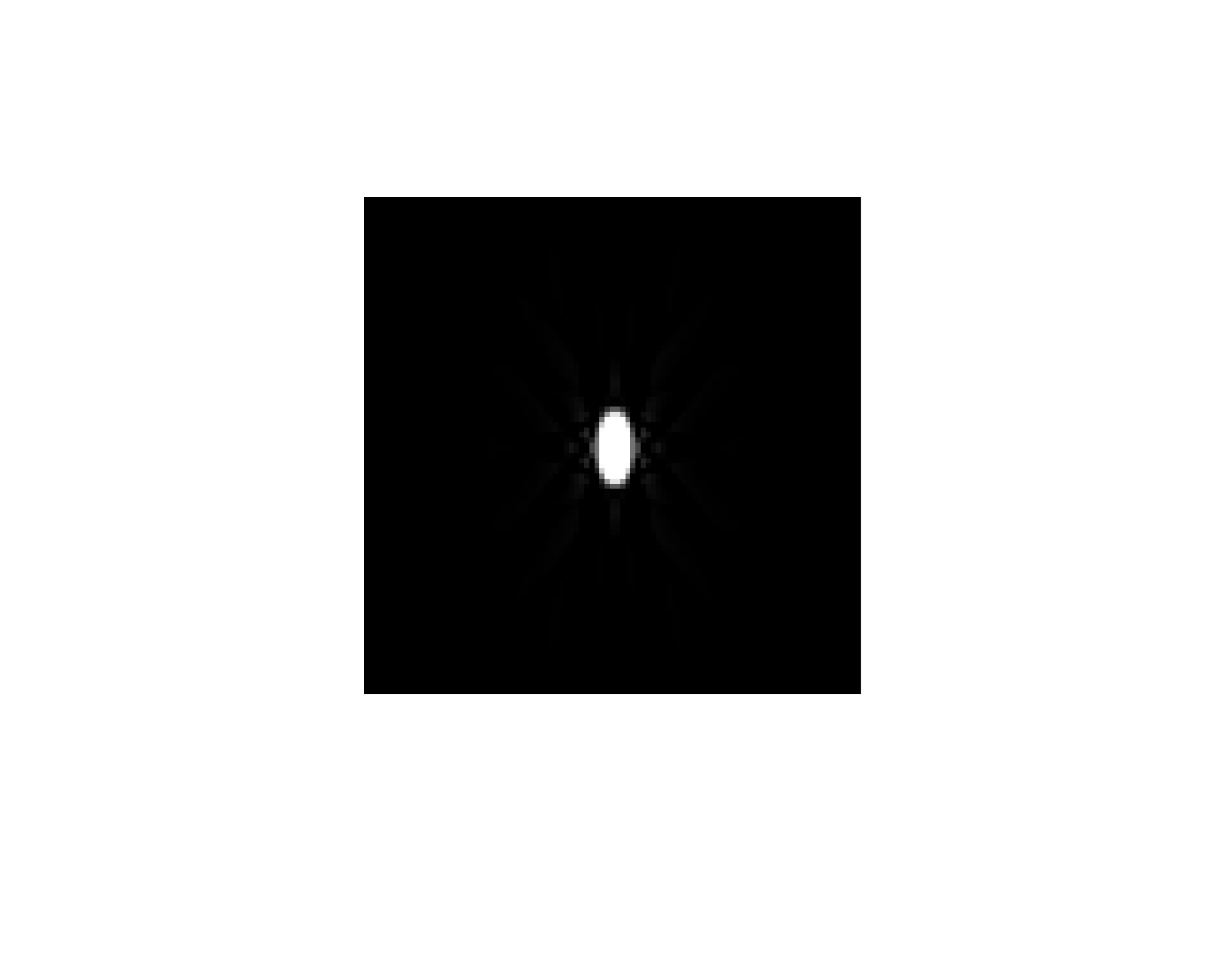}  \hspace{-6pt}& \includegraphics[width=123pt, bb = 70 42  160 134, clip]{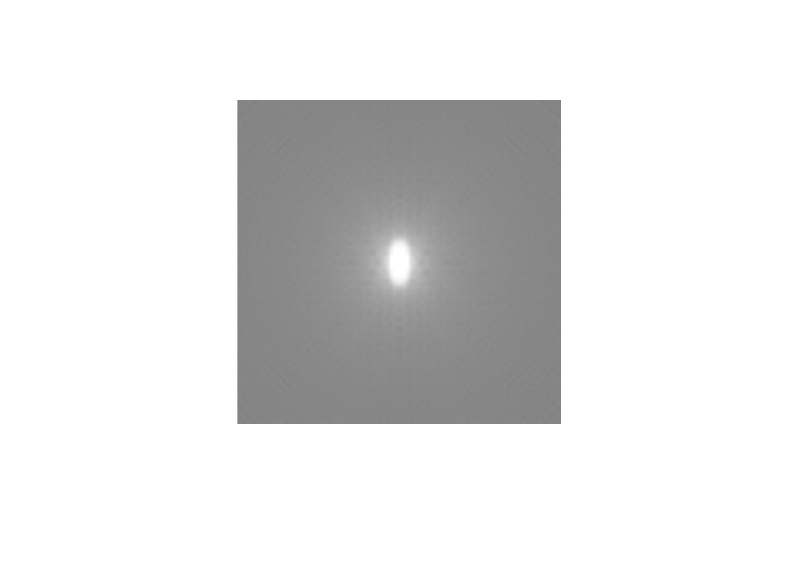} \\[-8pt]
 \includegraphics[width=123pt, bb = 69 43 161 136, clip]{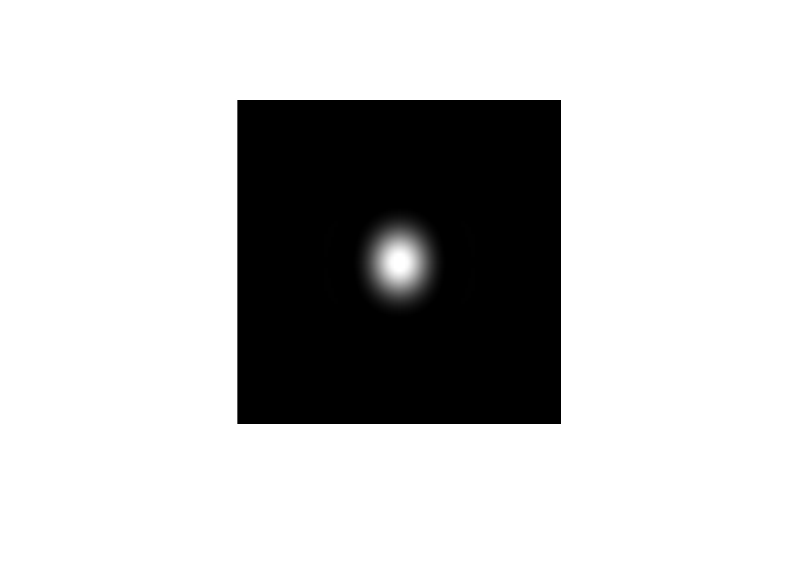}  \hspace{-6pt}& \includegraphics[width=123pt, bb = 188 156 443 412, clip]{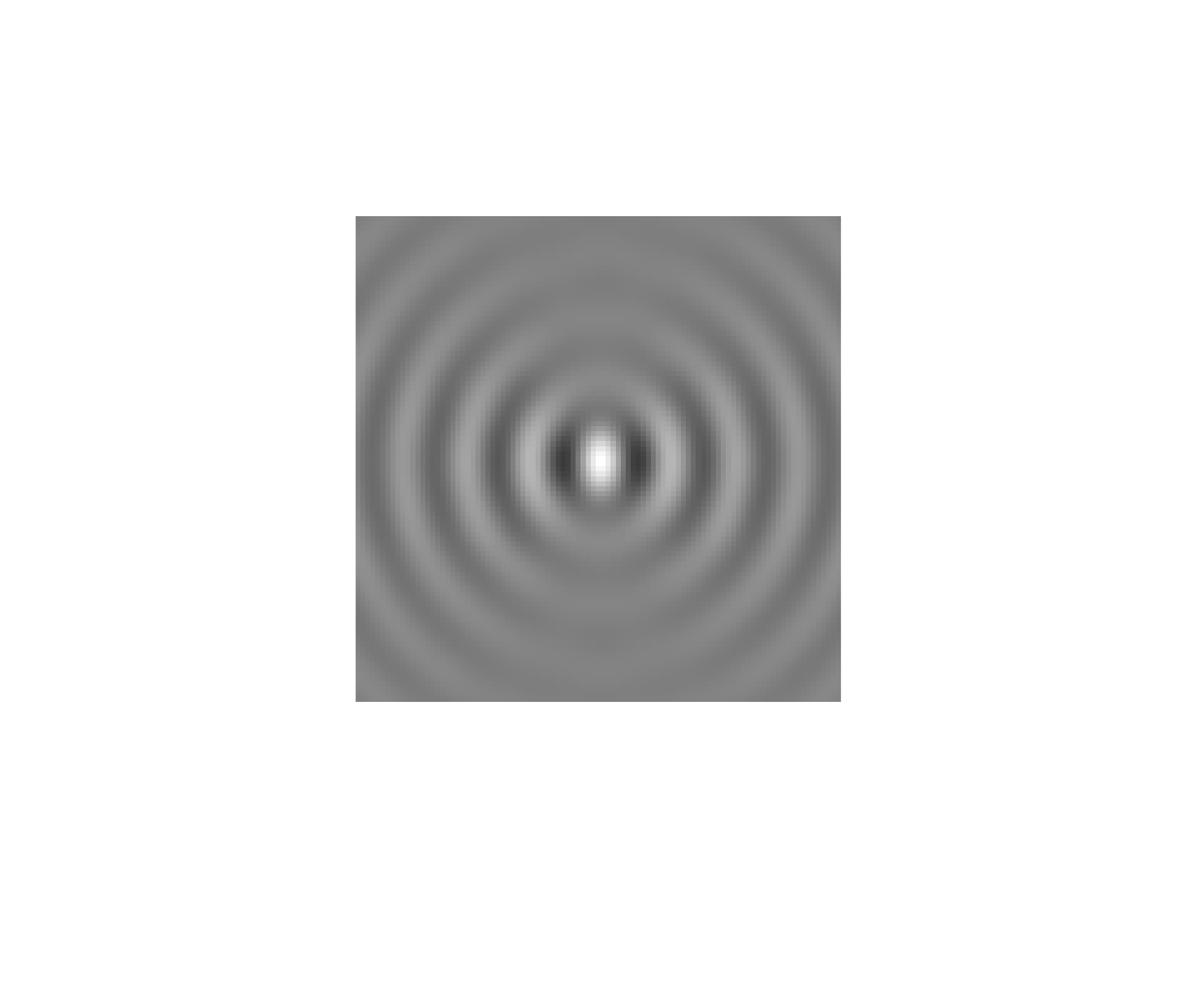} \\[-8pt]
\includegraphics[width=123pt, bb= 281 202 379 300, clip]{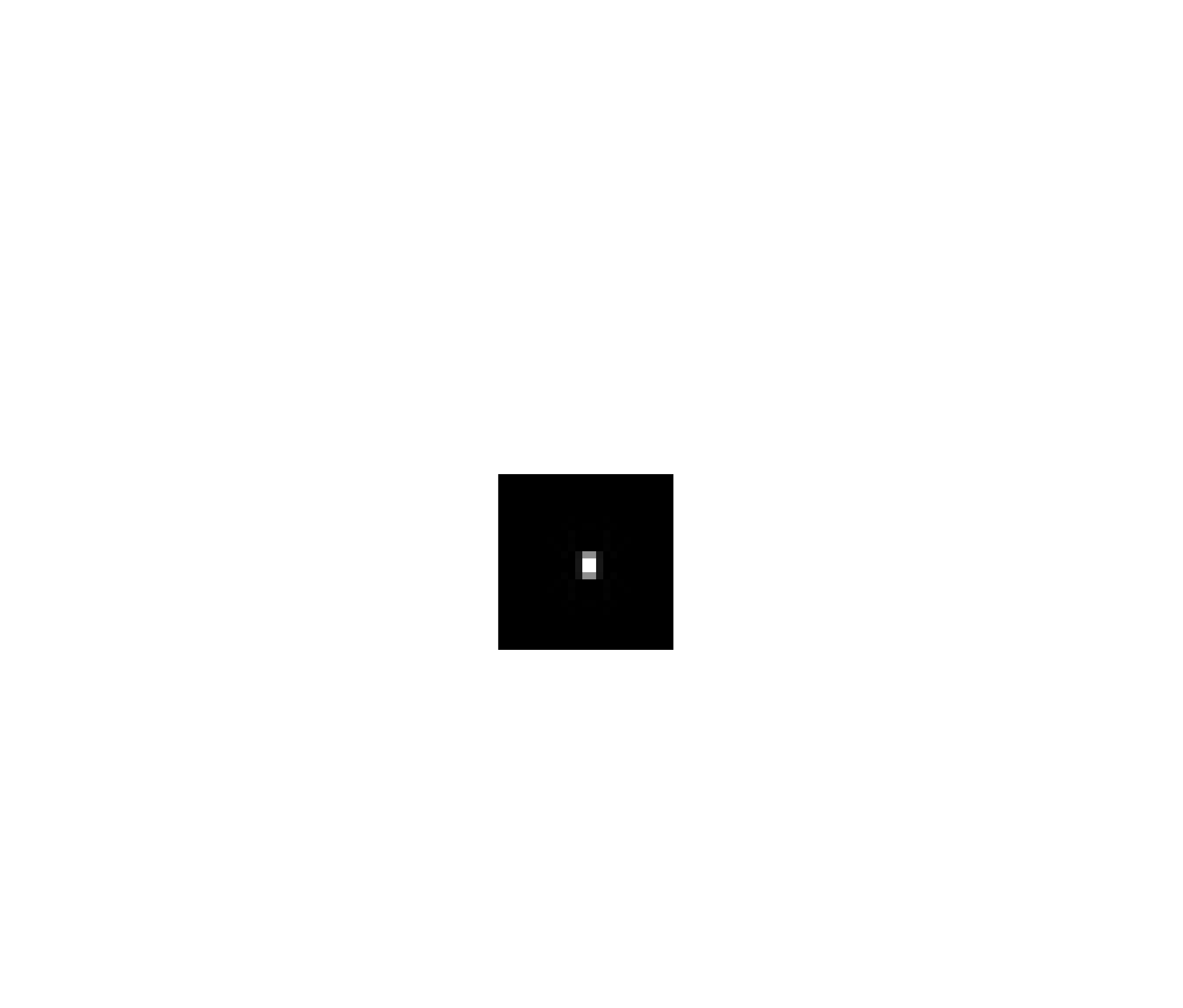}  \hspace{-6pt}& \includegraphics[width=123pt, bb = 70 44 93 67, clip]{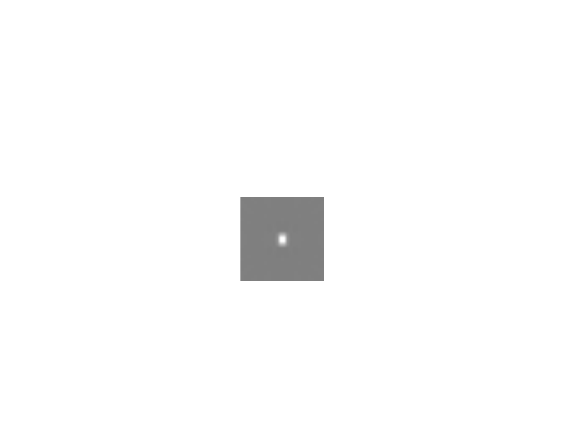} \\
\end{array}
$\end{center}
 \caption{Top Left: The reconstructed reference signal with ramp filtering and no added noise. Display window: 
[0 1.0] Visible artifacts are a result of the coarse level of discretization in the sinogram and image domains 
necessary to allow tractability of the relevant linear systems. 
Top Right: The Hotelling template for the reference reconstruction. Display window: [-0.2 0.2]
Middle Left: The reconstructed reference signal with Hanning filtering and no added noise. The window used here is narrower than that used in 
the experiment for printing purposes. Display window: [0 0.45]
Middle Right: The Hotelling template for the filtered reconstruction. Display window: [-1.7 1.7]
Bottom Left: The reconstructed reference signal with ramp filtering, increased pixel size
 and no added noise. Display window: [0 1.0]
Bottom Right: The Hotelling template for the large-pixel reconstruction. Display window: [-0.26 0.26] 
\label{sigs-temps}}
 \end{figure}

The reconstruction algorithm used was the FBP algorithm discussed in section \ref{recon}. Projection
 data for the reference case was acquired over 
a full $2\pi$ rotation at 1024 evenly spaced angles with 256 detector elements. The reconstruction was
 performed onto a $384\times 384$ pixel image 
grid with pixels roughly twice the size of a single detector element. The images were then cropped 
to a central $96\times 96$ pixel ROI, which was then displayed at $4\times$ magnification. Black and 
white lines were then superimposed on the image in order to aid the observers in localizing the signal 
at the center of the ROI. 
The reference images for the human
 and model observer study were reconstructed 
without regularization, i.e. with only the ramp kernel used for filtration. The ratio of the 
source-to-detector distance to the source's radius of rotation was $8:5$.
The reference reconstructed signal is shown on the top left of Fig. \ref{sigs-temps}. Visible artifacts in the 
reconstructed signal are a result of the discretization in the sinogram and image domains and the small size of the signal. 
This small signal size is desirable in this case because we can then expect greater variability in observer performance 
with respect to the reconstruction algorithm with the given noise model. The corresponding 
Hotelling template is pictured on the top right of Fig. \ref{sigs-temps}.

\subsubsection{Noise Model}
\label{noise}
\begin{figure}[h!]
\begin{center}$
\begin{array}{cc}
 \includegraphics[clip, width=123pt,trim = 5 5 5 5]{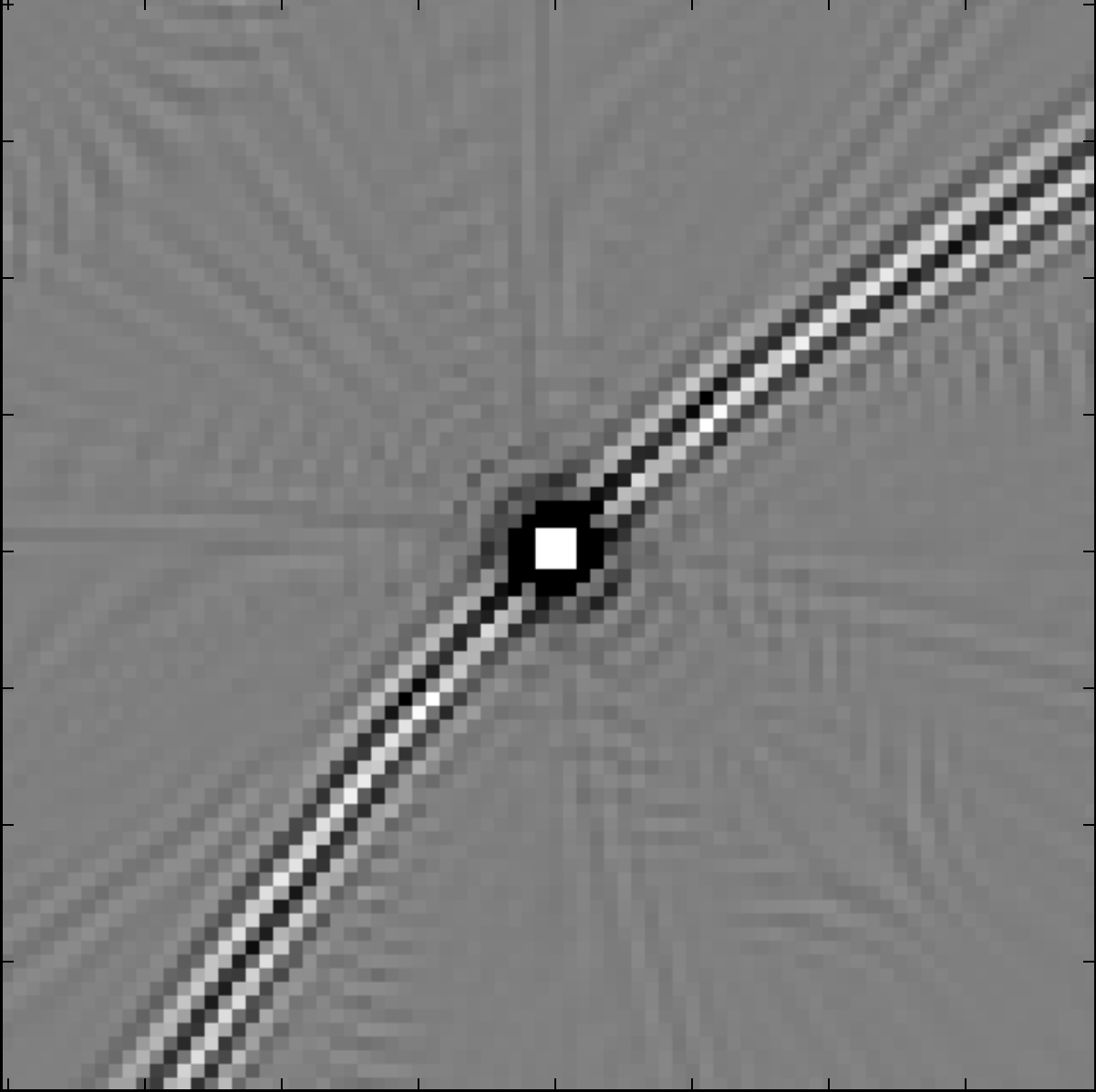} \hspace{-6pt} & \includegraphics[clip, width=123pt, trim = 5 5 5 7]{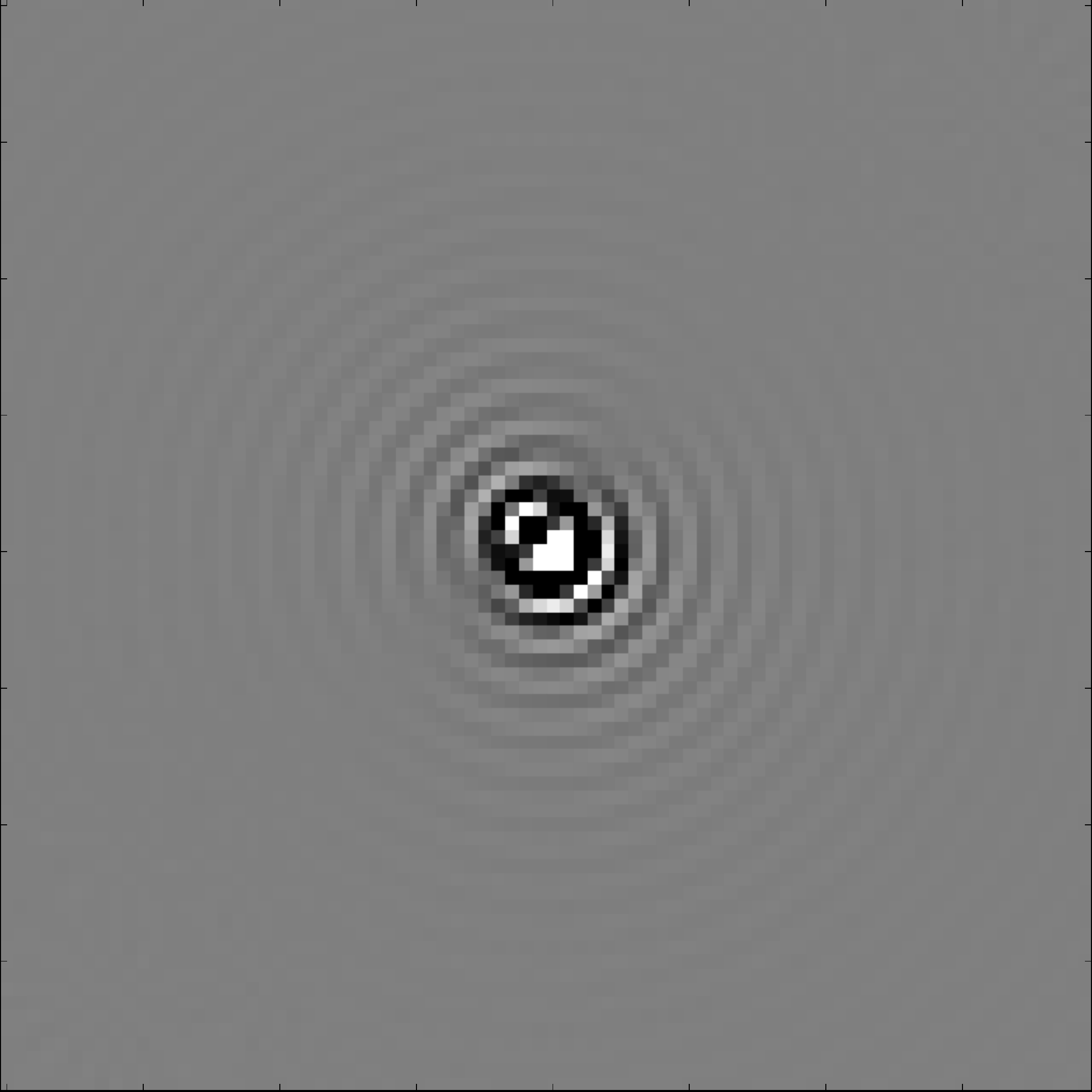} \\[-6pt]
 \includegraphics[clip, width=123pt,trim = 5 5 5 5]{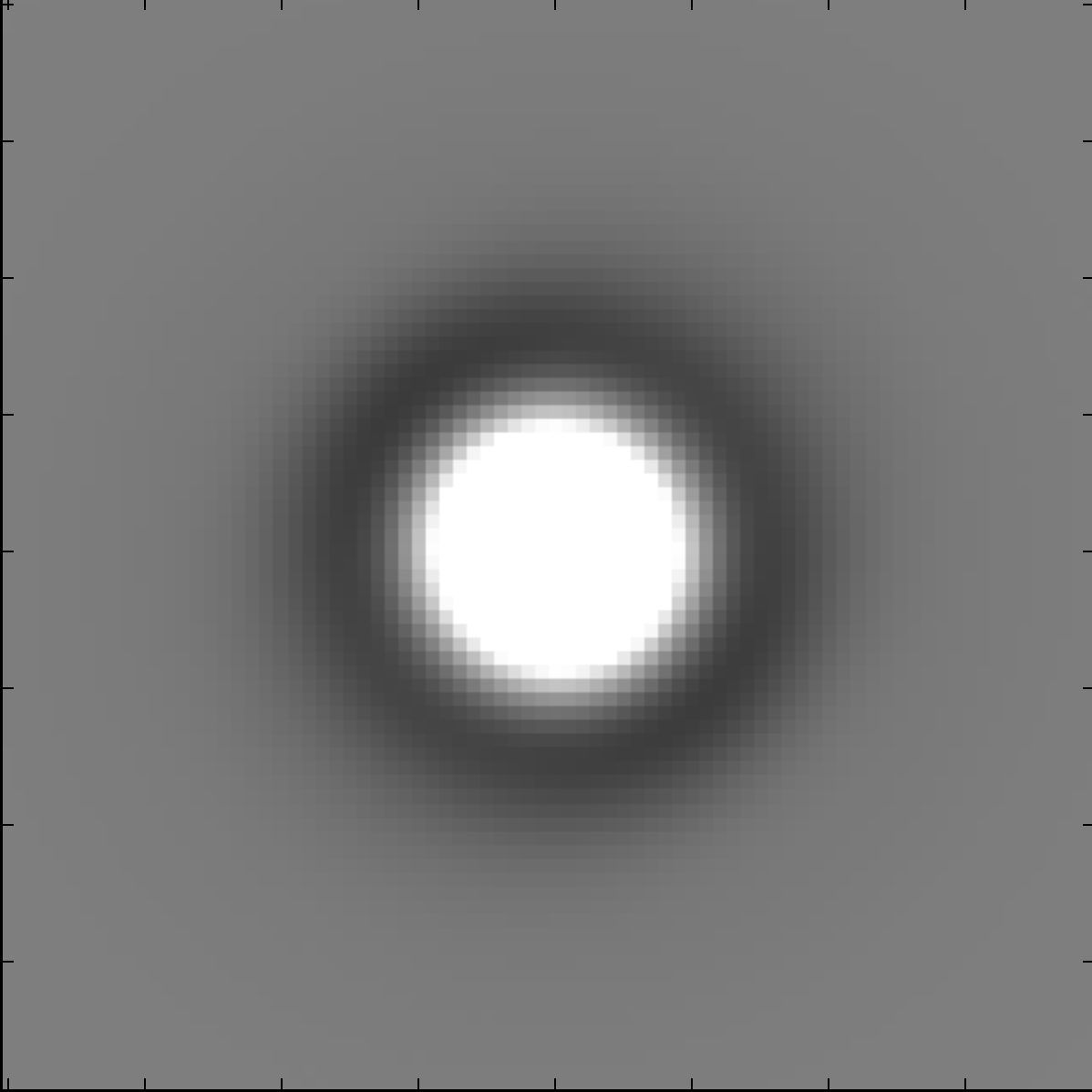} \hspace{-6pt} & \includegraphics[clip, width=123pt, trim = 5 5 5 4]{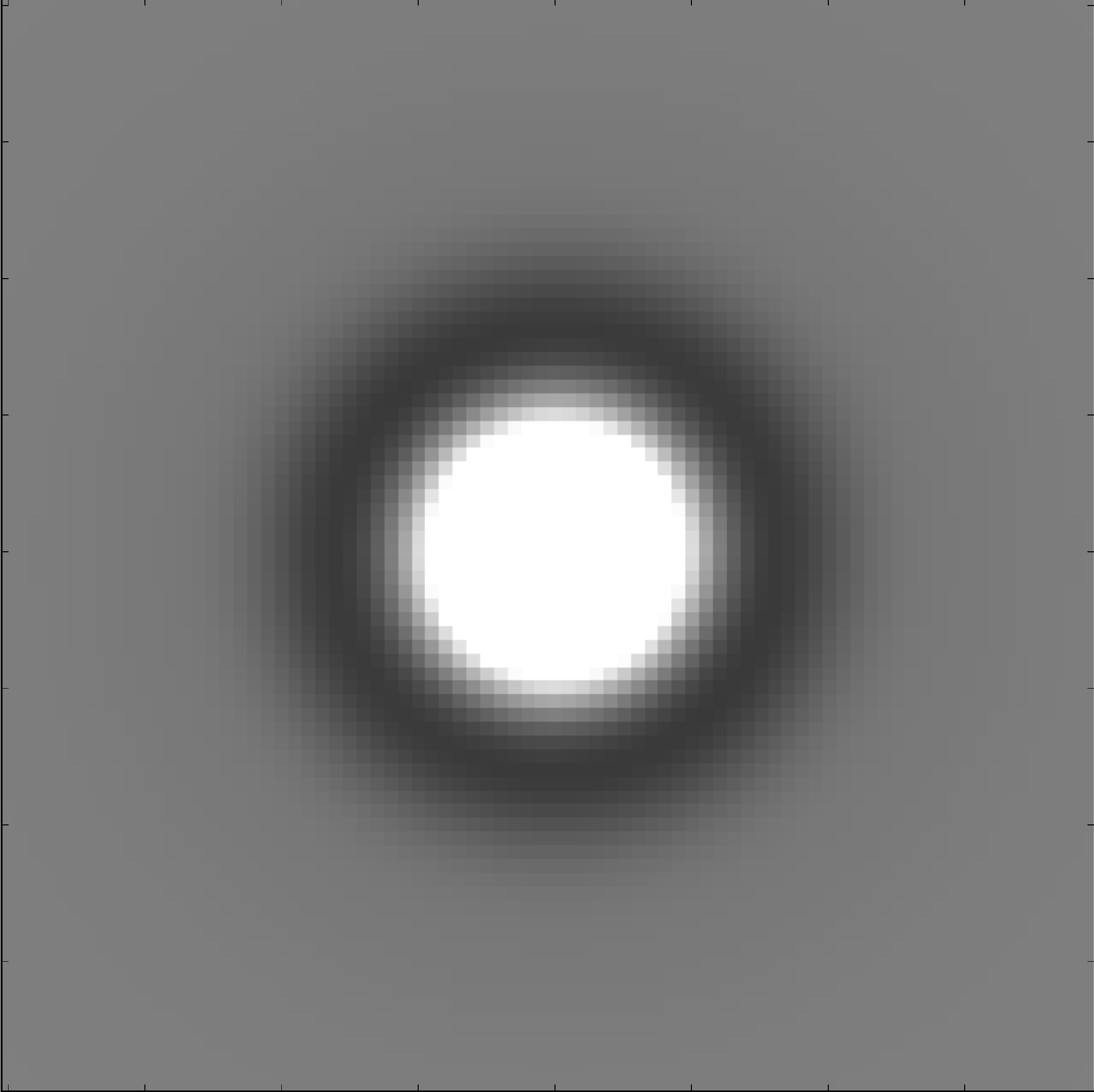} \\[-6pt]
 \includegraphics[clip, width=123pt,trim = 5 5 5 5]{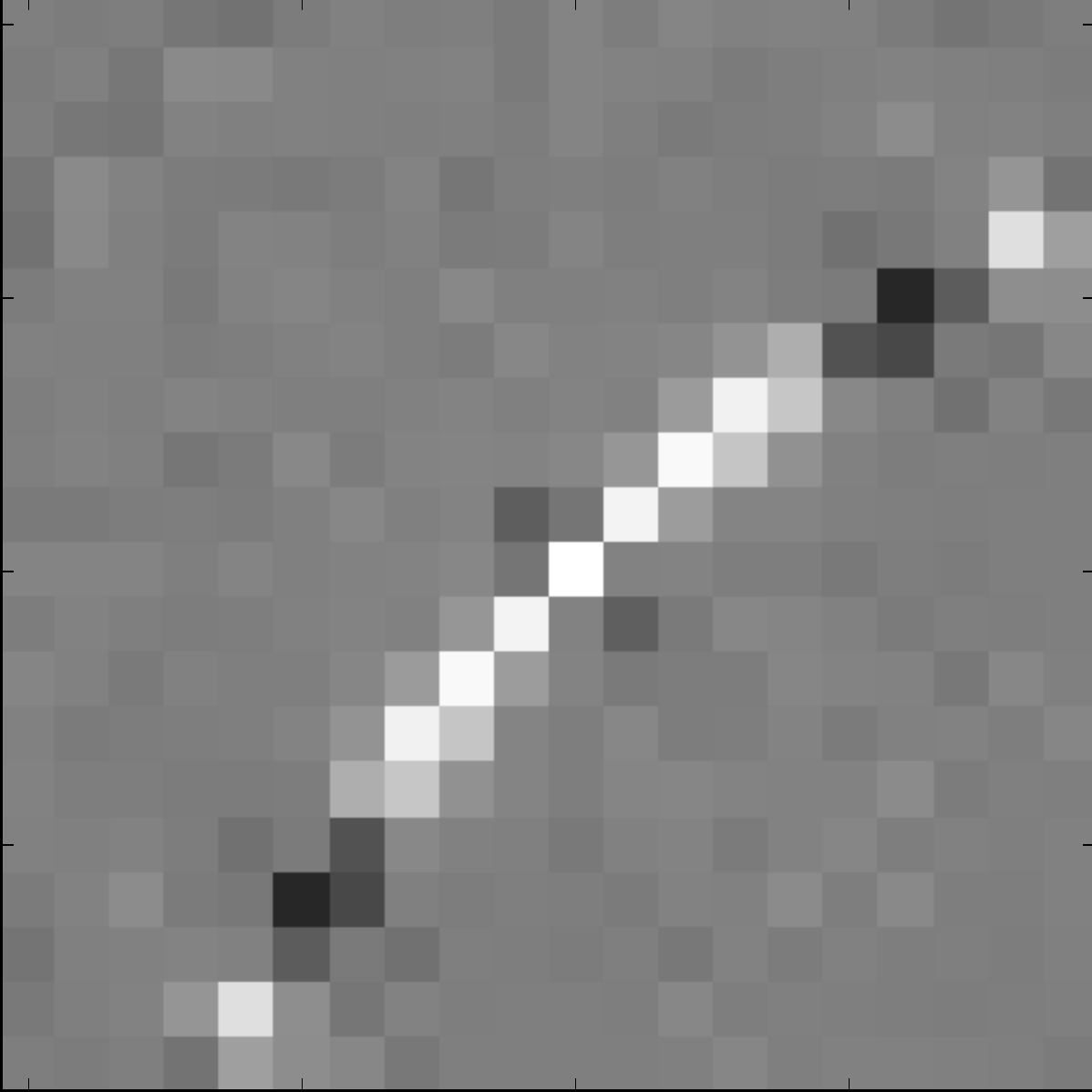} \hspace{-6pt} & \includegraphics[clip, width=123pt, trim = 5 5 5 5]{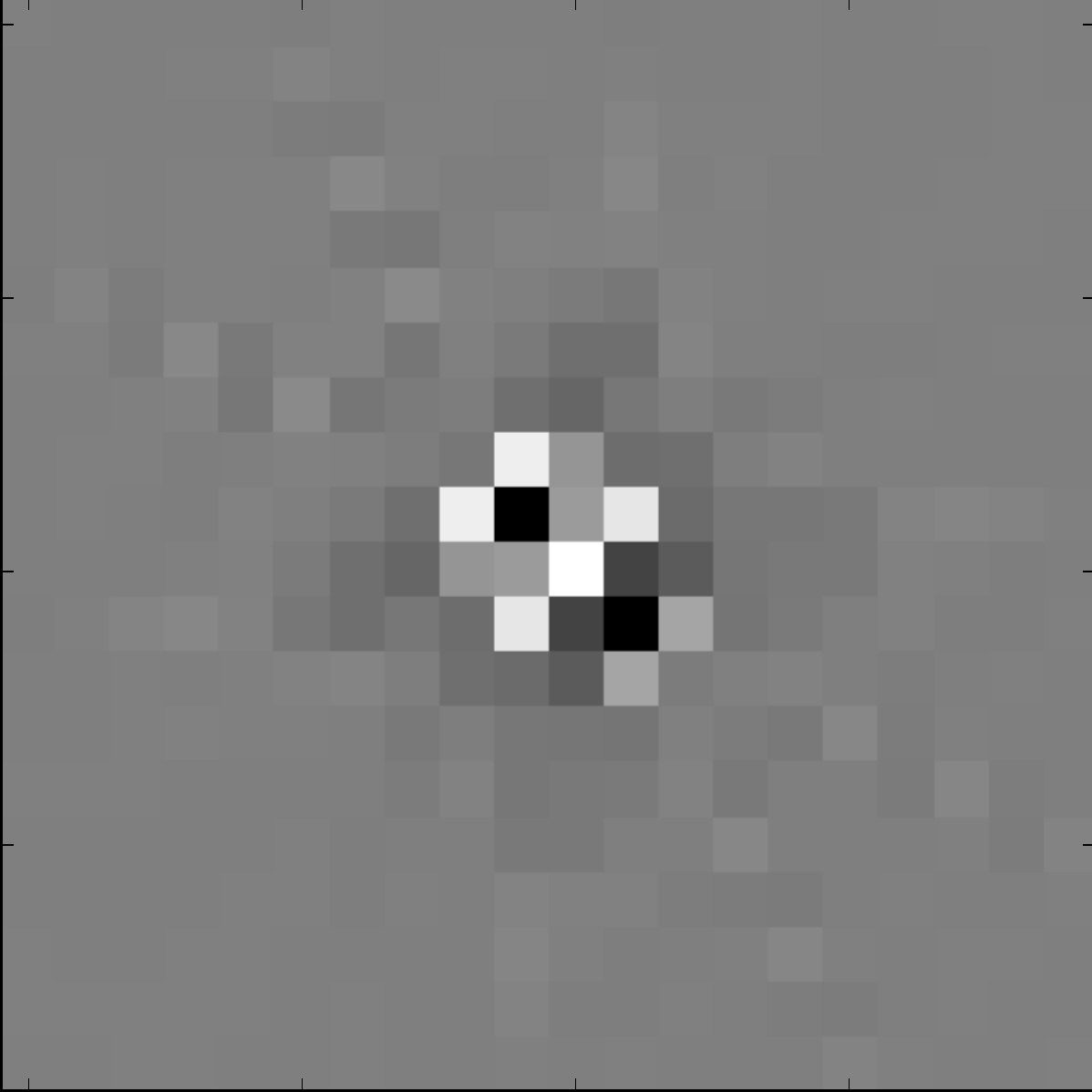} 
\end{array}
$\end{center}
 \caption{These images are the result of right multiplying the matrices $K_y$ with column vectors containing only a single non-zero element, i.e. images of a single non-zero pixel. 
The nonzero pixel (valued 1) was placed either halfway to the edge 
of the FOV (left) or in the center of the FOV (right). The images shown are cropped ROIs showing the pixels which covary with the given pixel. Therefore, each image represents a portion of a 
single row of the matrix $K_y$. The nonzero components away from the pixel of interest demonstrate 
the non-diagonality of $K_y$, while the change in shape between the images in the two columns 
demonstrate the shift-variance of image covariance. The images correspond to the reference FBP 
implementation (top), large image pixels (middle), and Hanning filtration (bottom). Note that the "Mexican hat" shape and near shift-invariance of the 
Hanning filtered case lead to ripples in the corresponding Hotelling template in Fig. \ref{sigs-temps}. The physical extent of each ROI is identical. 
The display window of the top and bottom rows is centered on 0 with a width equal to 0.04 times the maximum covariance element. The display window of the middle row is 
centered on 0 with a width equal to 0.4 times the maximum covariance element.
\label{Ky's}}
\end{figure}

After computing the discrete projection data as in Eqn. \ref{data_model}, zero-mean, independent, identically distributed Gaussian noise was added to the projection data. The standard deviation of the additive noise was 
uniform across the detector elements and equal to roughly ten times the maximum value of the signal 
in the projection data domain. This implies that the data covariance vector 
$K_g$ is diagonal such that
\begin{equation}
\label{data_cov}
\left(K_g\right)_{i,j} = 
	\begin{cases}
	\alpha  & : i = j \\
	0 	    & : i \not= j 
	\end{cases}
\end{equation}
where $\alpha$ is a constant.
We then have that the reconstructed image covariance matrix will be given by
\begin{equation}
\label{Ky}
K_y = AK_gA^T = \alpha AA^T
\end{equation}
where $A$ is the reconstruction matrix described in Section \ref{recon}, and $K_y$ is again the 
reconstructed image covariance matrix. Inspecting this expression for the image covariance matrix, it
becomes obvious why the inversion of this matrix is nontrivial. $K_y$ has dimensionality $M\times M$,
 where $M$ is the number of pixels in the reconstructed image. Further, various components of the 
matrix $A$ such as the matrix representing the weighted back-projection step in the FBP algorithm 
make $K_y$ non-diagonal. Selected components of the matrices $K_y$ for the three reconstruction 
implementations considered here are shown in Fig. \ref{Ky's}. 

The noise model was assumed to be
 invariant under the two 
hypotheses. This assumption is particularly reasonable in our case in light of the fact that the signal 
of interest is small, with size on the order of only a few detector bins. An example of a pair of 
images used in the human observer study is presented in the top row of Fig. \ref{all_examples}.

\subsubsection{Regularization and Image Domain Sampling}
As mentioned in Section \ref{recon}, regularization can be performed in fan-beam FBP by applying 
a multiplicative apodization window to the ramp kernel in the Fourier domain. A case of heavy 
regularization is considered here, wherein a Hanning window with cutoff frequency equal to $1/4$ of 
the Nyquist frequency is used. For this case, we expect a drop in HO SNR relative to the reference 
(unmodified ramp kernel) case since this multiplicative window will place high-frequency components 
of the weighted projection data in the null-space of the reconstruction operator. 
%
The mean reconstructed signal and Hotelling template for the regularized study are shown in the 
middle row of Fig. \ref{sigs-temps}.

\begin{figure}[h!]
\begin{center} $
\begin{array}{c}
\includegraphics[width=246pt]{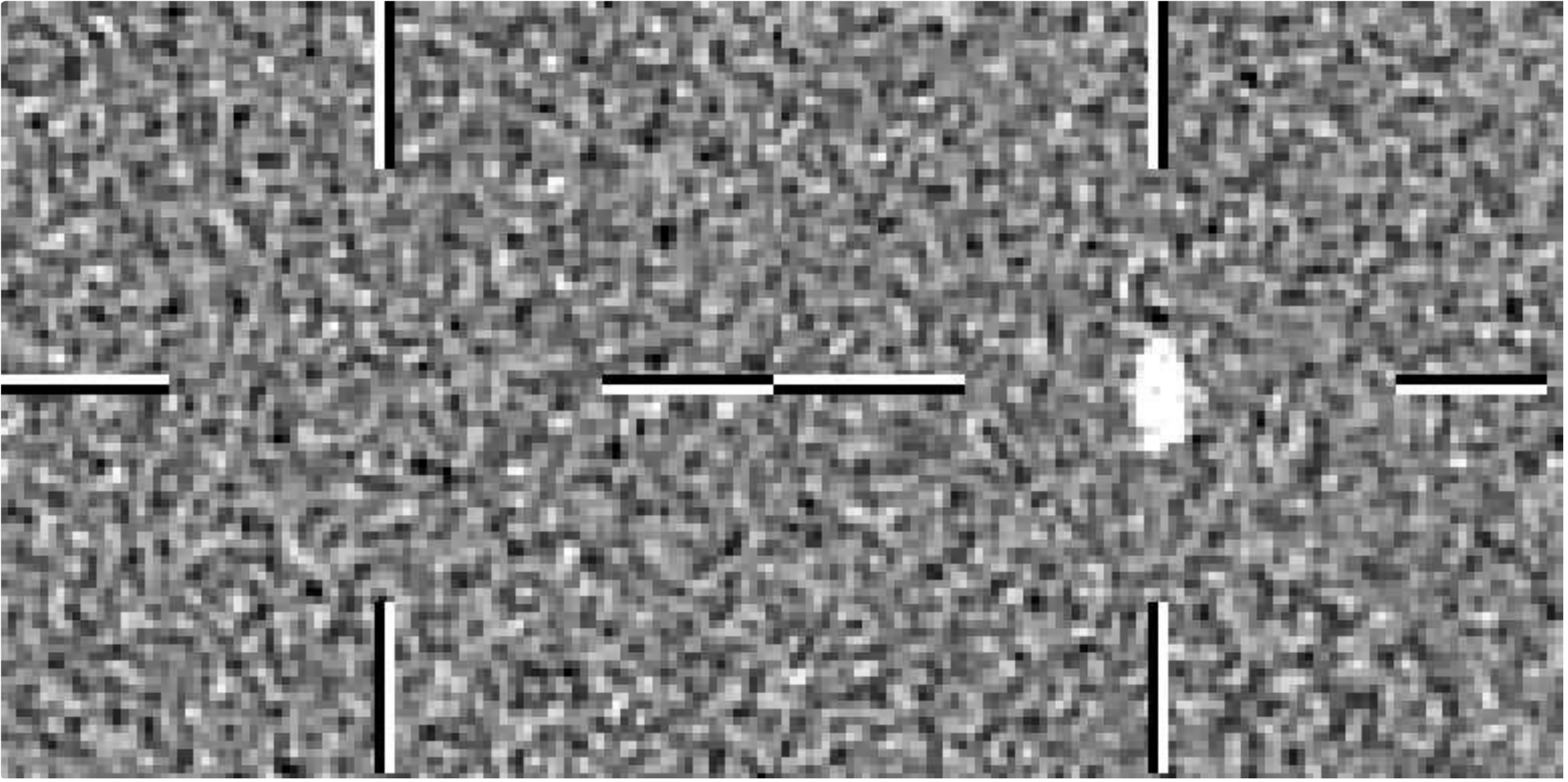} \\
\includegraphics[width=246pt]{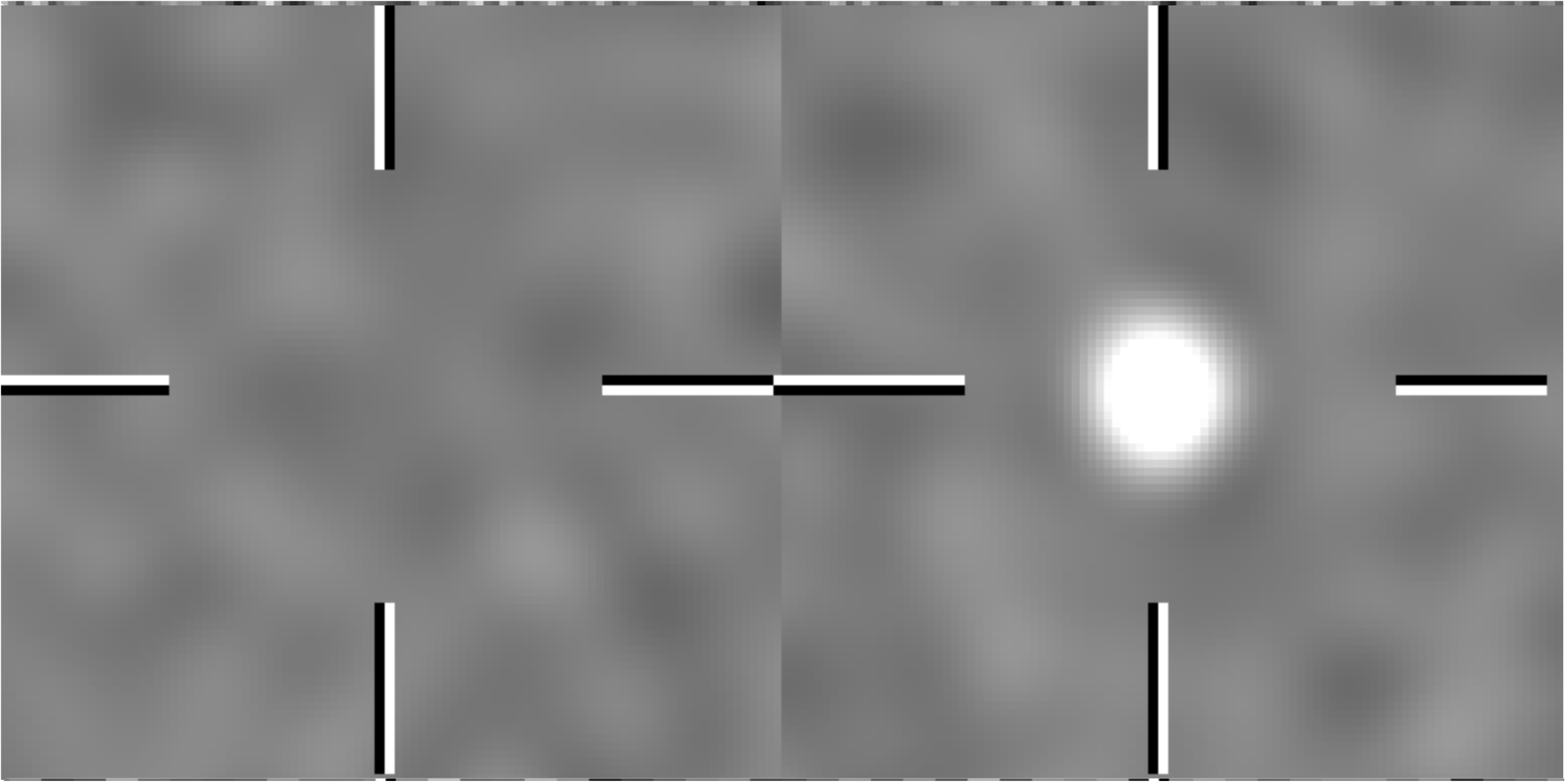} \\
\includegraphics[width=246pt]{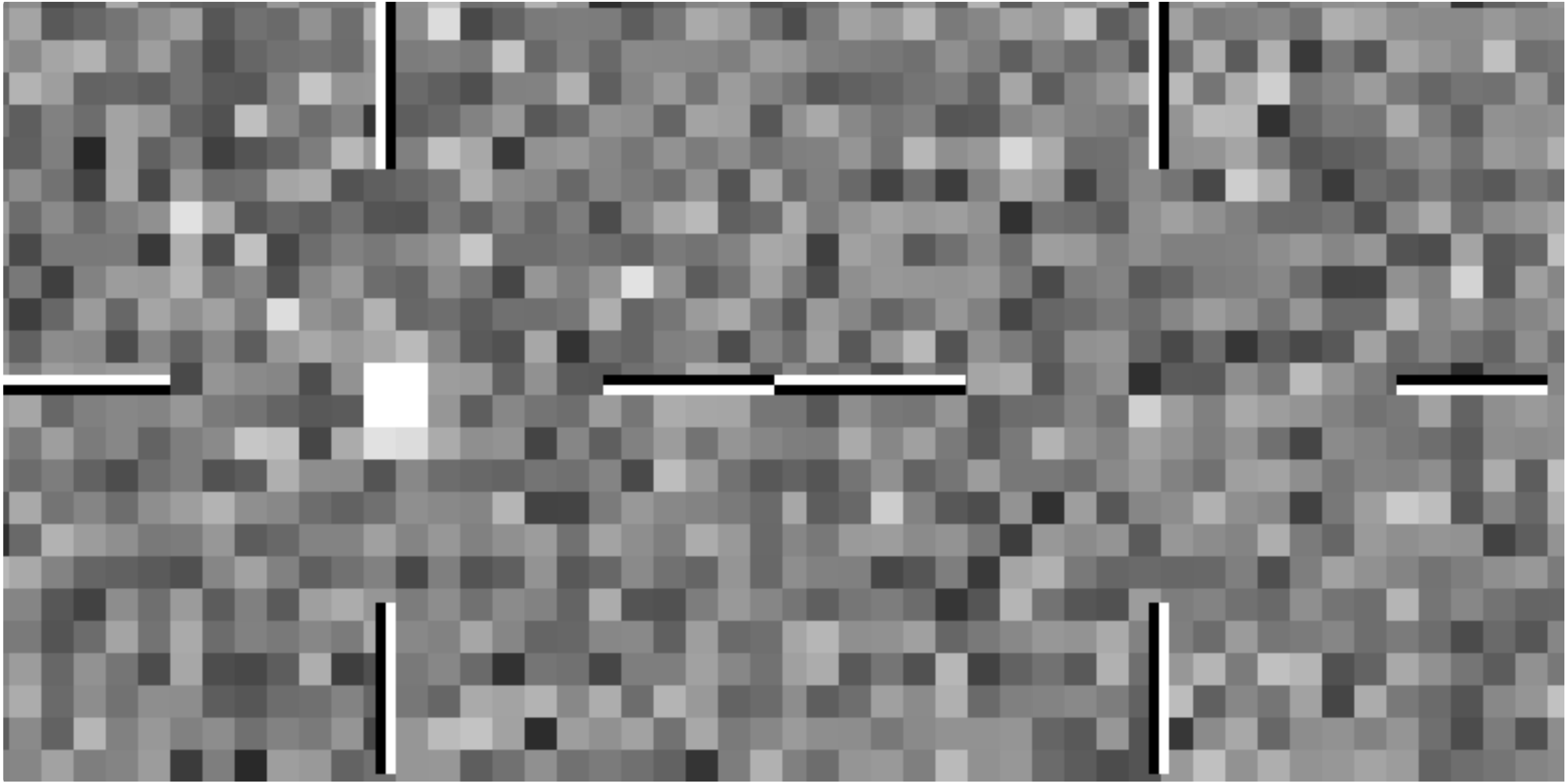} 
\end{array}
$
\end{center}
\caption{
Top: An example of a 2AFC trial used in the human observer study for the reference case of 
ramp filter FBP reconstruction. The signal amplitude here and in the other two sub-figures is 10 times greater than in the experiment for 
the sake of visualization. In addition to two images such as these, the reconstructed signal (left of Fig. \ref{sigs-temps}) was shown for each trial. Display window: [-2.5 2.5]
Middle: An example of a 2AFC trial used in the human observer study for the filtered case. Display window: [0.3 0.7]
Bottom: An example of a 2AFC trial used in the
 human observer study for the large pixel case. 
The window used here is narrower than that used in the experiment for printing purposes. Display window: [0.3 0.7]\label{all_examples}}
\end{figure}

Next, we considered the case of reconstruction onto pixels which are a factor of four larger than in 
the reference case, i.e. approximately eight detector bin widths square. Here, the expected loss in 
HO SNR arises from the fact that the reconstruction matrix $A$ has a more significant null-space due to 
the lower number of output image pixels relative to the number of input data elements. In other words, 
the matrix $A$ is now farther from being square, transforming an $N$-element data vector to an 
$M$-element image array where $M<N$. The mean reconstructed signal and Hotelling template for 
the study involving larger pixels is shown in the bottom row of Fig. \ref{sigs-temps}. 
Example pairs of 
images used in the regularization and large pixel studies are shown in the middle and bottom of 
Fig. \ref{all_examples}.

\subsection{Psychophysical Studies}
\label{human_study}
Seven human observers participated in 100 trials of a 2AFC procedure for each of the three reconstructions investigated: reference FBP, heavy regularization, 
and $4\times$ larger pixel size. In order to achieve improved confidence limits, 
an eighth human observer (hereafter referred to as Observer 8) performed 1500 trials of a 2AFC procedure for the same three experimental 
conditions. Observer 8 rested after every 250 images in order to ameliorate observer fatigue.  
In addition to the two image choices presented to the observers, the reconstructed signal 
was also shown accompanying each pair of images (i.e. the signals on the left of Fig. \ref{sigs-temps}). 
Ten trials were performed preceding the 100 (or 1500) recorded trials for 
each set of conditions as training for each observer. The monitor used for the psychophysical experiments 
was a standard grayscale-calibrated DICOM 14 mammography monitor, and the ambient light and monitor 
brightness were fixed across all observers and reconstruction implementations. The images were windowed and 
leveled so that the level was roughly centered on the mean pixel value and the window width roughly corresponded to 
two-thirds of the full range of pixel values. The exact window settings used were $[-10,10]$, $[-2.5,2.5]$, and $[-10,10]$, for the reference, filtered, and 
large pixel case, respectively. 
While the window and level were changed for each of the three experimental conditions, they were 
held fixed for the trials and observers within a given experimental condition. However, it should be noted that, in general, the results of preliminary 
investigation did not appear to be  sensitive to the window and level used. The software used 
for the image display and data acquisition was written in the Matlab programming language. 
None of the eight observers had received medical training. Each was a graduate student in medical 
physics with a basic understanding of the goals of the studies. 

For each observer and experimental 
condition, the proportion of correct decisions can be shown to be an unbiased estimator of 
the area under the ROC curve, $\widehat{\textbf{AUC}} = \frac{1}{n}\sum_{i=1}^n \lambda_i$, where $\lambda_i$ denotes the outcome of the $i$th trial (1 for a correct response and 0 for an incorrect response). In order to compute a confidence interval (CI) for the true AUC, we use Papoulis's\cite{papoulis_probability_2002} expression for estimating the CI when the variance is unknown 
and the number of trials is large enough that the central limit theorem applies (i.e. the estimator $\widehat{\textbf{AUC}}$ is approximately normal). 
Namely, we find that 
\begin{equation}
\label{CI}
\begin{split}
P\left\{\widehat{\textbf{AUC}}-\frac{s}{\sqrt{n}}z_{1-\delta / 2} < \text{AUC} < \widehat{\textbf{AUC}}+\frac{s}{\sqrt{n}}z_{1-\delta / 2}\right\} \\
> 1-\delta = \gamma \hspace{5cm}
\end{split}
\end{equation}
where $\delta$ is the confidence level, $\gamma$ is the 
confidence coefficient, $s$ is the sample standard deviation, $z_u$ denotes the $u$th percentile of the standard normal density, 
and n is the number of trials (always 100 except for Observer 8). The sample standard deviation $s$ is computed using the unbiased estimate of variance defined by
Eqn. 9-13 of Papoulis\cite{papoulis_probability_2002}. This estimate of standard deviation is also predicated upon the assumption that n is large enough for the central limit theorem to apply. 

Since by Eqn. \ref{AUC2SNR} we have that
\begin{equation}
\text{AUC} = \frac{1}{2}\left[\text{erf}\left(\frac{1}{2}\text{SNR}\right) + 1\right],
\end{equation}
we can rewrite Eqn. \ref{CI} as
\begin{equation}
\begin{split}
P\left\{\widehat{\textbf{AUC}}-\frac{s}{\sqrt{n}}z_{1-\delta / 2} < \frac{1}{2}\left[\text{erf}\left(\frac{1}{2}\text{SNR}\right) + 1\right] \right. \\
\left. < \widehat{\textbf{AUC}}+\frac{s}{\sqrt{n}}z_{1-\delta / 2}\right\} >  \gamma 
\end{split},
\end{equation}
or equivalently
\begin{equation}
\label{CI2}
\begin{split}
P\left\{ 2\text{erf}^{-1}\left[2\left(\widehat{\textbf{AUC}}-\frac{s}{\sqrt{n}}z_{1-\delta / 2}\right) -1\right] < \text{SNR} \right. \\
\left. < 2\text{erf}^{-1}\left[2\left(\widehat{\textbf{AUC}}+\frac{s}{\sqrt{n}}z_{1-\delta / 2}\right) -1\right]\right\} > \gamma.
\end{split}
\end{equation}
95\% confidence intervals for AUC and SNR for each observer and experimental condition were then found by setting $\gamma = 0.95$ and evaluating Eqns. \ref{CI} and \ref{CI2}. 

\subsection{Model Observer Studies}
For each of the three experimental conditions considered here, HO SNR was computed as in Eqn. 
\ref{HOSNR}, where the image covariance matrix $K_y$ is given by Eqn. \ref{Ky}. The Hotelling 
template was computed iteratively via the method of conjugate gradients. 
One important distinction between the HO SNR or AUC and the corresponding human AUC or $d_A$ is that, while the 
human observer estimation of human AUC and $d_A$ are statistically variable, and hence have 
associated uncertainties, the HO SNR is a deterministic quantity derived 
directly from the statistical properties of the two hypotheses. 

\section{Results}
\label{results}
\begin{figure}[ht]
 \includegraphics[height=\columnwidth,angle=-90]{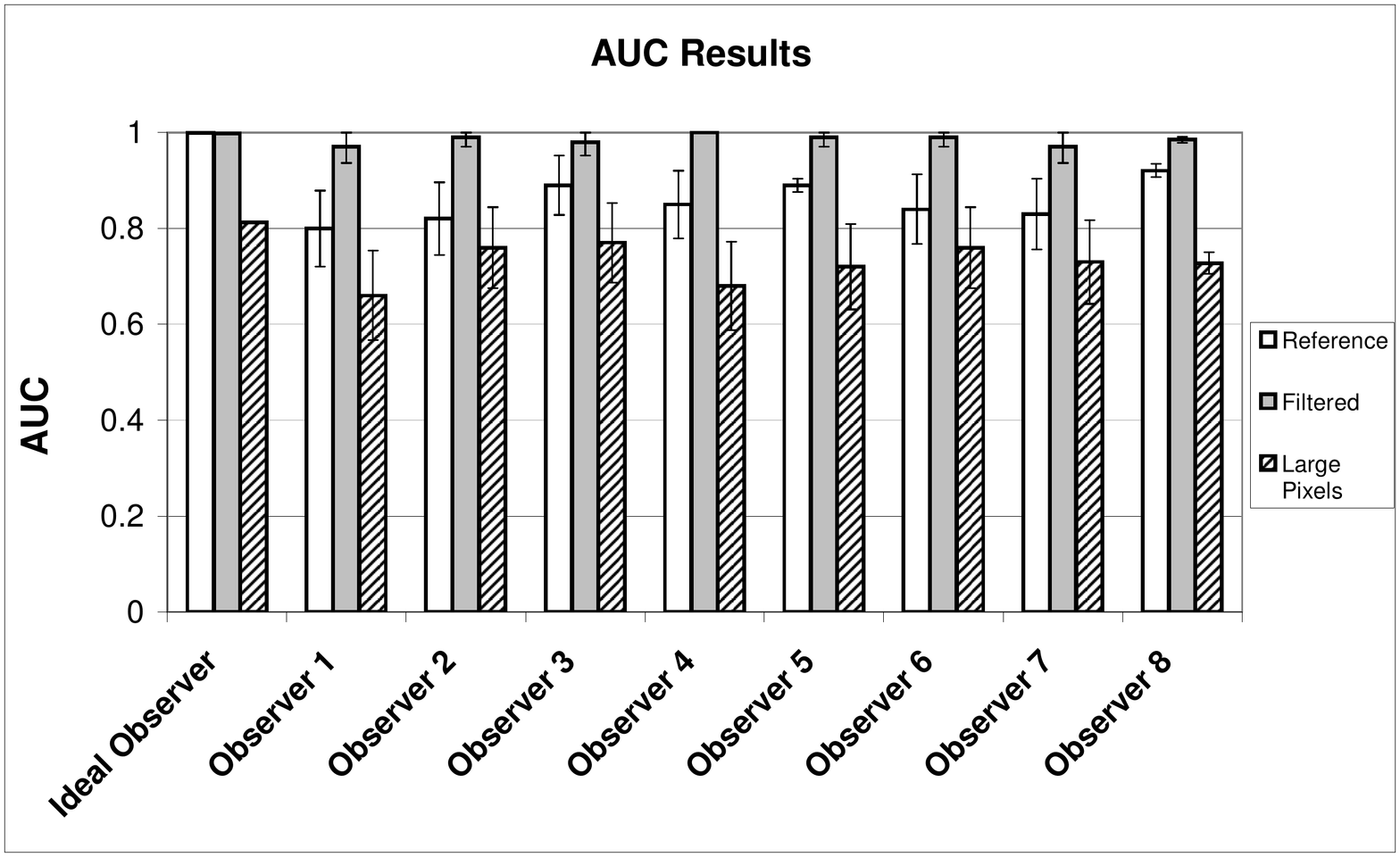}
 \caption{AUC results obtained for the eight human observers and the HO for each of the 
three experimental conditions: reference fan-beam FBP, FBP with Hanning filtration, and FBP with 
increased pixel size. The error bars shown are 95\% confidence intervals obtained through 
Eqn. \ref{CI}, however, the HO performance should be considered an absolute 
upper bound. The confidence interval for Observer 4 in the filtered case is unknown because this observer achieved 100\% correct on this 2AFC study.\label{AUC}}
 \end{figure}

\begin{figure}[ht]
 \includegraphics[height = \columnwidth,angle=-90]{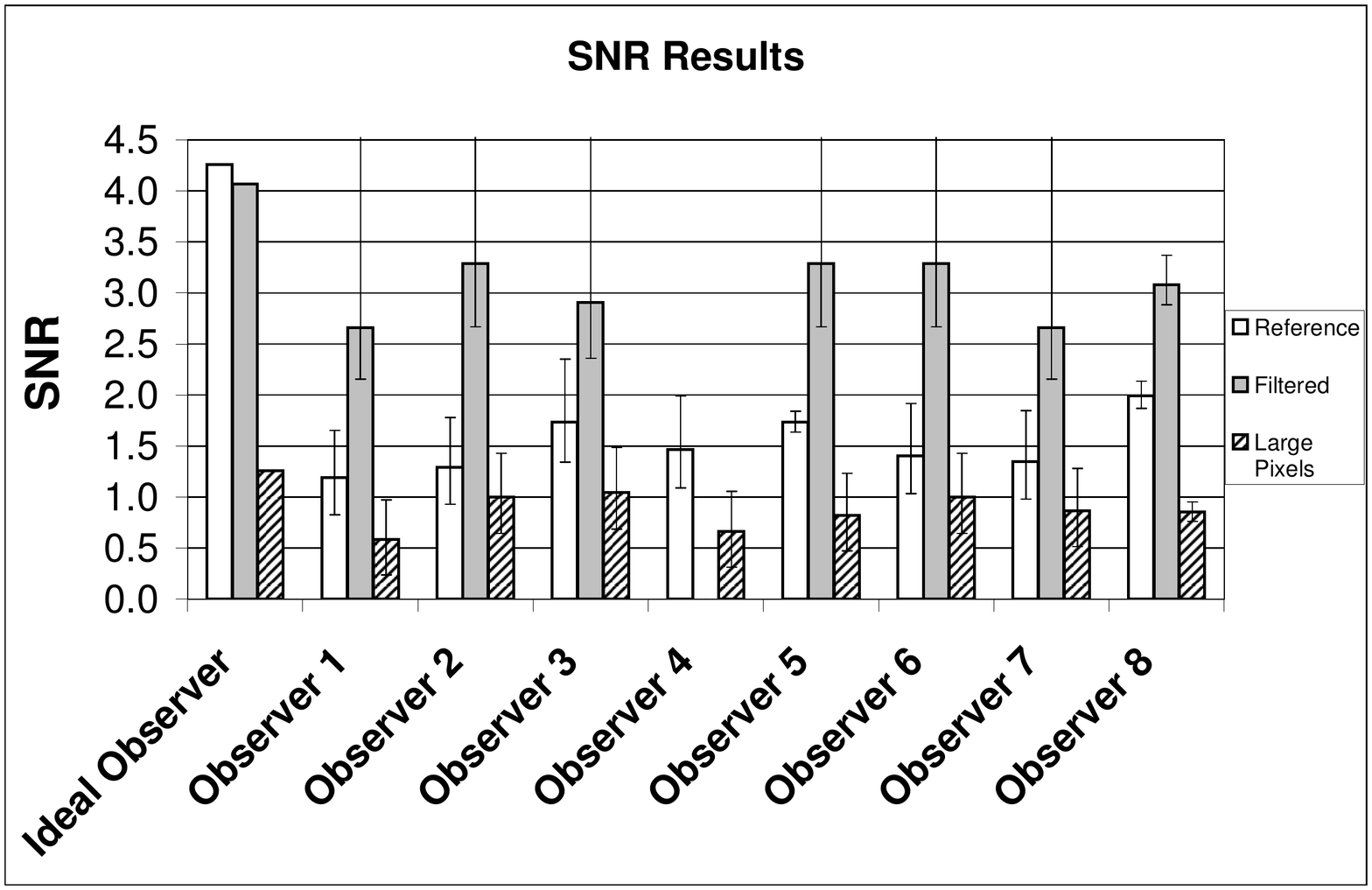}
 \caption{SNR results obtained for the eight human observers and the HO for each of the 
three experimental conditions: reference fan-beam FBP, FBP with Hanning filtration, and FBP with 
increased pixel size. The error bars shown are 95\% confidence intervals obtained through 
Eqn. \ref{CI2}, however, the HO performance should be considered an absolute 
upper bound. For instance, when the CI on AUC extends to 1, the SNR error bars extend to infinity, and are therefore truncated. 
Observer 4 has an unknown SNR for the filtered case because this observer achieved 100\% correct on this 2AFC study.\label{SNR}}
 \end{figure}

Fig. \ref{AUC} shows the AUC results obtained for the eight human observers and the HO for each of the 
three experimental conditions. The corresponding results in terms of SNR (or $\hat{d}_A$ for the 
human observers) are shown in Fig. \ref{SNR}. These results are also summarized in Tables \ref{AUCtable} and \ref{SNRtable}, 
respectively. Clearly, there is a marked improvement in human 
observer performance for the case of regularization, whereas the ideal observer performance 
degrades with regularization due to the loss of information through smoothing. By contrast, increasing 
the pixel size in the reconstructed image domain degrades both the human and Hotelling observers 
similarly. Note that Observer 4 achieved 100\% correct on the regularized 2AFC study, so that a reliable CI for AUC or estimate of $d_A$ 
is not possible. However, we include the results for Observer 4 under the other two experimental conditions for completeness.

The error bars shown are 95\% confidence intervals obtained through Eqns. \ref{CI} and \ref{CI2}. Although the upper limit of the confidence interval exceeds the 
deterministic HO SNR for the filtered case, the HO performance should be considered an absolute 
upper bound. The statistical error bars are only shown to provide intuition regarding the nonlinearity 
of the confidence intervals when converting from AUC to SNR or $d_A$. 


It should be noted that the result that regularization improves human observer performance to the 
extent that (in the low statistics case of 100 trials) it is statistically indistinguishable from the ideal 
observer performance is a consequence of the particular detection task investigated. In other words, 
in order to have any statistical power in the large pixel, human observer study, a signal was chosen that would 
necessarily be at a low-difficulty operating point for the humans in the regularization study. For this 
reason, the performance of Observer 8, the high statistics observer was investigated, clarifying the conclusions of the regularization study. 
Namely, in keeping with the findings of Abbey and Barrett \cite{abbey_human-_2001}, regularization was found to enable human observers to function at 
roughly 50\% efficiency with respect to $\text{SNR}^2$ of the Hotelling observer (the square of the values presented in Table \ref{SNRtable}. The fact that these results are 
at the high end of those found in Abbey and Barrett \cite{abbey_human-_2001} (efficiency of humans ranging from 40\% to 50\% of HO detectability)
 is likely due to the more complex noise model employed in both 
the Hotelling and human observer studies. In particular, whereas Abbey and Barrett \cite{abbey_human-_2001} approximate regularization in tomographic 
reconstruction by applying a discrete Fourier transform to a white noise image, apodizing, and applying the corresponding 
inverse transform, the noise structure in this study is the result of applying a full tomographic reconstruction algorithm. 
It is reasonable, therefore, that the precise efficiency results for humans detecting a signal in regularized noise could 
differ somewhat between the two investigations. 

While the results of the regularization study provide insight into a case where the upper bound provided by the HO 
moves in the direction opposite to the mean human observer performance, the case of large pixels, in some sense, 
provides a simpler result. Namely, decreasing the number of image pixels destroys information in the reconstruction by 
increasing the null-space of the reconstruction matrix $A$, and this loss of information appears to have no underlying benefit 
to the human observer. Rather, the impairment of the HO in the case of larger image pixels is mirrored by a similar impairment 
of human observer performance, suggesting that the modeling of human performance in this case may not need to be as 
complex as in the case of regularization.

\begin{center}
\begin{table}
\hspace{2.0cm}
\begin{tabular}{ | l|c|c|c|}
\hline
Observer & Reference AUC & Filtered AUC & Large Pixel AUC \\ 
\hline
Ideal      & 0.9987 & 0.9980 & 0.8126\\
Obs. 1    & $0.80 \pm 0.0791$ & $0.97 \pm 0.0337$ & $0.66 \pm 0.0937$\\
Obs. 2  & $0.82 \pm 0.0759$ & $0.99 \pm 0.0197$ & $0.76 \pm 0.0844$\\
Obs. 3   & $0.89 \pm 0.0619$ & $0.98 \pm 0.0277$ & $0.77 \pm 0.0832$\\
Obs. 4    & $0.85 \pm 0.0706$ & $1.0 \pm ?$ & $0.68 \pm 0.0922$\\
Obs. 5  & $0.89 \pm 0.0137$ & $0.99 \pm 0.0197$ & $0.72 \pm 0.0888$\\
Obs. 6  & $0.84 \pm 0.0725$ & $0.99 \pm 0.0197$ & $0.76 \pm 0.0844$\\
Obs. 7 & $0.83 \pm 0.0743$ & $0.97 \pm 0.0337$ & $0.73 \pm 0.0878$\\
Obs. 8 & $0.9207 \pm 0.0137$ & $0.9853 \pm 0.0061$ & $0.7273 \pm 0.0226$\\
\hline
\end{tabular}
\caption{AUC of the ideal observer and eight human observers under each of the three experimental conditions. The intervals shown are the 95\% confidence intervals.
The confidence interval for Observer 4 in the filtered case is unknown because this observer achieved 100\% correct on this 2AFC study. \label{AUCtable}}
\end{table}
\end{center}

\begin{center}
\begin{table*}
\footnotesize
\hspace{0.2cm}
\begin{tabular}{ | l|c|c|c|c|c|c|}
\hline
Observer & Reference SNR& Mean Val. & Filtered SNR & Mean Val.& Large Pixel SNR & Mean Val.\\ 
\hline
Ideal      & 4.259 & N/A& 4.070 & N/A& 1.255 & N/A\\
Obs. 1    & $(0.828,1.655)$& $1.190$ & $(2.156,\infty)$ & $2.660$ & $(0.236,0.970)$ & $0.583$\\
Obs. 2  & $(0.928,1.780)$  & $1.295$ & $(2.667,\infty)$ &$3.290$ & $(0.644,1.432)$& $0.999$ \\
Obs. 3   & $(1.339,2.353)$ & $1.735$ & $(2.358,\infty)$ & $2.904$ & $(0.688,1.485)$ & $1.045$\\
Obs. 4    & $(1.089,1.993)$ &$1.466$& unknown & unknown&$(0.314,1.055)$&$0.661$\\
Obs. 5  & $(1.636,1.843)$ & $1.735$ & $(2.667,\infty)$ & $3.290$ & $(0.474,1.235)$ & $0.824$\\
Obs. 6  & $(1.033,1.918)$ & $1.406$ & $(2.667,\infty)$ & $3.290$ & $(0.644,1.432)$ & $0.999$\\
Obs. 7 & $(0.979,1.848)$ & $1.349$ & $(2.156,\infty)$ & $2.660$ & $(0.515,1.283)$ & $0.867$\\
Obs. 8 & $(1.870,2.134)$ & $1.993$ & $(2.882,3.371)$ & $3.082$ & $(0.761,0.953)$ & $0.855$\\
\hline
\end{tabular}
\caption{SNR of the ideal observer and eight human observers under each of the three experimental conditions. The intervals shown are the 95\% confidence intervals.
Observer 4 has an unknown SNR for the filtered case because this observer achieved 100\% correct on this 2AFC study. \label{SNRtable}}
\end{table*}
\end{center}

\section{Internal Noise Methods}
\label{sec:int_noise}

In this section, we focus on addressing the question of approximating human performance. Specifically, we want to know if the HO is a 
good approximation of human performance. If so, then we gain another degree of significance to the HO, since it is then the ideal observer and
also carries direct information about human observer performance. 
While the preceding results have quantified the response of human and ideal observer performance to two forms of image 
regularization, showing their discrepancies in some cases, 
the question remains as to whether the ideal observer or Hotelling observer might be useful for system design and parameter optimization, 
even in cases where it is not predictive of human performance in absolute terms. This question has been addressed in 
some nuclear medicine applications, where objective assessment has a more well established role, by investigating 
modifications of the Hotelling observer \cite{barrett_foundations_2004}. We will look at one such modification, namely internal noise, in this section.

Here, we investigate both the HO and 
the non-prewhitening matched filter (NPWMF) \cite{barrett_foundations_2004} for a detection task. Like the HO, the NPWMF is a linear observer, however 
since the NPWMF does not use the Hotelling template, its performance will in general be lower than the HO. Like the HO, the NPWMF has been shown to 
well approximate humans in some cases \cite{beutel_handbook_2000}. These model observers 
are investigated with and without internal noise models and compared to human observer results for 
validation. In this context, the phrase \emph{internal noise} refers to a model for the inherent uncertainty a human has when making 
a decision. This is one common approach to degrading model observer performance in order to better predict humans \cite{beutel_handbook_2000}.

\subsection{Simulation Parameters}

While in the preceding sections we were concerned with characterizing overall trends of humans and the Hotelling 
observer and comparing the two in general terms, in this section we are attempting to bridge the gap between the two observers. 
In order to increase the relevance of our conclusions, we focus on a particular system and relevant clinical task, namely microcalcification 
detection in dedicated breast CT, which will be described in more detail in the chapters to come. For now, we will only provide the 
basic simulation parameters used. 
 
The CT system modeled in this work is based on the system described in Gong et al. \cite{gong_microcalcification_2004}.
To summarize, we model a flat-panel fanbeam
system with a source-to-detector distance of 80cm, and
a source-to-axis distance of 60cm. We consider 128 projection
views equally spaced over 2$\pi$ radians and an array of 1590
detector pixels spaced over a detector width of approximately
31.8cm, for a detector pixel width of 0.2mm. For image
reconstruction, the circular field-of-view is then inscribed in a
(1659 pixel)$^2$ image array, with square pixels of width 0.1mm.
Finally, a 0.4mm x-ray focal spot was modeled by convolving
the projection data with a rect function. FBP reconstruction
was used, along with a 2D Butterworth filter applied to the
reconstructed images. The Butterworth filter was of order 5.0,
with a cutoff of 0.25mm$^{-1}$.

We model a digital breast phantom as a uniform circle with a
7.0cm radius. The background is uniform with an attenuation
value midway between adipose and fibro-glandular breast
tissue, however preliminary results not shown here demonstrate that 
our results are consistent with somewhat more complex background models. 
The simulated microcalcifications are circularly symmetric
Gaussian functions with full widths at half-maximum defined
by the microcalcification size and peak attenuation values
equal to that of calcium. The attenuation is scaled with the
inverse of the microcalcification diameter for diameters less
than 1.0mm to account for linear partial volume averaging
within the 1.0mm thick slice. Finally, the location of the
microcalcification is set to the center of the field of view,
2.0cm from the center, and 4.0cm from the center, and the
corresponding human and model observer results were averaged across
signal locations.

\subsection{Psychophysical Studies}

In order to provide a reference for the model observer
results, a human volunteer performed a 2AFC experiment for
detection of microcalcifications using images simulated with
the same system, noise, and phantom parameters used in the
model observer studies.  
The human observer performed 300
2AFC trials for each of five microcalcification sizes ranging
from 100$\mu$m to 200$\mu$m.

\subsection{Model Observer Experiment}

The computation of the NPWMF performance is similar to HO performance computation, however the template used is not 
ideal, but instead is simply a matched filter. In other words for the NPWMF, $w$ becomes 
$\Delta\bar{y}$, and the resulting NPWMF SNR is given by
\begin{equation}
\text{SNR}^2_\text{NPW} = \frac{\left\| \Delta\bar{y} \right\|_2^2}{\Delta\bar{y}^TK_y\Delta\bar{y}},
\end{equation}
with the corresponding $P_C$ value again given by
\begin{equation}
P_C = \frac{1}{2} \left\{ 1+\mathrm{erf}\left(\frac{\mathrm{SNR}_\mathrm{NPW}}{2}\right)\right\}.
\end{equation}

In general terms, the difference between the HO and the NPWMF is that the HO accounts for 
noise correlations in the image by performing a prewhitening step before applying a matched 
filter. Humans have been shown to have varying ability to prewhiten images depending on the 
specific task, so that the appropriateness of the HO or NPWMF is highly task-dependent.

A somewhat more realistic noise model was employed for this study than the preceding study. In order 
to construct $K_y$ we first consider a projection data covariance matrix 
based on uncorrelated Gaussian projection noise with Poisson-like variance \cite{barrett_radiological_1996}:

\begin{equation}
\left(K_g\right)_{i,j}=
	\begin{cases}
	\frac{e^{\bar{g}_{i,j}}+1}{\bar{N}_0} &: i=j \\
	0 &: \text{else}.
	\end{cases}
\label{cov}
\end{equation}
This covariance is object-dependent, and for the purposes of this study, we consider the average of 
$K_g$ between the two hypotheses. We then consider the matrix $A$, which represents the  
discrete-to-discrete image reconstruction operator. The image covariance matrix is then given by
$K_y = AK_gA^T$, where $A^T$ is the transpose of $A$, 
and the Hotelling template $w_y$ is obtained by solving the linear equation
$K_yw_y = \Delta\bar{y}$ with the method of conjugate gradients. 

Finally, internal noise (observer uncertainty) is modelled by applying a constant factor less than 1 to the model 
observers' SNR values. In our case this is equivalent to applying zero-mean Gaussian noise to the 
test statistic with a variance that is constant across the different signal sizes. The magnitude of the 
internal noise is determined by a weighted least squares fit to the human data, implying that 
internal noise is the only presumed source of human inefficiency in this study. 

\section{Internal Noise Results}
\label{sec:int_noise_res}

\begin{figure}[h!]
\begin{centering}
\includegraphics[width=0.6\columnwidth, bb = 0 0 610 477, clip = True]{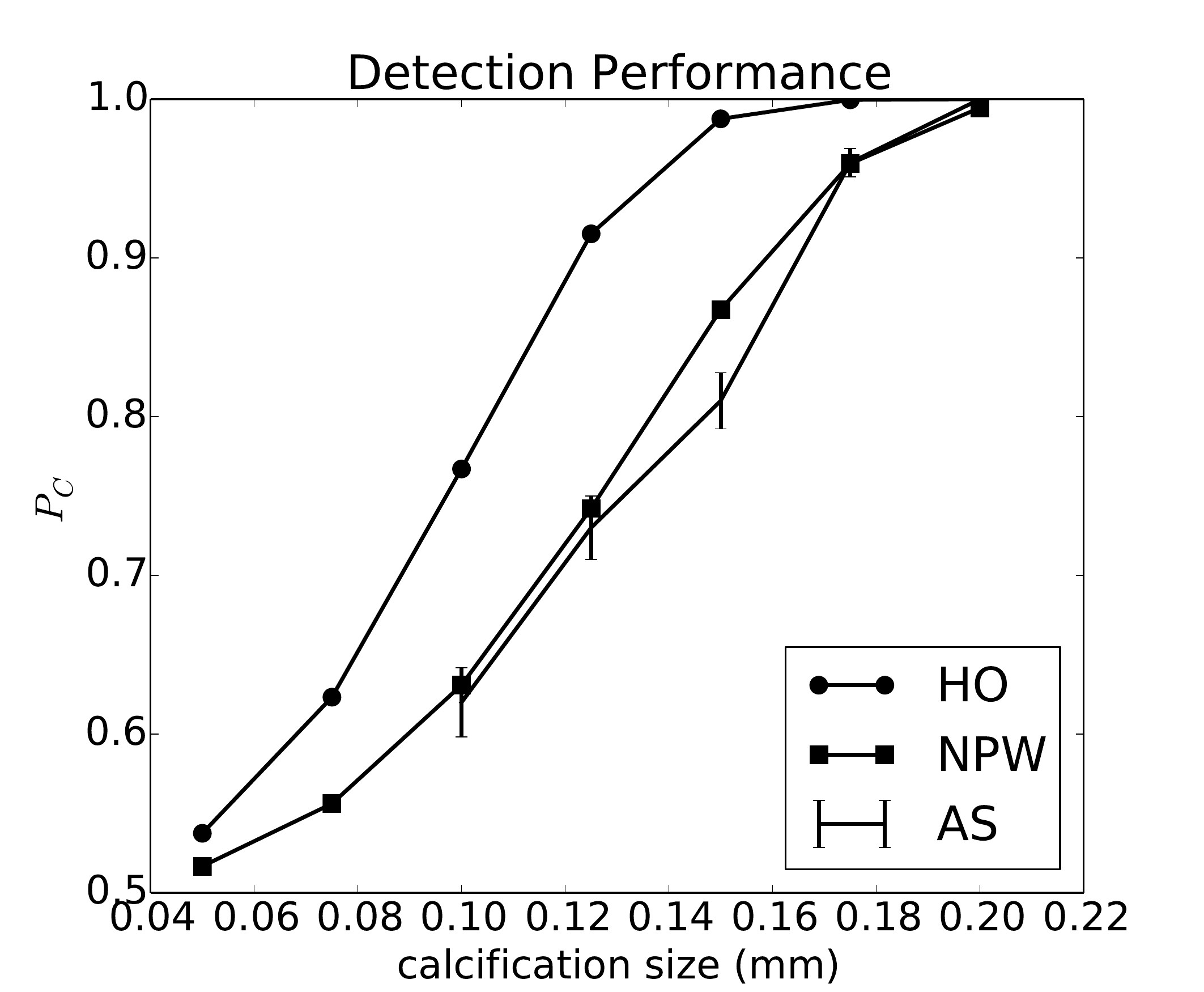}
\caption{Human observer results in terms of $P_C$ are shown for the 2AFC test of microcalcification detection, 
along with equivalent results for the HO and the NPWMF. The error bars represent 95\% confidence intervals.
\label{fig}}
\end{centering}
\end{figure}

\begin{figure}[h!]
\begin{centering}
\includegraphics[width=0.6\columnwidth,bb=0 0 610 477, clip=True]{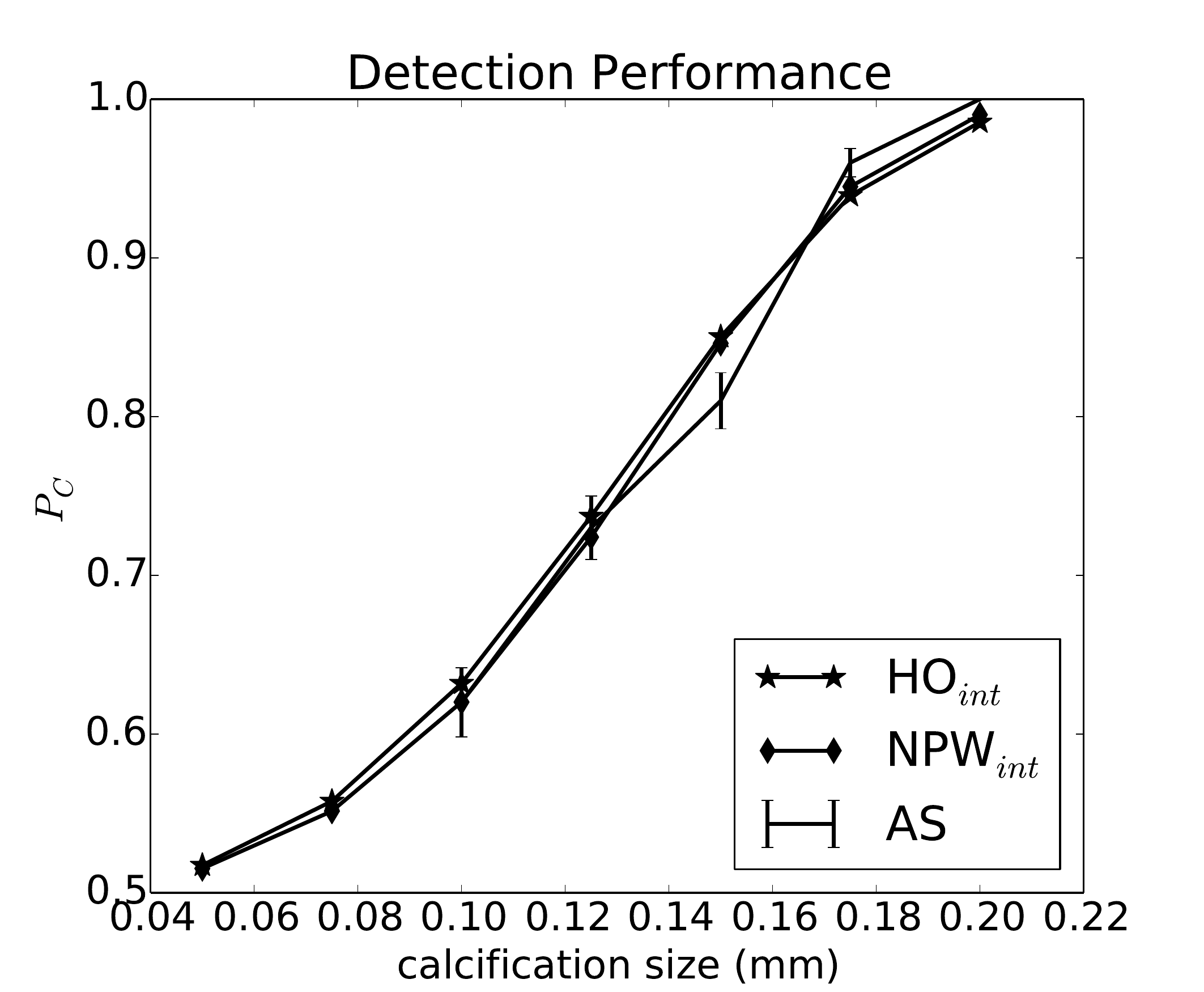}
\caption{Same as previous figure, but with the addition of internal noise to the model observers to account for 
human observer inefficiency.
\label{fig2}}
\end{centering}
\end{figure}

Figure \ref{fig} shows the results of the human observer experiment (labeled AS) along with the 
HO and NPWMF results. In terms of SNR$^2$, the human observer demonstrated an efficiency of 
0.84 relative to the NPWMF and an efficiency of 0.21 relative to the HO. Meanwhile Figure \ref{fig2}
shows the results of adding an internal noise model to each the HO and the NPWMF. Clearly both 
the HO and the NPWMF correlate well with the human results. It is interesting to note that the NPWMF 
is nearly predictive of absolute human observer performance, while the gap between the HO and the NPWMF
clearly illustrates that noise correlations play an important role in optimal task performance.

While both model observers correlate well with humans in this case, it is likely that alternative smoothing strategies 
in the image reconstruction will lead to qualitatively different predictions for the HO and the NPWMF, since 
noise correlations are at the heart of the difference between these two observers. 
Possible future work could also investigate the impact of modeling additional sources of human inefficiency, 
such as psychophysical channels, eye filters, and contrast sensitivity, however the present findings support the 
hypothesis that the HO could be successfully used to optimize reconstruction parameters for humans, since 
internal noise is sufficient to lead to good agreement between humans and the HO. In other words, since the HO performance 
with internal noise is a monotonic function of the HO performance without internal noise, an optimal system configuration for 
one model observer is optimal for the other. To the extent that the HO with internal noise is predictive of humans, the same system 
configuration is optimal for humans as well. 

\section{Conclusion}
\label{sec:conc}

The purpose of this chapter has been to evaluate human observer performance for a signal-detection 
task in fan-beam CT and to compare the results to the Hotelling observer (HO) performance. In order to 
establish a clear upper limit on detectability for each case, HO SNR values were computed, 
using knowledge of the statistics underlying the signal-present and signal-absent hypotheses. 
Meanwhile, human observer performance was computed using a 2AFC experiment, wherein approximate 
AUC values were obtained. These values were compared to the corresponding AUC values for the HO. 
Further, the detectability indexes $d_A$ for each observer and reconstruction implementation were 
computed from the approximate AUC values. These detectability estimates were then also related to 
Hotelling SNR. 

Three cases for reconstruction were considered. First, a reference case of unfiltered, ramp-spectrum 
noise was investigated. Next, heavy regularization was implemented by means of a Hanning filter in the 
fan-beam FBP algorithm. Finally, the reconstructed pixel size was increased. The last two implementations 
introduce signal detectability loss by virtue of the fact that they expand the null-space of the reconstruction 
matrix which constitutes the discrete implementation of the FBP algorithm. 

While the regularization impairs the performance of the HO, it was shown to result in a marked 
improvement for the human observers, making their performance statistically indistinguishable from that 
of the ideal observer. The general trend of this result is consistent with the results of Abbey and Barrett\cite{abbey_human-_2001}.
 Increasing the
reconstructed image pixel size, however, degraded both HO and 
human observer performance. Future work could investigate the optimum degree of regularization in 
the reconstruction algorithm with respect to human observer performance.


%% file: Chapter3/Chapter3.tex
\chapter{FBP Sampling Properties}
\label{ch:sampling}
\section{Introduction}

In this chapter, having provided evidence that the HO is likely to be useful for optimizing CT systems for quantum-noise-limited tasks involving small signals, 
we turn our attention to the primary practical barrier for HO implementation in CT with analytic algorithms. Specifically, we begin 
to address the dimensionality issue which arises from having many image pixels, many of which have long-range covariance with other image pixels. In short, we wish 
to characterize the image-domain sampling properties of the HO, so that the dimensionality of solving for HO SNR can be reduced by, for example, 
truncation of the image to a small ROI. 

By image-domain sampling, we refer to the discretization of the continuously-defined reconstructed image into pixels. 
Discretization of a CT image has the potential to introduce an effect known as noise aliasing, with the local noise power spectrum (NPS) being 
affected by the NPS from other locations in the image \cite{kijewski_noise_1987}. A similar effect is produced on the 
Fourier components of a reconstructed signal. The basic result of noise and signal aliasing is that high-frequency spatial features 
are ``aliased'' onto lower-frequency features, producing artificial low-frequency structures in the reconstructed image. The extent to which signal and
noise aliasing occurs depends both on the extent of the field of view and on the size of the image pixels. The
purpose of this chapter is to determine the subsequent impact of these parameters on HO performance for the 
task of detecting a small signal. 

Specifically, in the chapters which follow, we will make the assumption that restriction 
of an image to a region-of-interest (ROI) around a given signal does not affect our estimation of HO performance.
Although this assumption is rarely stated in the literature, it is implicit in any number of methods for model observer performance estimation which rely on 
local stationarity assumptions, or extraction of local noise properties from a series of noisy images.
In order to investigate the effect of limiting an observer to using an ROI and to give some intuition regarding the impact of ROI and pixel size, we will 
perform parameter sweeps of of each of these parameters and evaluate the resulting HO efficiency. 

\new{When varying pixel and ROI size, two principal issues must be considered. The first issue relates to the computational burden of computing large numbers of 
pixel values. In an effort to fully capture the relevant signal and noise components in the reconstructed image, one could simply decrease pixel size and increase ROI size 
\emph{ad infinitum}, with the hope that eventually, all of the important image content will be captured. However, this is not a practical approach because one quickly
obtains images whose covariance matrices are excessively large. Further, it is not even clear that perpetually extending the ROI would always capture all of the useful pixel values 
since for a finite number of projection views, the back-projection operator leads to reconstructed objects without finite support.} 

\new{The second sampling issue relates to the Hotelling template, which characterizes the optimal linear decision 
strategy for a given task. Some sampling configurations will result in high HO task performance, but lead to Hotelling templates with unintuitive structures or to templates which are not 
even uniquely defined. This is undesirable since it does little good to design an imaging system for which pixel values must be combined in unintuitive ways in order to 
perform simple tasks. We therefore would ideally like to find an image sampling paradigm which results in high HO performance metrics with simple and intuitive template structure.} 

\new{In the interest of clarity, we will now briefly review the relevant HO formalism with an eye toward characterizing sampling properties in parallel-beam CT. 
For the time being, we will consider a general linear 
reconstruction operator $\mathcal{R}$, which can be interpreted as either FBP or some other linear reconstruction method.
For instance, before investigating the 
x-ray transform directly, in section \ref{sec:dft} we will turn our attention to a simpler illustrative example employing a version of the inverse discrete Fourier 
transform (DFT). Like the FBP algorithm, our example illustrates a case where a continuously-defined image function can be sampled
at an arbitrary number of image pixel locations. However, unlike the FBP algorithm, the inverse DFT has sampling properties which are well 
understood and can provide some insight into the more difficult to characterize FBP sampling problem. Finally, in section \ref{sec:results} we will 
investigate the sampling properties of FBP.}

\subsection{HO Formalism}  \label{sec:methods}

In this chapter, we employ the HO and its 
associated metric, the HO SNR. Recall from Chapter 1, that the HO quantifies the degree to which an ideal observer with full knowledge of the relevant distributions can 
classify an image as belonging to one of two classes (e.g. signal-absent or signal-present), using only linear operations. Given a data vector $\mathbf{g}$, the HO will 
classify an image based on the outcome of a scalar value $t$, computed as 
\begin{equation}
\mathbf{t} = w_g^T \mathbf{g}.
\end{equation}
 If we take $\Delta\bar{g}$ to 
be the mean difference between data belonging to class 1 and data belonging to class 2, then the Hotelling template, $w_g$ is defined as
\begin{equation}
w_g = K^{-1}_g \Delta\bar{g},
\end{equation}
where $K_g$ is the data covariance matrix, whose ($i$,$j$)th component is given by Cov($g_i$,$g_j$). For the inverse DFT, for example, we assume uncorrelated additive 
Gaussian noise in the data domain with variance equal to one, so that $K_g = I_{M\times M}$ and $w_g = \Delta\bar{g}$. 
We then have that the HO SNR in the data domain (SNR$_g$) is given by
\begin{equation}
\textrm{SNR}_g^2 = w_g^T g.
\end{equation}

In order to evaluate the impact of image domain sampling, we consider classification in the reconstructed image domain, rather than 
in the projection data domain. The 
HO SNR in the image domain (SNR$_y$) is then given by 
\begin{equation}
\textrm{SNR}_y^2 = w_y^T \Delta \bar{y},
\end{equation}
where $w_y = K_y^{-1} \Delta \bar{y}$, $K_y = \mathcal{R}K_g\mathcal{R}^\dagger$ and $\dagger$ represents the Hermitian conjugate. 
Finally, we define the HO efficiency as $\varepsilon = \left(\frac{\textrm{SNR}_y}{\textrm{SNR}_g}\right)^2$.
We consider $\varepsilon$ a more useful parameter for algorithm optimization since for many CT applications, one would not operate at the threshold for observer 
performance within which SNR is a meaningful metric. Instead, $\varepsilon$ is a useful metric for algorithm and system evaluation that remains independent 
of the difficulty of a given task, while still retaining the exactness of computation inherent in HO performance calculation through the exact covariance matrix
$K_y$. 

\section{Inverse DFT Example}
\label{sec:dft}

\begin{figure}[h!]
   \begin{center}
   \includegraphics[width=0.47\columnwidth]{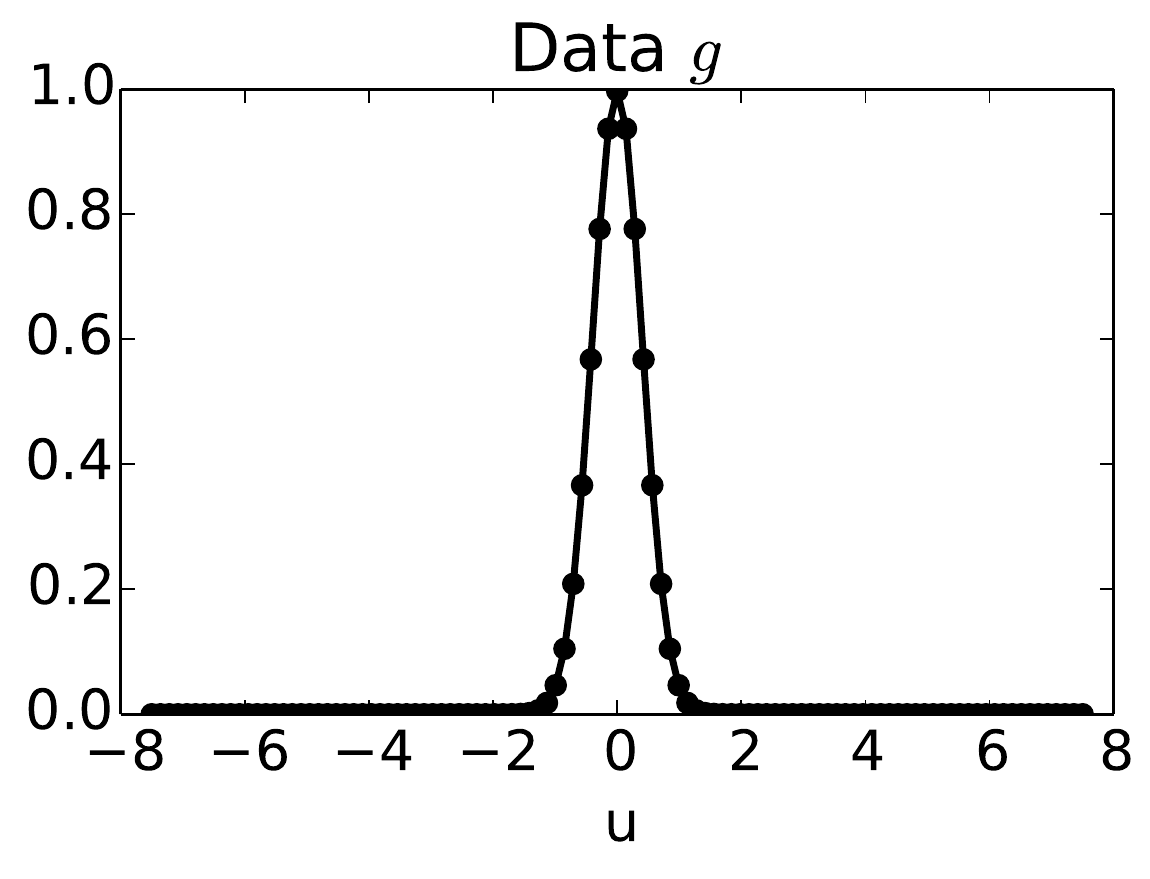} \includegraphics[width=0.51\columnwidth]{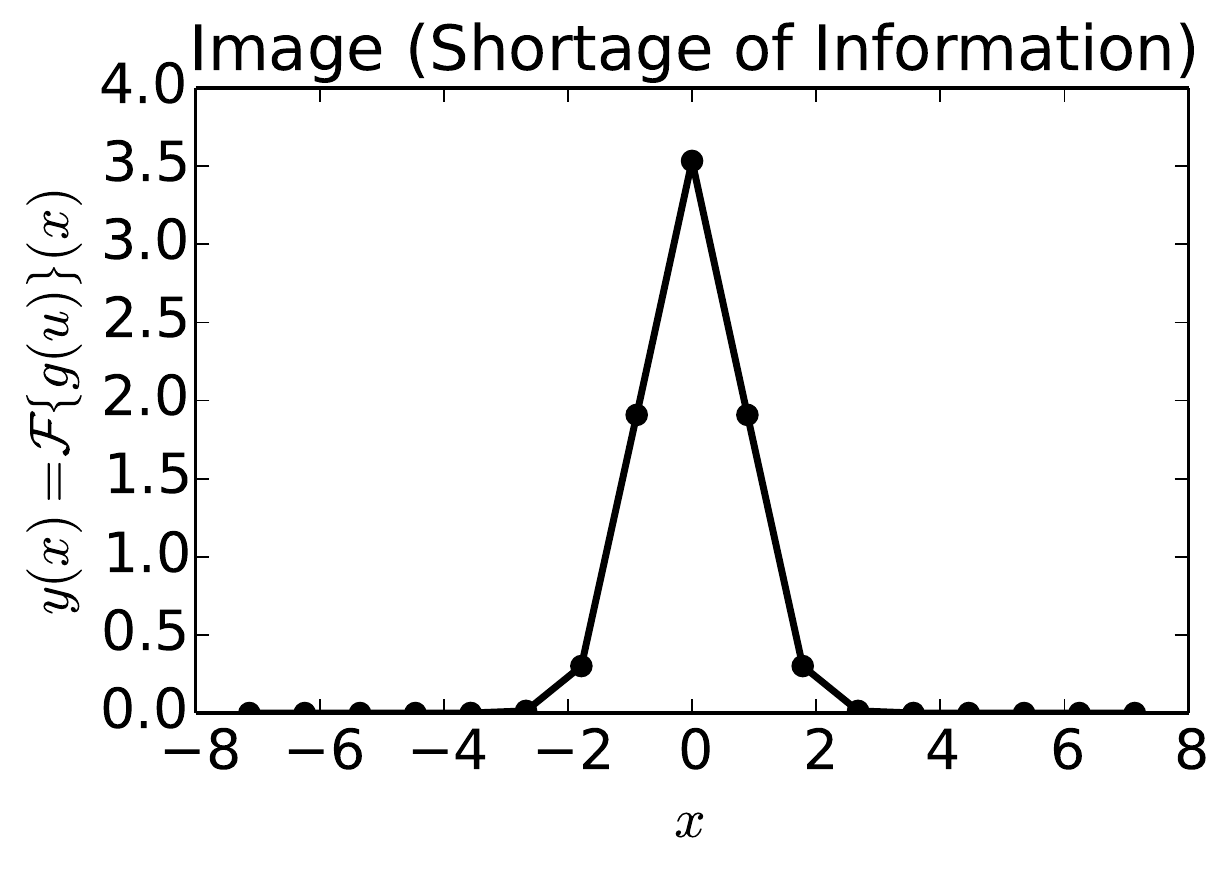} 
   \end{center}
   \caption{\textbf{Left:} The discrete data vector $g$ used to illustrate the effects of image-domain sampling for the DFT example. \textbf{Right:} The ``image'' vector corresponding 
   to the data in the left-hand figure. Note that inadequate sampling in the image vector leads to a loss of information, i.e. an non-invertible transformation has occurred. \label{fig:gaussians}}
   \end{figure}

In order to illustrate the interplay between pixel size and ROI size, consider a 1-dimensional discrete data vector $g \in \mathbb{R}^M$ which is a sampled Gaussian signal, 
as shown on the left of Figure \ref{fig:gaussians}. Next, we replace the FBP reconstruction operator with another linear operator, namely the inverse DFT. 
We will interpret the output of the inverse DFT as the reconstructed image $y = \mathcal{F}g \in \mathbb{R}^N$. Note that although the conventional DFT operator is square, with $M=N$, we 
will consider a more general case where the inverse DFT operator $\mathcal{F}$ is defined through
\begin{equation}
y_k = \sum_{n=-\left(\frac{M-1}{2}\right)}^{n=\frac{M-1}{2}} g_n \mathrm{exp}\left[ 2\pi ia\frac{nk}{M-1}\right], \hspace{0.5cm} k \in \left[-\left(\frac{N-1}{2}\right),\frac{N-1}{2}\right],
\end{equation}
where $M$ and $N$ are odd and $M$ is not necessarily equal to $N$. 
The output of this function for $N<M$ is shown in the right of Figure \ref{fig:gaussians}. 
Setting $N$ effectively controls the ROI size of the image in this case, since this parameter allows us to consider higher frequency 
components of the transform, extending the $x$-axis on the right side of Figure \ref{fig:gaussians}. Meanwhile, the scalar parameter $a$ controls the scale of the $x$-axis, 
which in this example corresponds to controlling the pixel size. 

\begin{figure}[h!]
   \begin{center}
   \includegraphics[width=0.6\columnwidth]{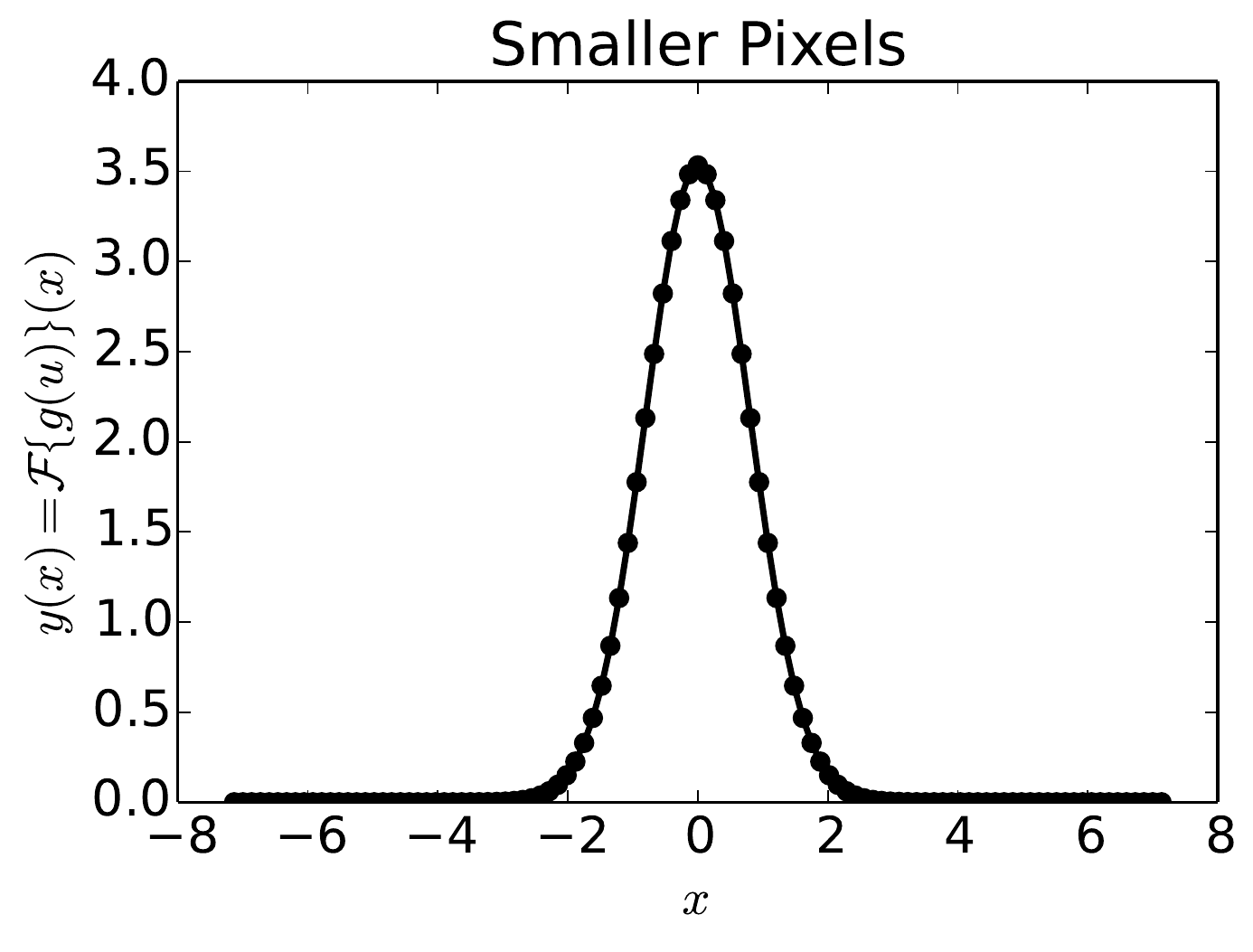}
   \end{center}
   \caption{ Shown here is one possible approach to recovering information which is not present in the right of the preceding figure. Namely, decreased sampling distance (pixel size) 
   has been used to recover the information in the original data. \label{fig:small_pix}}
   \end{figure}

The image-domain signal on the right of Figure \ref{fig:gaussians} corresponds to a non-invertible reconstruction, where information has been lost due to insufficient sampling.
By \emph{information}, we specifically mean that there are components of $g$ which lie in the null-space of the operator $\mathcal{F}$, with the result that HO performance 
is lower in the image domain than in the data domain. 
In general, there are two means of addressing the problem of preserving the information present in the original data $g$. The most intuitive approach is to modify the parameter 
$a$ so that smaller pixelization is used in the reconstruction. In order to preserve the ROI size, $N$ would then be increased appropriately. This strategy is illustrated in Figure 
\ref{fig:small_pix}, and it does indeed lead to a perfect preservation of information through the operator $\mathcal{F}$.  

\begin{figure}[h!]
   \begin{center}
   \includegraphics[width=0.6\columnwidth]{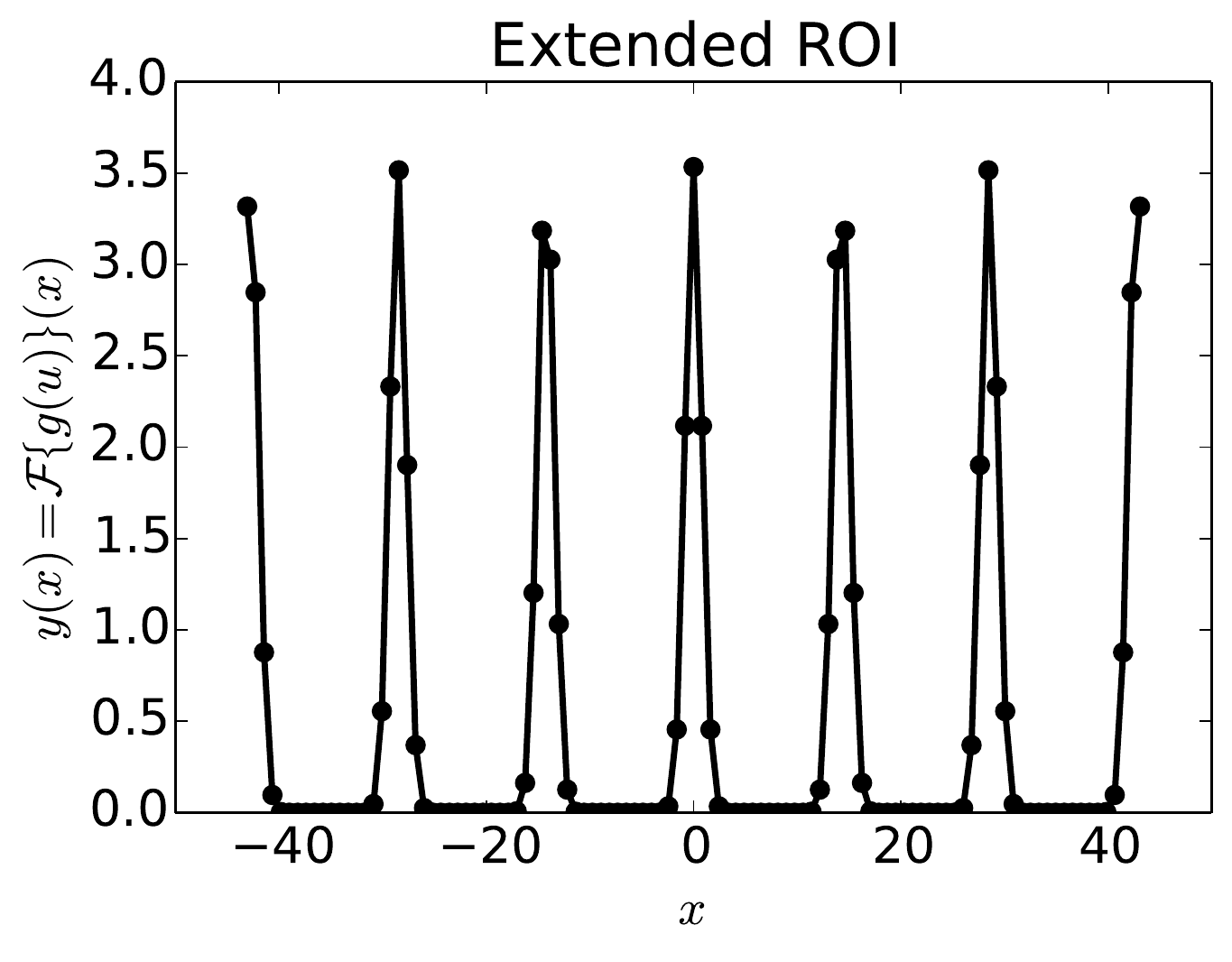}
   \end{center}
   \caption{For the DFT example, extending the ROI of the reconstructed image is exactly equivalent to increasing the sampling within the ROI. This procedure is illustrated here, where the full 
   information of the original data is also preserved perfectly. \label{fig:big_roi}}
   \end{figure}

Alternatively, one could fix the parameter $a$ but increase $N$. This extends the dimension of the ROI and recovers information by sampling higher-order aliases of the object in the 
reconstruction-domain. This approach is shown in Figure \ref{fig:big_roi}. For the case of the operator $\mathcal{F}$, these two approaches are mathematically equivalent. This is 
because image values outside of an ROI are linearly dependent on the values within the ROI. This dependence also exists for pixels in images reconstructed with FBP, since back-projection 
introduces long-range correlations. 
For the present DFT example, the reconstructed object is replicated in the image domain because of the discretization in the data domain, so it makes no difference whatsoever whether we interpret our sampled 
pixels as originating from one alias of the object or another. 

In order to illustrate the equivalence of these two approaches for the HO acting on an image produced by the inverse DFT, we will now directly compute HO performance for a range of values for $a$ and $N$, thereby varying pixel size 
and ROI extent. Later in this chapter, we will perform the equivalent experiment, replacing $\mathcal{F}$ with another discrete-to-discrete linear operator, the parallel-beam FBP algorithm.

\subsection{Decreasing Pixel Size vs. Increasing ROI Size}

   \begin{figure}[h!]
   \begin{center}
   \includegraphics[width=0.6\columnwidth]{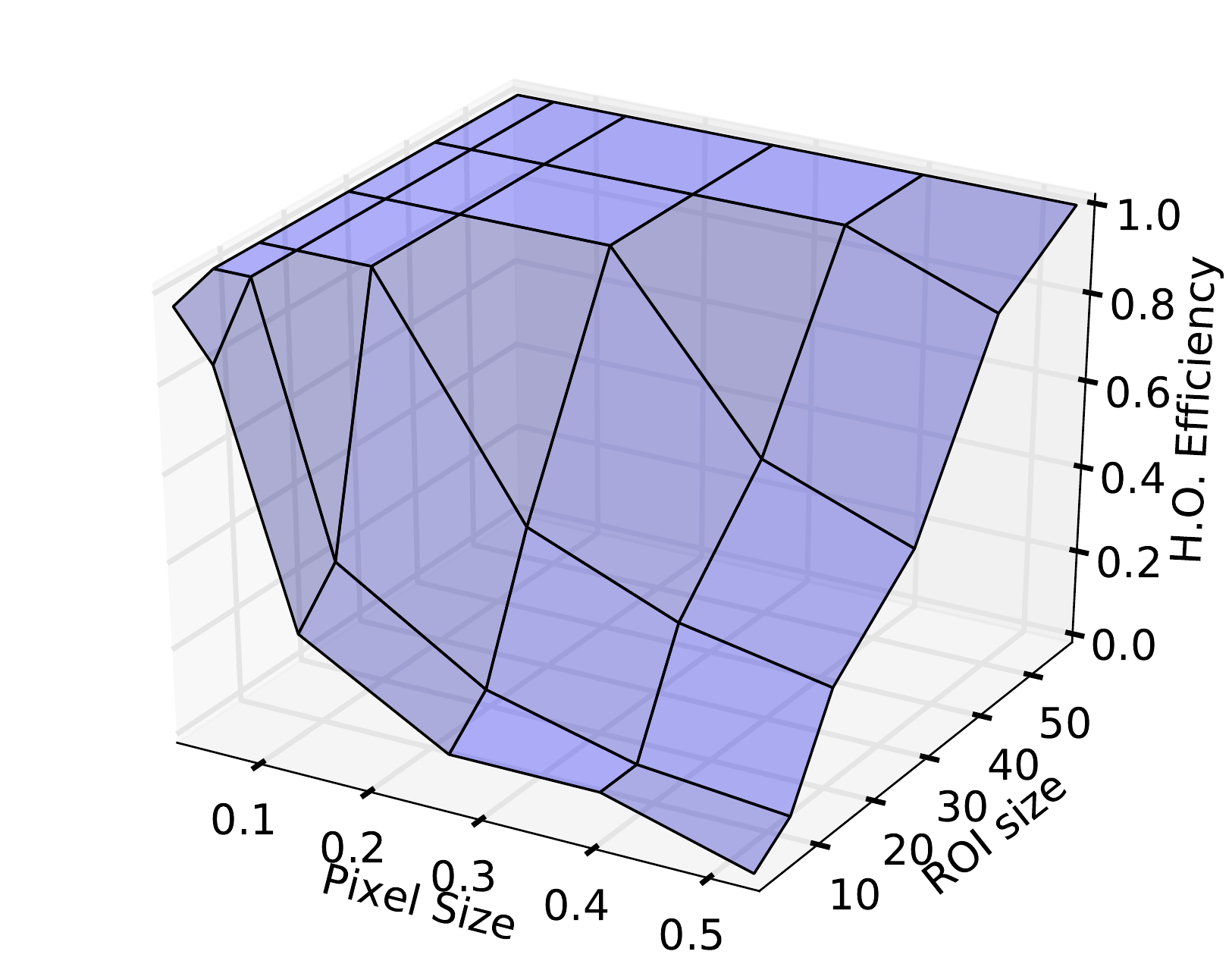} 
   \end{center}
   \caption{ This surface plot shows the HO efficiency for the illustrative inveres DFT sampling example. In this simple example, extending the image field-of-view and decreasing the image sampling distance (analogous to pixel size) 
   are mathematically equivalent means of preserving information from the data domain in the final 1-dimensional ``image.'' \label{fig:dft_surface}}
   \end{figure} 

For each combination of pixel size and ROI size used in this study, $K_y$ could be directly stored and 
inverted in computer memory. 
The resulting HO efficiency values for the inverse DFT experiment are shown in Figure \ref{fig:dft_surface}. 
As can be seen in the figure, equivalent information about the reconstructed object can be obtained either by increasing the image ROI, by decreasing image pixel size, 
or by some combination of the two. What may not be obvious from Figure \ref{fig:dft_surface} is that for ROI sizes less than 10 (arbitrary units) decreasing pixel size will not 
fully recover an efficiency of 1. Rather, a plateau of efficiency is reached as pixel size decreases, and computing smaller pixel values only provides redundant information to the HO. 
Note that this happens for ROI sizes which are substantially larger than the apparent width of the Gaussian signal.
The situation in FBP with small signals is similar, since, like a Gaussian distribution, the back-projection operation spreads small non-zero signal components throughout the image space.
For any finite number of projection views, these back-projections will produce signal components visible as angular under-sampling artifacts (streaks) far from the signal's central location. 
For parallel-beam geometries, increasing the number of projection views does not ever fully remove these slight signal components, but rather extends the radius at which the under-sampling 
artifacts become noticeable. Because of this property of FBP, we expect to see a similar phenomenon of saturating the HO efficiency to a value below 1 when the ROI is excessively restricted.

\subsection{The Hotelling Template for Two Sampling Schemes}

\begin{figure}[h!]
   \begin{center}
   \includegraphics[width=0.49\columnwidth]{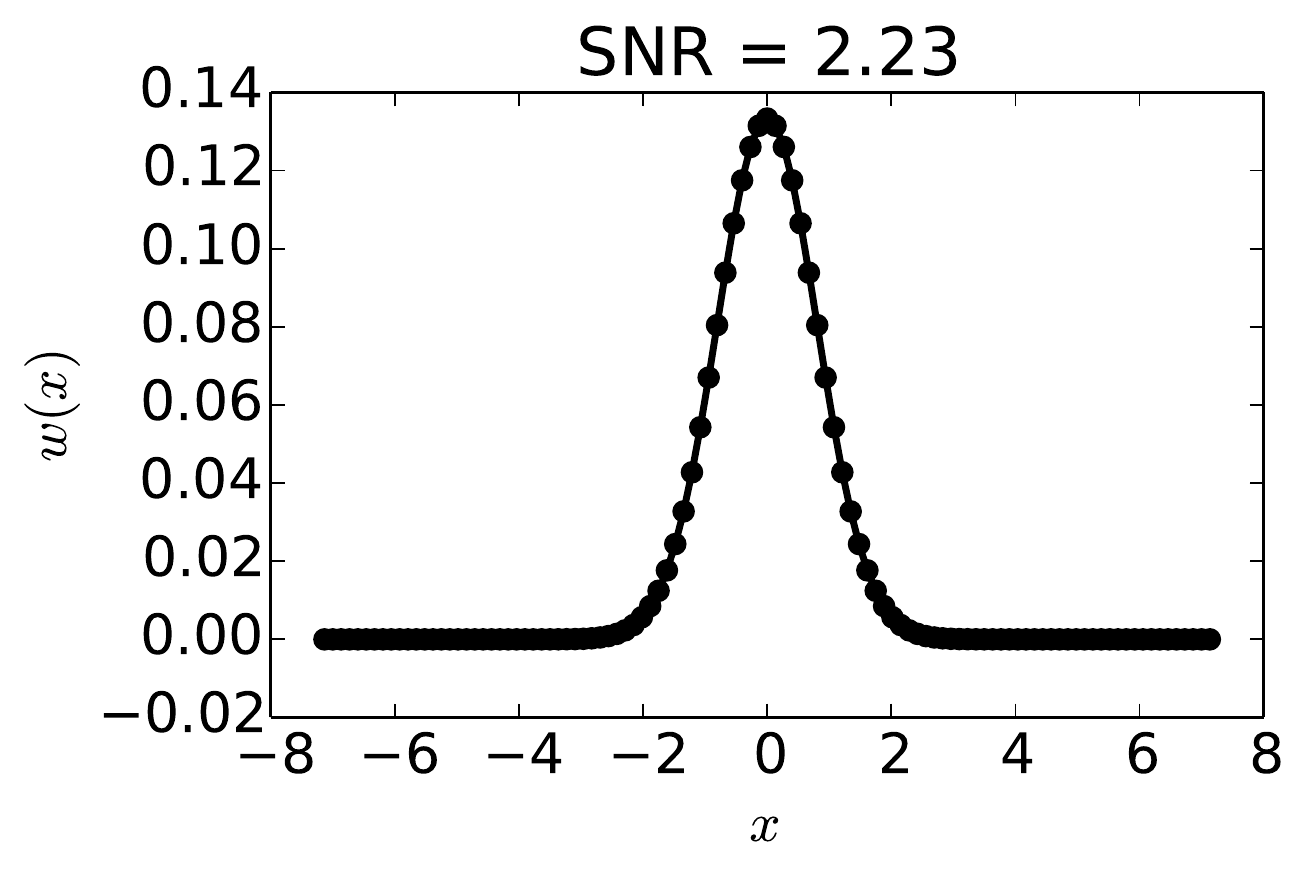} \includegraphics[width=0.49\columnwidth]{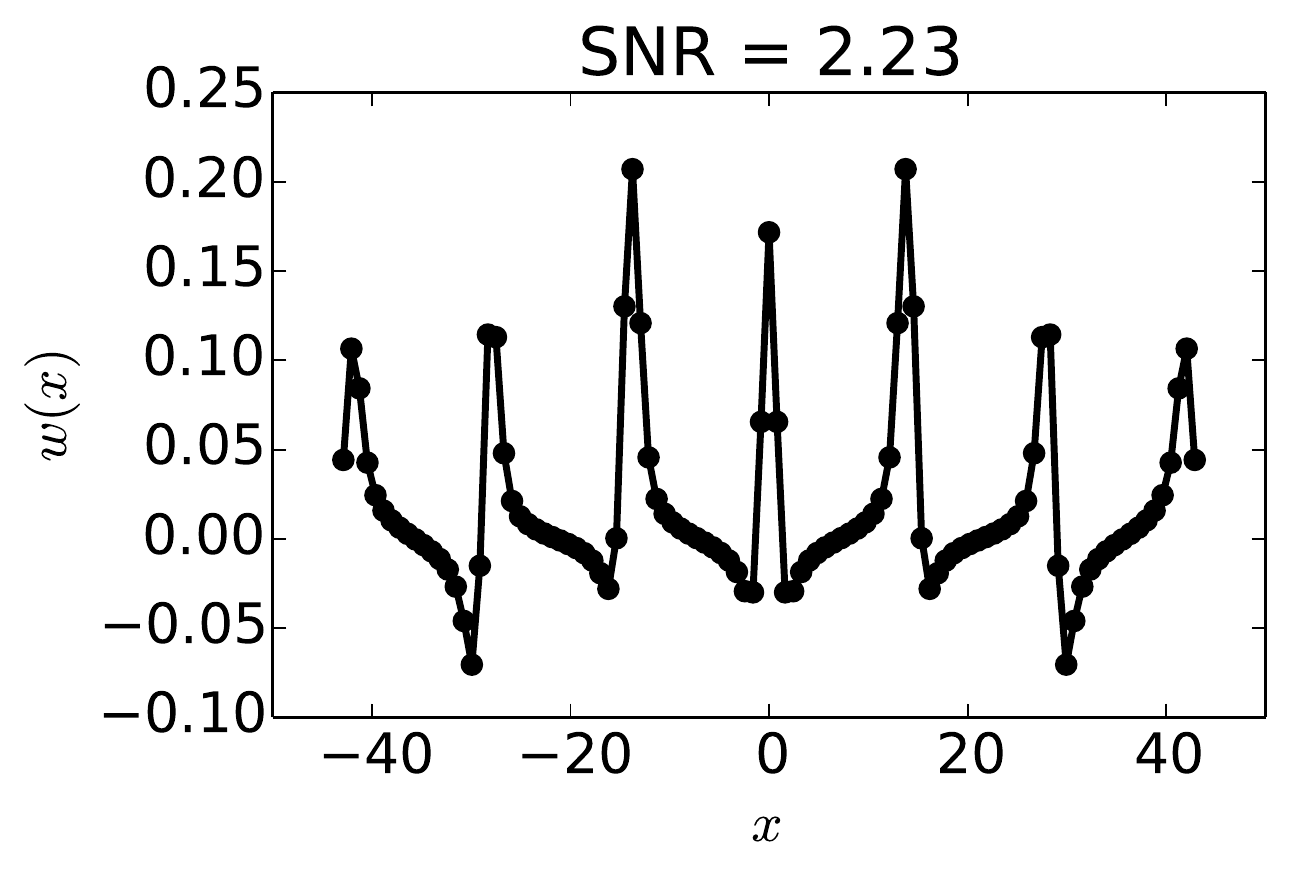} 
   \end{center}
   \caption{ \textbf{Left}: The Hotelling template when smaller pixels are used to recover the full HO SNR inherent in the original data. \textbf{Right}:
   The Hotelling template when the image ROI is expanded in order to sample higher-order aliases (replications) of the signal to recover HO SNR, resulting 
   in a HO efficiency of 1. \label{fig:two_approaches}}
   \end{figure}

Since we have stated that decreasing pixel size and increasing ROI size are two mathematically equivalent approaches to increasing HO efficiency, 
the question naturally arises: When designing an image-domain HO, which sampling approach is preferable? While the two approaches for the inverse DFT 
example are equivalent in terms of HO efficiency, they are not equivalent in terms of the Hotelling template. The HO's template reveals the optimal 
linear decision strategy for performing a task using pixel values. We would ideally like for the Hotelling template to resemble something intuitive, such as 
the signal itself. The Hotelling templates for the sampling approaches from Figures \ref{fig:small_pix} and \ref{fig:big_roi} are shown in Figure \ref{fig:two_approaches}. 
This figure highlights our motivation for restricting the HO to a small ROI in subsequent chapters. While signal components in the higher order signal aliases (replications) can contribute 
to HO efficiency, the optimal use of the corresponding pixels is unintuitive, even in this simplified example using the inverse DFT. Obviously replications constitute an extreme 
example of non-local signal components in the image domain, however even for FBP, these non-local components are present. Further in the case of CT reconstruction, there is even 
less intuitive justification for a model observer using non-local information to perform a task. One certainly would not expect this complex use of non-local information to be performed by humans. 
In terms of the Hotelling template, therefore, we consider truncation of the image to an ROI to be a reasonable approach for implementing the HO. 

\subsection{Matrix Rank Considerations}

\paragraph{Invertability of Reconstruction:}
One common misconception regarding the use of the HO for optimizing linear algorithms is that 
the HO cannot provide useful information since HO performance is invariant under any invertible 
linear operation. Specifically, since the HO is the optimal linear observer, any invertible reconstruction 
algorithm is simply ``undone'' by the HO. The key word in this statement is \emph{invertible}. In general,
reconstruction algorithms are not invertible processes. Particularly, HO efficiency values less than 1 occur 
whenever the linear reconstruction operator is not invertible, and we have demonstrated that 
it is not difficult to encounter sampling conditions where this is the case, even for the FBP algorithm (see 
Figure \ref{fig:fbp_surface}). 

The restriction of the HO to even a large ROI with small pixels can still result in sub-unity 
efficiency values. Further, dimensionality considerations are not enough to guarantee that the reconstruction 
is invertible, since, as discussed previously, one cannot decrease pixel size without limit in hopes of recovering 
HO efficiency lost due to a restricted field-of-view. In other words, even if there are more pixels than 
original data elements, this is not a guarantee that the reconstruction operation is invertible, and therefore 
does not guarantee that the HO efficiency is 1. Since image-domain sampling can introduce a null-space in the 
reconstruction operator (i.e. make it non-invertible), it is unclear how other reconstruction operations such as
regularization will impact HO performance. Subsequent chapters will demonstrate some interesting effects 
which arise from this situation, such as smoothing of the projection data counter-intuitively increasing 
HO performance.

\paragraph{Template Uniqueness and Covariance Rank:}

Based on our observations regarding HO efficiency and invertability of the reconstruction oparator, one 
could assume that the rank of the image covariance matrix might carry information regarding HO performance. 
For a generic, real reconstruction operator $R\in \mathbb{R}^{N\times M}$, invertability corresponds to $R$ having 
rank $M$ with $N \ge M$, i.e. $R$ is full-rank. In general, because it is difficult to know exactly 
which $N$ pixels will produce a full-rank reconstruction operator, 
we have that $N>M$. Next, notice that by the form of 
the covariance matrix $K_y = RK_gR^T$, that the image covariance has dimensions $N \times N$, but cannot 
exceed rank $M$. This implies that, in general, the image covariance matrix can be rank-deficient when the 
HO efficiency is equal to 1. Further, this situation can arise when HO efficiency is less than 1, if some pixels 
in the image are linearly dependent on other image pixels. The ``saturation" of HO efficiency with decreasing 
pixel size described earlier is one such case.

Since the Hotelling template is defined by $K_y w_y = \Delta \bar{y}$, rank-deficiency (i.e. non-invertability) of 
the covariance matrix is of practical concern for computing the Hotelling template. The HO template in this 
case is not unique, but fortunately any template which solves the corresponding linear equation will have the same 
HO SNR. Our main concern is then selecting \emph{which} template we wish to investigate. Ideally, we 
would like to use the Hotelling template to learn something about the optimal strategy for detecting a signal. 
This helps to inform our understanding of whether or not humans might be capable of optimal performance, and 
can also aid in the design of reconstruction algorithms when HO efficiency is not a sensitive metric. In this 
case, algoroithms resulting in intuitive templates, i.e. ones resembling the signal, would be prefered.

   \begin{figure}[h!]
   \begin{center}
   \includegraphics[width=0.49\columnwidth]{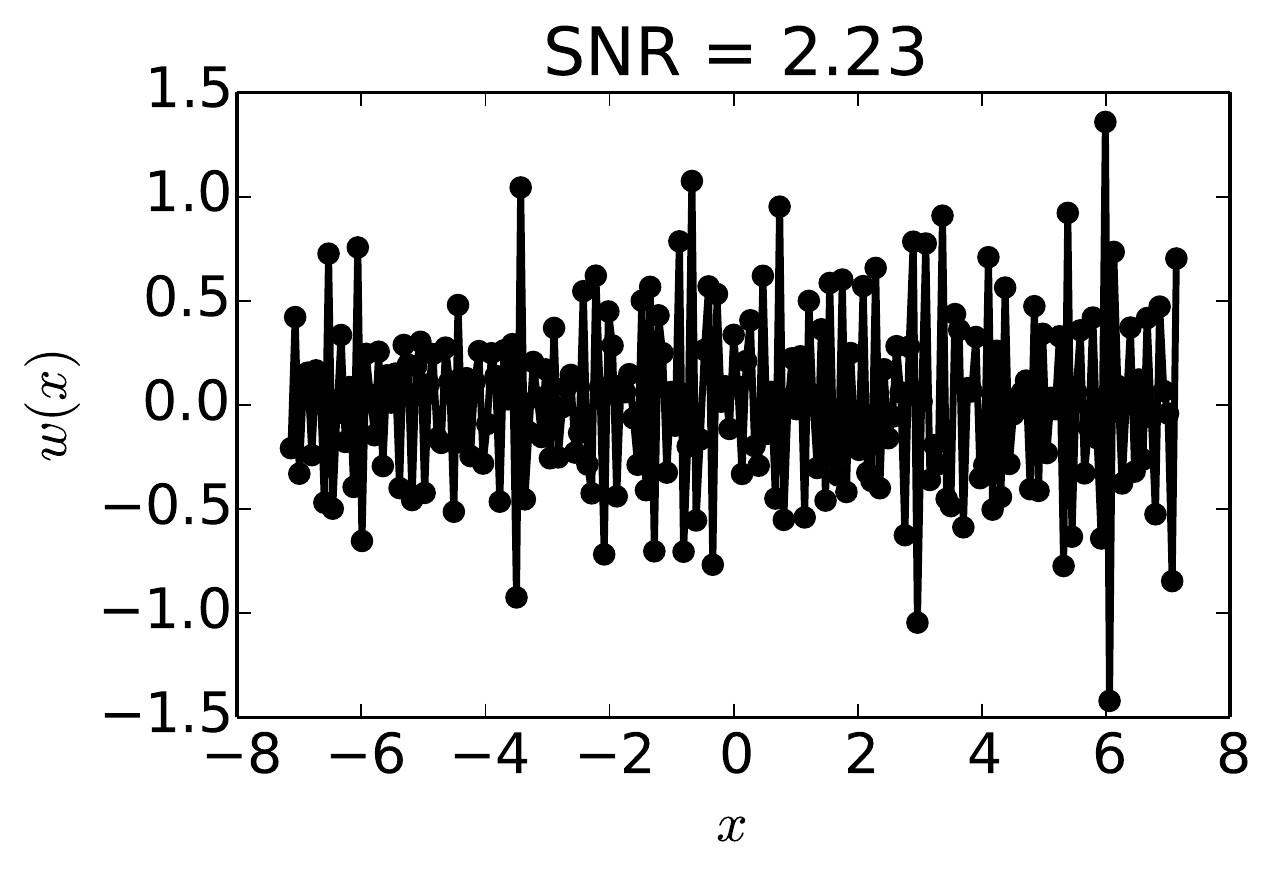} \includegraphics[width=0.49\columnwidth]{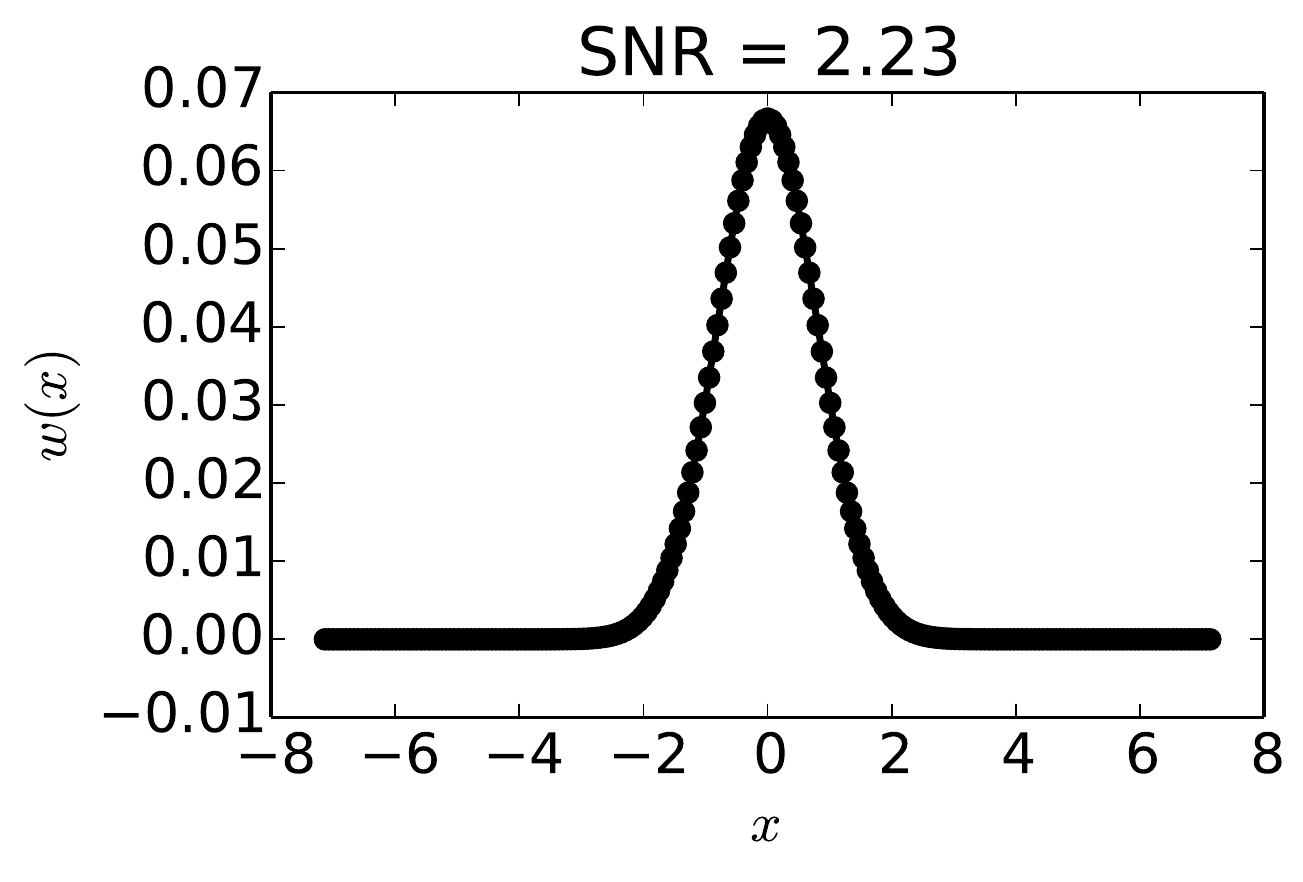} 
   \end{center}
   \caption{ Two Hotelling templates when the number of image pixels is unnecessarily large. \textbf{Left}: 
A Hotelling template with a large component in the null-space of $K_y$. \textbf{Right:} A Hotelling template obtained with slight Tikhonov regularization, so that it has no component in the null-space of $K_y$. This is the minimum-norm Hotelling template. \label{fig:regular_temp}}
   \end{figure} 

An example of this situation is illustrated in Figure \ref{fig:regular_temp}, where an excessive number of 
image pixels have been used in our inverse DFT example and the two HO templates shown 
have equal HO SNR. Clearly the template on the right is the one which conveys relevant information 
about an optimal decision strategy. The obvious approach to this problem is to employ some form of 
regularization in obtaining the Hotelling template. Tikhonov regularization, which penalizes the magnitude 
of the template, can be used in an iterative solution to the Hotelling linear equation, resulting in the 
template shown on the right of Figure \ref{fig:regular_temp}. This method is equivalent to the internal noise 
approach outlined in the previous chapter, but with a very small internal noise magnitude. Essentially, the 
regularization is intended to break the degeneracy of the solution space for $w_y$ while still obtaining a 
template which solves the Hotelling equation. 

\section{FBP Sampling} \label{sec:results}

We now turn our attention to replacing the inverse DFT operator with the FBP operator. 
For the FBP case, we still take $K_g$ to be diagonal, however we now replace the diagonal elements with an object-dependent data variance\cite{barrett_radiological_1996}:
\begin{equation}
\left(K_g\right)_{i,i} = \left(e^{\bar{g}_i}+1\right)/\bar{N}_0,
\end{equation}
where  $\bar{N}_0$ is a constant specifying the mean number of x-rays incident on the detector in a blank scan. Further, we replace $\mathcal{F}$ in 
the preceding equations with the matrix $A$, which represents the action of the FBP algorithm. 
The task investigated for the FBP case is detection of a small (0.15 mm) Gaussian signal in a uniform background. 
The scanning configuration is parallel-beam, and the detector pixel size is 0.8mm. 
A two-dimensional parameter sweep of pixel size and ROI extent was performed, and the resulting HO efficiency 
values computed. Our hypothesis can be described by stating that some information external to an ROI can be
recovered by finer pixelization within the ROI, and conversely, some information lost due to coarse pixelization can
be recovered through expansion of the field of view. 

   \begin{figure}[h!]
   \begin{center}
   \includegraphics[width=0.6\columnwidth]{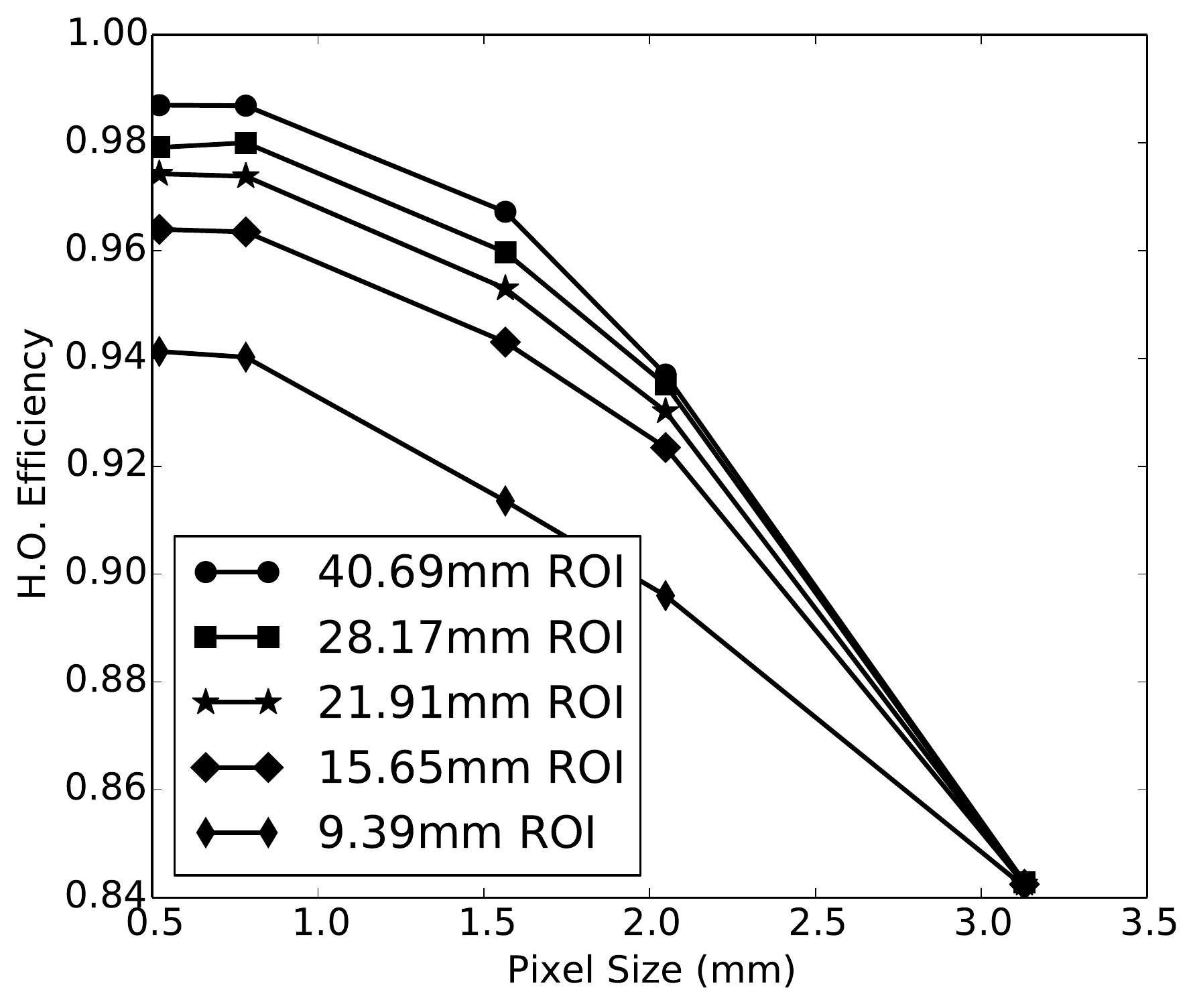} 
   \end{center}
   \caption{ \label{fig1}The HO efficiency $\varepsilon$ as a function of image pixel size for a range of ROI 
diameters}
   \end{figure}
   \begin{figure}[h!]
   \begin{center}
   \includegraphics[width=0.6\columnwidth]{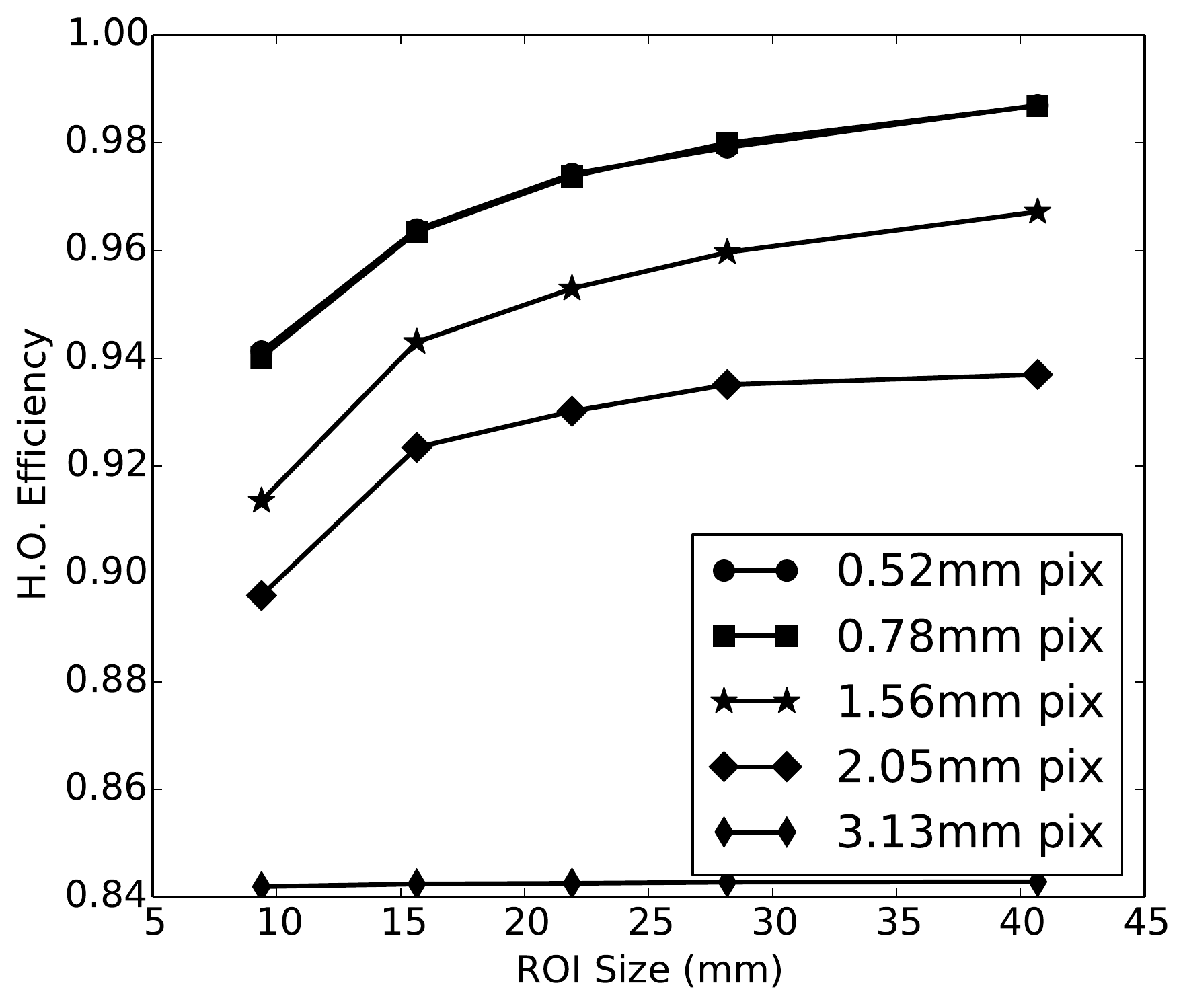} 
   \end{center}
   \caption{ \label{fig2} The same results as in figure \ref{fig1}, but plotted as a function of ROI size.}
   \end{figure} 

\begin{figure}[h!]
   \begin{center}
   \includegraphics[width=0.9\columnwidth]{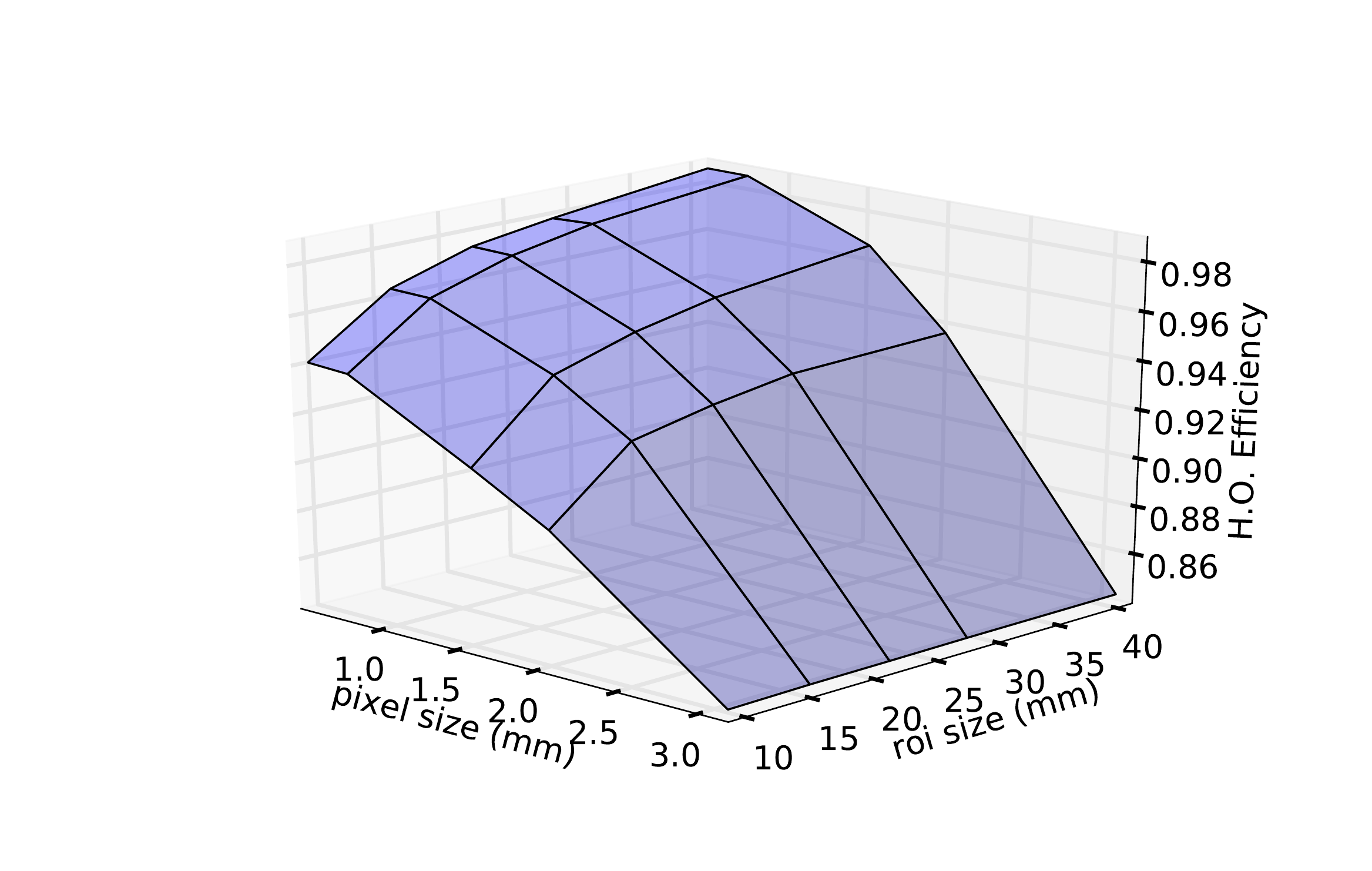} 
   \end{center}
   \caption{ \label{fig:fbp_surface} A surface plot of the HO efficiency $\varepsilon$ as a 2D function of image pixel size 
and ROI diameter. The highest efficiencies shown are roughly 0.99. Modest improvements in efficiency for larger
 ROI sizes may be possible for pixel sizes in the range 0.5-2.0mm, however the system used in this study would not 
accomodate the memory requirements for the covariance matrix $K_y$ in these cases.}
   \end{figure}

The results of the parameter sweeps for FBP are shown as line plots in Figures \ref{fig1} and \ref{fig2} and as a single surface plot 
in Figure \ref{fig:fbp_surface}. These results show that there is a maximum pixel size and minimum ROI size such that 
lost information cannot be recovered by ROI expansion or pixel size reduction, respectively. However, there 
does appear to be some redundancy in information beyond and within the ROI, since efficiencies near 1.0 
can be achieved either by increasing the ROI size or reducing the pixel size. The improved efficiency from 
increasing the ROI size has not fully plateaued in these results, however computational demands would require 
alternative methods such as iterative computation of $w_y$ in order to fully characterize the impact of information 
far from the signal location. It is encouraging, however, that the maximum ROI size necessary to fully capture the HO performance seems relatively insensitive to 
the image pixel size. This suggests a stronger dependence on the scanning configuration than on pixel size.

\begin{figure}[h!]
   \begin{center}
   \includegraphics[width=0.9\columnwidth]{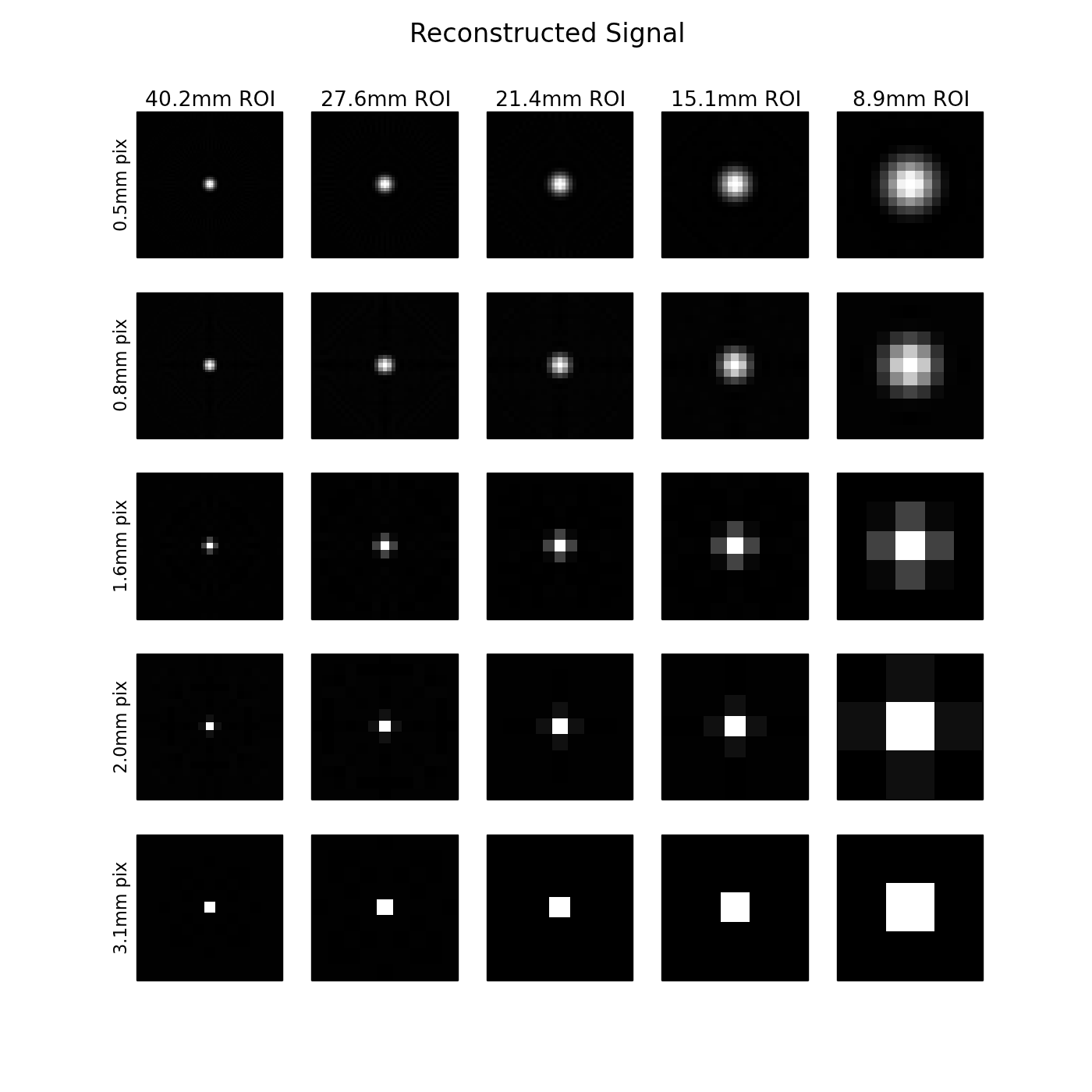} 
   \end{center}
   \caption{ The reconstructed mean difference between signals for the FBP case $\Delta \bar{y}$ is shown here for the full range of pixel and ROI sizes used. Note that for every ROI size, the full 
   physical extent of the signal is contained in the ROI. \label{fig:sig}}
   \end{figure}

\begin{figure}[h!]
   \begin{center}
   \includegraphics[width=0.9\columnwidth]{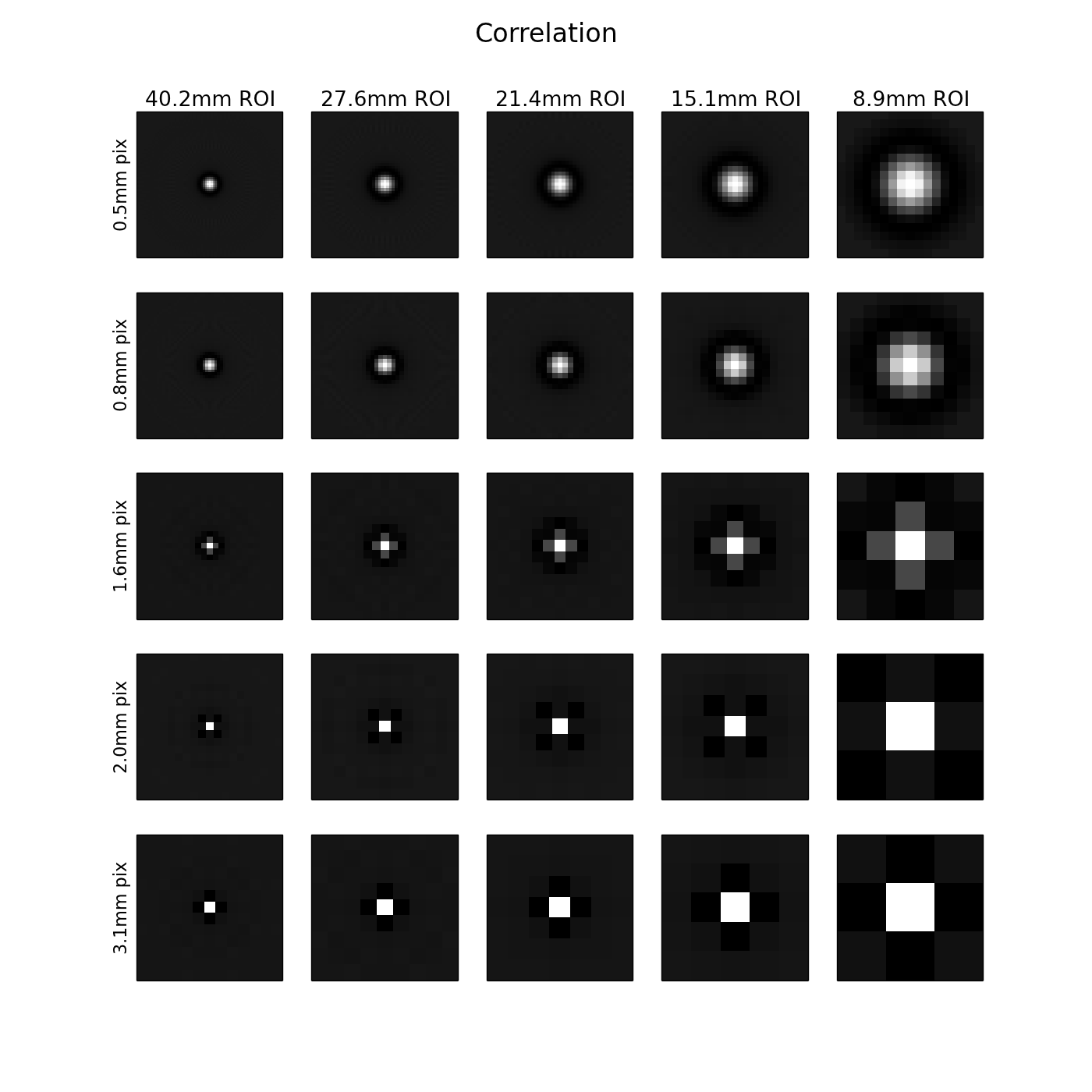} 
   \end{center}
   \caption{ Similar to the previous figure, this figure shows the correlation structure at the center of the ROI for each pixel size and ROI extent investigated. The correlation 
   structure is computed by extracting a single row of the image covariance matrix corresponding to the central pixel and reshaping the 1-dimensional matrix row as an image. Like with 
   $\Delta \bar{y}$ each ROI size adequately captures the predominant correlation extent. \label{fig:correlation}}
   \end{figure}

Figure \ref{fig:sig} illustrates the mean difference between reconstructed images $\Delta \bar{y}$ for the full range of ROI sizes and pixel sizes used to 
construct Figure \ref{fig:fbp_surface}. Inspection of the corresponding HO efficiency 
values in Figure \ref{fig:fbp_surface} demonstrates that the ROI extent necessary to preserve HO detectability cannot be immediately inferred from the 
physical extent of the signal. Clearly, the signal is fully encompassed by the ROI in each case, yet even the increase from 30mm to 40mm improves HO performance for the small 0.15mm signal. 
Note also that the use of an unapodized ramp filter leads to noise correlations which are also highly localized (Figure \ref{fig:correlation}). 
Therefore the extent of noise correlations is also inadequate to guide the choice of ROI size for dimensionality reduction 
of the HO linear system of equations. 

\begin{figure}[h!]
   \begin{center}
   \includegraphics[width=0.9\columnwidth]{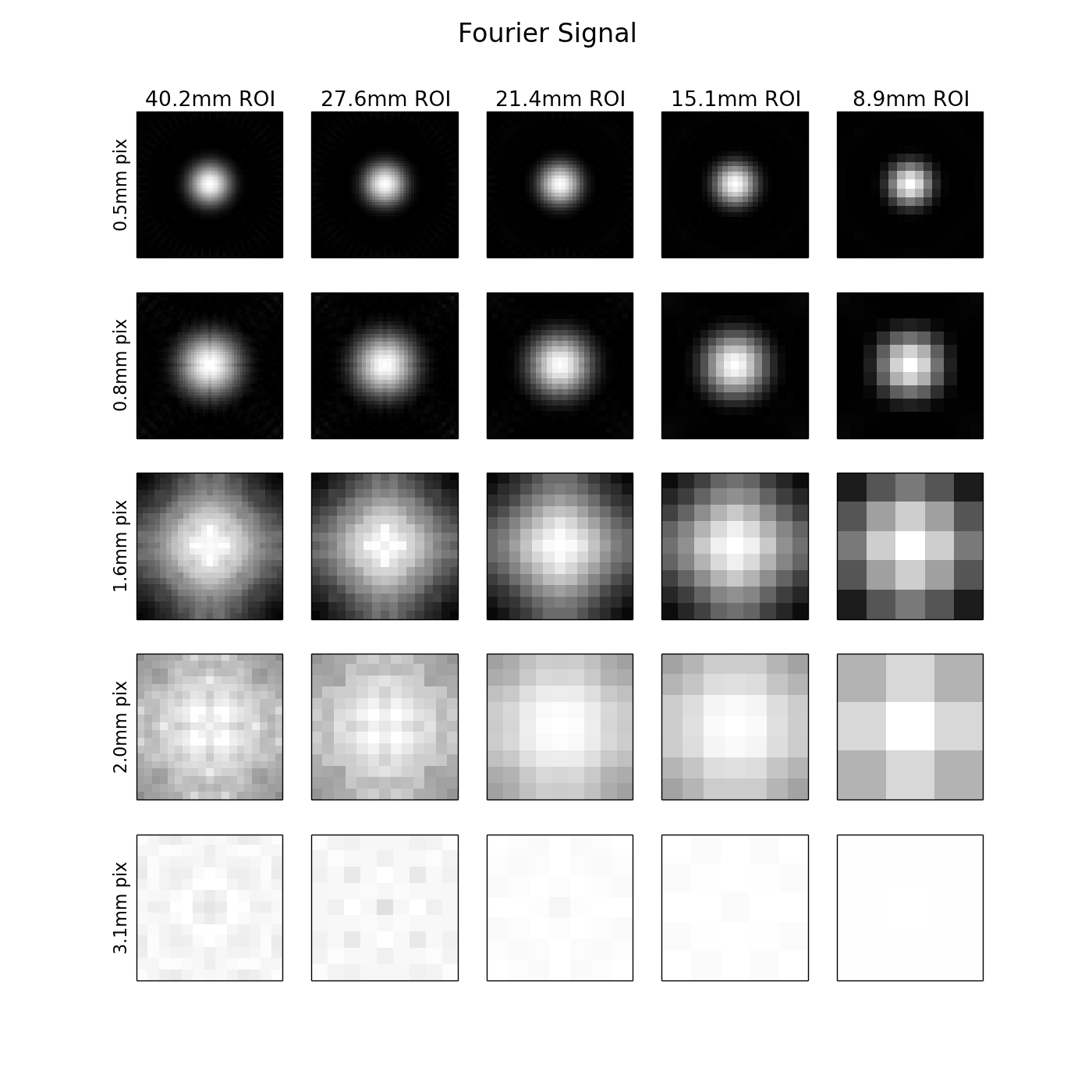} 
   \end{center}
   \caption{ Shown are the corresponding 2D discrete Fourier transforms of the images in Figure \ref{fig:sig}. It is possible that structure and extent in the Fourier domain is a 
   clearer indication of adequate image sampling than the spatial domain signal. \label{fig:fsig}}
   \end{figure}

\begin{figure}[h!]
   \begin{center}
   \includegraphics[width=0.9\columnwidth]{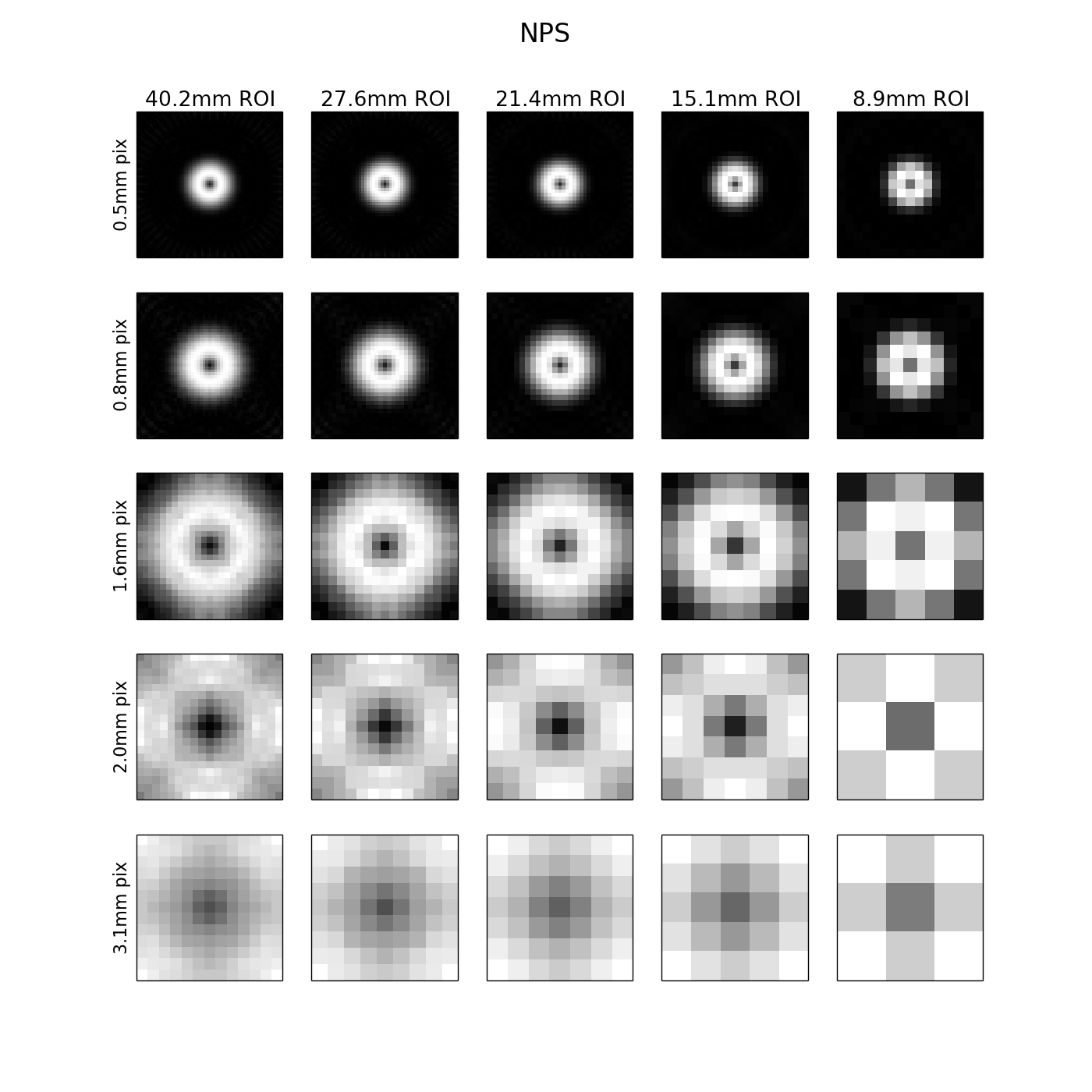} 
   \end{center}
   \caption{ Same as Figure \ref{fig:fsig}, but for the correlation structure. Under the approximation of local stationarity, the DFT of the autocorrelation is the same as the 
   noise power spectrum (NPS). \label{fig:NPS}}
   \end{figure}

The DFT-domain versions of the same reconstructed signals are shown in Figure \ref{fig:fsig}. Pixel size seems to 
have a straightforward effect on HO efficiency, with aliasing in the Fourier domain signal occurring roughly at the pixel size where HO efficiency begins to 
substantially decrease. Structure of the signal in the Fourier domain may occasionally provide insight into determination of an adequate ROI size. For example, beginning in the 
second row of Figure \ref{fig:fsig}, some structure is visible for large ROI sizes which becomes lost for ROI sizes less than 20mm. This approximately corresponds to the ROI size 
below which appreciable degradation of HO efficiency occurs. Therefore investigation of the structure and extent  of the reconstructed Fourier transform could be a more meaningful 
indication of necessary sampling parameters than inspection of the signal itself. A similar effect can also be seen in the local noise power spectra, shown in Figure \ref{fig:NPS} and computed 
as the 2D DFT of the image in Figure \ref{fig:correlation}. 

\begin{figure}[h!]
   \begin{center}
   \includegraphics[width=0.9\columnwidth]{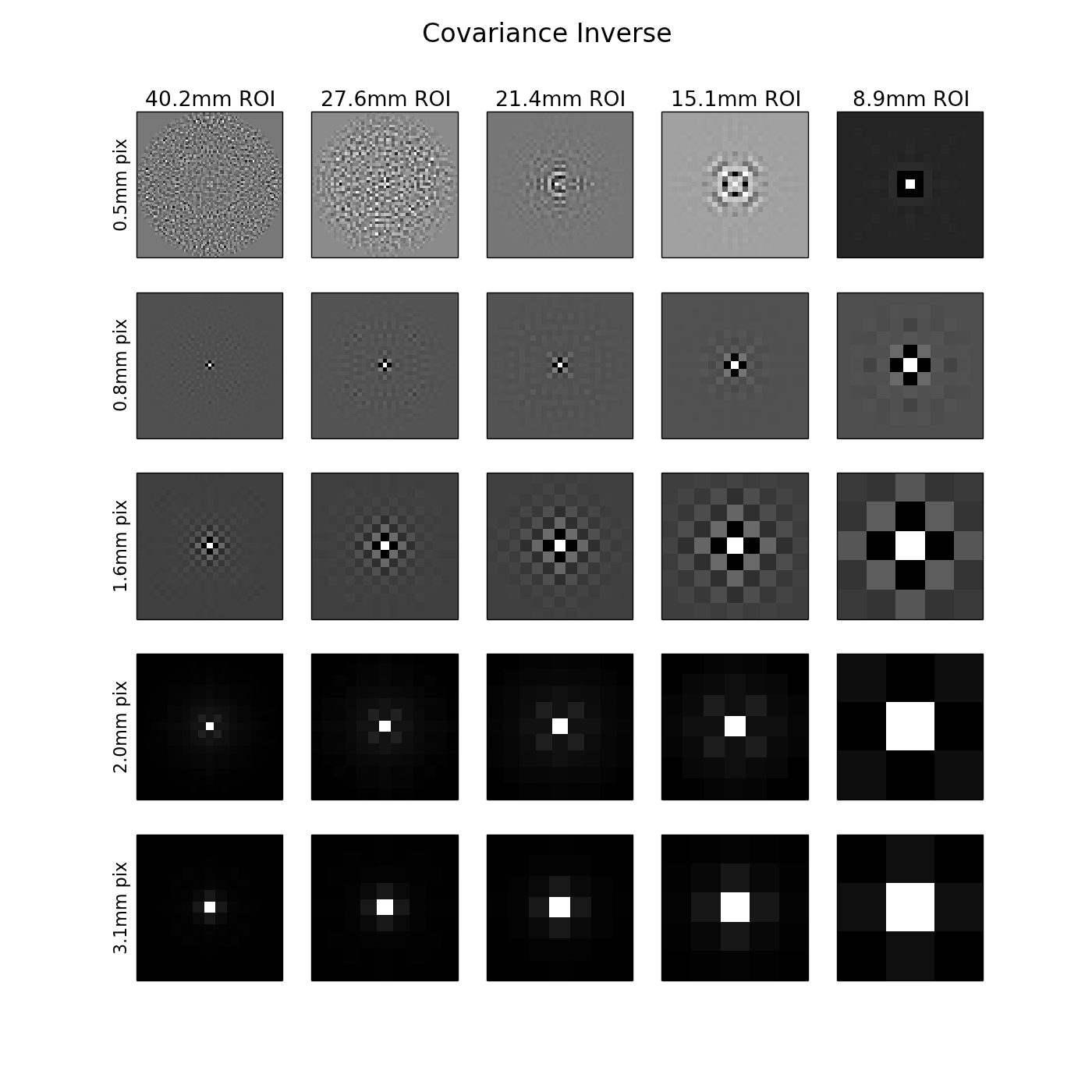} 
   \end{center}
   \caption{ Same as Figure \ref{fig:correlation}, but the images correspond to rows of $K_y^{-1}$ rather than $K_y$. Note that poor conditioning of $K_y$ leads to a lack of coherent structure in the 
   inverse covariance. \label{fig:cov_inv}}
   \end{figure}

Finally, since we have alluded to the fact that HO efficiency of 1 indicates that an invertible operation has been performed, this would seem to suggest a relationship 
between preservation of HO SNR and conditioning of the image covariance matrix $K_y$. While extracted rows of $K_y$ were shown in Figure \ref{fig:correlation}, 
we also present the corresponding rows of the inverse covariance $K_y^{-1}$ in Figure \ref{fig:cov_inv}. For singular $K_y$, the Moore-Penrose pseudo-inverse is computed. 
Careful inspection of the equation $K_y = AK_gA^T$ reveals that whenever linearly-dependent pixels are produced by $A \in \mathbb{R}^{N\times M}$, $K_y$ becomes singular, meaning 
no unique inverse exists. This occurs whenever extraneous pixels are evaluated (since these pixels are linearly related to other pixels), for example by reducing image pixel size excessively. 
In this case, the inverse of $K_y$ is not unique, and even obtaining a pseudo-inverse can be an ill-conditioned problem. This poor conditioning results in the unstructured $K_y^{-1}$ estimates 
shown in the top-left of Figure \ref{fig:cov_inv}.

\begin{figure}[h!]
   \begin{center}
   \includegraphics[width=0.9\columnwidth]{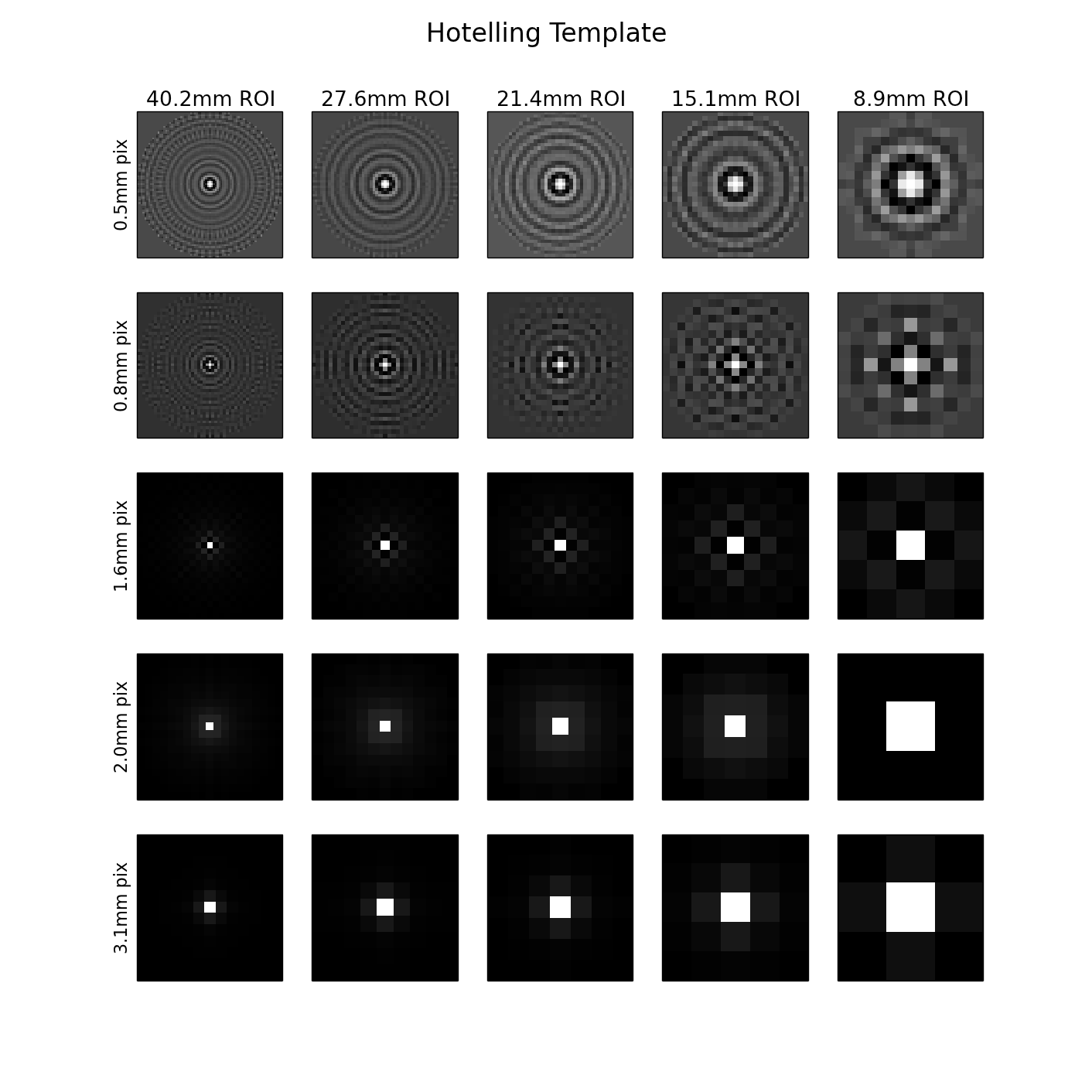} 
   \end{center}
   \caption{ Shown are the Hotelling templates corresponding to each sampling condition. The lack of structure in the inverse covariance (Figure \ref{fig:cov_inv}) is propagated into the Hotelling template. 
   The templates on the top left correspond to near perfect HO efficiency (see discussion in text). \label{fig:templates}}
   \end{figure}

   The structure of the inverse covariance is important because it impacts the structure of the Hotelling template, which indicates the optimal linear decision strategy for a given task. 
   The Hotelling templates corresponding to each sampling condition are shown in Figure \ref{fig:templates}. \new{Based on the inverse covariance matrix (Figure \ref{fig:cov_inv}), i}t is apparent that ill-conditioned or singular covariances \new{have the potential to} lead to unstructured 
   templates, which in turn tend to indicate high HO efficiency. This is because most or all relevant information for task performance is contained in the image pixels, and increasing image sampling only replicates information which is already present. 
This is the same effect that was observed in Figure \ref{fig:regular_temp}, which corresponded to a 
HO efficiency of 1. 
   Since $K_y$ has no unique inverse in these cases, the Hotelling template is also not unique (although HO SNR is). For this reason, various methods have been proposed which extract \emph{a} 
   Hotelling template which has reasonable structure. These include methods of internal noise \cite{beutel_handbook_2000, abbey_human-_2001} as well as regularization strategies \cite{sanchez_investigation_2013}.
   While this would seem to suggest inspection of the condition number of $K_y$ as a guide to selecting an appropriate ROI size, in practice we have not observed a stable trend between the 
   condition number of $K_y$ and HO efficiency. 
   \new{As in Figure \ref{fig:regular_temp}, the templates in Figure \ref{fig:templates} are obtained with a slight Tikhonov regularization which results in structured templates without substantially lowering the HO SNR.}
   
\section{Conclusions}

\new{
Through an illustrative example using the inverse DFT and through corresponding experiments with parallel-beam FBP, we have demonstrated 
the basic image-domain sampling considerations inherent in using the HO for CT system optimization. In order to efficiently compute HO metrics 
and ensure intuitive task performance (templates with intuitive structure), a restricted ROI should be used for HO computations. In the DFT example, 
the extent of the best ROI is defined by the first signal alias. Extending this result by analogy with FBP, we propose that the ROI should be restricted based on the 
radius at which signal components first produce angular under-sampling artifacts. In the next chapter, we will describe an efficient means of determining this radius 
based on a single reconstruction of the quantity $\Delta \bar{y}$. The resulting ROI size is then considered a fixed part of the HO model where we enforce a preference for 
structured Hotelling templates, which imply intuitive task-performance strategies. 
}

To summarize, direct and efficient computation of HO performance for small signals may be feasible in CT since the full set of image pixels 
appears to contain some redundant information. 
Future work in developing a precise mathematical model for the 
impact of image domain sampling will be necessary in order to perform a rigorous characterization of
the empirical sampling effects demonstrated in this work. For instance, the Fourier-based approach of Ref. \cite{kijewski_noise_1987} could be 
adapted to construct a rigorous framework for selecting necessary image ROI extent. Similarly, alternatives to Fourier descriptors could be developed, 
for example based on the continuous-domain SVD of the Radon transform. In subsequent chapters, however, we will adopt a different approach and 
consider the ROI extent to be an attribute of the model observer. While this modified HO will no longer strictly be equivalent to the ideal linear observer acting on the 
full reconstructed image, it 
will still be ideal under the restriction that only a subset of image pixels are used. This is a reasonable restriction since, in practice, human observers 
do not incorporate every image pixel when attempting to detect a small signal, but rather restrict their attention to a local neighborhood of pixels. 

%% file: Chapter4/Chapter4.tex
\chapter{Comparison of the ROI Hotelling Approach with Alternative Methods}
\label{ch:roi}
\section{Introduction} \label{purpose}

While objective assessment of image quality through task-specific metrics has a long history in medical imaging and is regarded by many
as ultimately being the most meaningful approach to medical image evaluation\cite{wagner_unified_1985, burgess_efficiency_1981, barrett_foundations_2004, metz_roc_1986},
the application of task-based assessment to x-ray CT is recent relative to its application to planar imaging modalities and nuclear medicine. One reason for this delay 
is that metrics based on the HO, such as those considered in this work, involve the image covariance matrix, and in CT this matrix is often
extremely large (well over $10^9$ elements), poorly conditioned, and possesses few if any simplifying structural properties. 
In order to address the challenge of large dimensionality, efficient channels have been proposed\cite{gallas_validating_2003, park_singular_2009, witten_partial_2010}, 
which essentially constitute a transformation 
of the image into a new basis, where the number of basis functions is substantially less than the number of image pixels. Another common means 
of circumventing the dimensionality problem is to assume noise stationarity, so that HO metrics can be obtained 
with relative computational efficiency through discrete Fourier transform (DFT) operations (see Section \ref{sec:methods}). Meanwhile, 
in order to address the somewhat unpredictable structure of the image covariance, various estimation strategies have been proposed which 
rely on samples of noisy images in order to construct an estimate for the image covariance when an analytic formulation of image covariance is 
impossible or infeasible\cite{chan_classifier_1999, wunderlich_new_2013}. 

In this chapter, we propose a formalism which, in certain cases, may be more appropriate than the use of efficient channels or estimation techniques,
while still addressing the issues of dimensionality and structural complexity of the image covariance matrix. 
Our approach differs from alternative methods in that the resulting metrics are non-stochastic, and we impose no restrictions 
on the structure of the signal or on the nature of the noise correlations. Instead, we only assume that the relevant noise correlations and 
signal extent can be restricted to a region-of-interest (ROI) in the image, as discussed in more detail in the previous chapter.
In order to investigate the impact and validity of 
various assumptions which simplify the computation of HO metrics, 
we herein compare the results of parameter optimization using our proposed method to two alternatives: a set of efficient (Laguerre-Gauss) channels, 
and a discrete-Fourier-transform (DFT) domain approach. Each of these approaches corresponds to a different assumption regarding the signal, the image noise, or both.
We then also investigate two statistical estimation strategies for HO metrics and compare these results with our proposed method.

The specific context in which our formalism is intended to be applied is detection or classification 
tasks where the object of interest is small (on the order of several pixels), and the reconstruction algorithm is a direct, linear algorithm such as filtered back-projection 
(FBP). In this case, we hypothesize that 
most of the relevant information for performing the detection or 
classification task is likely to be contained in pixels within 
a small region of interest (ROI) surrounding the signal. For direct linear algorithms, 
an analytic form of the image covariance matrix for a small ROI 
can be constructed and its elements can be 
stored directly in computer memory, so long as the ROI contains at most 
several thousand pixels. While we consider only 2D reconstruction in this work, the extension to 3D reconstruction is straightforward so long as the corresponding 3D ROI contains 
only several thousand voxels. 

Since the metrics evaluated in this work are task-specific, we restrict ourselves to two relevant tasks in a specific CT application, namely dedicated breast CT. The 
tasks considered are microcalcification detection and the Rayleigh discrimination task, which measures the ability of an observer to distinguish between a 
single object and two smaller objects. 
The specific metric we investigate is the HO efficiency, which is a summary scalar metric describing the preservation of relevant information from 
the projection data measurements to the reconstructed image. The mathematical construction of this metric is described in Chapter 1, while Section \ref{sec:methods} provides
a description of the two imaging tasks considered. Section \ref{sec:results} demonstrates the use of the proposed ROI HO method for the optimization of 
reconstruction filter width in the FBP algorithm. Section \ref{sec:results} also contains results for the application of a channelized HO (CHO), a DFT-domain HO, 
and two estimation approaches to the same parameter optimization. Finally, a discussion and brief conclusion is included in Section \ref{sec:conclusions4}.

\section{Methods}  \label{sec:methods}

\subsection{Task modeling}
\begin{figure}[h!]
   \begin{center}
   \begin{tabular}{cc}
   \includegraphics[width=0.6\columnwidth]{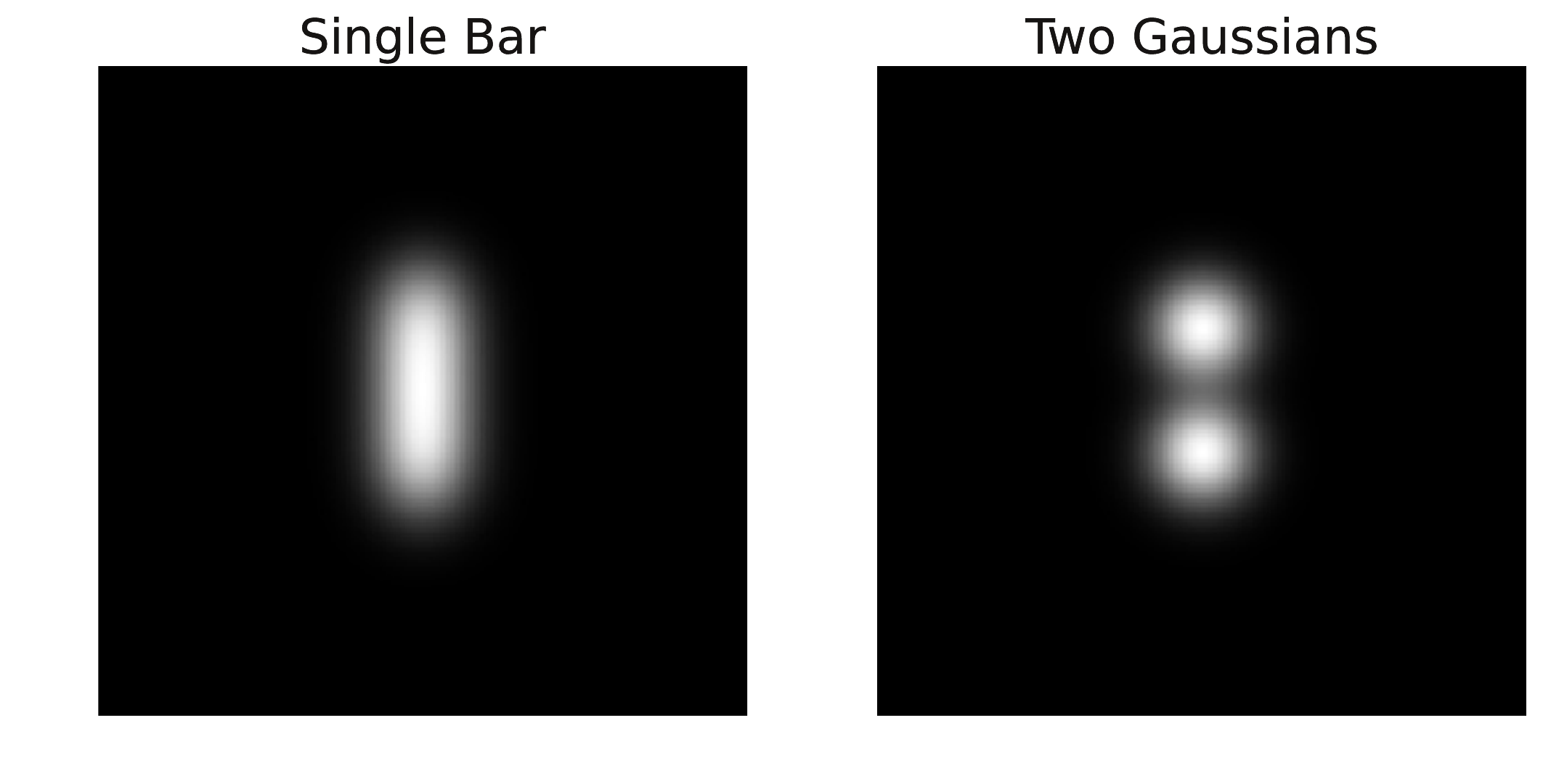}
   \end{tabular}
   \end{center}
   \caption{The two signals simulated for the Rayleigh discrimination task.
\label{fig:rayleigh_task_objects}}
   \end{figure}

Two tasks are considered in this work. The first is a Rayleigh resolution task, wherein an observer must classify an image as either an image of a single 
1.4mm bar convolved with a Gaussian with a full width at half maximum (FWHM) of 0.53mm or two distinct Gaussians of FWHM = 0.53mm, separated by a 0.8mm trough. 
These two signals were modeled on a discrete grid with pixels an order of magnitude smaller than the detector pixels back-projected to the center of the field-of-view. The two signals 
are shown in Fig. \ref{fig:rayleigh_task_objects}.

The second task is a signal detection task, where we model a microcalcification as a Gaussian with FWHM of 80$\mu$m. Here the two classes into which 
an observer classifies the image are a signal present class and a signal absent class. For both tasks, we operate under the signal-known-exactly-but-variable 
(SKEV) paradigm\cite{barrett_foundations_2004}, where three signal locations are considered: the center of the FOV, the position (2cm,2cm), and the position (4cm,4cm), however
results among the three signal locations are not substantially different. 

\subsection{Breast CT simulation}
Since the tasks considered in this work involve small signals, all of the relevant geometric and 
scanning parameters used in the breast CT simulations are based on 
Ref. \cite{kwan_evaluation_2006}, the purpose of which is to characterize spatial resolution properties of a dedicated breast CT scanner. In the 
following chapter, we fully explore parameter ranges for this system and compare with other work in optimization of breast CT systems. However,
we here investigate only a few parameter settings and instead focus on HO performance estimation methods. 
To summarize, we 
simulate a 40cm flat panel detector, rebinned to a sampling of 1024 detector bins. The source-to-isocenter distance is 45.8cm, and 
the source-to-detector distance is 87.8cm. The field-of-view (FOV) is restricted to 16.59cm, and this is sampled on a 512 $\times$ 512 image pixel grid, 
for an image pixel size of 0.324mm. Meanwhile, the detector element size, back-projected into the center of the FOV is approximately 0.2mm. 

The simulated x-ray spectrum corresponds to an 80kVp setting with added Be and Al filtration of 0.8mm and 2.5mm, respectively. Methods from Refs \cite{boone_accurate_1997, boone_molybdenum_1997, boone_spectral_1998}
 were used in simulation of the x-ray spectrum.
Finite detector bin size is modeled by sub-sampling each detector 
bin at 16 evenly-spaced locations. The peak attenuation value for both types of signal considered is 
that of calcium. The attenuation of the background medium is the mean attenuation of a breast composed of 50\% adipose tissue 
and 50\% glandular tissue, as determined in Ref \cite{johns_x-ray_1987}. The numerical phantom diameter used is 14cm, and the total 
photon fluence at isocenter necessary to achieve the same dose as two-view mammography for this diameter ($\approx 2\times 10^8$mm$^{-2}$) 
is obtained from Ref \cite{boone_technique_2005}. 
We restrict the dose in our simulation to the corresponding dose from mammography since breast CT is being investigated as an alternative screening modality.
In order to determine the 2-dimensional fluence necessary for our fan-beam simulations, we assume 
a slice thickness of 1mm. The total number of incident photons in our fan-beam simulations is then $\bar{N}_0 = 4.17\times 10^{10}$.
For the range of parameters investigated here, 50 projection views were sufficient to ensure that angular under-sampling artifacts remained farther from the signal of
interest than the typical correlation length of the image noise. For this reason, the HO optimization was insensitive to these artifacts. We therefore used only 50 projection 
views in order to further minimize the computation burden of the optimization.
The total fluence is divided equally among these 50 projection views. The reconstruction algorithm used is the FBP algorithm with a Hanning filter. The
optimization of the Hanning filter width is the system optimization task demonstrated in this work.

\subsection{ROI Selection Method}
\begin{figure}[h!]
   \begin{center}
   \begin{tabular}{cc}
   \includegraphics[width=0.9\columnwidth]{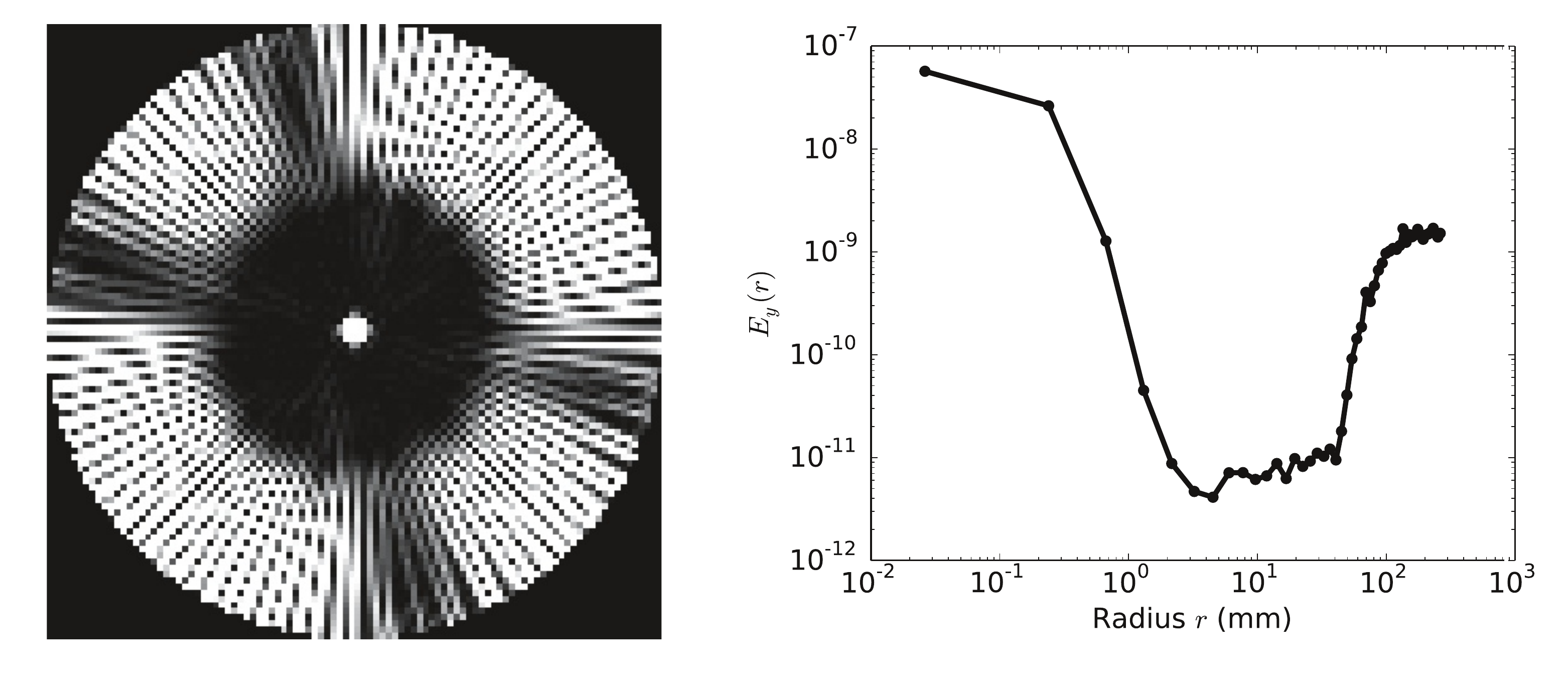}
   \end{tabular}
   \end{center}
   \caption{ Left: The integrand of Equation \ref{eqn:energy} for 
the microcalcification detection task.
Right: The reconstructed microcalcification signal energy as a function 
of radius from the signal, $E_y(r)$. In this case, the resulting ROI would have dimension 14$\times$14 pixels.
The results shown here correspond to 50 projection views
\label{fig:artifact}}
   \end{figure}

For each of the approaches to system optimization considered, we employ the Hotelling Observer (HO)\cite{barrett_foundations_2004} and its 
associated metric, the HO signal-to-noise ratio (SNR) (see Chapter 1). 
In order to allow for direct manipulation of the image covariance matrix $K_y$, we restrict 
the reconstruction to an ROI image. As stated previously, 50 projection views are simulated. 
This angular sampling is sufficient for accurate reconstruction in the immediate vicinity of the 
signal, however undersampling artifacts are visible elsewhere in the full image. We therefore 
restrict the ROI size based on the radius from the signal at which these artifacts begin to 
appear. Specifically, we consider the signal energy in the reconstructed image within an 
annulus of radius $r$ and 
width 2$\Delta r$, given by
\begin{equation}
\label{eqn:energy}
E_y(r) = \int_0^{2\pi}d\theta\int_{r-\Delta r}^{r+\Delta r}\left(\Delta\bar{y}(r^\prime,\theta)\right)^2r^\prime dr^\prime.
\end{equation}
The integrand of Equation \ref{eqn:energy} is shown on the left of Figure \ref{fig:artifact} for the 
microcalcification signal and a single Hanning filter width. In order to select the ROI
size, we approximate the above integral at evenly spaced values of $r$ and set the ROI radius to the value of $r$ which 
minimizes $E_y(r)$. An example of the function $E_y(r)$ is shown on the right of Figure \ref{fig:artifact}, 
corresponding to the image shown on the left side of the figure.

For the ROI sizes investigated here, ranging from 4$\times$4 to 40$\times$40, the reconstruction matrix $A$ ranges in size from 16$\times$4100 to 
1600$\times$4100. Meanwhile, the covariance matrix $K_y$ ranges from 16$\times$16 to 1600$\times$1600. The method proposed remains numerically 
feasible for ROI sizes up to roughly 100$\times$100. Hereafter, we refer to this approach as the ROI-HO.

\subsection{Approximation strategies}
One can show (see, for example, Ref. \cite{barrett_foundations_2004}, Section 7.4.4) that if one assumes that the matrix $K_y$ is circulant, the equation  
$K_y w_y = \Delta \bar{y}$ can readily be solved by application of the discrete-Fourier-transform (DFT) operator. The assumption that $K_y$ is Toeplitz corresponds 
to an assumption of noise stationarity, and the more restrictive assumption of a circulant $K_y$ further implies cyclic correlations at the boundaries of the 
image. In order to test the validity of these assumptions, and hence of the DFT approach to determining HO performance for the present task, we extract a single row of the 
matrix $K_y$, corresponding to a pixel near the center of the signal location, and base a circulant approximation of $K_y$ on this single row. We then repeat the 
optimization of filter width for microcalcification detection and Rayleigh discrimination 
under this approximation. 

An alternative approach suggested by Gallas and Barrett\cite{gallas_validating_2003, barrett_foundations_2004}, 
is the use of efficient Laguerre-Gauss (LG) channels for estimation of HO performance. 
These channels are discretizations of circularly symmetric basis functions, formed as the product of an exponential function and the Laguerre polynomials. They form an 
orthonormal basis for all circularly symmetric square-integrable functions in $\mathbb{R}^2$, and hence, they are an appropriate choice of channels for the task of 
detecting a circularly symmetric microcalcification. We therefore also apply 50 LG channels to the 
optimization of filter width for the microcalcification detection task. Rationale 
for this number of channels, as well as for the scale factor that determines the width of the Gaussian envelope of the LG channels is given in Section \ref{sec:results}. 

\subsection{Estimation strategies}
Yet another approach to the estimation of HO metrics is to train and then test a linear classifier on samples of noisy images from each of the classes being considered \cite{wagner_multivariate_1993,
fukunaga_effects_1989}. Using this approach, a set of training images from each class is used to compute a sample covariance matrix, along with sample mean images corresponding to 
each class. These are used to compute a template estimate which is then applied to a series of testing images. The outcomes of the test statistic for each testing image 
are then recorded and a Mann-Whitney U statistic is computed on the test statistic values, yielding an estimate of the HO AUC \cite{hanley_meaning_1982}. Recall that the HO AUC 
can be related to the HO SNR through the equation
\begin{equation}
\textrm{SNR} = 2\textrm{erf}^{-1}\left(2\textrm{AUC} - 1\right),
\end{equation}
where erf$^{-1}$ denotes the inverse error function\cite{beutel_handbook_2000}.

For this approach, the authors of Ref. \cite{gallas_validating_2003} suggest that, as a rule of thumb, 10-100 images are required for each row of the covariance matrix. 
For the case of a 30$\times$30 ROI, this corresponds to anywhere between roughly 9,000-90,000 images. While this is infeasible for a parameter optimization study in general,
we demonstrate the outcome of this approach for image training and testing sets ranging from 500 to 3,000 images
in order to demonstrate the general trend 
of results obtained using this method. Due to the large number of independent noisy data sets required, determination of HO performance for each reconstruction filter width is 
performed using the same sets of noisy data. Two approaches, commonly termed the hold-out method and resubstitution, are applied. In the first case, the training and testing 
phase use separate, independent sets of noisy data. In the second, the HO is tested on the same images with which it was trained. These two approaches yield estimates with 
negative and positive biases, respectively \cite{wagner_multivariate_1993, fukunaga_effects_1989}.

A more feasible approach based on sample images can be taken when prior knowledge of the mean images under 
each hypothesis is exploited in order to reduce the bias and variance of the estimator of HO performance, thereby reducing the necessary number of sample images. 
The construction of an unbiased estimator of HO SNR in this case is 
detailed in Ref. \cite{wunderlich_new_2013}. 
We demonstrate the application of this approach using 300 and 
700 noisy image samples (far fewer than the number of samples required for training and testing), and compare to the previous methods. 
Two scenarios are considered: (1) the same 300 or 700 noisy data realizations are used to reconstruct noisy images at each filter width considered and 
(2) a separate set of 300 or 700 data realizations is used to create noisy images at each filter width.

\section{Results} \label{sec:results}

   \begin{figure}[h!]
   \begin{center}
   \begin{tabular}{cc}
   \includegraphics[width=0.96\columnwidth]{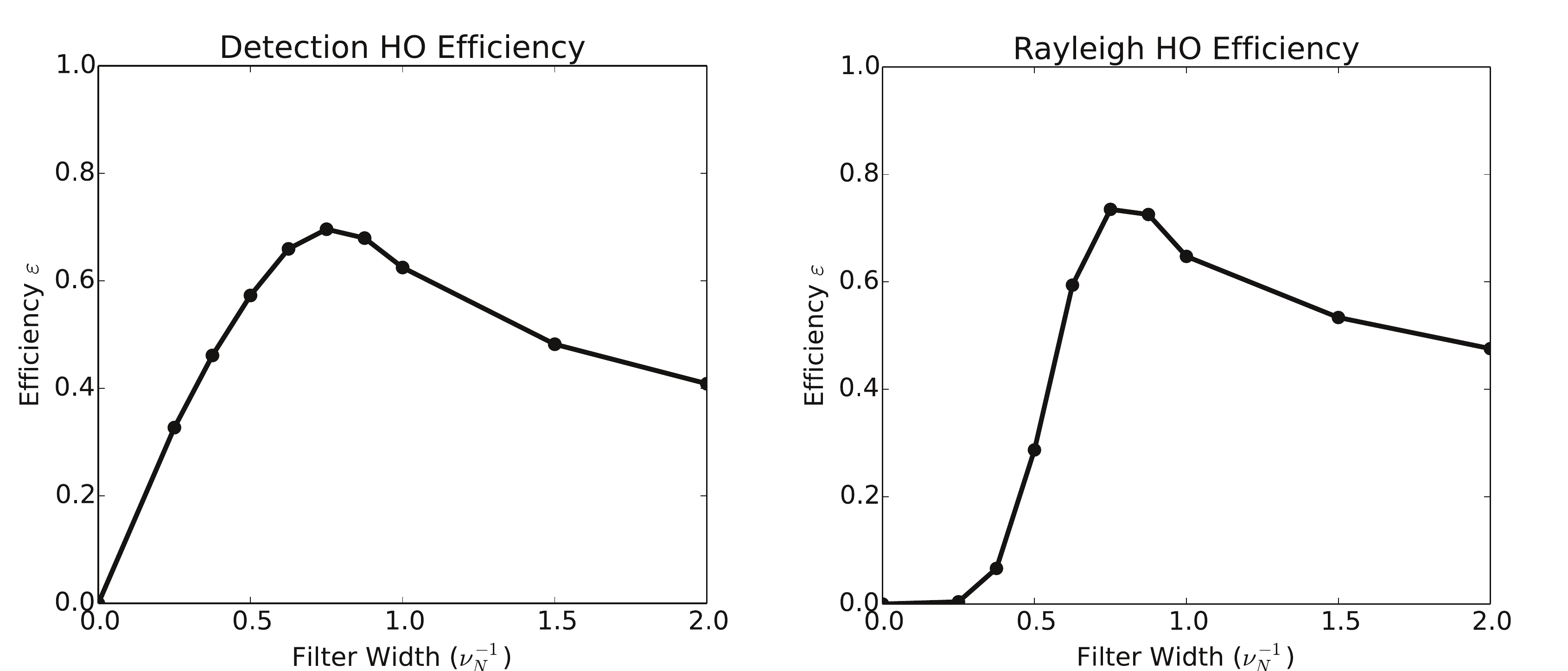}  
   \end{tabular}
   \end{center}
   \caption{ Efficiency values are shown for various views and Hanning window widths (relative to the Nyquist frequency on the detector)
 for both the Rayleigh task (left) and the 
detection task (right). The Nyquist frequency in this case is $\nu_N = \frac{1}{2\Delta u} \approx 1.3mm^{-1}$. While the efficiency values for moderate filtering 
to no filtering were seen to have a dependence on ROI size, the same trend pictured here was seen for ROI sizes up to roughly 1.5cm in diameter.\label{fig:plots} }
   \end{figure} 

The results of applying the ROI-HO for microcalcification detection and Rayleigh discrimination are shown in Figure \ref{fig:plots} for a range of Hanning filter widths. 
The ROI-HO is noticeably sensitive to the reconstruction filter width, showing a clear maximum in performance for Hanning windows in the range of 0.75$\nu_N$ to 
0.825$\nu_N$ for each task. However, this result should not be interpreted as giving a universally 
optimal filter width, but rather as a demonstration of the sensitivity of HO efficiency to relevant 
reconstruction algorithm parameters. 

\begin{figure}[h!]
   \begin{center}
   \begin{tabular}{cc}
   \includegraphics[width=0.96\columnwidth]{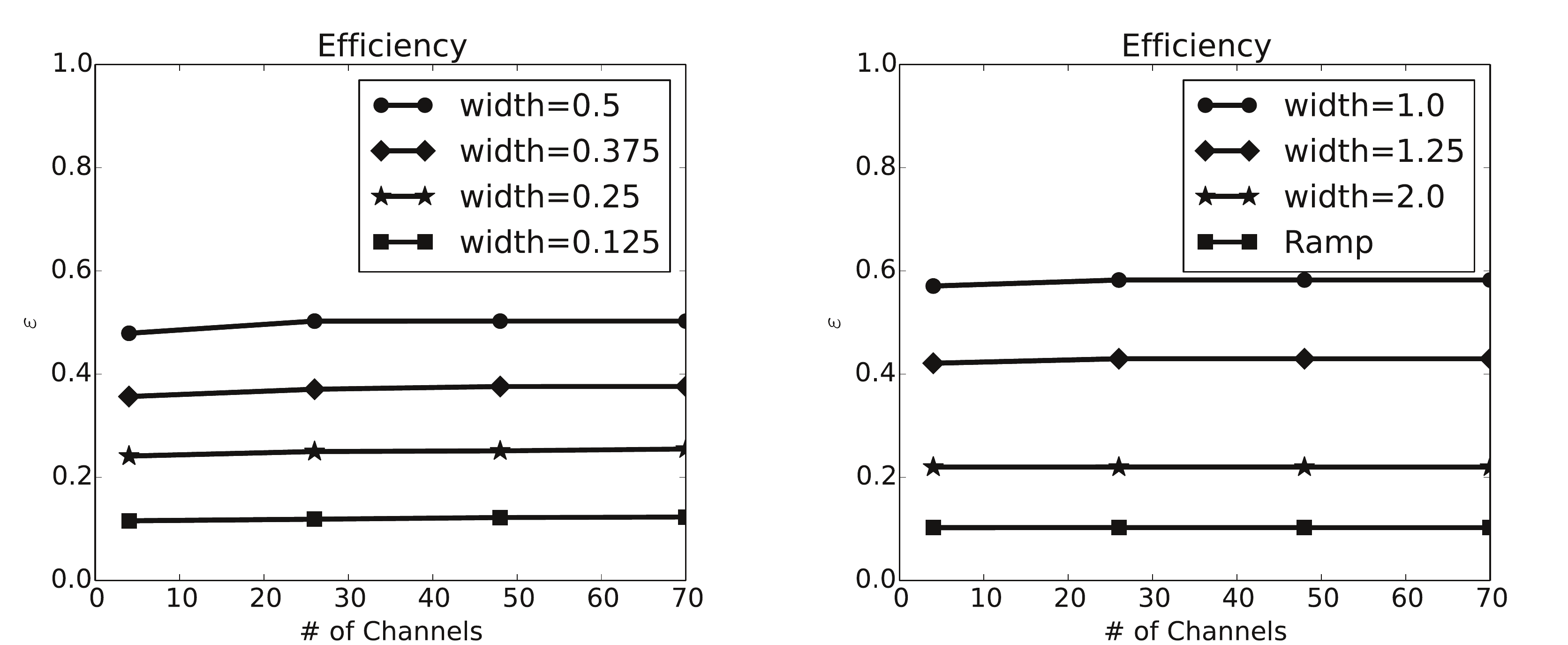}
   \end{tabular}
   \end{center}
   \caption{  Dependency of CHO efficiency on number of channels for a range of reconstruction filter widths. For most filter widths, the CHO 
efficiency stabilizes with 50 channels, however the performance estimates for this task and system 
are not sensitive 
to the number of channels used, in general. Subsequent results shown are for 50 channels.\label{fig:nchan}}
   \end{figure}
\begin{figure}[h!]
   \begin{center}
   \begin{tabular}{cc}
   \includegraphics[width=0.96\columnwidth]{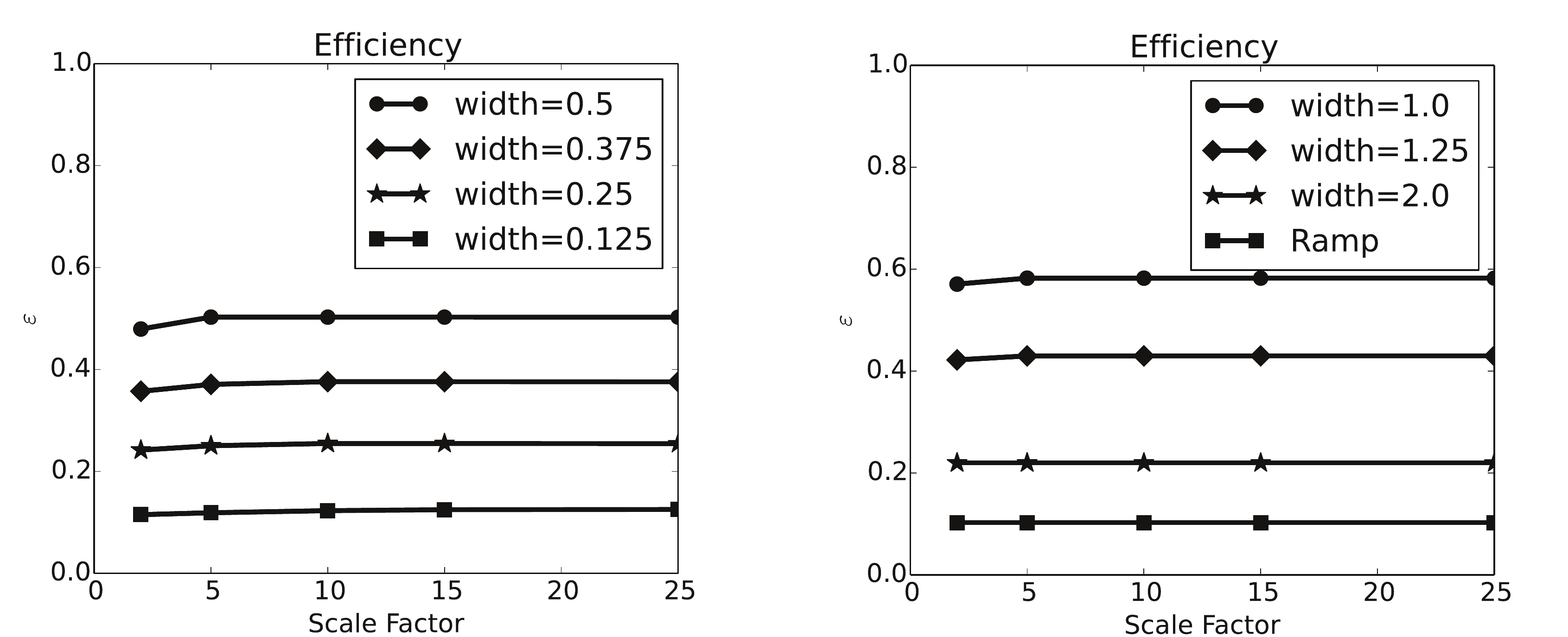}
   \end{tabular}
   \end{center}
   \caption{  CHO efficiency using 50 channels and 50 projection views for a range of Gaussian envelope widths in the LG channels. 
The x-axis is normalized to the 80$\mu$m microcalcification diameter. A scale factor of roughly 7 times the microcalcification width was determined 
to be optimal.\label{fig:scale}}
   \end{figure}

Next, we consider the use of channels for estimating HO performance in microcalcification detection. As with the ROI-HO, we restrict the CHO to an ROI, however due to the Gaussian envelope which 
modulates the LG channels, this only has an effect for the smallest ROIs used. 
For the CHO using LG channels, results correspond solely to the microcalcification task, since 
the Rayleigh task involves a signal which is not radially symmetric.
Figures \ref{fig:nchan} and \ref{fig:scale} demonstrate the dependence of CHO efficiency on the number of channels and the scale factor which modulates the width of the 
Gaussian envelope of the Laguerre-Gauss functions. For the majority of filter widths considered, the CHO efficiency estimate is completely stable above roughly 50 channels, while the optimum 
scale factor of the channels (full width at half maximum of the Gaussian) is roughly 7 times the 
width of the microcalcification diameter. 
The remainder of the results presented correspond to these CHO parameters. 
In our case, however, the dependence of the 
CHO performance estimates upon each of these parameters is weak, so that fewer channels 
or a slightly different scale factor could likely produce comparable results.

\begin{figure}[h!]
   \begin{center}
   \begin{tabular}{cc}
   \includegraphics[width=0.96\columnwidth]{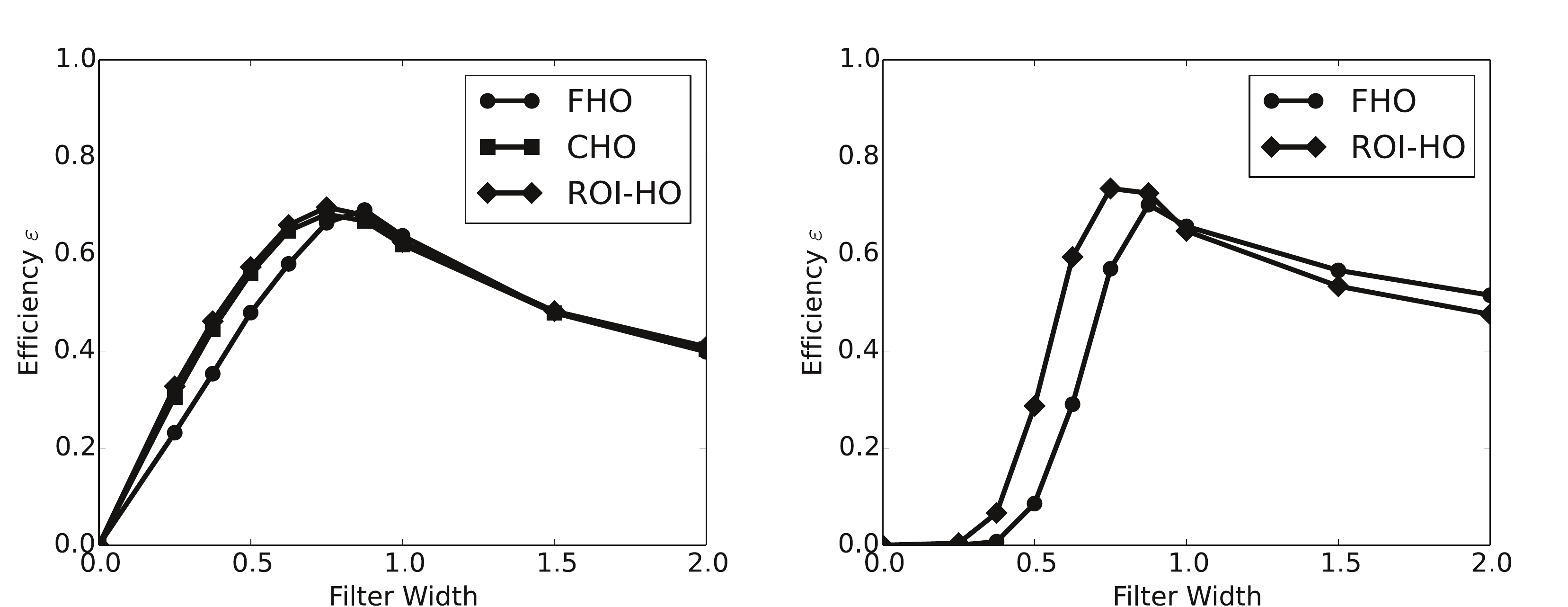} 
   \end{tabular}
   \end{center}
   \caption { A comparison of optimization of reconstruction filter width for 50 projection views. Shown are results obtained with the proposed 
ROI-HO, an approximate HO computed using the DFT (labeled FHO), and the CHO.\label{fig:approx}}
   \end{figure}

Figure \ref{fig:approx} compares three approaches to optimization of the reconstruction filter width parameter, namely the ROI-HO approach proposed in this work, 
the CHO with 50 LG channels, and the DFT-domain approach (labeled FHO). 
Since each of the methods utilizes the same ROI images, the ROI-HO, which computes the exact 
HO performance, should be taken as the ground truth for the sake of evaluating the alternative 
approaches. Clearly, for microcalcification detection, the CHO is an excellent approximation of the 
true HO. Further, the CHO allows for greater computation efficiency, since, in general, the ROIs used 
contained more than 50 pixels. However, the efficiency of this CHO comes with a loss of generality, 
as it is not applicable to general signals lacking radial symmetry, as in the Rayleigh discrimination 
task. 

Meanwhile, the DFT-domain HO was capable of reproducing the general trend of HO results, 
however the quantitative reliability of this method varied widely depending on the specific 
parameter setting used. For example, while wider filter widths allowed for accurate computation 
of detection task performance with the DFT-domain approach, 
the estimates of Rayleigh task performance using this method 
could vary from the HO by as much as 40\%, as seen in the right side of Figure \ref{fig:approx}. 
While this may not impact the optimization of filter width substantially, if absolute HO performance 
with a fixed filter width of, say, 0.625$\nu_N$ were of interest, the DFT-domain HO could be 
misleading.

\begin{figure}[h!]
   \begin{center}
   \begin{tabular}{cc}
   \includegraphics[width=0.96\columnwidth]{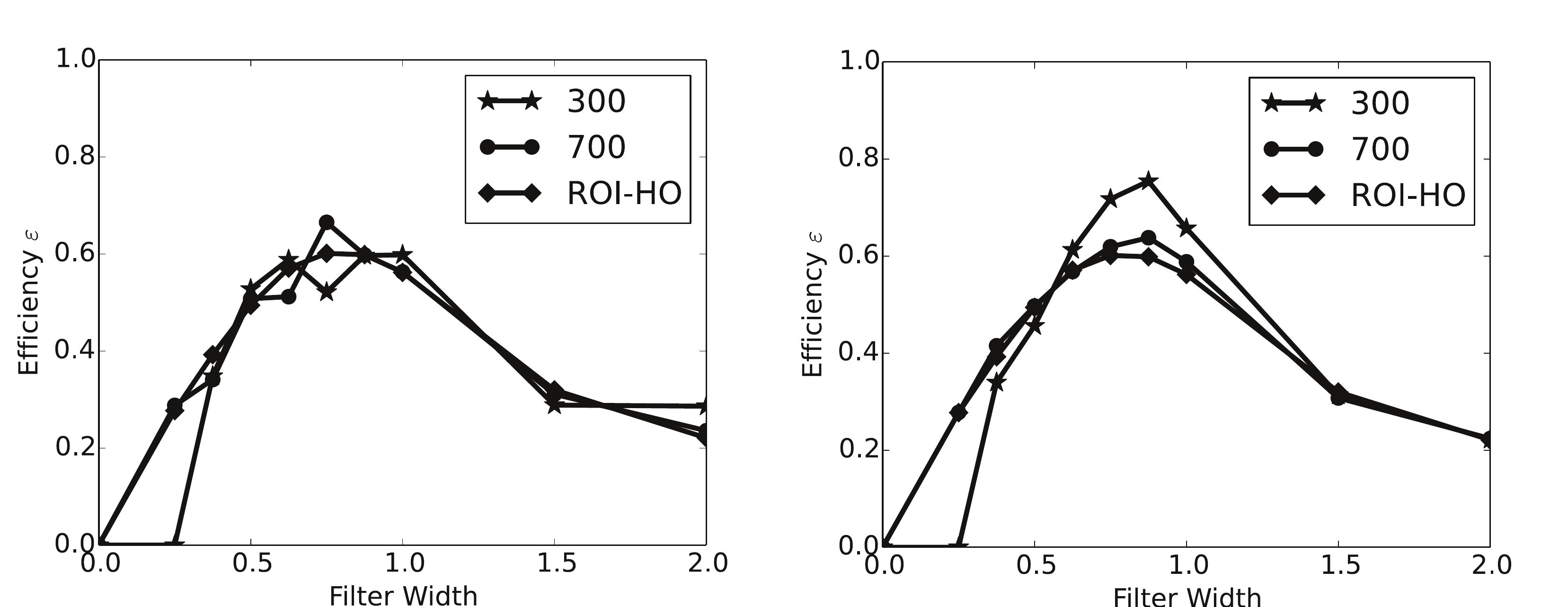} 
   \end{tabular}
   \end{center}
   \caption{ Results of estimating the HO efficiency from sample images using the method proposed by Ref. \cite{wunderlich_new_2013} with 300 
and 700 noisy sample images. Left: An independent set of images is used for each filter width. Right: The same noisy data realizations are used for each 
filter width. Error bars corresponding to two standard deviations of the estimates for 700 sample images are shown in Fig. \ref{fig:est_err}.}\label{fig:est} 
   \end{figure}

Figure \ref{fig:est} demonstrates the use of noisy sample images to estimate HO efficiency when the mean image under each hypothesis is known. In our case, since we consider
only linear image reconstruction algorithms, the mean image is produced simply by reconstructing an image without noise. The left and right plots correspond respectively to the use of 
independent noisy data sets for each filter width and reuse of the same simulated data for each filter width. 
The variability seen in the left-hand figure provides intuition as to the variance of the efficiency estimates, while the figure on the right illustrates that, 
although subject to the same variability, reusing the same data realizations correlates the estimates for different filter widths, potentially allowing for more reliable rank ordering 
of parameter settings. In other words, the estimated curves on the right could undergo vertical translation due to the variance of the estimator, but are more likely to 
preserve their shape than the curves shown on the left.

   \begin{figure}[h!]
   \begin{center}
   \begin{tabular}{cc}
   \includegraphics[width=0.96\columnwidth]{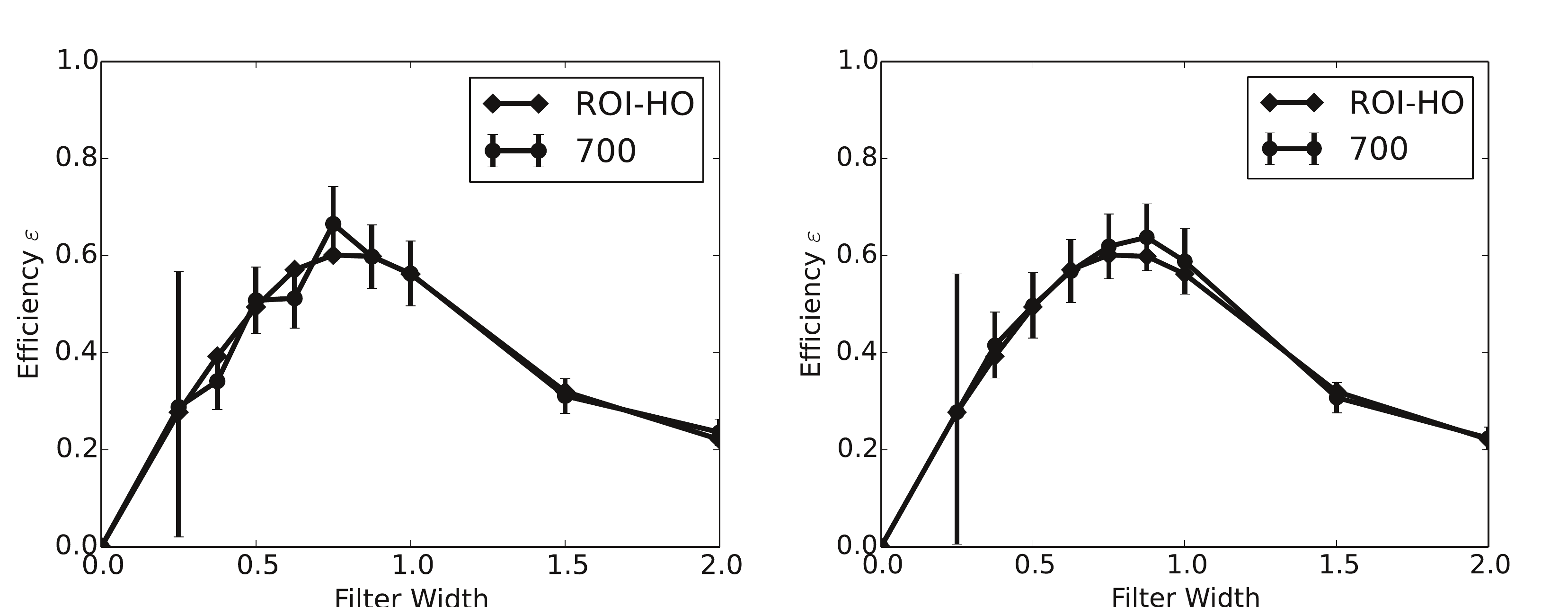} 
   \end{tabular}
   \end{center}
   \caption{ Results from Fig \ref{fig:est} for 700 noisy sample images are shown with error bars corresponding to two standard deviations derived from jackknifing.
   The error bars illustrate statistical variation, but do not account for inherent estimator bias, which can be seen by comparing with the analytically computed ROI-HO curves. It is possible
   that other sources of uncertainty not investigated here, aside from statistical variations, could influence these results. \label{fig:est_err} }
   \end{figure} 

\begin{figure}[h!]
   \begin{center}
   \begin{tabular}{cc}
  \includegraphics[width=0.96\columnwidth]{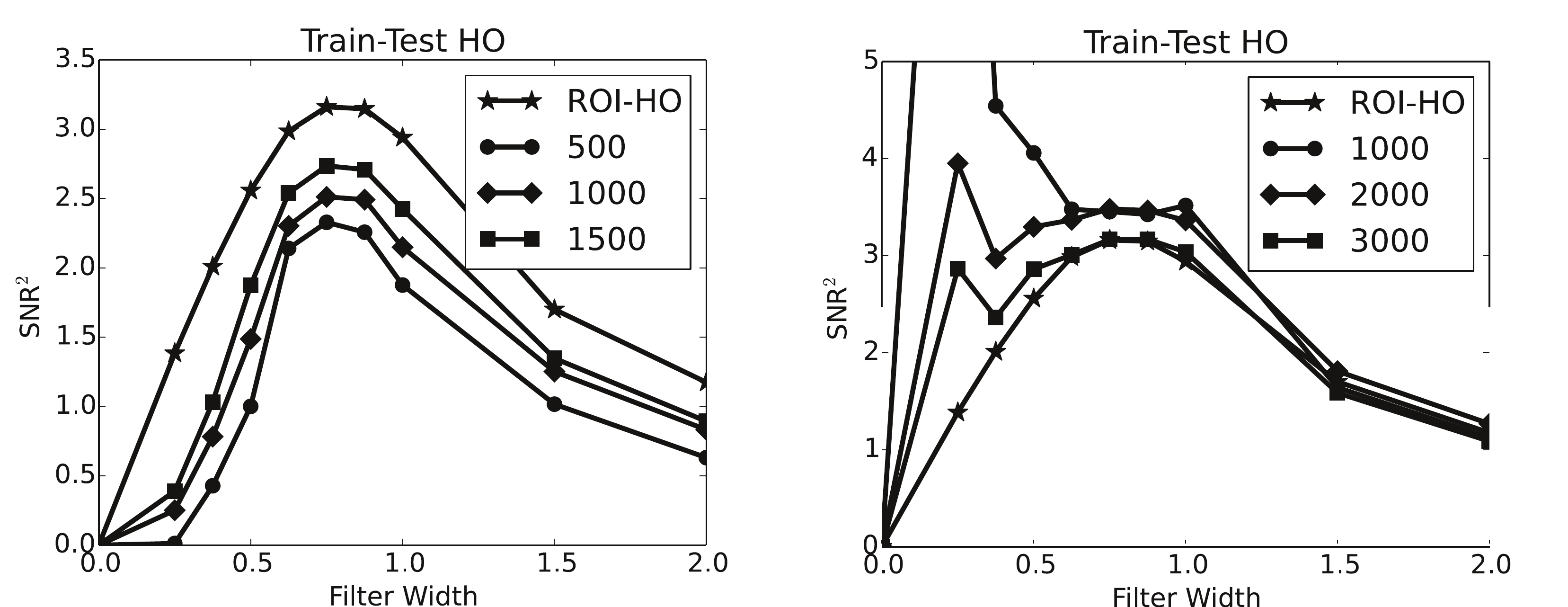} 
   \end{tabular}
   \end{center}
   \caption{ Left: HO SNR$^2$ estimates from training and testing performed using the hold-out approach for 500, 1000, and 1500 training images, with an equal number of testing images. 
The prevalence of images from each class is also equal. 
Right: HO SNR$^2$ estimates resulting from training and testing performed using resubstitution for 1000, 2000, and 3000 total images. The bias and variance of the estimates is worst for 
narrow filter widths, where the size of the ROI used is largest. Variance of the estimates is illustrated in Fig. \ref{fig:tterr} through 95\% confidence intervals derived from bootstraping.\label{fig:tt}}
   \end{figure}

Figure \ref{fig:tt} illustrates the use of linear classifier training and testing in order to estimate HO 
performance. In this case, prior knowledge of the mean images is not exploited, 
and sample estimates of the images are computed instead. 
The left-hand plot corresponds to the hold-out approach, where independent image sets are used for 
the training 
and testing phases, while the right-hand plot is generated using re-substitution. As seen in the figure, 
these two approaches introduce negative and positive biases in the estimates, respectively. 
Note that thousands of images are required in order to construct quantitatively meaningful estimates. 
However, the general trend of the results can be seen 
for comparatively few samples. This suggests that for certain tasks, such as rank-ordering of only a few 
options for algorithm implementation, the training and testing approach could be useful 
if one lacks an adequate model of the image covariance or class means. 

   \begin{figure}[h!]
   \begin{center}
   \begin{tabular}{cc}
  \includegraphics[width=0.96\columnwidth]{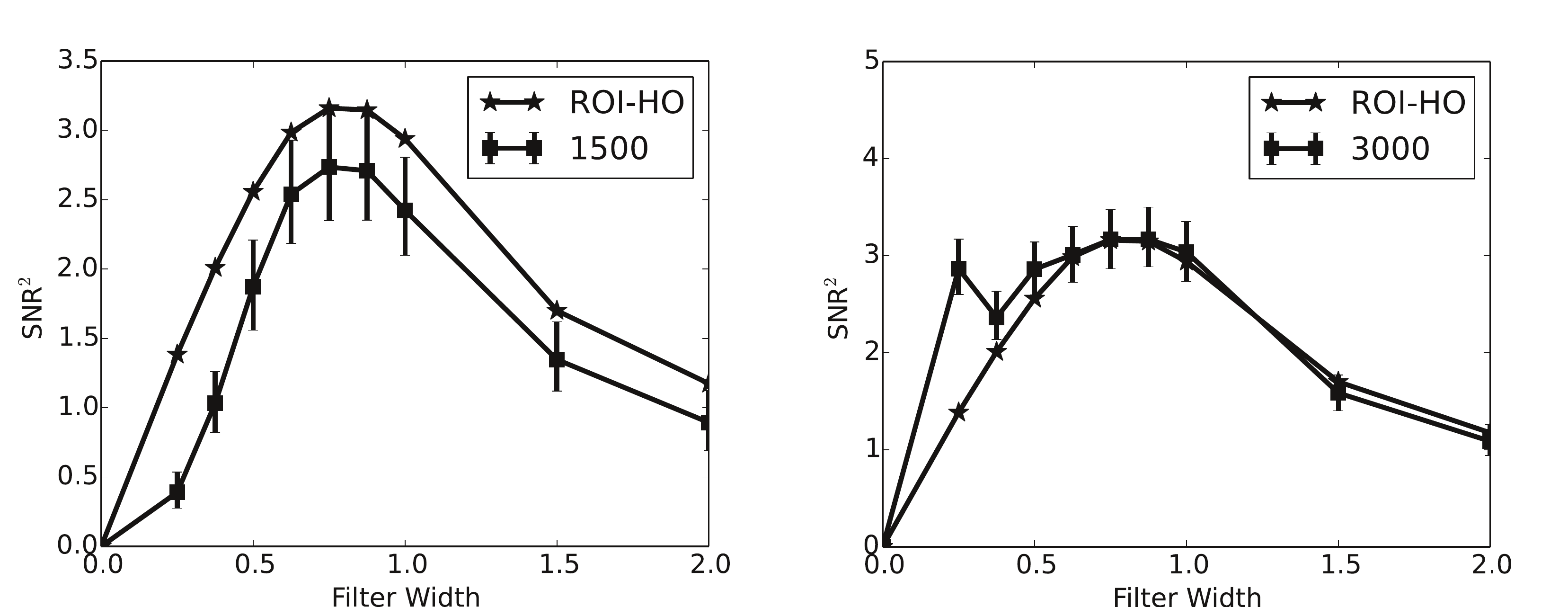} 
   \end{tabular}
   \end{center}
   \caption{Shown here are the results from Fig. \ref{fig:tt} for the largest number of sample images with error bars denoting 95\% confidence intervals derived from 1000 bootstrap samples.
   These errors derive from the variance of the training and testing estimator, while inherent bias in the estimator contributes additional error, especially for small filter widths where the number of ROI pixels 
   is large. As in Fig. \ref{fig:est_err}, the error bars here only convey statistical uncertainties, and bias can be inferred by comparison with the analytically computed ROI-HO curves. \label{fig:tterr}}
   \end{figure} 
\section{Conclusions}
\label{sec:conclusions4}
We have demonstrated that for classification tasks involving small signals in CT, the HO performance computed within an ROI constitutes
an efficient and objectively meaningful approach to reconstruction algorithm and system optimization. Specifically, we have employed the HO to 
optimize reconstruction algorithm filter width for two tasks: microcalcification detection in dedicated breast CT, and
Rayleigh discrimination.  

We performed a comparison 
of this approach with several alternatives for objective assessment. In broad terms, these alternatives fall into two categories: methods 
based on computation of statistical estimates and methods based on approximations regarding the signal and/or image noise. While our proposed
methodology assumes that information beyond a small ROI is irrelevant to the classification task, the CHO imposes assumptions of rotational 
symmetry on the signal and Hotelling template, and the Fourier-domain approach assumes stationary noise. Strictly speaking, none of these assumptions
are valid in CT for the tasks considered here, however each approach possesses strengths and weaknesses. The Fourier-domain HO is the most 
computationally efficient method for large ROIs, but is less accurate than the ROI-HO or CHO approach. Meanwhile, the CHO with LG channels is limited to cases when the signal of interest and Hotelling template possess 
radial symmetry. 

The statistical estimation approaches investigated possess an attractive degree of flexibility, in that they do not rely on any assumptions regarding the image covariance matrix or, 
in the case of the training/testing approach, the signals themselves. However, this flexibility comes at the expense of computational efficiency, as 
this family of approaches requires excessively large numbers of noisy samples, making any extensive system or algorithm optimization prohibitively 
time-consuming. Finally, while we have illustrated some of the trade-offs, benefits, and 
shortcomings of several 
methods for objective assessment in CT, ultimately, future work will be necessary to ensure that the 
proposed methodology yields metrics which correlate with improved performance of humans for the 
given tasks.

%% file: Chapter5/Chapter5.tex
\chapter{Using the Hotelling Observer to Optimize a Breast CT System}
\label{ch:breastct}
\section{Introduction}

In this chapter, we demonstrate the application of the ROI-based HO developed in the preceding chapters to the optimization of a range of system and image reconstruction parameters. 
We focus on optimization of dedicated breast CT, which  is currently being investigated as a potential screening and diagnostic tool for breast cancer at several institutions
\cite{boone_dedicated_2001, chen_cone-beam_2002,yang_dedicated_2007,gong_microcalcification_2004,mckinley_development_2012}.
Breast CT is an attractive technology, largely because it has the potential to provide improved visualization of breast lesions in cases 
where conventional mammography lacks sensitivity \cite{chen_cone-beam_2002,gong_computer_2006,glick_breast_2007}. 
Examples include young women (less than 50 years old) and those with dense breast tissue, wherein the 
overlay of anatomic breast structures in a mammographic projection image can obscure lesions and impede diagnostic accuracy \cite{pisano_diagnostic_2005}. 

While breast CT can potentially improve the utility of breast screening relative to mammography, the individual projection views 
of the CT acquisition are limited in terms of x-ray fluence due to dose concerns, resulting in considerable quantum noise in the projection data. 
Further, dedicated breast CT cannot currently match the ``in plane'' resolution of
full-field digital mammography \cite{boone_dedicated_2001, lindfors_dedicated_2008}, 
leading to potential shortcomings in terms of microcalcification 
detection or visualization of tumor margins, two tasks which can help to indicate malignancy of breast lesions.

In this chapter, we demonstrate the impact of several system and object
parameters on image quality. First, we investigate the effects of patient breast diameter 
for both fixed total exposure and fixed mean glandular dose. Since patient breast diameter cannot be controlled, the impact of breast size on 
task performance is important to ascertain. We therefore demonstrate that our proposed formalism is capable of qualitatively reproducing previous findings
regarding the impact of breast size on image quality. Next, we investigate the impact 
of signal location within the breast, which is important since it leads to the conclusion that accounting for location variability can meaningfully 
impact the outcome of the image quality assessment. In order to demonstrate the use of HO metrics to address a question of system parameters, 
we next evaluate the effects of varying the number of projection views while keeping the total radiation exposure fixed. The trade-off between 
noise in the individual projection views and adequate angular sampling provides a meaningful evaluation of the HO metrics' ability to optimize acquisition parameters. 
Finally, since the reconstruction algorithm itself connotes some potential loss in information, we demonstrate optimization of two of the most ubiquitous 
parameters in the conventional FBP algorithm: the image pixel size and the reconstruction filter width. Image quality evaluation in each case was performed 
with respect to two tasks: the detection of a simulated microcalcification 
and the resolution of two adjacent, high-contrast objects. We choose these tasks because 
the tradeoff between image noise and resolution is quite sensitive to CT system parameters.

The noise/resolution tradeoff also depends strongly on the noise model covariance and the form of object being viewed in
the detection/discrimination task. 
The HO combines this complex image quality information into a single meaningful metric for a given task.
In order to illustrate this advantage of the HO, we present several realizations of reconstructed noisy images for subjective assessment.
We include several examples of images 
created with reconstruction filters and view numbers deemed optimal by the HO. Also shown are images obtained from view numbers or reconstruction 
filter widths deemed suboptimal by the HO. The numerical phantoms used to create these images do not correspond exactly to the tasks for which these parameters 
were optimized, but instead are either anthropomorphic breast phantoms or phantoms intended to mimic clinical quality assurance phantoms. This comparison 
is important because it illustrates the extent to which these two approaches to image quality assessment agree, demonstrates how such images provide 
a useful check for HO optimization, and highlights specific limitations of each approach. 

The outline of the chapter is as follows. Section \ref{sec:bg} provides
background in image quality evaluation in dedicated breast CT, along with motivation for the evaluation 
formalism we propose. In sections \ref{sec:imaging_system} and \ref{sec:phantom} we 
describe the simulated breast CT system and numerical phantoms. In sections \ref{sec:data} and \ref{sec:noise_model} we describe the data and noise 
models in the projection data and image domains, while in sections \ref{sec:tasks} and \ref{sec:HO} we outline the task-based assessment formalism. 
Next, we describe the application of the formalism to ascertaining the impact of a range of imaging parameters, some of which are 
inherent to the breast being imaged, section \ref{sec:patient}, and others which are controllable through system design or reconstruction algorithm implementation, sections \ref{sec:system} and
\ref{sec:recon_parms}.
Results of the application of the methodology are shown in section \ref{sec:results1}, with 
corresponding discussion in section \ref{sec:discussion1}. Examples of reconstructed images generated
using the determined optimal system parameters are provided in section \ref{sec:results2} and 
discussed in section \ref{sec:discussion2}. Finally, we 
provide conclusions in section \ref{sec:conclusions}.

\section{Background}
\label{sec:bg}
Since in this chapter we have decided to focus on dedicated breast CT, we shall focus our discussion of CT image quality metrics to those which 
have most commonly been applied to that modality.  
A variety of approaches have been adopted to aid in image quality evaluation for dedicated breast CT, some of which have been applied to the optimization of 
the physical system, technique factors, or algorithm implementation. Boone \etal \cite{boone_dedicated_2001} performed image quality evaluation 
by computing contrast-to-noise (CNR) and corresponding signal-to-noise ratio (SNR) values for lesions in breast cadavers. Comparison of these values to the 
Rose criterion \cite{rose_human_1974} was then performed to establish feasibility of low-dose dedicated breast CT. A similar approach 
was also used more recently by Shen \etal \cite{shen_suei02:_2013} in order to ascertain the impact of projection view number with fixed total exposure.

Meanwhile, Refs. \cite{kwan_evaluation_2006, yang_computer_2007, yang_noise_2008} explored the impact of an extensive range of imaging 
parameters on image resolution (Refs. \cite{kwan_evaluation_2006, yang_computer_2007}) and noise (Ref. \cite{yang_noise_2008}) by 
computing Fourier-based metrics. Specifically, modulation transfer functions (MTFs) and noise power spectra (NPSs) were measured experimentally by repeated imaging 
of phantoms with varied imaging parameters. In the case of resolution measurement, a high-contrast wire was used to mimic a point source, so that empirical point-spread 
functions could be seen in the reconstructed images. Meanwhile, uniform cylindrical polyethylene phantoms were used to generate mean-subtracted noise images, whose 
discrete Fourier transform yielded empirical NPSs. These metrics have the benefit of ubiquity across many medical imaging applications.

A task-based approach to breast CT image quality evaluation was pursued by Lindfors \etal and Lai \etal in Refs. \cite{lindfors_dedicated_2008} and \cite{lai_visibility_2007}, respectively. The 
former study used human observers to subjectively classify a variety of lesion types as visible or not visible, for the purpose of comparing lesion detectability in dedicated breast CT 
to screen-film mammography. The authors allude to performing human observer ROC studies, which indeed would provide a very meaningful metric for comparison of these 
two modalities, but would be too costly and time-consuming to perform investigation of the impact of many system parameters in isolation. 
Meanwhile, Lai \etal \cite{lai_visibility_2007} also performed 
a subjective image-quality evaluation by having human observers count the number of microcalcifications in various microcalcification clusters which were clearly visible. The clusters were 
in known locations and had fixed, known locations for individual microcalcifications. The purpose of this study was to investigate the effects of x-ray tube voltage and radiation dose.

Moving toward methods which are more robust against the subjectivity of human observer responses, the results of Refs. \cite{gong_microcalcification_2004} and
 \cite{gong_computer_2006} are particularly significant in that they help to establish the capability of dedicated breast 
CT in microcalcification detection through assessment of human performance in the form of full human ROC studies \cite{metz_basic_1978}. 
Meanwhile, the methods proposed in Ref. \cite{glick_evaluating_2007}
overcome some of the limitations of using humans by computing ideal observer performance to study the impact of x-ray spectral shape. The authors' approach in that work is similar to that which we 
propose, however the assumptions upon which each formalism is based are different. Glick \etal \cite{glick_evaluating_2007} apply a parallel cascade approach to studying 
the signal and noise properties of dedicated breast CT by modeling it as a linear shift-invariant (LSIV) system with stationary Gaussian noise (both anatomical and quantum). They then analyze 
lesion detectability in the projection domain, rather than optimizing with respect to the reconstructed image. 

As an alternative to simulation of the breast CT system, Packard et al. used sample images to obtain mean signal profiles and breast power spectra, in turn constructing a mathematical 
model observer, the pre-whitening matched filter, based on these quantities derived from samples \cite{packard_effect_2012}. The goal of that work was to ascertain the effect of 
slice thickness on lesion detectability. 
More recently, Chen et al.\cite{chen_association_2013} have used similar methods to investigate an association between lesion detectability in breast imaging and spectral components of anatomic noise.

We propose an assessment methodology based on the Hotelling observer (HO). By considering the HO's performance on a relevant task within a small ROI, summary metrics can be 
computed quickly enough to sweep over a range of imaging parameters and evaluate their impact on image quality. This approach has the benefit of being task-based and non-stochastic, 
meaning that it does not rely on a collection of noisy images or on the assessment of human performance. Further, the methodology we propose does not assume an LSIV system or stationary noise. 
Instead of making assumptions regarding the noise and signal properties in the reconstructed images, our approach is based on 
assumptions of noise and signal properties in the original projection data, which, in the absence of object variability, are simpler to model accurately.
We then exactly account for the action of the reconstruction operation on HO task performance, including the noise correlations and 
signal bias which reconstruction introduces. 
The details of the noise model assumed in this chapter are given in section \ref{sec:noise_model}, and the resulting scalar image quality metrics are described in section \ref{sec:HO}.

\section{Methods}

\subsection{Imaging System Geometry}
\label{sec:imaging_system}

The simulated CT system in this chapter was modeled based on that from Refs. \cite{kwan_evaluation_2006} and \cite{yang_noise_2008}. Those studies investigated
noise and resolution properties, respectively, of the same dedicated breast CT system. This system utilizes a pendant geometry and a flat-panel CsI detector to acquire cone-beam CT data of the 
breast. We chose to model that system here in order to facilitate a meaningful comparison of our 
task-based assessment approach to more standard approaches, namely performing measurements of the MTF and NPS, pursued by the authors of those works.  

The chosen detection/discrimination tasks are performed viewing a single slice of the reconstructed volume,
and in order to simplify the HO presentation we restrict the observer model computation to the mid-plane of the circular
cone-beam scanning configuration. As a result we consider only the two-dimensional circular fan-beam configuration for the
mid-plane.
The axial position of
the mid-plane for the simulated detector is approximately 2 cm below the top edge of the detector face. 
The simulated detector had a width of approximately 40 cm and consisted of 2048 detector pixels, each 0.194 mm in width. 
The distance from the x-ray source to the axis of rotation is 458 mm, and the distance from the x-ray source to the 
detector is 878 mm. The resulting geometric magnification for an object at the axis of rotation is therefore 1.92. 
The system's field-of-view (FOV) is restricted to 165.9 cm in diameter, and this circular FOV is inscribed in a 1024 $\times$ 1024 image pixel grid, with a corresponding 
image pixel size of 0.162 mm $\times$ 0.162 mm. Meanwhile, the detector pixel size, back-projected to the axis of rotation, is 0.101 mm. Finally, 
a 0.4 mm focal spot is modeled, as described in section \ref{sec:data}.

\subsection{Breast Phantom}
\label{sec:phantom}
The digital breast phantom used in this work is circular with a diameter of 14 cm. The phantom composition is modeled as being 50\% adipose tissue and 
50\% fibroglandular tissue, uniformly distributed throughout the breast (see Figure \ref{fig:tasks} for ROIs from the phantom). The corresponding attenuation values are taken from Ref. \cite{johns_x-ray_1987}.
Simulated microcalcifications are modeled as being circularly symmetric Gaussians of peak x-ray attenuation equal to that of calcium. However, the attenuation value is decreased in order 
to simulate the first-order partial volume averaging in a finite slice thickness of 0.5 mm containing a microcalcification.

Whenever the simulated microcalcification has a diameter less than the slice thickness of 0.5 mm, the modified x-ray attenuation for the microcalcification is computed as
\begin{equation}
\mu_\text{calc} = \frac{d_\text{calc}}{0.5\text{mm}} \left(\mu_\text{Ca} - \mu_\text{br}\right) + \mu_\text{br}
\end{equation}
where $d_{calc}$ is the diameter of the calcification, defined as the full width at half maximum of the Gaussian signal, 
$\mu_{Ca}$ is the x-ray attenuation of calcium, and $\mu_{br}$ is the attenuation of the 50/50 breast phantom. 

\subsection{Projection Data Model}
\label{sec:data}

In this section, we will repeat the formulation of our data model, as we have included some physical effects which were not incorporated into the 
model presented in Chapter 1. 
We again model the CT data after the point where the transmitted X-ray intensity is processed by the negative logarithm, using the
standard line integral formulation as the mean:
\begin{equation}
\label{eqn:data}
g_i=\int_L f\left(s_i + l \hat{\theta}_i \right) dl,
\end{equation}
where $g_i$ is the sinogram data corresponding to the $i$th detector element, and $f$ is the numerical phantom value at a point which is $l$ distance from the 
source position $s_i$ in the direction $\hat{\theta}_i$. The line integral is performed along all values within the compact object support in the direction $\hat{\theta}$, 
and projection through 
the discrete numerical phantoms is performed by considering the path-length of a given ray through a square pixel. The pixel size of the numerical phantoms is 
always set to be at least an order of magnitude smaller than the pixel size of the final reconstructed image. 

The simulated x-ray spectrum corresponds to an 80kVp setting with added Be and Al filtration of 0.8mm and 2.5mm, respectively. Methods from Refs \cite{boone_accurate_1997, boone_molybdenum_1997, boone_spectral_1998}
 were used in simulation of the x-ray spectrum.
In order to model the finite size of the detector pixels, we perform 16-fold subsampling, so that for each detector pixel, 16 equally spaced rays within that 
pixel are simulated.
We do not perform subsampling along the trajectory of the x-ray source, modeling a step-and-shoot acquisition.
A 0.4 mm 
focal spot is modeled by convolution of a rect function with the subsampled line integral data. The width of the convolution kernel (rect function) includes geometric magnification and is computed 
as
\begin{equation}
\label{eqn:fspot}
d = d_\text{fspot}\frac{SDD-r}{r}
\end{equation}
where $d_\text{fspot}$ is the width of the focal spot (0.4 mm here), $SDD$ is the source-to-detector distance (878 mm), and $r$ is the angle-dependent distance 
from the focal spot to the center of an ROI containing the signal of interest. 
For the present model we do not include other physical factors such as a full tube spectrum model, x-ray scatter, and spread of optical photons in the CsI detector.
Although inclusion of these physical factors may alter the presented results, they do not
alter the proposed assessment methodology outlined in section \ref{sec:HO}.

\subsection{Noise Model}
\label{sec:noise_model}
We view images and sinograms as one-dimensional vectors where the index scans through the respective
data structure in lexigraphical order.
We then consider additive Gaussian noise applied to the line integral data model of Eq. \ref{eqn:data}:
\begin{equation}
\textbf{g}_i=\int_L f\left(s_i + l \hat{\theta}_i \right) dl + \textbf{n}_i
\end{equation}
where bold font indicates a random variable. 
In particular, we consider the noise term $\textbf{n}_i$ to 
be zero-mean and have variance determined by
\begin{equation}
\text{Var}\left\{\textbf{n}_{i}\right\} = \frac{\sigma_e^2 + \text{Var}\left\{\textbf{I}_{i}\right\}}{\bar{I}_i^2},
\end{equation}
where $\textbf{I}_{i}$ is the number of optical quanta incident on the imaging array at location $i$ and converted to electrons, $\bar{I}_{i}$ is the mean of $\textbf{I}_{i}$, 
and $\sigma^2_e$ is the variance due to electronic noise. This noise model is a generalization of that put forward by Barrett and Swindell \cite{barrett_radiological_1996}, with the addition of the 
electronic noise term. Following Gong et al. \cite{gong_computer_2006}, we take $\sigma_e = 2200$, the optical collection efficiency (OCE) to be 0.64, 
the detector pixel fill factor to be 0.57, and the x-ray-to-optical conversion of the scintillator to be 58/keV. The OCE, fill factor, and optical conversion factor can all be combined to a single 
gain factor $\Omega=0.64\times0.57\times58$, so that the x-ray spectrum incident on the detector can be related to $\textbf{I}_{i}$ via
\begin{equation}
\textbf{I}_i = \Omega\sum_{k=1}^{N_E}E_k \textbf{I}^\text{x-ray}_{i,k},
\end{equation}
where $E_k$ is the energy in keV of the $k$th energy bin, and $\textbf{I}^\text{x-ray}_{i,k}$ is the incident number of x-rays at energy $E_k$ in location $i$. We take 
each $\textbf{I}^\text{x-ray}_{i,k}$ to be Poisson distributed, with Var\{$\textbf{I}^\text{x-ray}_{i,k}$\} = $\overline{I^\text{x-ray}_{i,k}}$, where the mean x-ray 
count data  $\overline{I^\text{x-ray}_{i,k}}$ can be obtained through noiseless simulation.
The resulting model for the line integral data variance is then given by
\begin{equation}
\left(K_g\right)_{i,j}=
	\begin{cases}
	\frac{\sigma^2_e + \Omega^2\sum_k E_k^2 \overline{I^\text{x-ray}_{i,k}}}{\left(\Omega\sum_k E_k \overline{I_{i,k}^\text{x-ray}}\right)^2} &: i=j \\
	0 &: \text{else}.
	\end{cases}
\label{cov}
\end{equation}
where $\left(K_g\right)_{i,j} = Cov\left\{g_i,g_j\right\}$.
In order to simulate a fluence which would result in the same mean glandular dose as two-view mammography, we approximate the total incident number of x-rays
based on the results for the photon fluence at the axis of rotation given in Ref. \cite{boone_technique_2005}, which range from approximately $3\times 10^7$ to 
$4 \times 10^8$ photons/mm$^2$ for breast diameters between 10 cm and 16 cm. This photon fluence is equally distributed among each projection view. 

While Eq. \ref{cov} describes the modeled additive noise in the projection data domain, the noise in the image domain is correlated by the reconstruction algorithm. 
In our case, we employ the FBP algorithm, whose action can be fully described by multiplication with a matrix $A$. The image covariance matrix is then 
given by (see Eq. 8.50 in Ref. \cite{barrett_foundations_2004}):
\begin{equation}
\label{imcov}
K_y = AK_gA^T,
\end{equation}
where $A^T$ denotes the matrix transpose of the reconstruction operator (not the discretization of the continuous adjoint operator). 
Recall that the image and sinogram can be represented by one-dimensional vectors. In this way, the image covariance and reconstruction operator can be coded as two-index matrices.

For large systems, where $A$ cannot directly be computed and stored,
$A$ represents the action of all the linear steps in the FBP algorithm and a separate
implementation for $A^T$ is needed. For the present studies, the ROI is small enough
that $A$ can be directly computed column-by-column and stored in computer memory.
Depending on the ROI size used, the dimensions of $A$ in this study range roughly from 16 $\times$ 400
to 7,000 $\times$ 80,000, with the number of columns varying because we consider only the shadow of the ROI projected onto the detector.
With the actual matrix $A$ available $A^T$ is trivial to compute and 
the covariance $K_y$ is obtained directly by matrix multiplication.
In general $K_y$ is rank-deficient and we use its Moore-Penrose inverse, $K_y^\dagger$.
Computation of $K_y^\dagger$ can be obtained by singular value decomposition (SVD).
With the SVD, the covariance is written as
\begin{equation}
\label{SVD}
K_y = U \Sigma V^T,
\end{equation}
where $U$ and $V$ are orthogonal matrices, and for the present case, where $K_y$ is always symmetric and positive semi-definite, $U=V$. The matrix $\Sigma$ is diagonal containing the singular values of $K_y$.
The Moore-Penrose pseudo-inverse becomes
\begin{equation}
\label{SVDi}
K_y^\dagger = V \Sigma^\dagger U^T,
\end{equation}
where $\Sigma^\dagger$ is diagonal, containing the multiplicative inverse of the non-zero singular values
and zero elements of $\Sigma$ remain zero in $\Sigma^\dagger$.

\begin{figure}
\centering
\includegraphics[width=0.6\columnwidth]{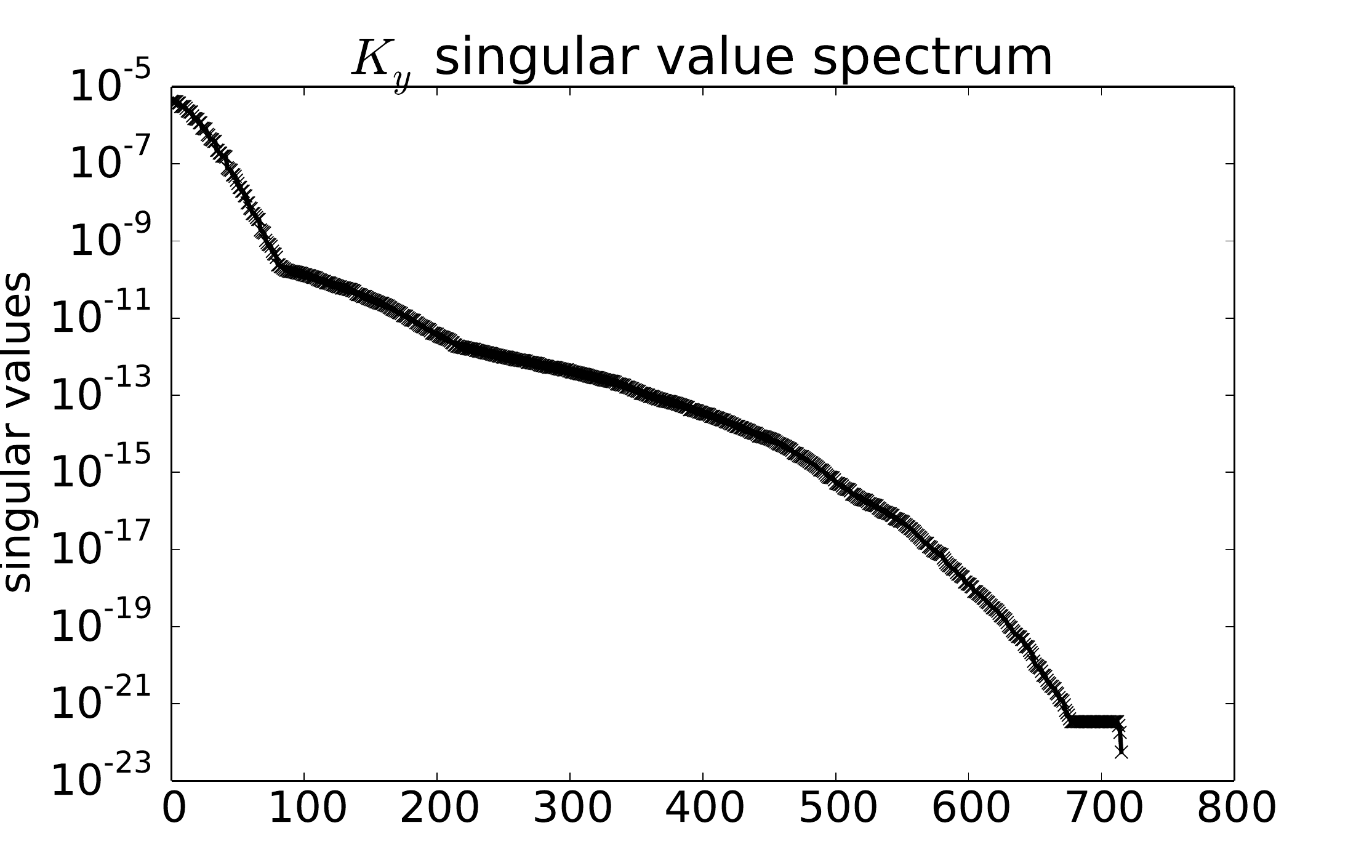}
\caption{The singular value spectrum for the matrix $K_y$ is shown for one combination of parameters investigated. The smallest singular values shown are within numerical precision of 0, and $K_y$ is, therefore, rank-deficient. \label{svdfig}}
\end{figure}

For some configurations investigated, rank-deficiency enters into $K_y$ through $A$, which can be rank deficient when image pixel size is mismatched to the size of the detector pixels or 
through a filtering operation on the data. An example of the singular value spectrum for $K_y$ is shown in Fig. \ref{svdfig} for a 5 $\times$ 5mm$^2$ ROI located 5.7cm from the center of the field of view.
The plateau of small singular values are within numerical precision of 0.

\subsection{Classification Tasks}
\label{sec:tasks}
\begin{figure}[h]
\centering
\includegraphics[bb = 105 111 762 440, width=0.4\columnwidth, clip=True]{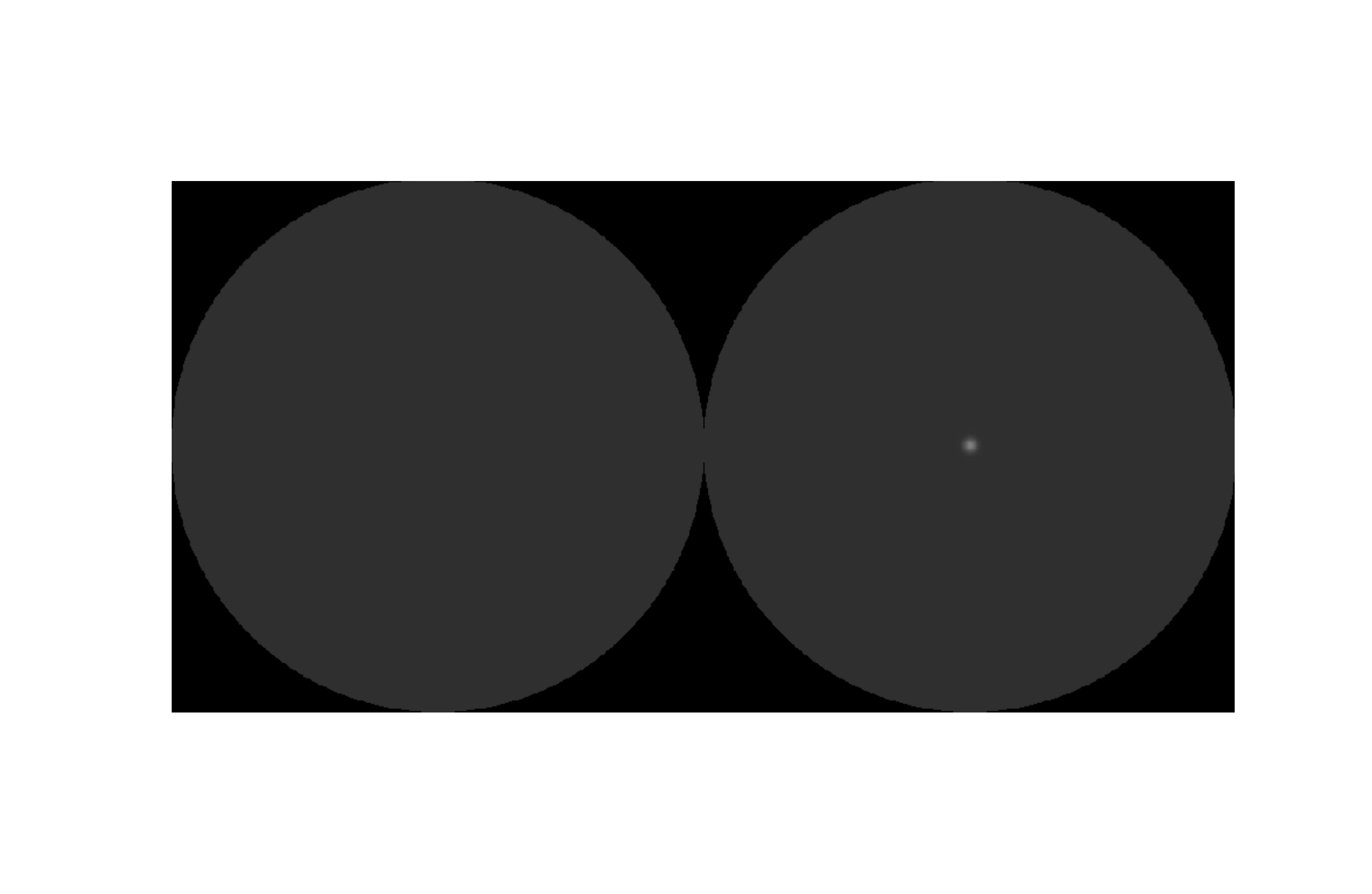} \\
\includegraphics[bb = 73 70 524 345, width=0.4\columnwidth, clip=True]{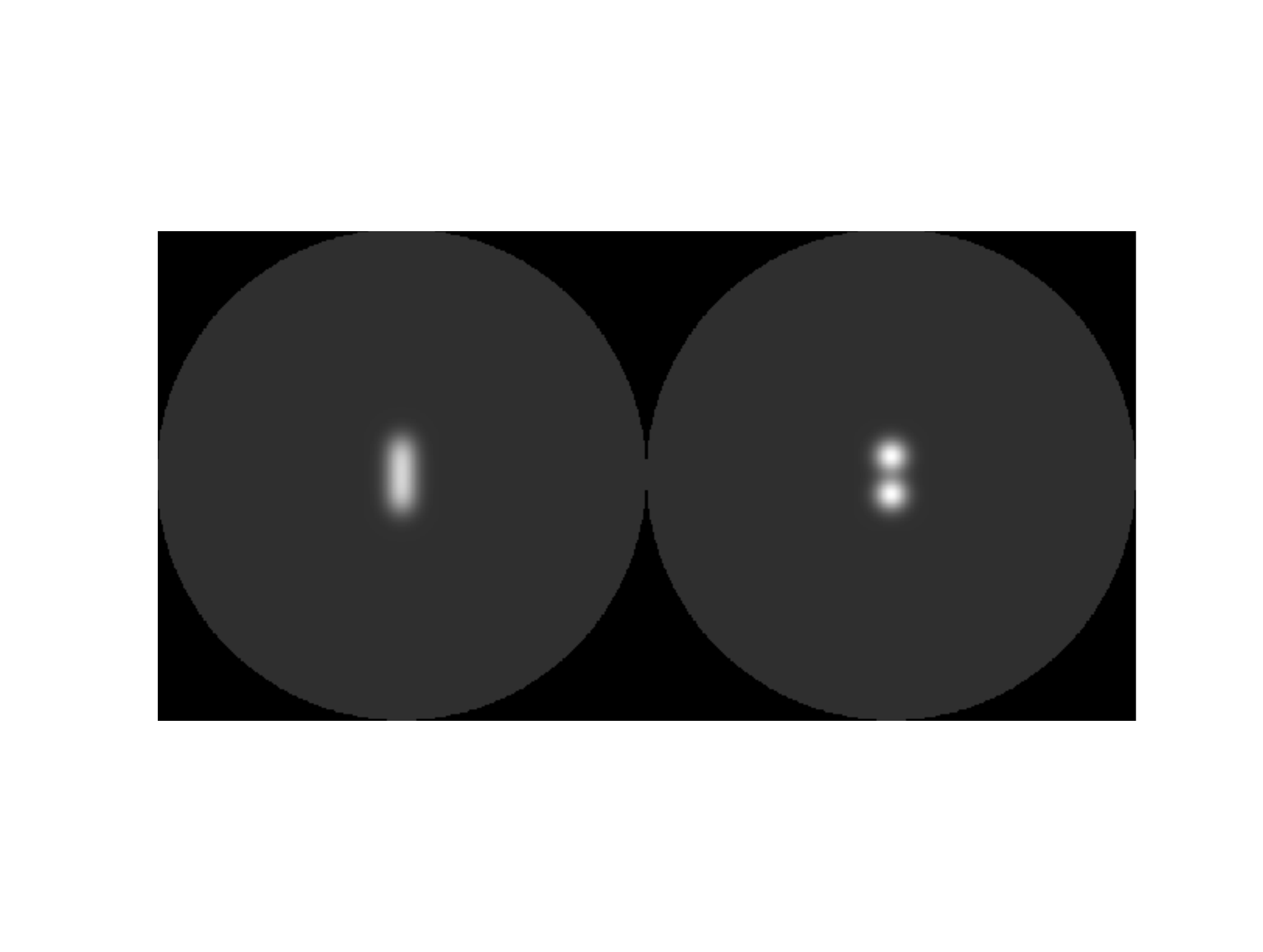}
\caption{Top Row: The numerical phantoms corresponding to the two hypotheses for the signal detection task, namely ``signal-absent'' and ``signal-present''. 
Bottom Row: The two hypotheses for the Rayleigh task, ``one object'' and ``two objects''. The width of each image is approximately 5 mm. 
Note that the simulated microcalcification is much smaller than the Rayleigh signals because the Rayleigh task must span several pixels in order to be feasible, whereas 
the microcalcification can be detected by the HO when it spans only one or two pixels. The display window is [0, 0.122] mm$^{-1}$.
\label{fig:tasks}}
\end{figure}

As in the preceding chapter, we focus on optimizing parameters with respect to 
two relevant tasks: the detection of microcalcifications and the resolution of two high-contrast signals in close proximity, known as the Rayleigh discrimination task
\cite{wagner_multiplex_1981, hanson_rayleigh_1991, myers_rayleigh_1998}.
Both of these tasks address different forms of an image noise-resolution trade-off, and they involve single slice viewing.
This latter point facilitates studies relating human observers to the HO, because observer studies can be conducted with single
or pairs of images and there is no need to consider slice scrolling or volume rendering, which would be needed to present
volume data in a human consumable format.
Each of these tasks can be seen as a two-class classification task, where one attempts to classify an image 
as corresponding to one of two hypotheses. In the case of microcalcification detection, the two hypotheses correspond to ``signal-absent'' and ``signal-present.'' In the 
case of the Rayleigh task, the two hypotheses are ``one elongated object'', namely a blurred line, or ``two distinct objects.''  
The blurred points and line used in the Rayleigh discrimination task are assigned a peak attenuation 
value equal to that of calcium. The sizes of the Rayleigh task signals are made proportional to those described in Ref. \cite{myers_rayleigh_1998}.
The numerical phantoms corresponding to each hypothesis are shown in Fig. \ref{fig:tasks} for 
each of the two tasks. The difference in scale between the two tasks is a result of the fact that the Rayleigh task must span several pixels so that the ``trough'' between 
the high-contrast signals is visible. Meanwhile, a detectable microcalcification can be much smaller. 

\subsection{Hotelling Observer Metrics}
\label{sec:HO}

The HO model used is the same ROI-HO model presented in the previous chapter. The HO defined within an ROI is the optimal linear observer if we assume that only the pixels within the ROI are known. Equivalence between the HO in an ROI and the HO for the full image depends on whether the ROI contains the full signal extent as well as relevant noise correlations. Regarding the former point, while under-sampling artifacts are present in $\Delta \bar{y}$, we intentionally exclude this region of the image in the task model so that the HO is not performing the task based on artifacts. As for the latter concern, we have consistently found that the ROIs determined using the proposed method are larger than the typical pixel correlation length.

In the following, we consider the number of projections to be at most 500 views. Therefore, for the ROI sizes investigated, ranging from 4$\times$4 to 84$\times$84, the reconstruction matrix $A$ ranges in size up to roughly
7,000$\times$80,000. Meanwhile, the covariance matrix $K_y$ ranges from 16$\times$16 to 7,000$\times$7,000. The method proposed remains numerically 
feasible for ROI sizes up to roughly 100$\times$100.
We point out that generalization to volume imaging is straight-forward;
the ROI becomes a volume of interest (VOI) and the detector data at each view become two-dimensional. The HO theory
remains the same, but the scale of the computation increases.

\noindent We briefly explain each of the studies performed:

\subsubsection{Patient Parameter Studies}
\label{sec:patient}
In order to illustrate the use of HO $P_C$ as a metric for image quality, we examine the impact on task performance of two relevant physical parameters 
which affect the original projection data: signal location within the breast, and breast diameter. Three radial positions are considered: 0 cm, 2.8 cm, and 5.7 cm from 
the axis of rotation, and curves of 
HO $P_C$ were swept out for a range of microcalcification sizes and Rayleigh task separations. The phantom size for the radial position study is 14 cm in diameter, 
and a Hanning filter
of width 1.0$\nu_N$ is used in reconstruction onto pixels of width 0.162 mm. The Hanning filter 
is defined as
\begin{equation}
\begin{split}
H(\nu) = &0.5 + 0.5\text{cos}\left(\nu/\nu_C\right) \hspace{0.75cm} 0\le |\nu| \le \nu_C, \\
= &0  \hspace{4.3cm} \text{otherwise}
\end{split}
\end{equation}
where $\nu_C$ is the cutoff frequency defining the filter's width.

The impact of breast diameter on image quality is investigated by computing HO $P_C$ for a range of microcalcification sizes and Rayleigh separations for breasts 
between 10 cm and 16 cm in diameter. The results are generated first for a fixed exposure level, and subsequently for a fixed average glandular dose, where 
x-ray fluence for a fixed average glandular dose is again obtained from Ref. \cite{boone_technique_2005}. The values of HO SNR$^2$ are computed at 
the three different signal locations mentioned above and averaged, so that the final $P_C$ is computed from the average HO SNR$^2$ for the three locations. 

\subsubsection{Impact of View Number}
\label{sec:system}
In order to study the impact of the number of projection views (at a fixed total dose), the size 
of the microcalcification is set to 100 $\mu$m, and the Rayleigh task separation is set to 0.3 mm. These signal sizes are chosen since they were typically shown 
to result in a HO $P_C$ value of near 95 \%, corresponding to a possible but not trivial task. 50, 100, 200, and, 500 view cases are investigated, and for each of these 
cases, a range of reconstruction filter widths are considered (see below). 

\subsubsection{Reconstruction Parameters}
\label{sec:recon_parms}
\begin{figure}[h]
\centering
\includegraphics[width=.8\linewidth]{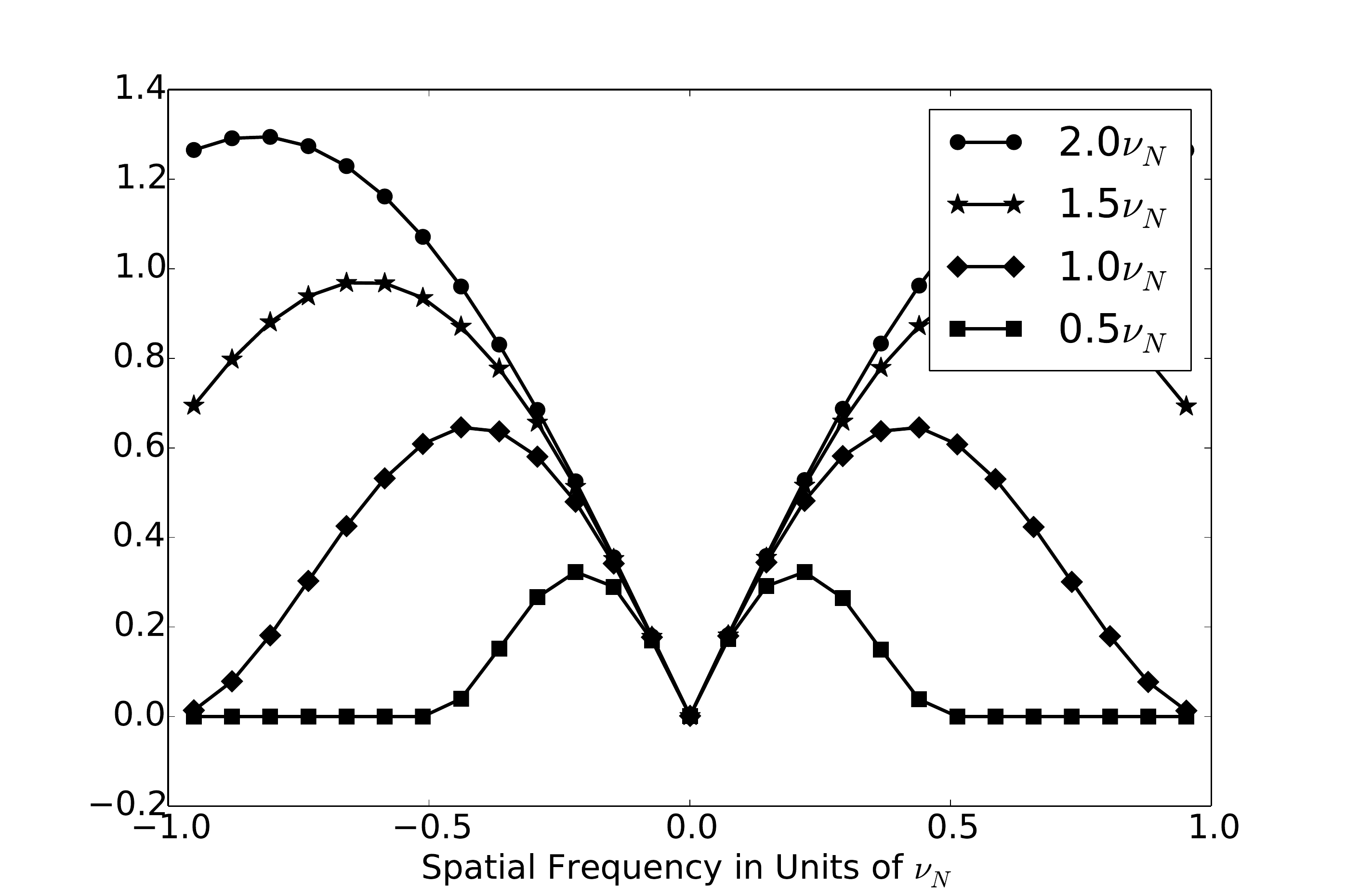}
\caption{
Examples of the impact of various Hanning filter cutoff frequencies on the apodized ramp filer (frequency domain).
\label{fig:kernels}}
\end{figure}

The Hotelling efficiency $\varepsilon$ is used to determine the image quality impact of two reconstruction parameters: image pixel size
and Hanning window width. HO efficiency is computed for a range of signal sizes for two image pixel
widths, 0.162mm and 0.384mm. For this study and the view number study, we consider the  performance of the HO averaged at three locations within the breast.
 The HO efficiency is then computed for 
a Hanning cutoff ranging between 0.125 $\nu_N$ and 2.0 $\nu_N$, where $\nu_N$ is the Nyquist frequency. The Fourier domain 
representation of the corresponding apodized ramp kernels is shown in Fig. \ref{fig:kernels}.

\subsubsection{Comparison to Subjective Assessment}
After using the HO methodology to determine an optimal set of acquisition and reconstruction parameters, we reconstruct simulated images of several numerical 
phantoms: a simulated breast phantom containing microcalcifications, a breast phantom with a Rayleigh signal, and simulated high- and low-contrast resolution 
quality assurance phantoms, based roughly on the \emph{Catphan} phantom (The Phantom Laboratory, Salem, NY). Example reconstructions are also computed 
for reconstruction filter widths and view numbers which are non-optimal, so that the scale of refinement achieved through HO optimization can be seen qualitatively. 

\section{Results - Patient and System Parameters}
\label{sec:results1}

\subsection{Breast diameter study}

\begin{figure}[h]
\centering
\includegraphics[width=.48\columnwidth]{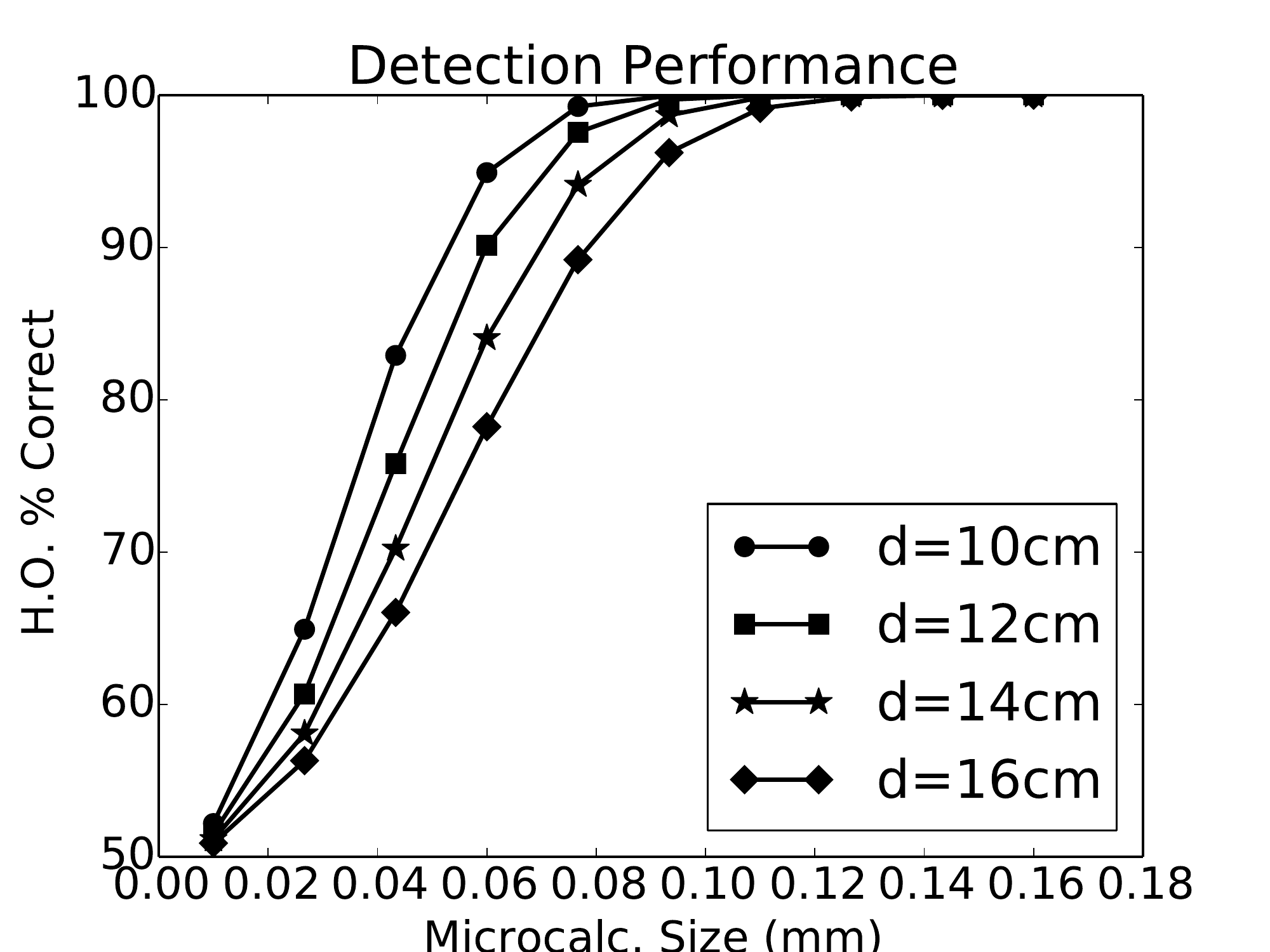}   \includegraphics[width=.48\columnwidth]{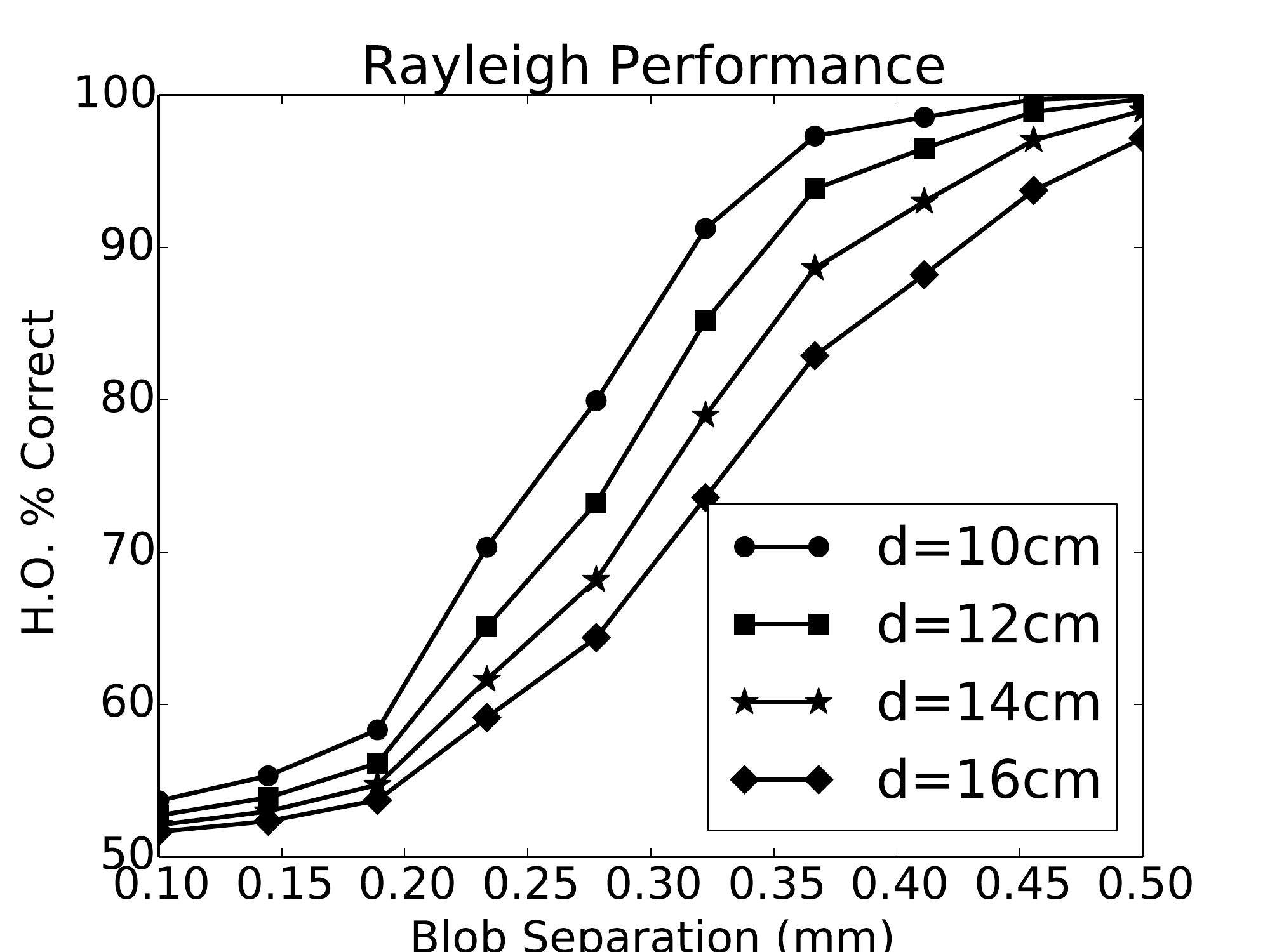}
\caption{\textbf{Left:} The HO $P_C$ as a function of microcalcification diameter for a range of breast sizes and a constant x-ray exposure. The quantity d in the legend 
is the breast diameter simulated. In a 2AFC trial, the 95\% correct performance level of the HO would range from microcalcifications of roughly 60 $\mu$m 
for a 10cm breast, versus roughly 100 $\mu$m for a 16cm breast. 
\textbf{Right:} The HO $P_C$ as a function of the separation between the blobs in the Rayleigh task. As expected, for reasonable task performance, the scale of the 
Rayleigh signal is substantially larger than that of a detectable microcalcification, as 
reflected in the left-hand figure. For the breast diameters studied here, the separation at 95\% correct ranges from roughly 0.35mm to 0.5mm.  
\label{fig:1}}
\end{figure}

Figure \ref{fig:1} illustrates the impact of breast size on HO performance for the two tasks considered in this work, microcalcification detection (Left) and 
Rayleigh discrimination (Right). Exposure for these results is fixed so that the mean glandular dose to the 14cm diameter breast is roughly 
equal to that of two-view planar mammography. 
Clearly, since exposure is fixed in these results, 
larger breasts lead to poorer task performance, as an increasing amount of noise is inherent in the projection data. This finding is in keeping with the results of Yang et al. \cite{yang_noise_2008}.
For a 14 cm breast diameter, these results indicate that 90$\mu$m microcalcifications are the smallest microcalcifications 
which can be detected with 95\% accuracy, while microcalcifications larger than 120$\mu$m can likely be detected with near perfect accuracy. Gong et al. found that using a similar system 
in simulation resulted in a human observer performance of greater than 95\% correct in detection of microcalcifications with diameter equal to or greater than 175$\mu$m 
\cite{gong_microcalcification_2004}. Given that the HO can be expected to establish an upper bound on human observer performance, a 95\% accuracy in detection of 90$\mu$m 
microcalcifications 
is reasonable, particularly in light of the differences in system simulation and observer experiment methodology between the current study and the study by Gong et al. Specifically, 
one would expect the inclusion of optical light spread, which was not considered in our work, to lessen the gap between the two results. 

The results of the Rayleigh discrimination experiment illustrate that for a 14cm diameter breast, two high-contrast objects can be well resolved by the HO (with 95 \% accuracy)
when their centers are separated by roughly 0.45mm. This separation is significantly larger than the smallest detectable microcalcification, simply because the Rayleigh task is 
limited by pixelization in the image, whereas microcalcification detection is still possible even when the image pixel size is somewhat larger than the microcalcification to be detected. 
The image pixel size for these results is 0.162mm, and the HO can only consistently perform the Rayleigh discrimination task once the separation between objects 
spans several pixels.

\begin{figure}[h]
\centering
\includegraphics[width=.48\columnwidth]{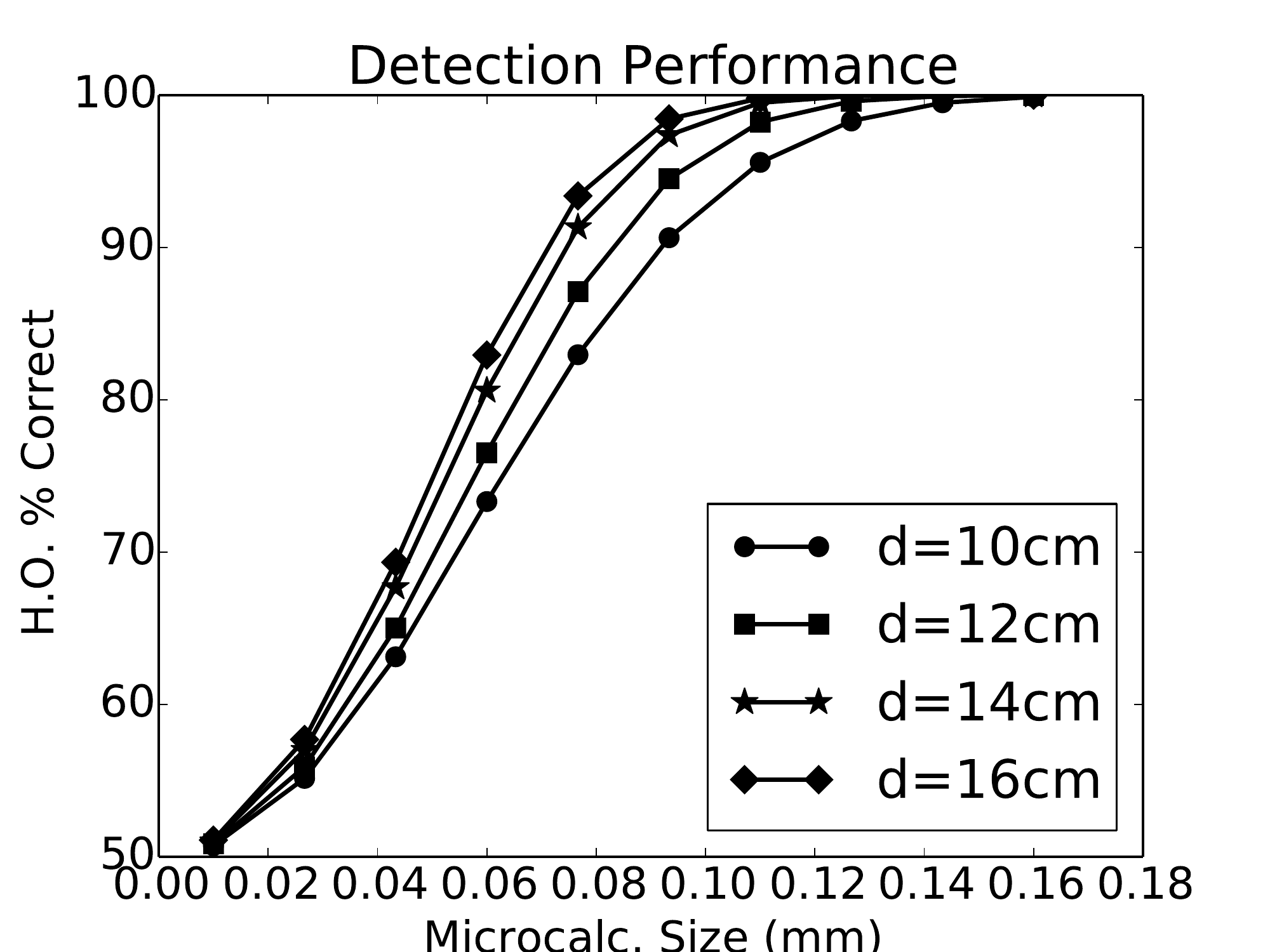}   \includegraphics[width=.48\columnwidth]{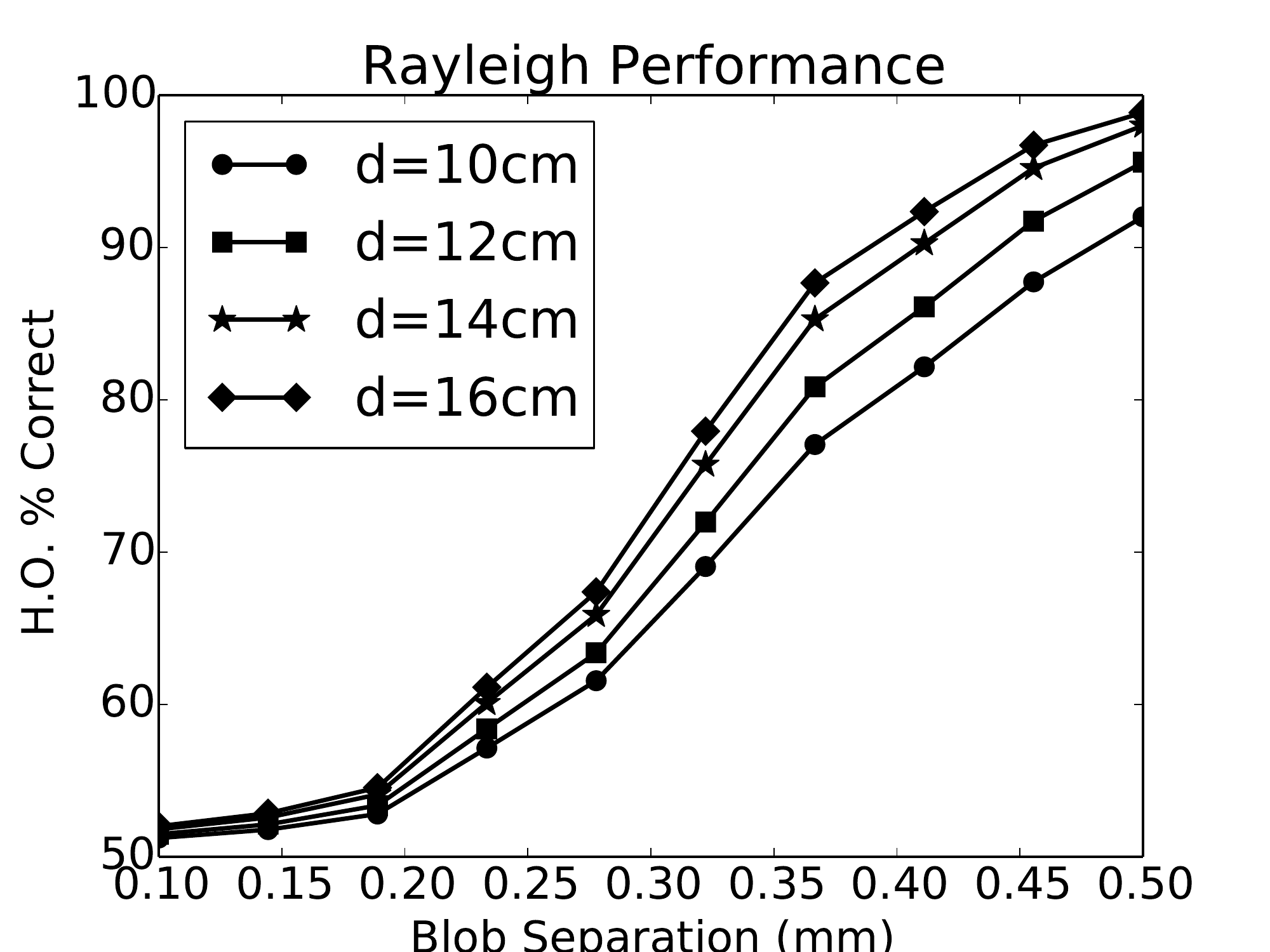}
\caption{\textbf{Left:} The HO $P_C$ as a function of microcalcification diameter for a range of breast sizes and constant breast AGD. These results are 
inverted relative to the corresponding results for constant exposure. This is likely due to the increased noise power for smaller breasts 
at constant dose which, as described in \cite{kwan_evaluation_2006}, is a consequence of the physical relationship between photon flux 
and breast diameter at a constant AGD. 
\textbf{Right:} Shown is HO $P_C$ as a function of the separation between the blobs in the Rayleigh task for various breast diameters with AGD held 
constant. These results mirror those shown in the left-hand figure.
\label{fig:2}}
\end{figure}

Figure \ref{fig:2} illustrates the impact of breast size on task performance when the radiation dose is held constant, equal to a two-view mammogram. 
Again, these results correspond closely to those from Yang et al. \cite{yang_noise_2008}, who demonstrated an increase in image noise power with 
decreasing breast size, which would lead to a decrease in task performance. That work demonstrated a noise level for smaller breasts which was 
nearly an order of magnitude higher for a 10.4cm breast compared to an 18.4 cm breast. While our results do reproduce the same trend observed
in that work, the noise penalty for smaller breast size does not appear to be as dramatic in our simulations.

\subsection{Signal location study}

\begin{figure}[h]

\centering
\includegraphics[width=.48\columnwidth]{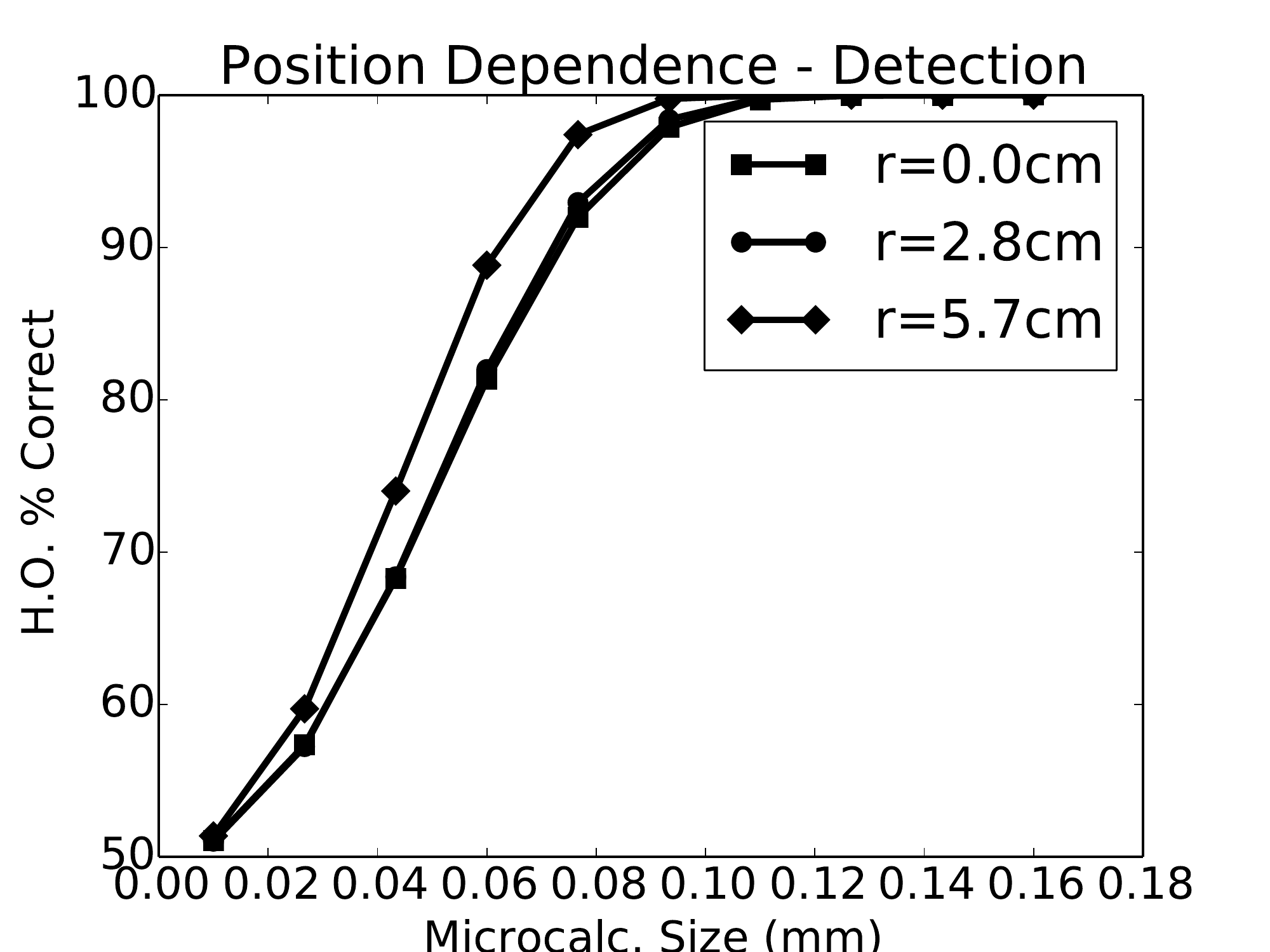} \includegraphics[width=.48\columnwidth]{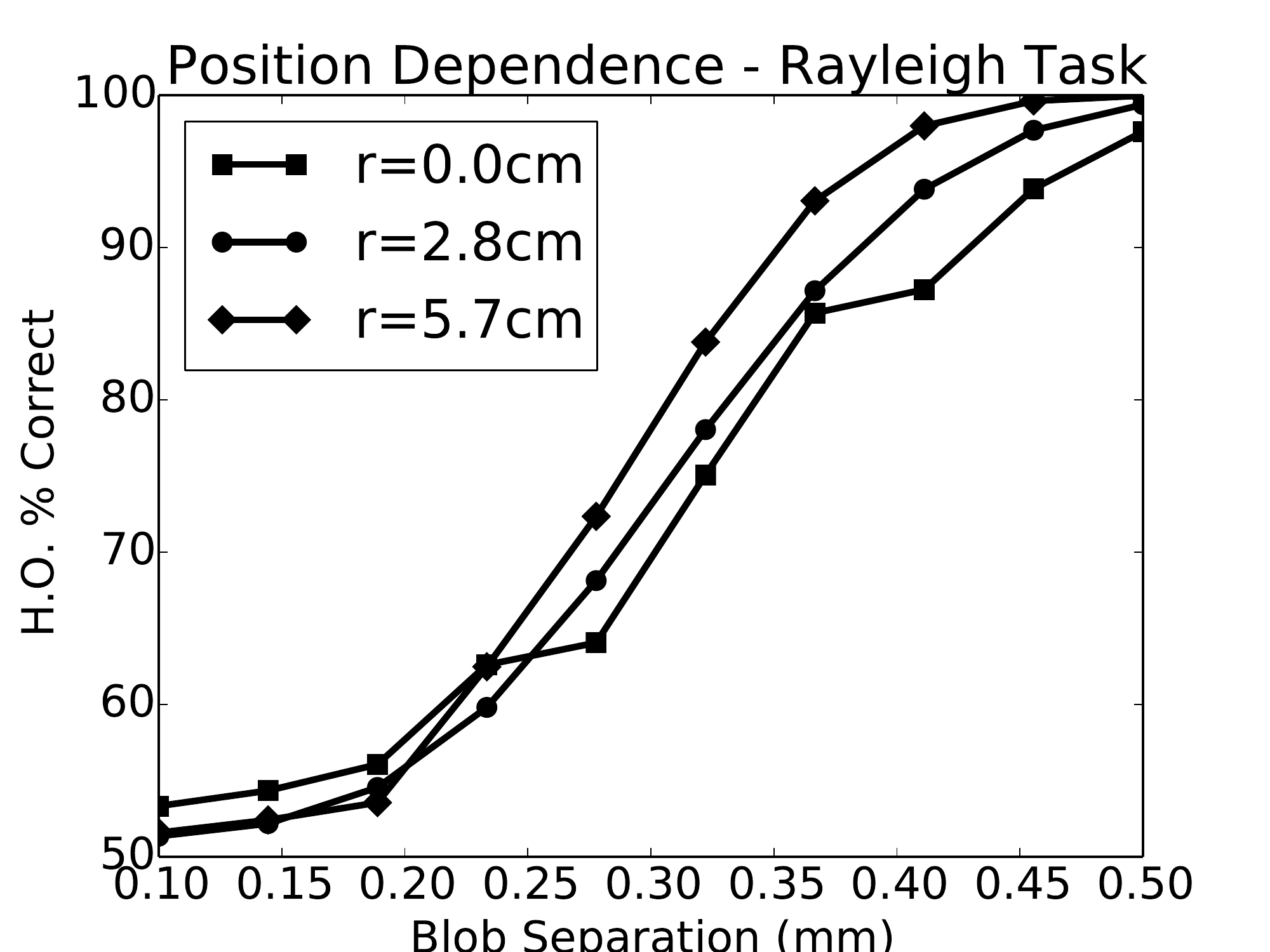}
\caption{\textbf{Left:} Plots of HO $P_C$ as a function of calcification size for a range of signal locations.
The radial position $r$ denotes the distance of the signal from the axis of rotation. Note that a bowtie
filter was not simulated, and the noise at the periphery of the phantom is therefore reduced relative
to the center of the FOV.
\textbf{Right:} Corresponding results to the left hand figure, but for Rayleigh discrimination.
\label{fig:location}}
\end{figure}

Figure \ref{fig:location} illustrates the effect of radial distance from the axis of rotation on the HO task performance for microcalcification detection (Left) and 
Rayleigh discrimination (Right). A slight increase in performance is seen for signals located away from the axis of rotation for the microcalcification and Rayleigh tasks, and this effect is seen 
for the full range of signal sizes investigated here. 
Kwan et al. \cite{kwan_evaluation_2006} demonstrated a significant reduction in resolution measure via the MTF 
with increasing radial distance from the axis of rotation. For microcalcification detection, our result therefore demonstrates that the location-dependent image noise 
properties have a stronger impact on task performance than image resolution. Note that a bowtie filter was not simulated in our results, so that image noise is lower in the 
periphery of the image.


\subsection{Image pixel size}

\begin{figure}[h]
\centering
\includegraphics[width=.48\columnwidth]{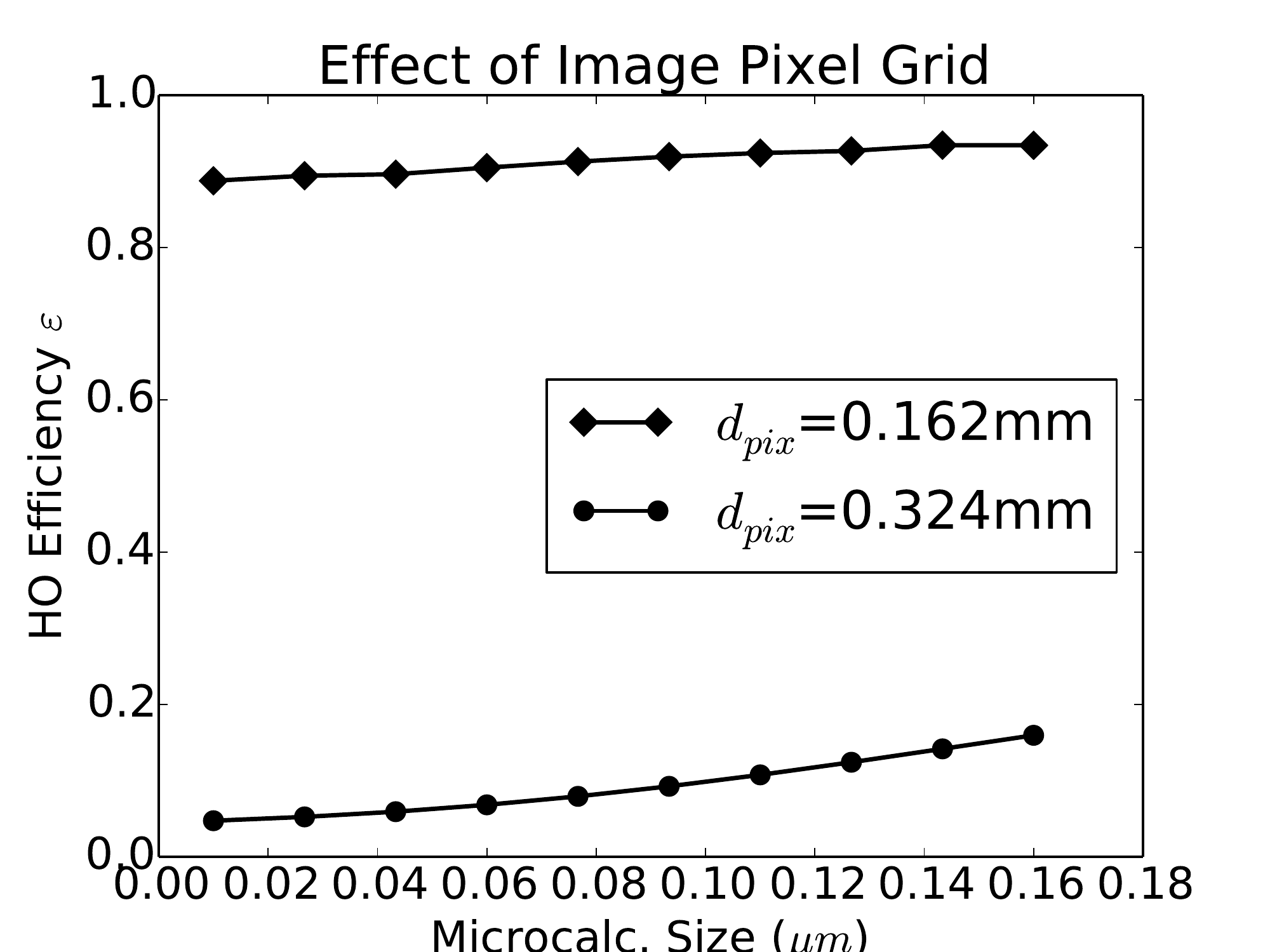} \includegraphics[width=.48\columnwidth]{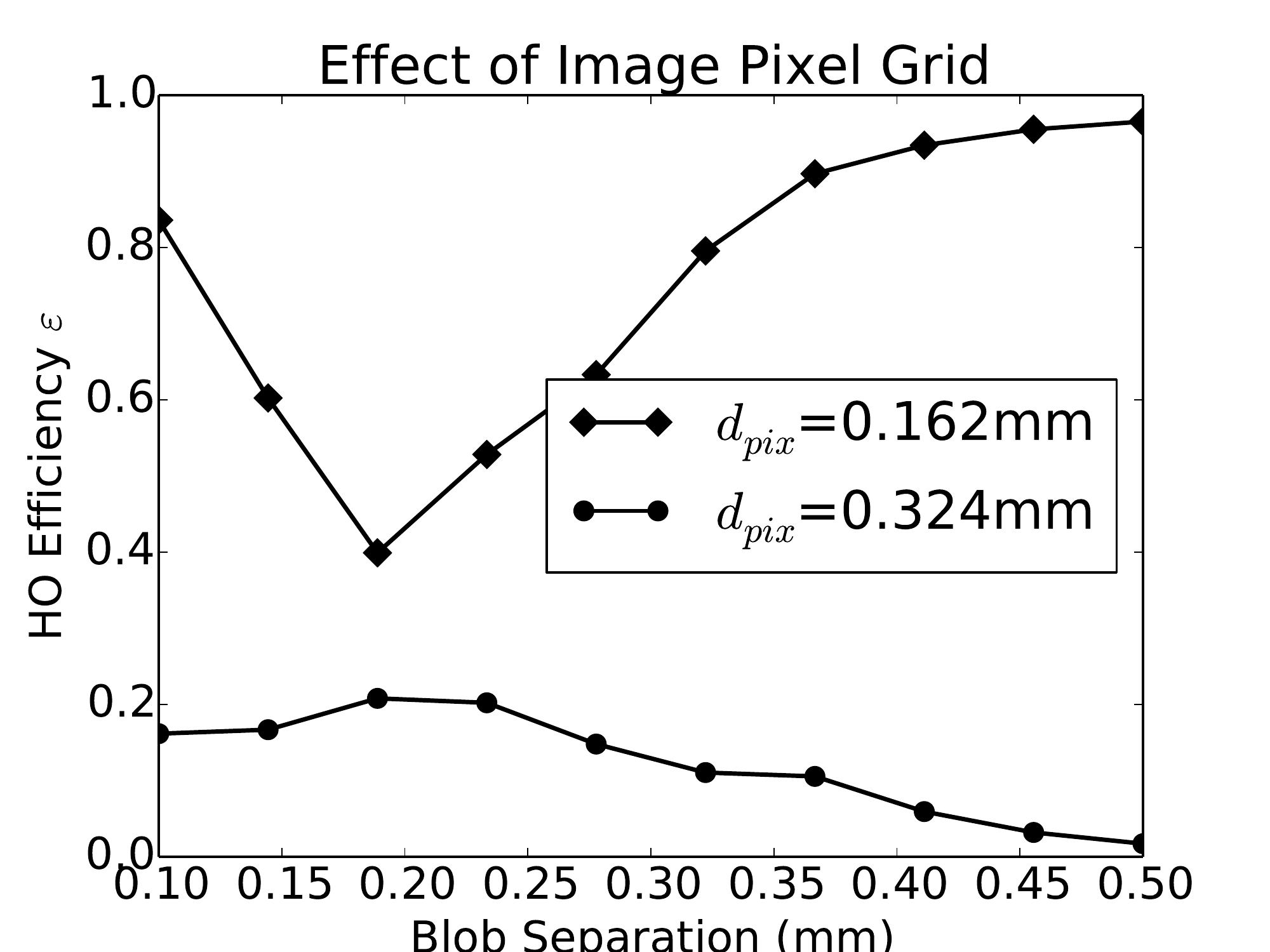}
\caption{\textbf{Left:} The HO efficiency $\varepsilon$ is shown as a function of microcalcification diameter for two image pixel sizes: 0.324mm and 
0.162mm. In general, smaller pixels result in greater algorithm efficiency in preserving detectability. Further, as the signal decreases in 
spatial extent, the benefit of finer pixelization is diminished since for these small signals, in either case the majority of the signal's intensity is contained in a single pixel.
\textbf{Right:} Shown are the HO efficiency for two image pixel grid sizes for the Rayleigh discrimination task. See text for discussion.
\label{fig:4}}
\end{figure}

\begin{figure}[h]
\centering
\includegraphics[width=.48\columnwidth]{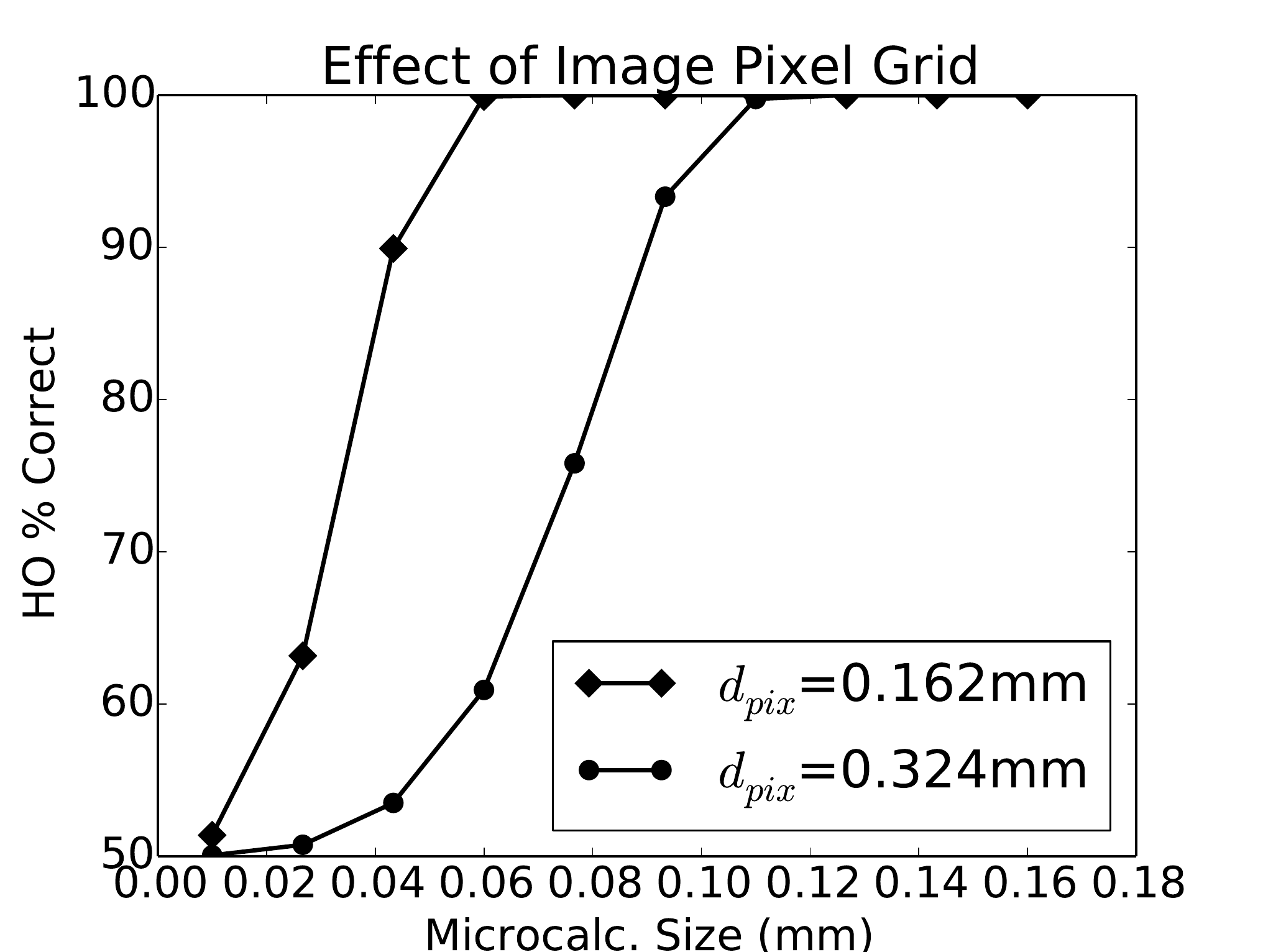} \includegraphics[width=.48\columnwidth]{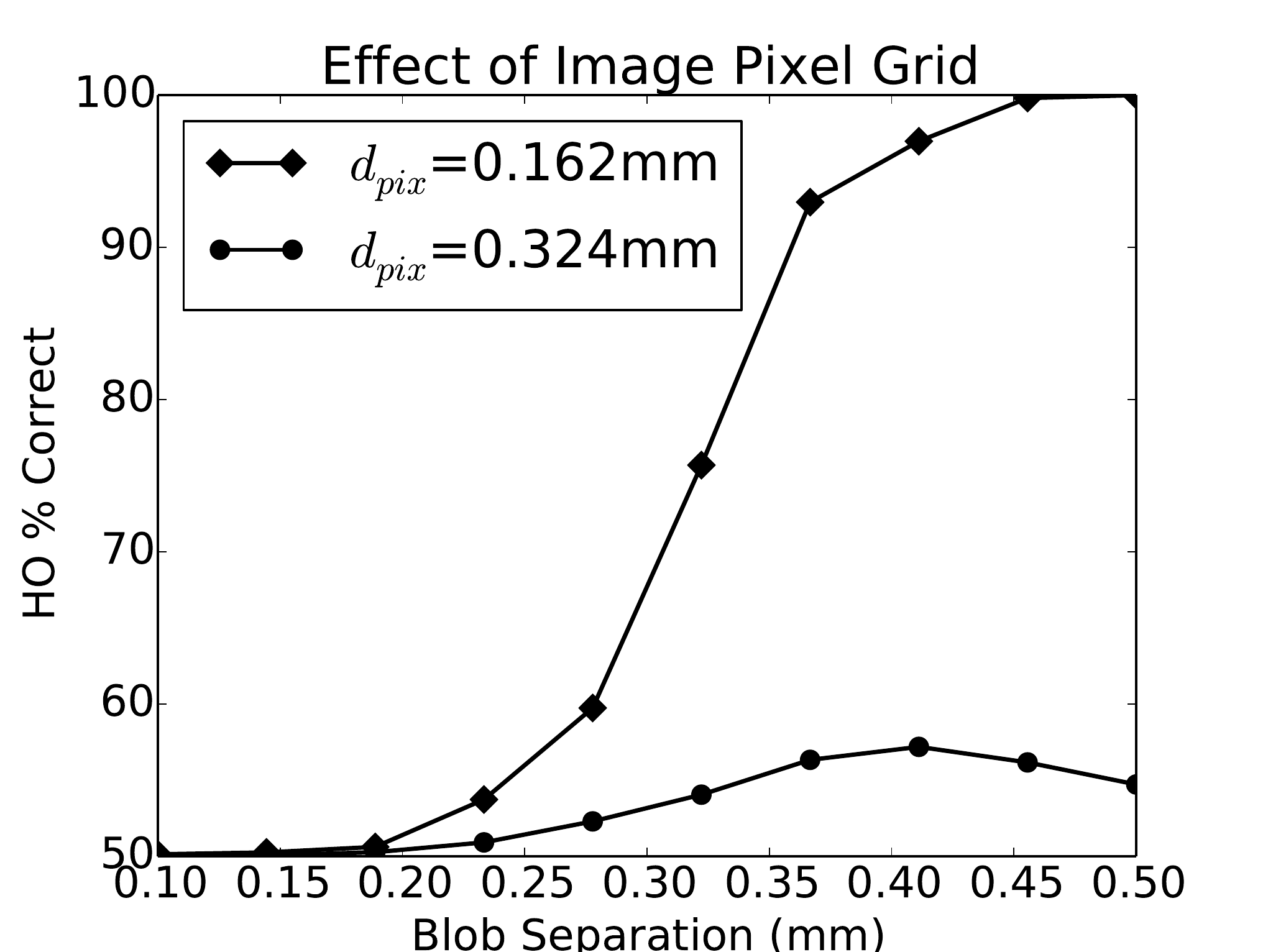}
\caption{\textbf{Left:} The HO percent correct is shown for the same microcalcification sizes and pixel sizes as in Fig. \ref{fig:4}
\textbf{Right:} Same as left hand figure, but for the Rayleigh discrimination task.
\label{fig:4b}}
\end{figure}

Figure \ref{fig:4} illustrates the impact of image pixel size on HO efficiency for a range of detection and Rayleigh tasks. Clearly, finer pixelization in the image grid 
results in a substantial benefit in HO efficiency for the full range of investigated microcalcification sizes (Left). 
For the Rayleigh discrimination task, there is also a pronounced benefit for smaller pixels, which is to be expected since the separation of the Rayleigh signals is less 
than 2 pixels when 0.324mm pixels are used. 
The jump in efficiency for object separations 
below 0.15mm on 0.162mm pixels is an artifact of the nature of the Rayleigh discrimination task changing once the two objects are in adjacent pixels. 
While the algorithm regains some efficiency 
in this signal regime, this is immaterial since the HO SNR which is being preserved from the projection domain is low to begin with. In other words, although the algorithm 
becomes slightly more efficient in this domain, the task is essentially impossible regardless. For an illustration of this point, see Figure \ref{fig:4b}, which shows the HO 
percent of correct decisions for the same image pixel sizes and tasks as are shown in Figure \ref{fig:4}. 

These results stand in contrast to those from Kwan et al. \cite{kwan_evaluation_2006}, 
who evaluated the impact of image pixel size using MTF measurements. That study found that selection of a 512$\times$512 image pixel grid (0.324mm pixels) versus a 1024$\times$1024 grid 
(0.162mm pixels) had minimal 
impact on image resolution. 
Modeling of additional physical factors in our simulated signal, such as spread of optical photons in the CsI flat-panel detector, could account for the discrepancy in these findings. It is also possible that directly modeling the nickel-chromium wire used by Kwan et al. instead of a microcalcification could lead to somewhat different results. Finally, the presence of correlated noise in the images could impact task performance for the HO, facilitating improved performance with smaller pixels. This effect would not be captured by MTF measurements.

\subsection{View Number Study}
\begin{figure}[h]
\centering
\includegraphics[width=.48\columnwidth]{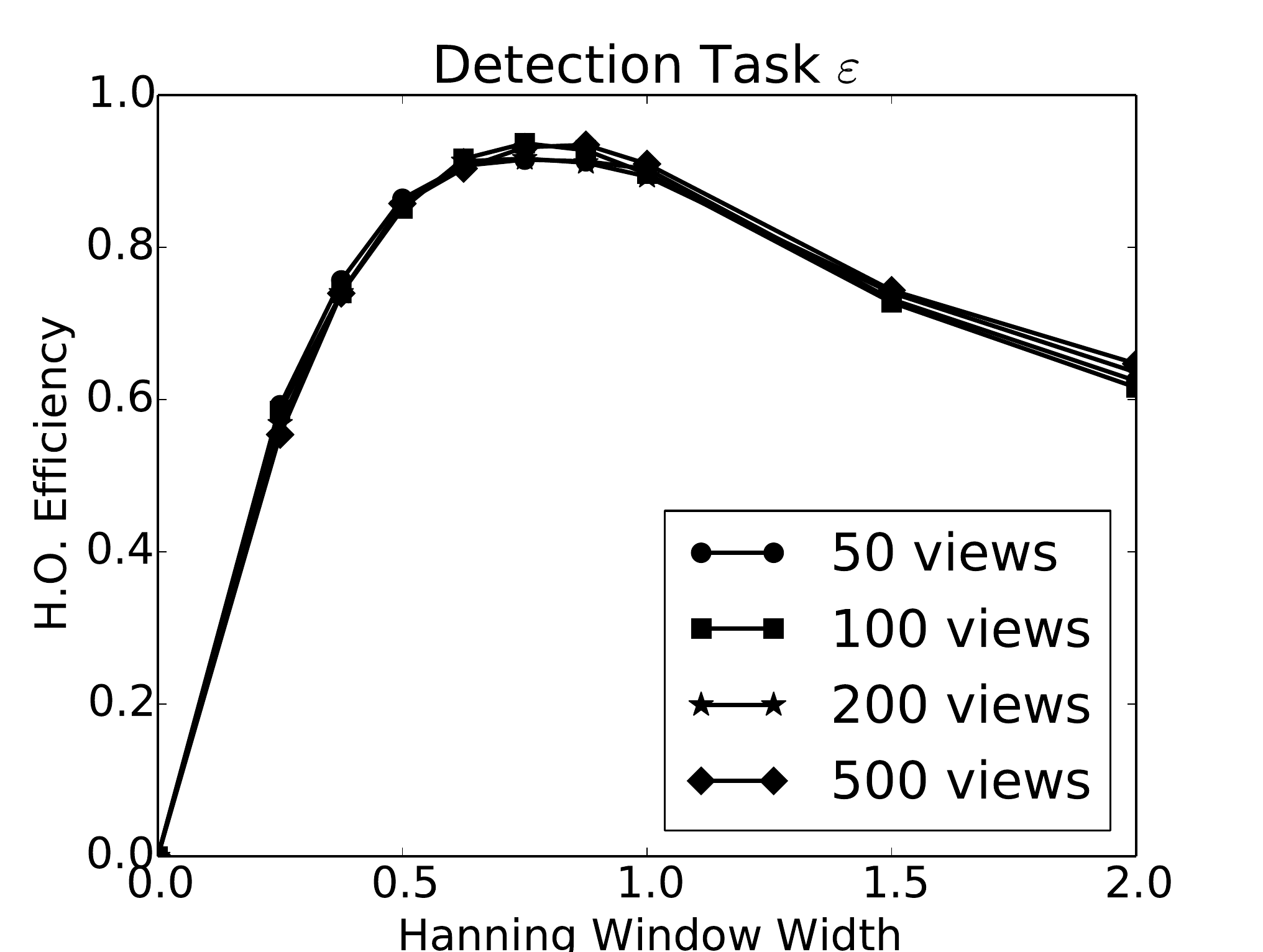} \includegraphics[width=.48\columnwidth]{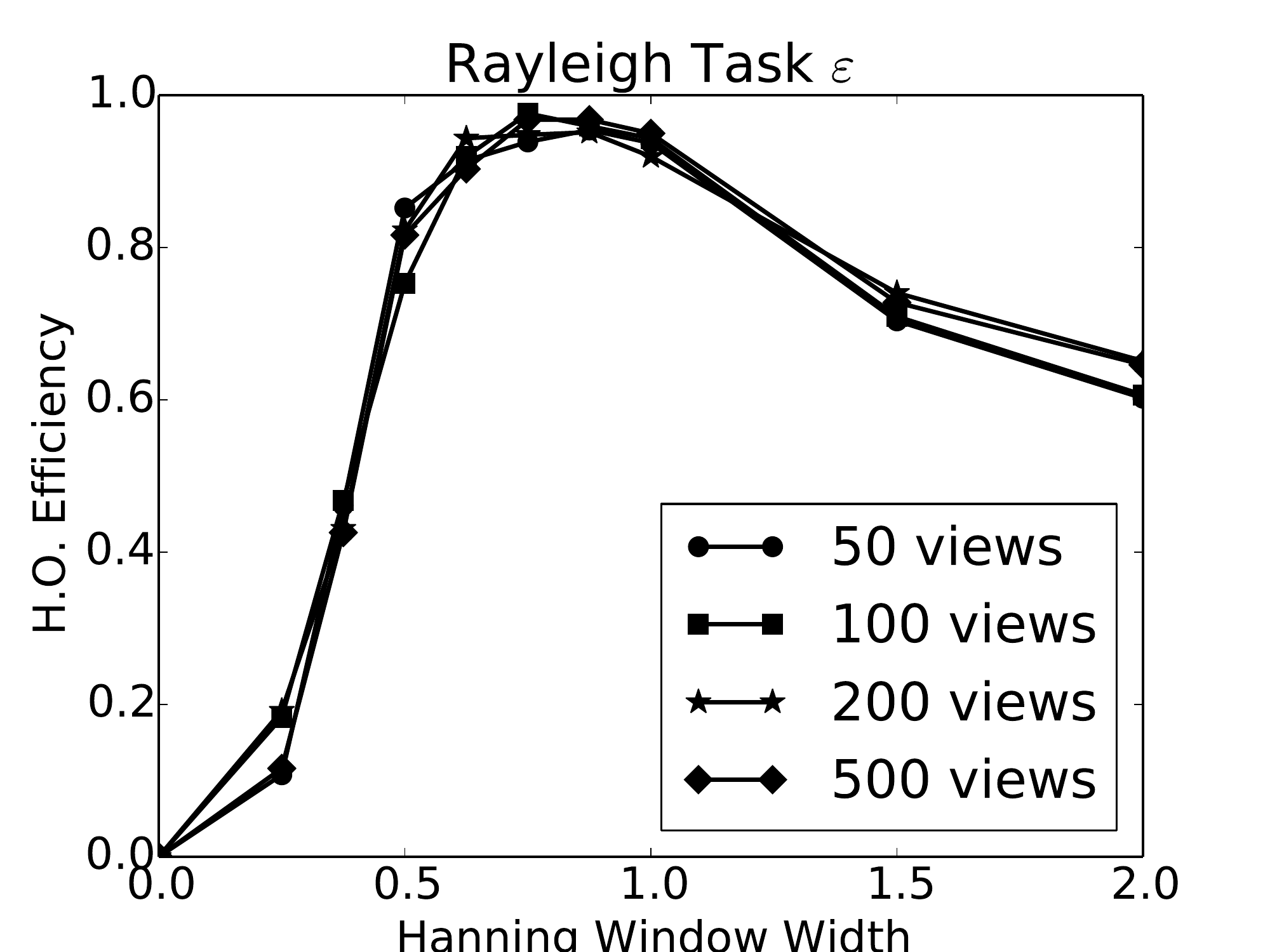}
\caption{
\textbf{Left:} Detection task HO efficiency $\varepsilon$ as a function of Hanning window width for a range of projection views from 50 to 500 and 0.162mm pixels
image pixels. All plots between 50 views and 500 essentially are overlaid. The total image dose is held constant.
\textbf{Right:} Rayleigh task HO efficiency versus the width of the Hanning window for 0.162mm image pixels. As 
before, there is little discernible difference for view numbers ranging from 50 to 500. 
\label{fig:5}}
\end{figure}

Figure \ref{fig:5} shows the impact of the number of projection views on the HO efficiency. The size of the microcalcification signal is held at 100$\mu$m, 
and the separation of the Rayleigh signal is set to 0.3mm. The total exposure for each acquisition is held fixed, so that the exposure per projection 
view varies. Kwan et al. predict improved resolution properties for larger numbers of views \cite{kwan_evaluation_2006}, while Yang et al. predict decreased image noise for 
fewer projection views \cite{yang_noise_2008}, so that the impact of view number on task performance in this case is unclear without a task-based approach to image quality 
evaluation. We find that when the signal of interest is at the center of the FOV, the noise benefits of restricting view number and the resolution benefits of increasing view number 
roughly cancel and lead to fairly uniform task performance for a wide range of projection view numbers. By contrast, as the 
signal moves toward the periphery of the FOV, a slight benefit is seen for lower view numbers. After averaging 
performance over the three signal locations, this effect is only noticeable for the Rayleigh task using 50 views and a wide 
Hanning filter. Since view number appears to only negligibly impact task performance in this case, the remaining results in this work correspond to 100 projection views in order to facilitate computational efficiency.

It is worth noting that our simulations neglect x-ray source motion, so that the 
under-sampling artifacts of an x-ray 
source operating in continuous mode are not captured in these results. In a study not presented here we find 
that such blurring from source motion becomes the dominant factor in task performance for 
signals near the FOV periphery. In a real system, this effect could be controlled somewhat by either using a step-and-shoot acquisition or a pulsed x-ray source.


\subsection{Hanning Filter Width Study}

Figure \ref{fig:5} also illustrates the effect of a Hanning filter on the performance of the two tasks. 
Moderate smoothing via the Hanning filter results in a modest improvement in HO efficiency for each task. Meanwhile, over-filtering impairs HO performance for the Rayleigh task more significantly than for the detection task, however the optimal filter widths for the two tasks seem to correlate closely. 
However, as with the investigation of view number, electronic noise could impact the outcome of this 
conclusion.  

\section{Discussion - Patient Parameters and System Optimization}
\label{sec:discussion1}
The results illustrated in Figures \ref{fig:1} and \ref{fig:2} illustrate the impact of patient breast diameter on two tasks: detectability of microcalcifications and the Rayleigh task. 
The performance of each of these tasks is limited not only by the spatial frequency content in the images, but also by the presence of noise, and the latter effect is clearly what 
is modified by a change in breast diameter. The study by Yang et al. which investigated the impact of breast diameter on image noise illustrated a degradation in image quality 
for decreased breast diameter when the mean radiation dose to the breast is kept constant. The same conclusion is reproduced here, but through a task-based approach that 
also quantifies an upper bound for task performance for a range of signal sizes and breast diameters. 

The effect of signal location within the breast CT scanner's field of view appeared to impact task performance somewhat less than previous studies would suggest, however the 
basic trend that the detection task is somewhat more difficult closer to the axis of rotation in the absence of a bowtie filter is shown in Figure \ref{fig:location}. Use of the HO efficiency metric 
demonstrated that for all investigated signal sizes, performance was improved through use of smaller image pixels (see Figure \ref{fig:4}). While the magnitude of this 
effect depends somewhat on the signal size, the improvement was relatively constant for signal sizes near the threshold of reliable task performance, roughly 100$\mu$m for 
microcalcification size, and 0.3mm for Rayleigh blob separation. 
Figure \ref{fig:5} also shows the impact of reconstruction implementation by illustrating the impact of width of a Hanning window in reconstruction.
For each task considered, a slow improvement in performance was seen with increasing filtering until a shallow maximum 
followed by a sharp drop in performance for over-filtering.

Fig. \ref{fig:5} also illustrates the impact of view number with fixed total radiation dose. 
The number of projection views was shown to have a minimal impact on task performance, with a slight improvement for 
few-view acquisitions (50 projection views). The lack of impact from view number is likely due to a combination of the small size of the ROIs investigated and the step-and-shoot 
model for the x-ray source motion. The small ROIs tended to remain free from angular under-sampling artifacts. This highlights the relevance of
task modeling, in that since we were only interested in improvement of classification for small signals, global changes to the image quality did not affect the metrics used. 
Clearly, if a fluoroscopic x-ray mode were modeled, wherein the x-ray source is constantly emitting x rays, there would also be a degradation in task performance due to blurring 
in the projection data from the moving x-ray source, and 50 projection views would be inadequate. Pulsed x-ray sources present an intermediate regime between step-and-shoot and 
fluoroscopic acquisitions, and investigation of the impact of view number with these sources requires further study.

\section{Results - Optimized Reconstructions}
\label{sec:results2}
\begin{figure}[h]
\centering
\includegraphics[width=0.7\columnwidth]{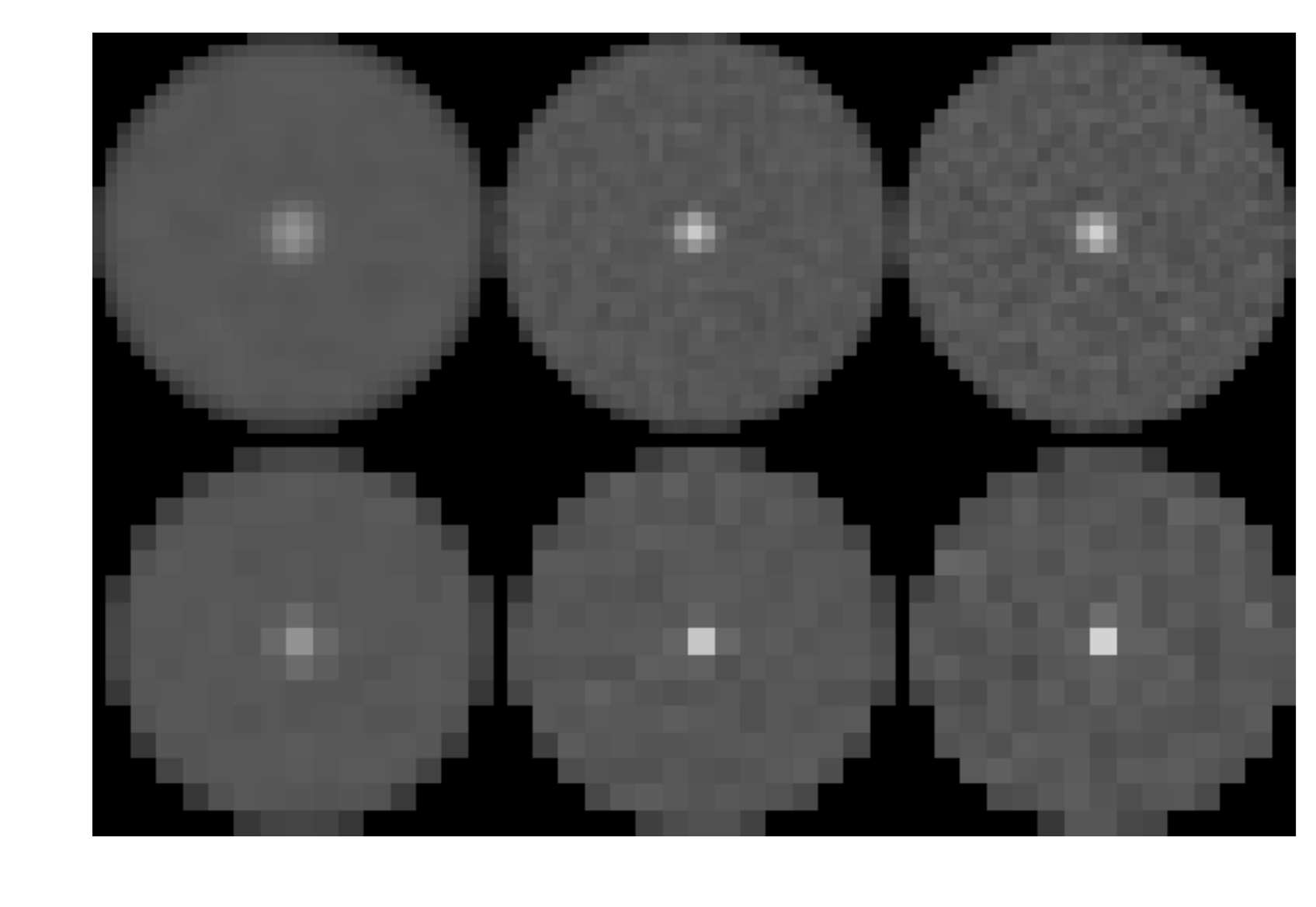}
\caption{
Noisy reconstructed images are shown with microcalcification signals. The microcalcifications are 100 $\mu$m in diameter and artificially amplified. The 
top center image was reconstructed using a Hanning window width of 1.0$\nu_N$, where $\nu_N$ is the Nyquist frequency. For all images, 50 projection views were used. The top row images correspond to a 1024 $\times$ 
1024 image pixel grid (0.162mm pixels). The left column of images was reconstructed with heavier smoothing using a 0.5$\nu_N$ Hanning window, and the right column used a 1.5$\nu_N$ 
Hanning window. The bottom row demonstrates reconstruction using a 512$\times$512 image pixel grid (0.324mm pixels). The diameter of the circular ROI is approximately 5mm, and the 
display window used is [0, 0.07] mm$^{-1}$. \label{fig:examples}}
\end{figure}

The foregoing image quality studies demonstrated that a Hanning window of width between 0.75$\nu_N$ and 1.0$\nu_N$, along with a 
1024$\times$1024 image pixel grid (0.162 mm pixel width) yields optimal HO performance 
for detection of 100$\mu$m microcalcifications for the imaging system model used here. As validation of this result, noisy reconstructed images with artificially amplified microcalcifications 
are shown in Fig. \ref{fig:examples}. The top center reconstruction was generated with the optimized parameters, 1.0$\nu_N$ Hanning width and 0.162mm pixels. 
The left column used a Hanning window width of 0.5$\nu_N$, while 
the right column was generated with a Hanning window width of 1.5$\nu_N$. The bottom row images were all reconstructed using a 512$\times$512 image pixel grid (0.324mm pixels). Each image was 
reconstructed from 50 equally spaced projection views. 

From inspection of these noisy reconstructed images, one sees that the HO metrics lead to a set of reconstruction parameters that produce a reasonable image, with enough smoothing 
from the Hanning filter to limit the background noise, but not so much smoothing that the signal itself became substantially attenuated. With the artificially amplified microcalcifications, 
it is not immediately clear that the larger pixels would result in decreased task performance. This illustrates that the HO enables the determination of image quality information which 
cannot be reliably obtained through subjective inspection of a single simulated image by a human.

\begin{figure}[h]
\centering
\includegraphics[width=0.45\columnwidth,bb = 77 159 550 632, clip=True]{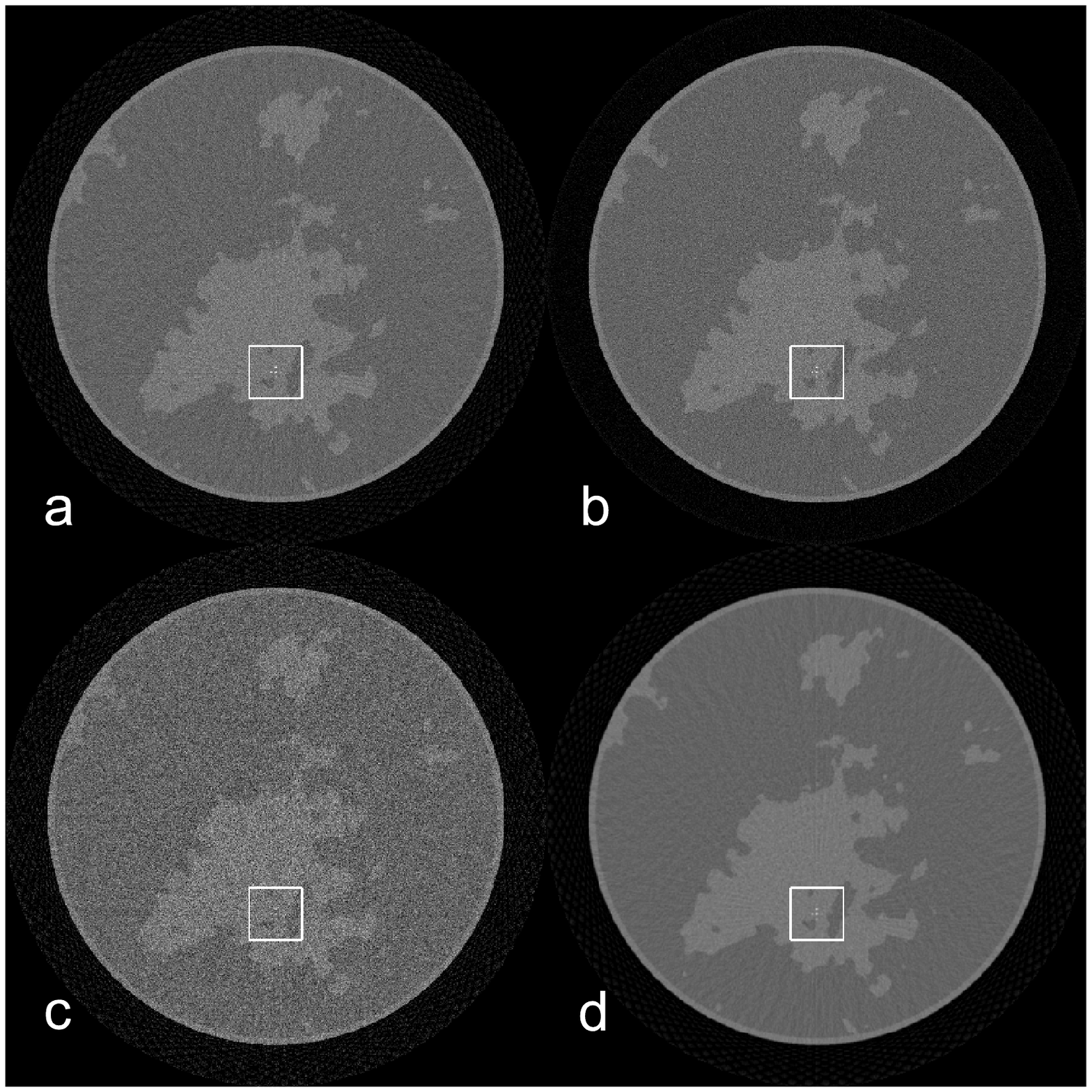} \includegraphics[width=0.45\columnwidth,bb = 77 159 550 632, clip=True]{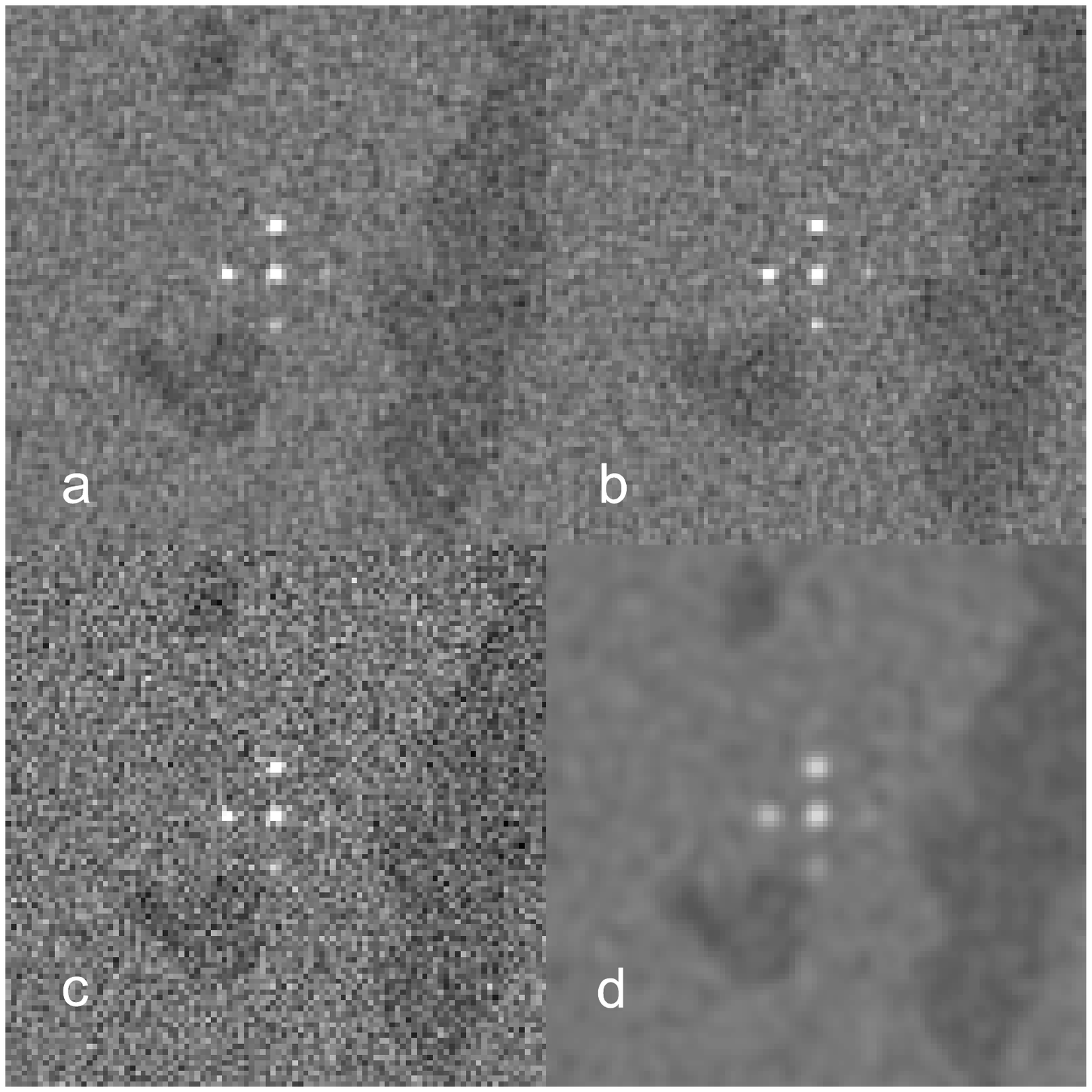}
\caption{Example noisy reconstructions from a simulated breast phantom with microcalcification cluster. The ROIs outlined in white are shown in the images on the right. 
a: a reconstruction from 100 views with reconstruction parameters determined 
to be optimal for microcalcification detection by the HO. b: the same phantom reconstructed from 500 views. c: a reconstruction with a wider-than-optimal (2.0$\nu_N$) Hanning filter. 100 
views were acquired for this image.
d: a 100-view reconstruction with a narrower-than-optimal (0.375$\nu_N$) Hanning filter. The display window is [0, 0.05] mm$^{-1}$. 
\label{fig:noisy_recons}}
\end{figure}

\begin{figure}[h]
\centering
\includegraphics[width=0.45\columnwidth,bb = 77 159 550 632, clip=True]{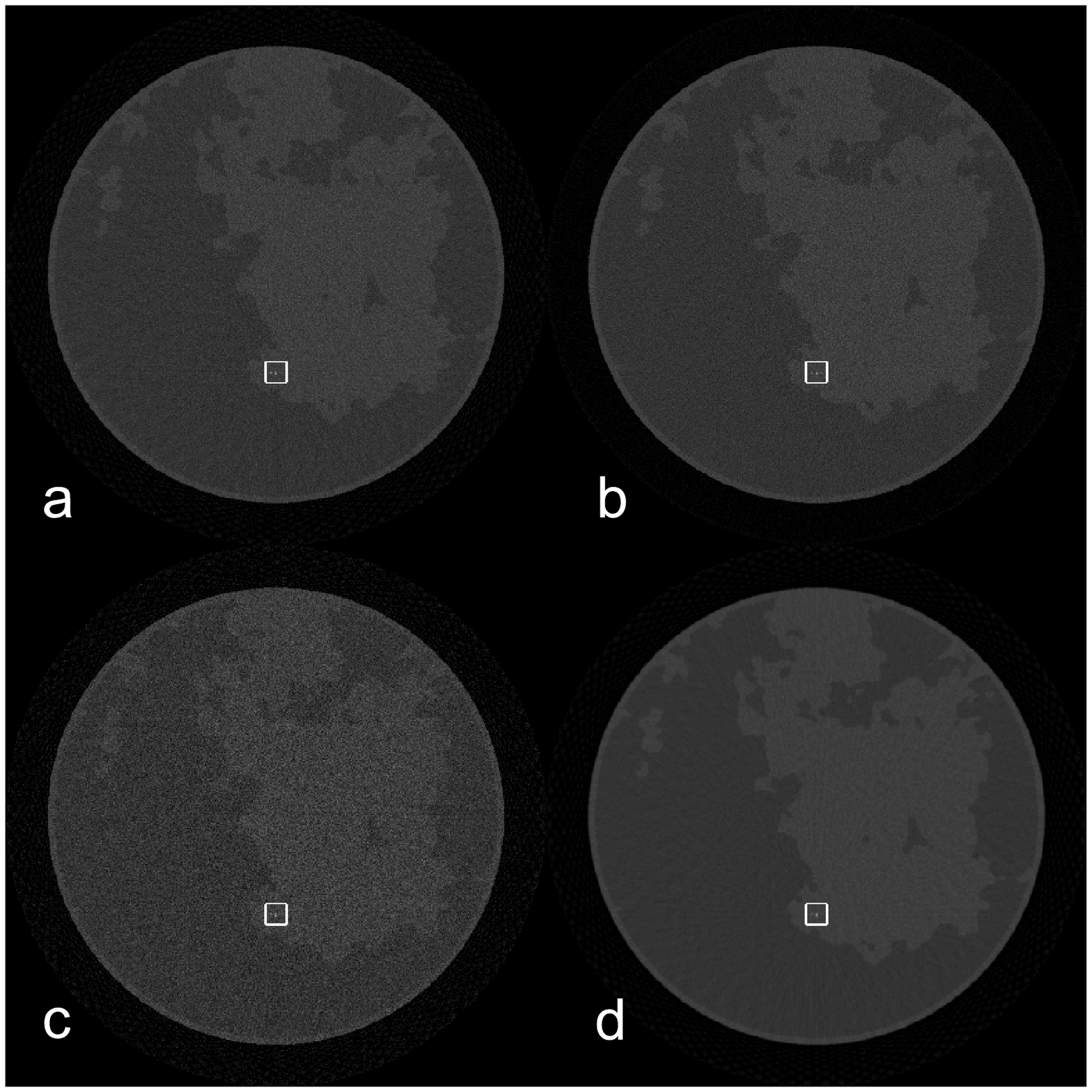} 
\includegraphics[width=0.45\columnwidth,bb = 77 159 550 632, clip=True]{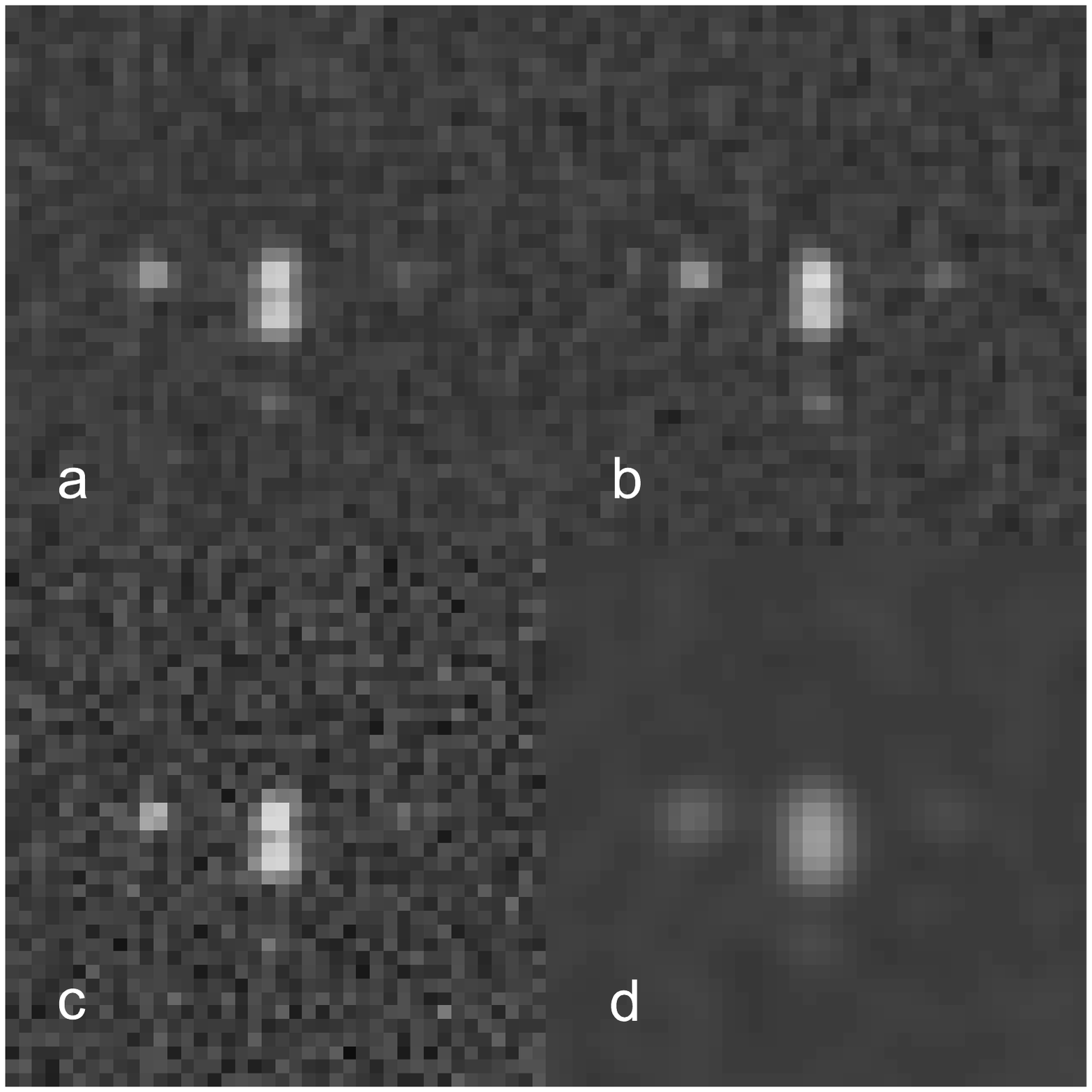}
\caption{Noisy reconstructed images from a breast phantom including two barely-resolvable microcalcifications to emulate the Rayleigh task. a: reconstruction from 100 views and optimal 
reconstruction parameters. b: reconstruction from 500 views. c: reconstruction with a wide (2.0$\nu_N$) Hanning filter. d: reconstruction with a narrow (0.375$\nu_N$) Hanning filter. 
The display window is [0, 0.1] mm$^{-1}$. \label{fig:noisy_recons_Rayleigh}}
\end{figure}

By the nature of the metrics used in this work, there is a clear dependence of the outcome of system parameter optimization
on the task being modeled, i.e. on the objects being imaged and noise correlations. 
This dependence is important to understand for two reasons. First, any task which is modeled for assessing model observer performance will necessarily 
be a simplification of the more realistic task which one hopes to ultimately perform.
In order to help ensure that this simplification does not lead to unreasonable conclusions for system optimization, inspection of simulated images of more realistic tasks provides a useful check for the derived optimal system parameters. In this way, far from solely relying on a noisy simulated image for system design, simulations constitute a useful supplement to the more fundamental objective assessment methods advocated in this work. 
Figures \ref{fig:noisy_recons} and \ref{fig:noisy_recons_Rayleigh} 
address this point by showing noisy reconstructed images of a numerical breast phantom with simulated microcalcifications. A reconstructed image with determined optimal 
system parameters is shown, along with images corresponding to two suboptimal Hanning filter widths. Since the HO appeared insensitive to the number of projection views 
with fixed total dose, an image corresponding to 500 projection views is shown in order to ensure that using 100 views indeed does not negatively impact the microcalcification 
detection task. The microcalcifications are not artificially enhanced in this case, but instead are made large enough to be visualized with relative ease.

\begin{figure}[h]
\centering
\includegraphics[width=0.45\columnwidth,bb = 77 159 550 632, clip=True]{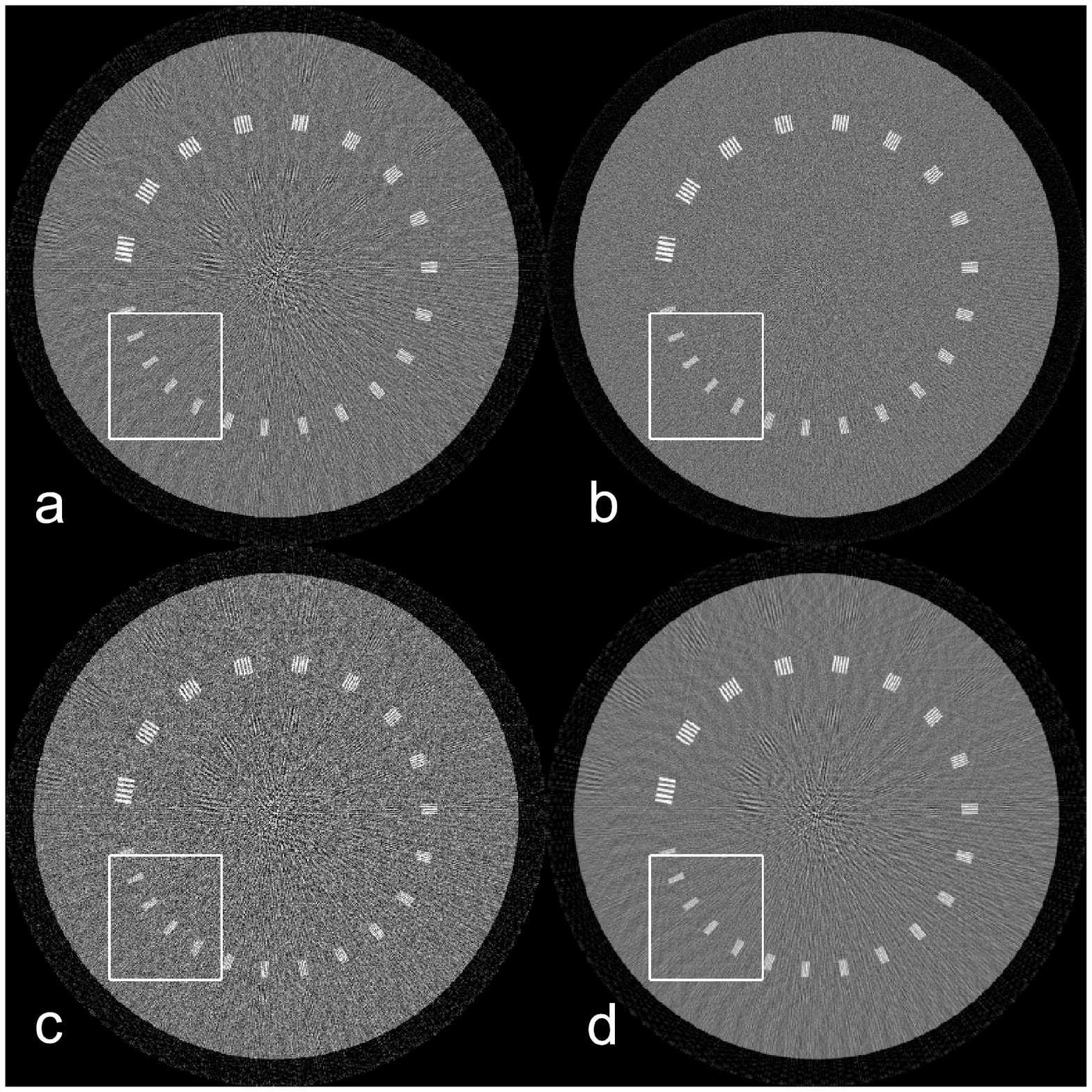} \includegraphics[width=0.45\columnwidth,bb = 77 159 550 632, clip=True]{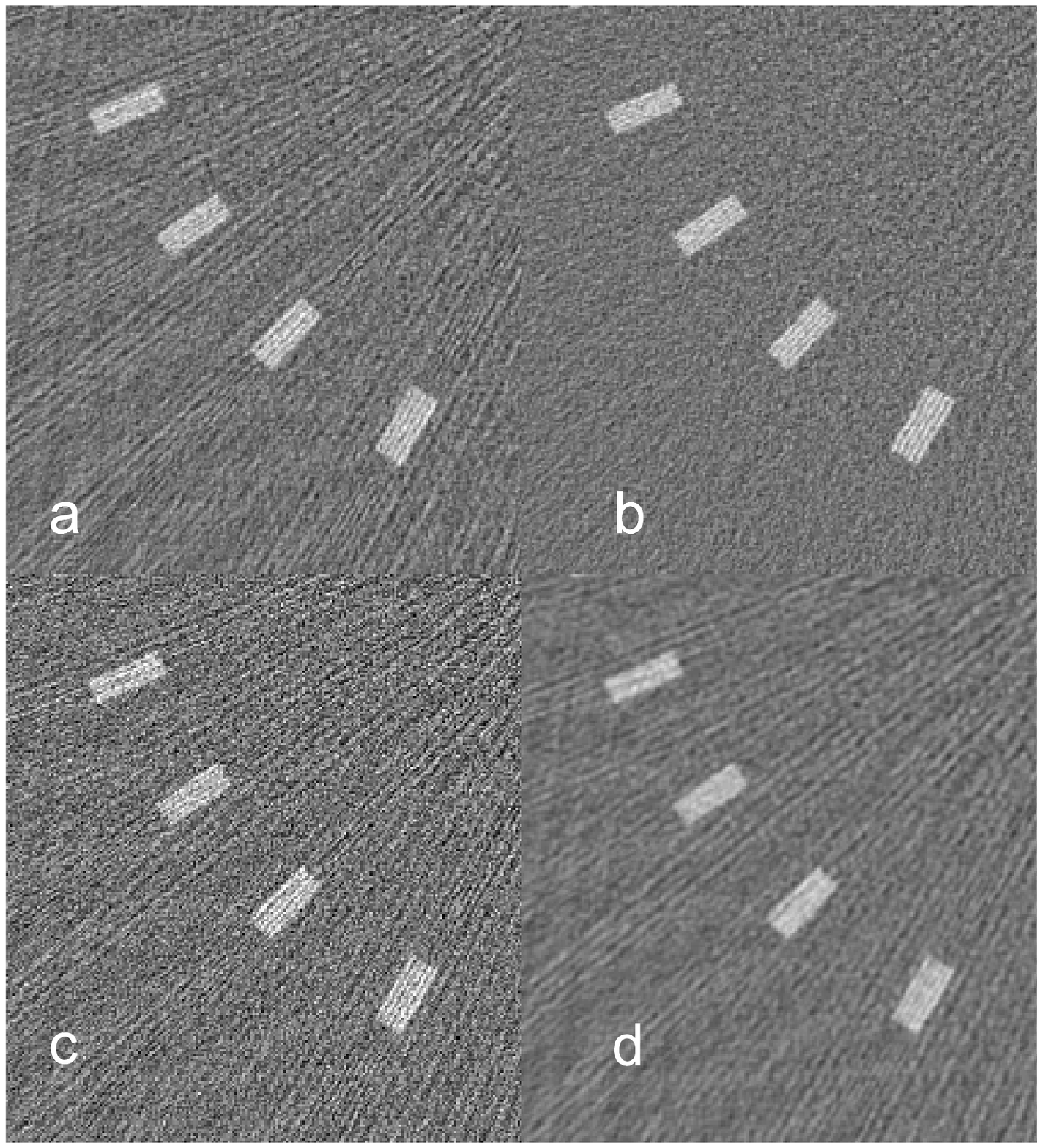}
\caption{Reconstructed noisy images from a bar phantom. The bars shown in the ROI image are of separations ranging from 0.173 mm to 0.22 mm. a: reconstruction from 100 views with 
parameters optimal for Rayleigh discrimination. b: reconstruction from 500 views. c: reconstruction with wider Hanning filter (2.0$\nu_N$). d: reconstruction with narrower Hanning filter
(0.375$\nu_N$). The 0.22 mm separation bars are resolvable in all images except d.  The display window is [0, 0.0544] mm$^{-1}$. 
\label{fig:noisy_cats}}
\end{figure}

\begin{figure}[h]
\centering
\includegraphics[width=0.45\columnwidth,bb = 77 159 550 632, clip=True]{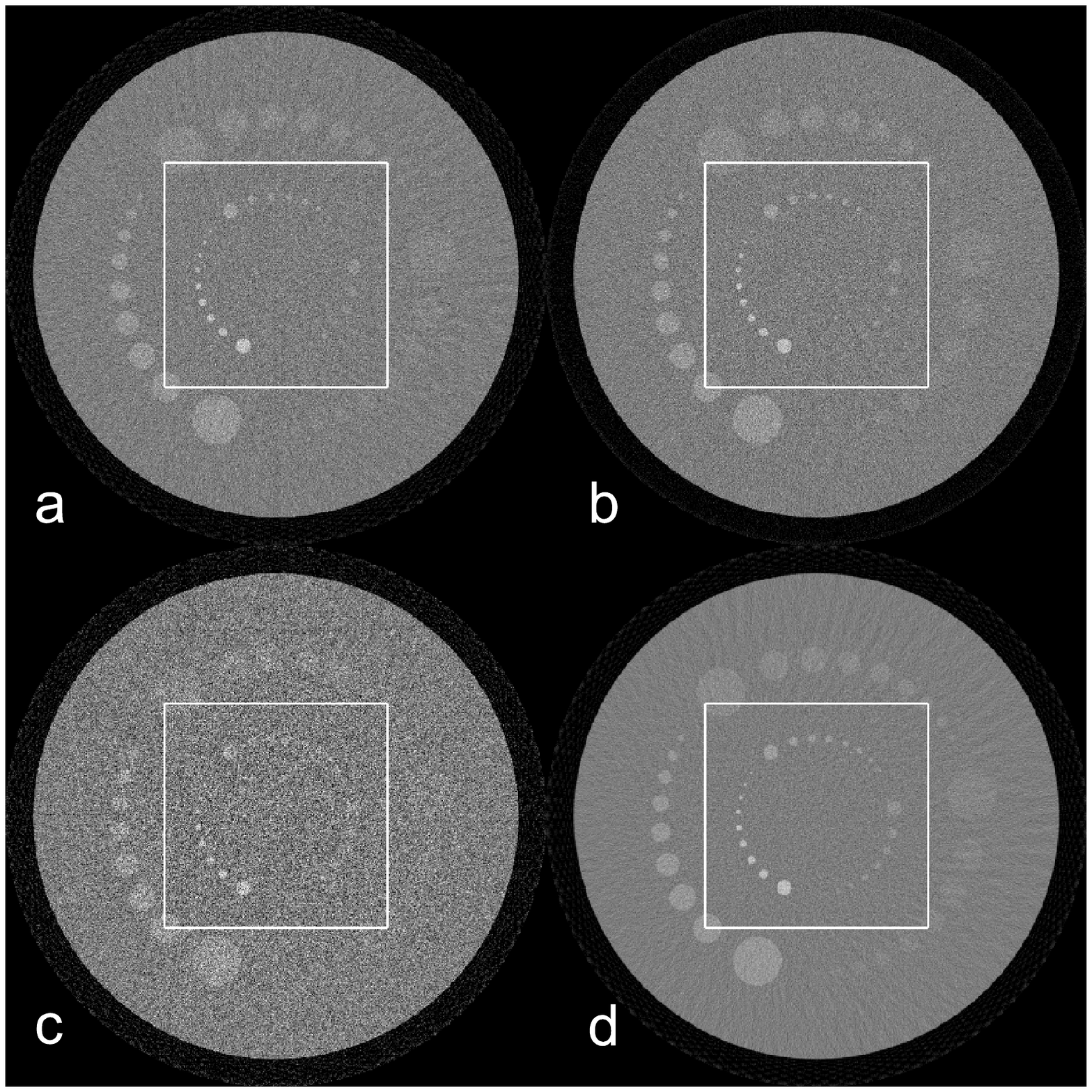} \includegraphics[width=0.45\columnwidth,bb = 77 159 550 632, clip=True]{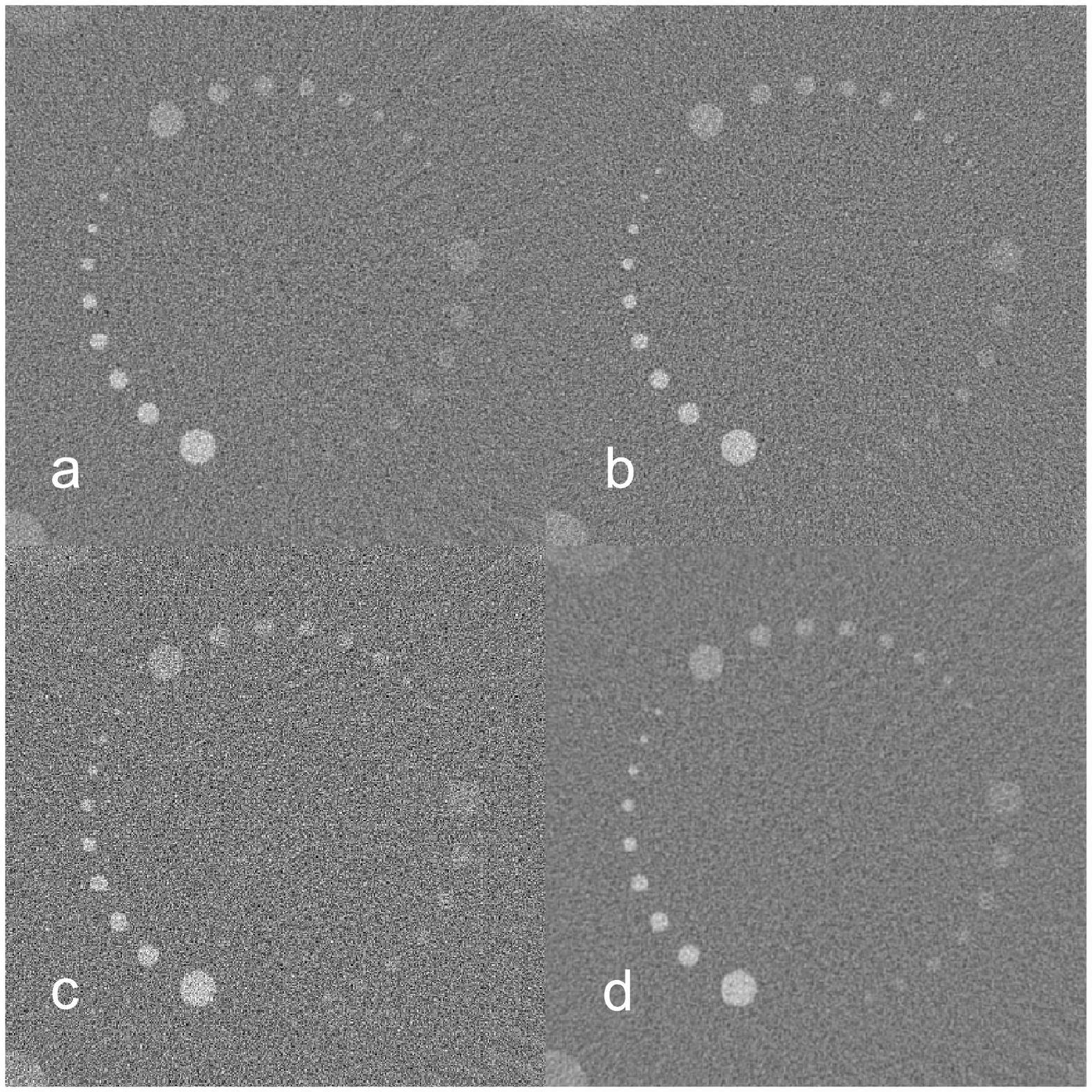}
\caption{Noisy reconstructed images from a low-contrast resolution numerical phantom. a: reconstruction with optimal parameters from 100 views. b: reconstruction from 500 views. 
c: 2.0$\nu_N$ Hanning width. d: 0.375$\nu_N$ Hanning width. The display window is [0, 0.05] mm$^{-1}$. 
\label{fig:noisy_cats_lc}}
\end{figure}

Another sense in which the form of the object being imaged is important is 
that stylized phantoms commonly used for quality assurance in CT may not be appropriate for task-specific 
system optimization. Figures \ref{fig:noisy_cats} and \ref{fig:noisy_cats_lc} address this point by showing noisy reconstructions of a phantom based on 
the \emph{Catphan} 
phantom for the same system parameters as Figures \ref{fig:noisy_recons} and \ref{fig:noisy_recons_Rayleigh}. We outline the specifics of each of the four phantoms used 
in this study:

Figure \ref{fig:noisy_recons} shows noisy reconstructions from a simulated breast phantom with a microcalcification cluster. Five microcalcifications of sizes 0.125, 0.15, 0.175, 0.2, and 
0.225 mm were included, with the smallest microcalcification located to the right, and the largest in the center. The image labeled \emph{a} was reconstructed from 100 views with 
parameters found to be optimal by the HO. The image labeled \emph{b} was reconstructed from 500 views, but the reconstruction parameters were unchanged. Images \emph{c} and 
\emph{d} were reconstructed from 100 views with Hanning filters that were 2.0$\nu_N$ and 0.375$\nu_N$ wide, respectively. These widths were chosen because they each correspond
to HO efficiency of approximately 75-80\%, compared to a maximal HO efficiency of 90-95\%. 

Figure \ref{fig:noisy_recons_Rayleigh} shows noisy reconstructed images from a similar numerical breast phantom, but with the addition of a pair of simulated microcalcifications 
in close proximity to simulate the Rayleigh task. The top-left image was reconstructed from 100 views using parameters deemed optimal by the HO for the Rayleigh task. Images 
\emph{b}--\emph{d} correspond to a 500-view reconstruction, a 2.0 $\nu_N$-wide Hanning filter, and a 0.375$\nu_N$-wide Hanning filter, respectively.   

Figure \ref{fig:noisy_cats} shows reconstructions from a bar phantom based on the \emph{Catphan} phantom. The ROI shown to the right contains bars with separations of 0.173, 0.186, 0.20,
and 0.22 mm. The image labeled \emph{a} is reconstructed from 100 views with parameters optimal for performing Rayleigh discrimination. The other subfigures are, as before, either from 
500 views, reconstructed with a wide Hanning filter, or reconstructed with a narrow Hanning filter. The 0.22mm bars appear distinct in all but image \emph{d}, while the 0.20 mm bars are only
 questionably resolved in each image. Resolution of smaller bars seems impossible based on these images, regardless of the parameter choices. 

Finally, Fig. \ref{fig:noisy_cats_lc} demonstrates noisy reconstructions of a low-contrast resolution phantom, roughly based on the \emph{Catphan} phantom. The optimal image, \emph{a}, was 
reconstructed using parameters which were optimal for microcalcification detection. The other images were generated as in the preceding figures. Faint undersampling artifacts are visible in 
the images reconstructed from 100 views (\emph{a}, \emph{c} and \emph{d}). 
Low contrast visibility is consistently improved with heavier filtering, while the higher-contrast inserts are arguably equally visible in each image. 

\clearpage

\section{Discussion of Optimized Reconstructed Images}
\label{sec:discussion2}
Figs. \ref{fig:noisy_recons}--\ref{fig:noisy_cats_lc} show examples of noisy reconstructed images for a variety of numerical phantoms. In each case, an optimal Hanning filter setting and 
view number is used, along with two suboptimal filters and a larger number of views. Based on these few image realizations, in each case, increasing the view number from 100 to 500 views appears to have
minimal 
impact on the resulting image within the selected ROIs. For the microcalcification detection phantom (Fig. \ref{fig:noisy_recons}), under-smoothing seems to result in the smallest microcalcification being overwhelmed by noise, 
while over-filtering increases the noise correlation distance to approximately the size of the microcalcification. Therefore, the smallest microcalcification seems marginally easier to visualize for the 
optimized Hanning filter width. Meanwhile, the Rayleigh signal shown in Fig. \ref{fig:noisy_recons_Rayleigh} appears approximately equivalent among all except the over-smoothed image. 
This and the preceding figure demonstrate that subjective assessment is not necessarily as sensitive to parameter choices or acquisition settings as the objective HO assessment. However,
subjective assessment of these particular simulated images tends to corroborate the HO optimization by demonstrating that the optimal parameters do, in fact, produce reasonable images. 

Figs. \ref{fig:noisy_cats} and \ref{fig:noisy_cats_lc} highlight the importance of task-based assessment in that while each of these two phantoms is intended to summarize image quality, 
neither provides results which agree with the more relevant HO optimization or with the subjective assessment in Figs. \ref{fig:noisy_recons} and \ref{fig:noisy_recons_Rayleigh}. In the 
case of Fig. \ref{fig:noisy_cats}, only the over-smoothed reconstruction resulted in a noticeable difference in resolution of the bar patterns. Meanwhile, the results in Fig. \ref{fig:noisy_cats_lc} seem to 
contradict the conclusion of the HO and of Fig. \ref{fig:noisy_recons}, with the over-smoothed image more clearly showing the small disks. 

\section{Conclusions}
\label{sec:conclusions}
We have investigated the use of a model observer based approach to image quality evaluation for the purpose of guiding design and use of a dedicated breast CT system. The use 
of the proposed methodology, based on the Hotelling observer, was demonstrated through investigation of performance in two tasks: microcalcification detection and Rayleigh discrimination.
 Breast diameter, signal location, image grid size, reconstruction filter, and projection view number were all considered, and the impact of each 
parameter on the HO metrics was computed. The proposed approach to system optimization and evaluation is applicable for any linear reconstruction algorithm, and 
the results of this study indicate that the HO and its associated metrics are a versatile and easily implemented tool for characterizing breast CT image quality and guiding decisions in 
system design and optimization. The value of this approach is further increased by the fact that it does not rely on human observers, repeated experimental scans, or assumptions of 
shift-invariance or stationarity in the reconstructed images. Further work is needed to include additional realism in the simulations, such 
as electronic noise modeling, background variability, and realistic anatomic background structure. 

%% file: Chapter6/Chapter6.tex
\chapter{Generalization to Iterative Image Reconstruction}
\label{ch:tv}

\section{Introduction}

To this point, our discussion of system optimization and image quality assessment has focused on analytic image reconstruction, specifically on the FBP algorithm. In this 
chapter, we will take an important first step in investigating the feasibility of extending this method to images reconstructed iteratively with penalties involving total variation. 

\subsection{Background}
Over the past decade, use of regularization based on total variation (TV) has had a profound impact on iterative image reconstruction (IIR) research in x-ray computed tomography (CT). 
In particular, TV-based regularizers have been motivated by their ability to accurately reconstruct objects whose gradient magnitude images are sparse under data sampling conditions where 
conventional algorithms fail \cite{sidky_accurate_2006,sidky_image_2008,defrise_algorithm_2011,jensen_implementation_2012,jorgensen_quantifying_2013,ramani_splitting-based_2012}. 
In short, use of the 
image TV has the potential to reduce the data sampling requirements necessary for recovery of a useful image, thereby limiting the necessary 
number of x-ray projections for performing reconstruction. 
This potential has been demonstrated in real systems, enabling a range of practical applications including sparse-view CT acquisitions \cite{chen_prior_2008,bian_evaluation_2010,sidky_image_2010,ritschl_improved_2011,han_algorithm-enabled_2011, han_optimization-based_2012} and improved dynamic CT \cite{chen_prior_2008,song_sparseness_2007,chen_temporal_2009,bergner_investigation_2010,ritschl_iterative_2012,kuntz_constrained_2013}. 
\new{Additionally, TV and related penalties have been motivated purely for their edge-preserving properties for situations other than sparse sampling. \cite{rudin_nonlinear_1992, delaney_globally_1998,vogel_iterative_1996,
panin_total_1999, persson_total_2001}}

While image TV and similar penalties are now widely accepted as potentially beneficial components in IIR, 
research in the formulation of novel algorithms, cost functions, and penalties has vastly outpaced the development of sound image quality metrics 
suitable to the task of evaluating this new class of images. Specifically,  images obtained from IIR do not satisfy the 
assumptions inherent in application of conventional metrics such as the modulation transfer function (MTF)\cite{rossmann_measurement_1964} or noise power spectrum (NPS)\cite{dainty_image_1974}. Therefore any
assessment of image quality in IIR based on these or related metrics requires an initial evaluation of assumptions such as local stationarity or shift invariance, with the resulting assessment being 
meaningful only insofar as these assumptions are valid. 

Alternatively, one can avoid the issue of limiting assumptions by collecting a sample of noisy images, so that image quality metrics based on statistical properties can be estimated. However, these methods 
often require many images, imposing a significant burden in terms of computation and, in the event real data is used, in terms of data acquisition.
Further, the metrics that result from these sample-based strategies are stochastic and possess a finite statistical variability.
The purpose of the present work is to aid in the development of image quality metrics for TV-based IIR by characterizing the noise properties of images reconstructed with TV penalties. 
Specifically, we seek to develop a framework for accurate computation of image pixel variance and covariance which does not rely on assumptions such as stationarity and does not 
require the collection of noisy sample images which leads to stochastic estimates. We therefore combine the concept of fixed-point covariance estimation with a noise propagation approach by employing 
the iteratively-reweighted least-squares (IRLS) algorithm \cite{lawson_contributions_1961, beaton_fitting_1974, wohlberg_iteratively_2007, chartrand_iteratively_2008, sidky_constrained_2014} in order to overcome the difficulty of nonlinearity in TV-penalized IIR.

\subsection{Motivation for Image Covariance Computation}

The full covariance matrix for an image $\mathbf{f}$ is defined as $K_\mathbf{f} = \textrm{E}\left\{\left(\mathbf{f}-\bar{\mathbf{f}}\right)\left(\mathbf{f}-\bar{\mathbf{f}}\right)^T\right\}$, where $E$ 
denotes the expectation operator and
$\bar{\mathbf{f}}=\textrm{E}\left\{\mathbf{f}\right\}$. The diagonal of this matrix contains the variance of each image pixel, while the off-diagonal elements contain covariances between separate pixels.
\new{Koehler and Proksa \cite{koehler_noise_2009} observed a strong object-dependence in image variance for TV-based reconstruction, with object edge locations corresponding to high pixel variances relative 
to uniform regions. This implies a need to understand the interplay between TV-based reconstruction and image noise, since
image pixel variance lies at the heart of a wide array of image quality metrics such as signal-to-noise ratio (SNR). Further,} there are many reasons for additionally considering pixel \emph{covariances} in the assessment of image 
quality in IIR. First, TV-based reconstruction can lead to a characteristic noise texture which is often anecdotally described as ``plastic'' or ``waxy'' \cite{leipsic_adaptive_2010, vorona_reducing_2011, ren_comparison_2012}.
 Noise texture is essentially the visualization of higher-order noise 
statistics (such as covariance) in a single realization. By accurately characterizing image covariance, a more rigorous analysis of TV-based reconstruction's noise texture can be performed.

Secondly, beyond qualitative subjective impacts on image texture, image covariance has a demonstrable effect on performance of many relevant clinical tasks, such as lesion detection \cite{burgess_human_2001, abbey_human-_2001}. 
This naturally leads to implications 
for more sophisticated image quality assessment, such as the use of model observers \cite{barrett_foundations_2004, barrett_model_1993}, 
but even direct analysis of the correlation structure of image noise can provide valuable information for assessing image quality. 
For example, one purpose of this work is to investigate the validity of the assumptions of local stationarity and object-independence of image noise. These assumptions are implicit whenever Fourier-based image 
quality metrics are constructed or whenever stylized phantoms are used to optimize a system employing IIR.

Finally, the eventual goal of this research direction would be the efficient and non-stochastic construction of objective image quality metrics such as Hotelling observer performance \cite{barrett_foundations_2004, fiete_hotelling_1987}, 
for directly relevant clinical tasks.
Accurate estimation and inversion of the image covariance matrix are the major barriers in applying the Hotelling observer to image quality assessment in CT. The nonlinearity of IIR can make image statistics difficult to 
describe, and the size of the images involved in a typical CT scan presents an additional difficulty. It is the goal of the present work to begin to address the former of these challenges, while maintaining awareness of
the latter.

\subsection{Relation to Prior Work}

Nearly all work to date in characterizing the effects of noise in images obtained through IIR has relied on the concept of linearizing perturbations in the image pixels about a mean image. 
If, for the purposes of noise analysis, the reconstruction algorithm behaves linearly, then the image covariance has a straightforward analytical expression based on a linear approximation 
to the reconstruction operation and the noise properties of the original data. Within this broad framework, two complementary approaches to noise estimation have emerged. The first is 
based on approximating noise at each iteration of a reconstruction algorithm, thereby propagating the original data noise through each step of the algorithm until the final image is obtained.
The second major approach is based on analysis of a convex objective function near its optimum in order to compute noise properties of a mathematically converged solution directly. The vast 
majority of work in both approaches has been applied to imaging in nuclear medicine rather than CT. This is likely due to the reduced dimensionality of images in PET and SPECT, which makes direct 
analysis of image covariance more feasible than in CT, where a typical 2-dimensional image can contain 1024$^2$ pixels. However, for the most part, prior work in PET and SPECT can generalize to 
problems in CT, so will now provide a brief overview of these methods.

Propagation-based methods were first developed and validated for the expectation maximization (EM) algorithm by Barrett \emph{et al.} \cite{barrett_noise_1994} and Wilson \emph{et al.}\cite{wilson_noise_1994}, and 
were subsequently generalized to block-iterative algorithms and maximum \emph{a-posteriori} (MAP) - EM algorithms by Soares \emph{et al.} \cite{soares_noise_2000,soares_noise_2005} and Wang and Gindi \cite{wang_noise_1997}, respectively.
This approach was further generalized by Qi \cite{qi_unified_2003}, who constructed a framework for propagation-based noise analysis with preconditioned gradient-ascent algorithms. Additionally, the same work provided
asymptotic analysis of noise properties so that the propagation-based approach could be bridged to the other major method, which we refer to as a fixed-point method, since it analyses noise properties 
at the fixed point of the objective function.
The basic approach for fixed-point noise analysis was presented by Fessler \cite{fessler_mean_1996}, with subsequent development and application still ongoing 
\cite{qi_fast_1999,qi_resolution_2000, zhang-oconnor_fast_2007, li_noise_2011, chun_noise_2013, dutta_quantitative_2013}. In particular, Refs. \cite{li_noise_2011} and \cite{dutta_quantitative_2013} 
clarify the relationship between iteration-based methods and fixed-point methods. Additionally, Ref. \cite{dutta_quantitative_2013} provides examples of the sort of system and reconstruction optimization 
which noise analysis can facilitate. 

A common limitation of most existing work is that non-quadratic regularization is not well handled since first-order approximations to these penalties can be inaccurate. One notable exception is the work of Ahn and Leahy \cite{ahn_analysis_2008},
which employs efficient Monte-Carlo methods to estimate noise and resolution properties rather than relying on analytic approximations. In this work, we adopt a different approach, by employing the IRLS algorithm 
for performing TV-penalized reconstruction. Our approach combines elements of both iteration-based and fixed-point methods. Like many EM algorithms, IRLS can be interpreted as a fixed-point iteration which iteratively 
solves a series of surrogate optimization problems. Similar to Li \cite{li_noise_2011}, we assume that each surrogate problem is solved to convergence and compute noise estimates at each iteration. The novel 
contribution of the work is that we employ the IRLS algorithm to address the use of the non-quadratic TV penalty. The method we propose also has a straightforward generalization to noise analysis of non-convex TpV 
minimization as discussed by Sidky \emph{et al.} \cite{sidky_image_2007,sidky_constrained_2014}.

This chapter is organized as follows: Section \ref{fessler_recap} provides necessary background on fixed-point methods for noise estimation; Section \ref{irls} describes the IRLS algorithm and its application to TV minimization; in Sections
\ref{linear_irls} -- \ref{implementation}, we apply fixed-point noise analysis to the IRLS algorithm; Section \ref{validation_method} describes the methods and results of the validation of our proposed method, while Section 
\ref{sec:large_system} applies the methodology to address several pertinent questions in TV-based IIR; finally Section \ref{sec:conclusion} provides a brief discussion and conclusions.


\section{Methods}
\subsection{Covariance of an Implicitly Defined Estimator}
\label{fessler_recap}
In this section, we will briefly introduce a method from Fessler \cite{fessler_mean_1996} which plays a key role in most approaches to covariance approximation in IIR, and which constitutes one component
of our proposed method as well. 
We begin by repeating the result for the covariance of an implicitly defined estimator. 
Here and elsewhere in this chapter, capital Latin letters denote matrices and bold, lowercase Latin letters denote vectors.
In general terms, one is interested in obtaining an image $\mathbf{f}^o\in \mathbb{R}^N$, which can be written as the optimizer of a cost function $\Phi$ that depends both on the image and data:
\begin{equation}
\mathbf{f}^o = \argmin_\mathbf{f} \Phi\left(\mathbf{f},\mathbf{g}\right),
\end{equation}
where $\mathbf{g} \in \mathbb{R}^M$ represents the projection data in x-ray CT. We restrict ourselves, for the time being, to considering functions $\Phi\left(\mathbf{f},\mathbf{g}\right)$ which possess a unique global optimizer that can be found by 
zeroing the partial derivatives of $\Phi$ with respect to $\mathbf{f}$, i.e.
\begin{equation}
\label{eqn:optimum}
\left. \frac{\partial \Phi\left(\mathbf{f},\mathbf{g}\right)}{\partial \mathbf{f}_i} \right|_{\mathbf{f}=\mathbf{f}^o}= 0 \hspace{1.0cm} \forall i \in \left[ 1, N\right],
\end{equation}
where $\mathbf{f}_i$ denotes the $i$th image pixel.

The basic idea for constructing an approximation to the image covariance matrix $K_\mathbf{f}$ relies on linearizing the implicitly defined image function $\mathbf{f}(\mathbf{g})$ with respect to the data $\mathbf{g}$. In other words, 
we wish to obtain the Jacobian matrix $J \in \mathbb{R}^{N\times M}$ whose $(i,j)$th entry we define by
\begin{equation}
J_{i,j} := \frac{\partial \mathbf{f}_i}{\partial \mathbf{g}_j},
\end{equation}
where $\mathbf{f}_i$ is the $i$th image pixel, and $\mathbf{g}_j$ is the $j$th sinogram (projection data) element. Specifically, we use the Jacobian to linearly model small perturbations of the image (i.e. noise) about the mean image $\bar{\mathbf{f}}$. 
In this work, we rely on the assumption that $\bar{\mathbf{f}}$ is well approximated by reconstruction of noise-free data, i.e.
$\bar{\mathbf{f}}\approx \mathbf{f}\left(\bar{\mathbf{g}}\right)$. In the examples investigated here, we have consistently observed this to be the case. We then use this 
information to construct an approximation of $J(\bar{\mathbf{f}})$. This leads to the approximation 
\begin{equation}
\begin{split}
\label{eqn:linearization}
&\mathbf{f}^o \approx \bar{\mathbf{f}} + \Delta \mathbf{f}, \\ 
&\Delta \mathbf{f} = J\left(\bar{\mathbf{f}}\right) \Delta \mathbf{g}, \\
&\Delta \mathbf{g} := \mathbf{g} - \bar{\mathbf{g}}
\end{split}
\end{equation}
where $\Delta \mathbf{f}$ and $\Delta \mathbf{g}$ define noise perturbations about the noise-free image and data, and $\bar{\mathbf{g}}$ denotes the noise-free data. Statistical variability 
enters into $\mathbf{f}^o$ only through $\Delta \mathbf{f}$, so that
\begin{equation}
\label{eqn:1stcov}
K_\mathbf{f} \approx K_{\Delta \mathbf{f}} = J \left(\bar{\mathbf{f}}\right) K_\mathbf{g} J^T \left(\bar{\mathbf{f}}\right),
\end{equation}
where the superscript $T$ denotes the matrix transpose (or Hermitian adjoint in the event that the entries are complex).

Since the image $\mathbf{f}^o$ is defined implicitly, the Jacobian $J$ cannot be constructed directly. Instead we can differentiate both sides of Eqn. \ref{eqn:optimum} with respect to $\mathbf{g}$ and 
apply the chain rule, yielding
\begin{equation}
H_{\mathbf{ff}}J + H_\mathbf{fg} = \mathbf{0}
\end{equation}
where $\mathbf{0} \in \mathbb{R}^{N\times M}$ is a matrix of all zeros, and we have defined the Hessian $H_\mathbf{ff} \in \mathbb{R}^{N\times N}$ such that its ($i$,$j$)th element 
\begin{equation}
\left[H_\mathbf{ff}\right]_{i,j} = \frac{\partial^2 \Phi \left(\mathbf{f},\mathbf{g}\right)}{\partial \mathbf{f}_i \partial \mathbf{f}_j}. 
\end{equation}
Similarly, we define the mixed Hessian $H_\mathbf{fg} \in \mathbb{R}^{N\times M}$ such that 
\begin{equation}
\left[H_\mathbf{fg}\right]_{i,j} = \frac{\partial^2 \Phi \left(\mathbf{f},\mathbf{g}\right)}{\partial \mathbf{f}_i \partial \mathbf{g}_j}.
\end{equation}
The resulting Jacobian is then given by
\begin{equation}
\label{eqn:jac1}
J = -H_\mathbf{ff}^{-1} H_\mathbf{fg}.
\end{equation}
The matrix inverse requires that $-H_\mathbf{ff}$ be positive definite; however, in order to construct our covariance approximation, we need only evaluate the Jacobian at $\bar{\mathbf{f}}$. Therefore, 
we only require  $-H_\mathbf{ff}$ to be positive definite when evaluated at $\bar{\mathbf{f}}$. As Fessler points out \cite{fessler_mean_1996}, this corresponds to requiring $\Phi \left(\mathbf{f},\mathbf{g}\right)$ to be locally strongly 
convex near the optimum for noise-free data. This is generally not true for objectives involving a TV term, however we will address this difficulty below. The final covariance approximation is then given by combining Eqns. \ref{eqn:1stcov} and \ref{eqn:jac1}:
\begin{equation}
K_\mathbf{f} = H_\mathbf{ff}^{-1}\left(\bar{\mathbf{f}}\right) H_\mathbf{fg}\left(\bar{\mathbf{f}}\right) K_\mathbf{g} H^T_\mathbf{fg}\left(\bar{\mathbf{f}}\right) H_\mathbf{ff}^{-1}\left(\bar{\mathbf{f}}\right).
\end{equation}

\subsection{TV-penalized IRLS Reconstruction}
\label{irls}

In this work, we consider the unconstrained form of TV-penalized image reconstruction, where given a data set of $M$ elements, $\mathbf{g} \in \mathbb{R}^M$, the reconstructed 
image is defined via the following optimization program:
\begin{equation}
\label{eqn:tv_obj}
\mathbf{f}^o = \argmin_\mathbf{f} \left\{ \lambda \| \left( | \nabla \mathbf{f} | \right)\|_1 + \|X\mathbf{f} - \mathbf{g}\|_2^2  \right\},
\end{equation}
where $\lambda$ is a free parameter controling the weight of regularization, $X \in \mathbb{R}^{M\times N}$ is a linear model of forward projection, and 
$\mathbf{f}^o \in \mathbb{R}^N$ is the image vector composed of reconstructed image pixel coefficients.  In this work, we consider $\mathbf{f}$ to be a 2D image, however 
the bulk of the formalism presented here can be trivially generalized to 3D, albeit with a corresponding increase in computational burden. Further, a weighting factor could be added to the 
least-squares term to generalize our approach, but this was not included in the present study. The argument of 
the $\ell_1$ norm is the pixel-wise magnitude of the image spatial gradient, where $\nabla \in \mathbb{R}^{2N \times N}$ is the discrete 
gradient operator for two-dimensional images, constructed as
\begin{equation}
\nabla = \binom{\nabla^x}{\nabla^y},
\end{equation}
where $\nabla^x$ and $\nabla^y$ represent forward difference operators in the $x$- and $y$-dimension of the image, respectively.

\new{The TV term in Eqn. \ref{eqn:tv_obj} lies at the heart of the difficulty of computing image noise properties accurately. Since the TV penalty is 
not smooth, its partial derivatives are not defined everywhere, and the foregoing approximation for image covariance is not well defined. 
Further, the lack of smoothness in the TV penalty is essential in encouraging gradient magnitude sparsity in the reconstructed image, and should 
therefore be considered in our noise approximation.
In this work, we propose analysis of image noise properties by propagation of noise through an iteratively reweighted least-squares (IRLS) algorithm applied 
to the problem in Eqn. \ref{eqn:tv_obj}. The aforementioned covariance approximation is then well defined at each iteration of the IRLS algorithm, and image noise properties can 
be propagated through the reconstruction process, which converges to the solution of the non-smooth TV objective. We summarize this algorithm below.}

The first iteration of IRLS involves the solution of a least-squares objective with a quadratic roughness penalty:
\begin{equation}
\label{eqn:1stiter}
\mathbf{f}^o_{1} = \argmin_\mathbf{f} \left\{ \lambda \|\nabla \mathbf{f} \|_2^2 + \|X\mathbf{f} - \mathbf{g}\|_2^2  \right\}
\end{equation}
Next, each subsequent iteration has two steps. First, a vector of weights is computed from the previous iterate:
\begin{equation}
\label{eqn:w}
\mathbf{w}_n = \left( \frac{1}{\sqrt{\eta_n^2+| \nabla \mathbf{f}^o_n |^2}}\right)
\end{equation}
where $\eta_n = 10^{-n}$ and $n$ denotes the iteration number. The parameter $\eta_n$ is a continuation parameter that addresses potential singularities in the definition of the weights. Following Chartrand and Yin \cite{chartrand_iteratively_2008}, 
we begin with a relatively large value of this parameter and rapidly shrink it with subsequent iterations. Next, the following iterate is computed from another quadratic optimization problem:
\begin{equation}
\label{eqn:new}
\mathbf{f}^o_{n+1} = \argmin_\mathbf{f} \left\{ \lambda \| \sqrt{\mathbf{w}_n} | \nabla \mathbf{f}|  \|_2^2 + \|X\mathbf{f} - \mathbf{g}\|_2^2  \right\}.
\end{equation}
Alternating applications of Eqns. \ref{eqn:w} and \ref{eqn:new} are then applied. Each cycle of computing weights and solving a resulting quadratic objective constitutes a single iteration of IRLS. This algorithm has been shown elsewhere to efficiently solve 
a general class of problems with Eqn. \ref{eqn:tv_obj} as a special case \cite{wohlberg_iteratively_2007, chartrand_iteratively_2008, rodriguez_efficient_2009,rodriguez_comparison_2012}.

\subsection{Linearization of IRLS}
\label{linear_irls}

In order to compute linear approximations for each iterate of IRLS, we begin by constructing $H_\mathbf{ff}$ and $H_\mathbf{fg}$ for $\Phi\left(\mathbf{f},\mathbf{g}\right)$ given in Eqn. \ref{eqn:1stiter}.
Expanding the matrix multiplication, we can rewrite the objective function as
\begin{equation}
\Phi_1\left(\mathbf{f},\mathbf{g}\right) = \lambda \mathbf{f}^T\nabla^T\nabla\mathbf{f} + \mathbf{f}^TX^TX\mathbf{f} - 2\mathbf{g}^TX\mathbf{f} + \mathbf{g}^T\mathbf{g},
\end{equation}
where the subscript denotes that this is the objective function defining the first iterate of the IRLS algorithm. By inspection, we then have that 
\begin{equation}
H_\mathbf{ff} = 2\left(\lambda \nabla^T\nabla + X^TX\right)
\end{equation}
and
\begin{equation}
H_\mathbf{fg} = -2X^T.
\end{equation}
The Jacobian and image covariance matrix are therefore,
\begin{equation}
J_1 = \left(\lambda \nabla^T\nabla + X^TX\right)^{-1} X^T
\end{equation}
and 
\begin{equation}
K_{\mathbf{f}_1^o} = \left(\lambda \nabla^T\nabla + X^TX\right)^{-1} X^T K_\mathbf{g} X \left(\lambda \nabla^T\nabla + X^TX\right)^{-1}.
\end{equation}
The expression above for the covariance of the first iterate is exact, since the objective function is quadratic and all higher order derivatives in the expansion of $\Phi\left(\mathbf{f},\mathbf{g}\right)$ 
are zero. 

For all subsequent iterates, we have
\begin{equation}
\Phi_n\left(\mathbf{f},\mathbf{g}\right) =  \lambda \mathbf{f}^T\nabla^T\mathrm{diag}\left(\mathbf{w}_{n-1}\oplus \mathbf{w}_{n-1}\right)\nabla\mathbf{f} + \mathbf{f}^TX^TX\mathbf{f} - 2\mathbf{g}^TX\mathbf{f} + \mathbf{g}^T\mathbf{g},
\end{equation}
where diag($\mathbf{x}$) denotes a diagonal matrix with $\mathbf{x}$ along the diagonal and $\oplus$ denotes concatenation of vectors. While the Jacobian for the first iteration $J_1$ has 
no dependence on the data $\mathbf{g}$, Jacobians for subsequent iterations will. We will therefore compute the quantities $H_\mathbf{ff}$ and $H_\mathbf{fg}$ at the location 
$\mathbf{f}^o_n=\bar{\mathbf{f}}^o_n \approx \mathbf{f}^o_n\left(\bar{\mathbf{g}}\right)$. Similarly, we evaluate these Hessians at the location $\mathbf{w}_n = \bar{\mathbf{w}}_n$.
The assumption $\bar{\mathbf{f}}^o_n \approx \mathbf{f}^o_n\left( \bar{\mathbf{g}}\right)$ is common and is not a limiting factor in the accuracy of our approximation. However the equivalent approximation for $\bar{\mathbf{w}}$ leads to inaccuracies in the 
noise model. This is particularly true at higher iterations of IRLS and in nearly uniform regions. 
An improved, non-linear estimate of $\bar{\mathbf{w}}_n$ based on an approximate cumulative distribution function for $\mathbf{w}_n$
is provided in Section \ref{wbar}.

By inspection, we now have
\begin{equation}
H_\mathbf{ff} = 2\left(\lambda \nabla^T\mathrm{diag}\left(\bar{\mathbf{w}}_{n-1} \oplus \bar{\mathbf{w}}_{n-1}\right)\nabla + X^TX\right),
\end{equation}
however constructing $H_\mathbf{fg}$ requires some care since $\mathbf{w}_{n-1}$ is a function of $\mathbf{g}$. Here, we introduce the shorthand notation for a Jacobian matrix
$\frac{\partial \mathbf{f}}{\partial \mathbf{g}} \in \mathbb{R}^{N \times M}$, whose entries are all partial derivatives of elements of $\mathbf{f}$ with respect to elements of $\mathbf{g}$.
We then have
\begin{equation}
H_\mathbf{fg} =\left. \left[ \frac{\partial}{\partial \mathbf{g}} \left( 2\lambda \nabla^T\mathrm{diag}\left(\mathbf{w}_{n-1} \oplus \mathbf{w}_{n-1}\right)\nabla \mathbf{f} - 2X^T\mathbf{g}\right) \right] 
\right|_{\mathbf{g}=\bar{\mathbf{g}}}.
\end{equation}
Rearranging the first term and applying the chain rule, we obtain
\begin{equation}
\begin{split}
H_\mathbf{fg} &= 2\left( \lambda \nabla^T \mathrm{diag}\left(\nabla \bar{\mathbf{f}}^o_n\right)\left. \left[\frac{\partial \left(\mathbf{w}_{n-1} \oplus \mathbf{w}_{n-1} \right)}{\partial \mathbf{g}} \right]\right|_{\mathbf{g}=\bar{\mathbf{g}}}
- X^T \right) \\
&= 2\left( \lambda \nabla^T \mathrm{diag}\left(\nabla \bar{\mathbf{f}}^o_n\right)\left. \left[ \frac{\partial \left(\mathbf{w}_{n-1} \oplus \mathbf{w}_{n-1} \right)}{\partial \mathbf{f}^o_{n-1}} \frac{\partial \mathbf{f}^o_{n-1}}{\partial \mathbf{g}} \right] \right|_{\mathbf{g}=\bar{\mathbf{g}}}
- X^T \right) \\
&= 2\left( \lambda \nabla^T \mathrm{diag}\left(\nabla \bar{\mathbf{f}}^o_n\right) \left. \left[ \frac{\partial \left(\mathbf{w}_{n-1} \oplus \mathbf{w}_{n-1} \right)}{\partial \mathbf{f}^o_{n-1}}\right]\right|_{\mathbf{g}=\bar{\mathbf{g}}} 
 J_{n-1} - X^T \right).
\end{split}
\end{equation}
In order to evaluate $\frac{\partial \left(\mathbf{w}_{n-1} \oplus \mathbf{w}_{n-1} \right)}{\partial \mathbf{f}^o_{n-1}}$ we note that
\begin{equation}
|\nabla \mathbf{f}^o_{n-1}|^2 = \left(\nabla^x \mathbf{f}^o_{n-1} \right)^2 + \left(\nabla^y \mathbf{f}^o_{n-1} \right)^2.
\end{equation}
Therefore, recalling Eqn. \ref{eqn:w} and applying the chain rule,
\begin{equation}
\label{eqn:wbar}
\begin{split}
\left. \frac{\partial \left(\mathbf{w}_{n-1} \right)}{\partial \mathbf{f}^o_{n-1}} \right|_{\mathbf{g}=\bar{\mathbf{g}}} =& -\left. \left[\frac{1}{2} \mathrm{diag}\left( \left\{ \eta_{n-1}^2+ | \nabla \mathbf{f}^o_{n-1} |^2 \right\}^{-\frac{3}{2}} \right)
\frac{\partial }{\partial \mathbf{f}^o_{n-1}} \left(  \left(\nabla^x \mathbf{f}^o_{n-1} \right)^2 + \left(\nabla^y \mathbf{f}^o_{n-1} \right)^2 \right) \right] \right|_{\mathbf{g}=\bar{\mathbf{g}}}\\
=&  -\left[\mathrm{diag}\left( \left\{ \eta_{n-1}^2+ | \nabla \bar{\mathbf{f}}^o_{n-1} |^2 \right\}^{-\frac{3}{2}} \right)
 \left(  \mathrm{diag}\left(\nabla^x \bar{\mathbf{f}}^o_{n-1} \right)\nabla^x + \mathrm{diag}\left(\nabla^y \bar{\mathbf{f}}^o_{n-1} \right)\nabla^y \right) \right].
\end{split}
\end{equation}
For the concatenated vectors, we then have
\begin{equation}
\frac{\partial \left(\mathbf{w}_{n-1} \oplus \mathbf{w}_{n-1} \right)}{\partial \mathbf{f}^o_{n-1}} = 
\begin{bmatrix}
\frac{\partial \left(\mathbf{w}_{n-1} \right)}{\partial \mathbf{f}^o_{n-1}} \\
\frac{\partial \left(\mathbf{w}_{n-1} \right)}{\partial \mathbf{f}^o_{n-1}}
\end{bmatrix}
\end{equation}
For compactness, we will denote the above matrix, evaluated at $\mathbf{g}=\bar{\mathbf{g}}$, as $W_{n-1}$, so that
\begin{equation}
H_\mathbf{fg}= -2\left( \lambda \nabla^T \mathrm{diag}\left(\nabla \bar{\mathbf{f}}_n^o\right) W_{n-1} J_{n-1}
+ X^T \right)
\end{equation}
The Jacobian matrix $J_n$ can then be defined recursively for $n>1$ as 
\begin{equation}
\label{eqn:jac}
J_n = \left(\lambda \nabla^T \mathrm{diag}\left(\bar{\mathbf{w}}_{n-1} \oplus \bar{\mathbf{w}}_{n-1}\right)\nabla + X^TX\right)^{-1}  \left( \lambda \nabla^T \mathrm{diag}\left(\nabla \bar{\mathbf{f}}_n^o\right) W_{n-1} J_{n-1}
+ X^T \right).
\end{equation}
The resulting Jacobian matrix fully describes the linearization of the TV-penalized reconstruction obtained with $n$ IRLS iterations. 
The resulting covariance matrix for the image is then
\begin{equation}
\label{eqn:cov}
K_{\mathbf{f}_n^o} = J_n K_\mathbf{g} J_n^T.
\end{equation}

\subsection{An Improved Estimate of $\bar{\mathbf{w}}$}
\label{wbar}

As stated previously, the mean value of the $n$th weight vector, $\bar{\mathbf{w}}_n$ cannot be accurately obtained through reconstruction of noise-free data.
\new{Further, accurate image covariance approximation relies on obtaining at least a rough approximation of $\bar{\mathbf{w}}_n$.
Without access to the full probability density function for $\mathbf{w}_n$, the approach we take is to assume that the image at each iteration of IRLS can be modeled 
as a multivariate Gaussian distribution, with covariance given by the method described in Section \ref{linear_irls}. Since the covariance of the first iterate, $K_{\mathbf{f}_1}$, 
does not depend on a weight vector, we can construct an approximate distribution function for each $\mathbf{w}_n$ one at a time, extracting the average values $\bar{\mathbf{w}}_n$
along the way.}
Note that for compactness we will temporarily drop the subscript $n$, as the following analysis 
holds for every iteration.
We begin by observing that each element of $\mathbf{w}$ depends on the previous iterate only through the corresponding elements of $\nabla_x \mathbf{f}$ and $\nabla_y \mathbf{f}$. 
We then define $u:=\left(\nabla_x \mathbf{f}\right)_i$ and $v:=\left(\nabla_y \mathbf{f}\right)_i$ as the $i$th pixel values of the $x-$ and $y-$ gradient images. We begin by 
considering the cumulative distribution function $F_{\mathbf{w}_i}\left(w\right) = P\left( \mathbf{w}_i < w \right)$. Inserting the definition of $\mathbf{w}_i$, we have 

\begin{equation}
\begin{split}
\mathbf{w}_i<w &\implies \frac{1}{\sqrt{\eta^2 + u^2 + v^2}} < w \\
&\implies \frac{1}{w} < \sqrt{\eta^2 + u^2  + v^2}\\
&\implies \frac{1}{w^2} - \eta^2 <  u^2 + v^2. 
\end{split}
\end{equation}
Then we can write the cumulative distribution function of $\mathbf{w}_i$ as
\begin{equation}
\begin{split}
F_{\mathbf{w}_i}(w) &= 1 - P\left\{ u^2 + v^2 < \frac{1}{w^2} - \eta^2 \right\} \\
&= 1 - \bigints_{-\sqrt{\frac{1}{w^2}-\eta^2}}^{\sqrt{\frac{1}{w^2}-\eta^2}}\left[ f_1(v)dv\bigints_{-\sqrt{\frac{1}{w^2} - \eta^2 -v^2}}^{\sqrt{\frac{1}{w^2} - \eta^2 -v^2}} f_2(u) du \right],
\end{split}
\end{equation}
where $f_1(v)$ and $f_2(u)$ are the probability density functions of $u$ and $v$. While the final probability density function of $\mathbf{w}$ is highly non-Gaussian, in our experience the gradient magnitude image 
is comparatively well approximated as a multivariate Gaussian distribution. Following this assumption, $u$ and $v$ can be described by Gaussian distributions with means $\mu_u$ and $\mu_v$ and variances 
$\sigma^2_u$ and $\sigma_v^2$, respectively.

We then have
\begin{equation}
\begin{split}
F_{\mathbf{w}_i}(w) =&  1 - \bigints_{-\sqrt{\frac{1}{w^2}-\eta^2}}^{\sqrt{\frac{1}{w^2}-\eta^2}}\left[ \frac{dv}{\sigma_v \sqrt{2\pi}}e^{-\frac{\left(v-\mu_v\right)^2}{2\sigma_v^2}}
\bigints_{-\sqrt{\frac{1}{w^2} - \eta^2 -v^2}}^{\sqrt{\frac{1}{w^2} - \eta^2 -v^2}}\frac{1}{\sigma_u \sqrt{2\pi}}e^{-\frac{\left(u-\mu_u\right)^2}{2\sigma_u^2}} du \right] \\
=& 1 - \bigints_{-\sqrt{\frac{1}{w^2}-\eta^2}}^{\sqrt{\frac{1}{w^2}-\eta^2}}\left[ \frac{dv}{\sigma_v \sqrt{2\pi}}e^{-\frac{\left(v-\mu_v\right)^2}{2\sigma_v^2}}
\frac{1}{2} \left\{ \mathrm{erf}\left( \frac{\sqrt{\frac{1}{w^2} - \eta^2 -v^2}-\mu_u}{\sqrt{2}\sigma_u}\right)  \right.\right.\\
& \left. \left. -\mathrm{erf}\left( \frac{-\sqrt{\frac{1}{w^2} - \eta^2 -v^2}-\mu_u}{\sqrt{2}\sigma_u}\right)  \right\} \right].
\end{split}
\end{equation}
Recall (see, for example, Section 5-3 of \cite{papoulis_probability_2002}) that for a random variable $x$ taking only positive values, $E\left\{x\right\} = \int_0^\infty dx^\prime\left[1-F_x\left(x^\prime\right) \right]$. We can then write $E\left\{\mathbf{w}_i\right\}$ as
\begin{equation}
\label{eqn:wapprox}
\begin{split}
E\left\{\mathbf{w}_i\right\} =& \bigints_0^{\frac{1}{\eta}} dw \bigints_{-\sqrt{\frac{1}{w^2}-\eta^2}}^{\sqrt{\frac{1}{w^2}-\eta^2}}\left[ \frac{dv}{\sigma_v \sqrt{2\pi}}e^{-\frac{\left(v-\mu_v\right)^2}{2\sigma_v^2}}
\frac{1}{2} \left\{ \mathrm{erf}\left( \frac{\sqrt{\frac{1}{w^2} - \eta^2 -v^2}-\mu_u}{\sqrt{2}\sigma_u}\right)  \right.\right.\\
& \left. \left. -\mathrm{erf}\left( \frac{-\sqrt{\frac{1}{w^2} - \eta^2 -v^2}-\mu_u}{\sqrt{2}\sigma_u}\right)  \right\} \right],
\end{split}
\end{equation}
where we have replaced the upper limit on the outer integral with the $1/\eta$, as this is the maximum value of $w$.

Given pixel variance estimates for a previous iteration of IRLS, the approximation of $\bar{\mathbf{w}}$ for the subsequent iteration can be calculated numerically using Eqn. \ref{eqn:wapprox} and any 
number of available numerical integration packages to perform the 2-dimensional integration for each image pixel. In practice we have found this computation to be feasible, even for large systems. Two 
points are important in order to speed the computation of $\bar{\mathbf{w}}$: First, since the numerical integration is performed pixel-by-pixel, it is trivially parallelizable. Second, while Eqn. \ref{eqn:wapprox}
relies on knowledge of the covariance matrix of the previous iterate (through $\sigma_u$ and $\sigma_v$), we have observed that a rough approximation of the pixel variances is sufficient for a good 
approximation of $\bar{\mathbf{w}}$. For large images, we have implemented this approximate solution by loosening the convergence criteria for the various matrix inversions of Eqn. \ref{eqn:jac}, which are implemented as 
linear solves for large images, while for smaller systems, we store the full covariance matrix at each iteration directly. More 
sophisticated methods exist for efficiently computing approximate variance maps, such as those in Ref. \cite{zhang-oconnor_fast_2007} employing Fourier methods, and 
these could also potentially be implemented to speed pre-computation of $\bar{\mathbf{w}}$ for each iteration.

\subsection{Implementation of $K_{\mathbf{f}_n^o}$ }
\label{implementation}

Even after pre-computing estimates of $\bar{\mathbf{f}}_n$ and $\bar{\mathbf{w}}_n$, 
it is not immediately obvious that the expression for the Jacobian given in Eqn. \ref{eqn:jac} enables one to perform efficient computation of inner products with $K_{\mathbf{f}_n^o}$.
If one were to expand that equation, the number of large matrix inversions would be seen to grow geometrically with the number of IRLS iterations. However, proper implementation 
allows only a linear dependence on the number of IRLS iterations. Further, in our experience fewer than 10 iterations was always sufficient to reasonably approximate the converged 
TV-penalized reconstruction, and as few as 5 or 6 iterations was commonly sufficient. 
In order to illustrate the implementation of an inner product with $K_{\mathbf{f}_n^o}$, the procedure is outlined in Algorithms \ref{alg1} and \ref{alg2}.
\begin{algorithm}                      
\caption{Calculate $\mathbf{y} =J_n\mathbf{x}$}          
\label{alg1}                           
\begin{algorithmic}                    
\State SOLVE$\left(A\right.$,$\left.\mathbf{b}\right)$ is any algorithm to solve a linear system $A\mathbf{x}=\mathbf{b}$ for $\mathbf{x}$.
\State Obtain $\bar{\mathbf{w}}_i$ and $\bar{\mathbf{f}}_i$ for $i=1\dots n$ by reconstruction of noise-free data and the procedure outlined in Section \ref{wbar}.

\State $\tilde{\mathbf{f}} \gets X^T\mathbf{x}$
\State $\mathbf{y}_1 = \mathrm{SOLVE}\left(\lambda \nabla^T\nabla + X^TX,\tilde{\mathbf{f}}\right)$
\For{$i  = 1 \dots n-1$} 
	\State $\tilde{\mathbf{y}}_{i} \gets \lambda \nabla^T \mathrm{diag}\left(\nabla\bar{\mathbf{f}_i}\right)W_i \mathbf{y}_i + \tilde{\mathbf{f}}$
	\State $\mathbf{y}_{i+1} = \mathrm{SOLVE}\left(\lambda \nabla^T \mathrm{diag}\left(\bar{\mathbf{w}}_{i} \oplus \bar{\mathbf{w}}_{i}\right)\nabla + X^TX,\tilde{\mathbf{y}}_i\right)$
\EndFor
\State \Return $\mathbf{y}_n$
\end{algorithmic}
\end{algorithm}

\begin{algorithm}                      
\caption{Calculate $\mathbf{x} =J_n^T\mathbf{y}$}          
\label{alg2}                           
\begin{algorithmic}                    
\State SOLVE$\left(A\right.$,$\left.\mathbf{b}\right)$ is any algorithm to solve a linear system $A\mathbf{x}=\mathbf{b}$ for $\mathbf{x}$.
\State Obtain $\bar{\mathbf{w}}_i$ and $\bar{\mathbf{f}}_i$ for $i=1\dots n$ by reconstruction of noise-free data and the procedure outlined in Section \ref{wbar}.
\State $\tilde{\mathbf{y}}_n \gets \mathbf{y}$
\State $\textbf{x} \gets \mathbf{0}_{M\times 1}$
\For{$i  = n-1 \dots 1$} 
	\State $\tilde{\mathbf{g}}_i \gets \mathrm{SOLVE}\left(\lambda \nabla^T \mathrm{diag}\left(\bar{\mathbf{w}}_{i} \oplus \bar{\mathbf{w}}_{i}\right)\nabla + X^TX,\tilde{\mathbf{y}}_{i+1}\right)$	
	\State $\textbf{x} \gets \textbf{x} + X\tilde{\textbf{g}}_i$
	\State $\tilde{\mathbf{y}}_i \gets \lambda W_i^T\mathrm{diag}\left(\nabla\bar{\mathbf{f}}_i\right)\nabla \tilde{\mathbf{g}}_i$
\EndFor
\State $\tilde{\mathbf{g}}_0 = \mathrm{SOLVE}\left(\lambda \nabla^T\nabla + X^TX,\tilde{\mathbf{y}}_1\right)$
\State $\mathbf{x} \gets \mathbf{x} + X\tilde{\mathbf{g}}_0$
\State \Return $\mathbf{x}$

\end{algorithmic}
\end{algorithm}

\section{Validation on a Small System}
\label{validation_method}
\subsection{Method for Validation}

The most direct means of validating the proposed covariance approximation is through comparison to sample statistics estimated from many noise realizations. 
In order to compute accurate sample statistics for covariances, however, the number of realizations obtained must be at least roughly on the order of the number of image pixels.
Therefore, in order to investigate a range of reconstruction parameter values, we perform validation using 2,500 noise realizations of a 32 $\times$ 32 pixel image for a variety of 
regularization parameters $\lambda$ and total IRLS iterations. The covariance estimates obtained from the noise realizations are the maximum likelihood estimates of image covariance. The 
noise model used is a Gaussian noise model with variance inversely proportional to the photon flux incident on the detector. The noise level simulated is characteristic of typical noise in dedicated breast CT, 
which is relatively high compared to many other CT applications.

\begin{figure}
 \includegraphics[width=0.98\columnwidth, trim= 30 0 10 0 ]{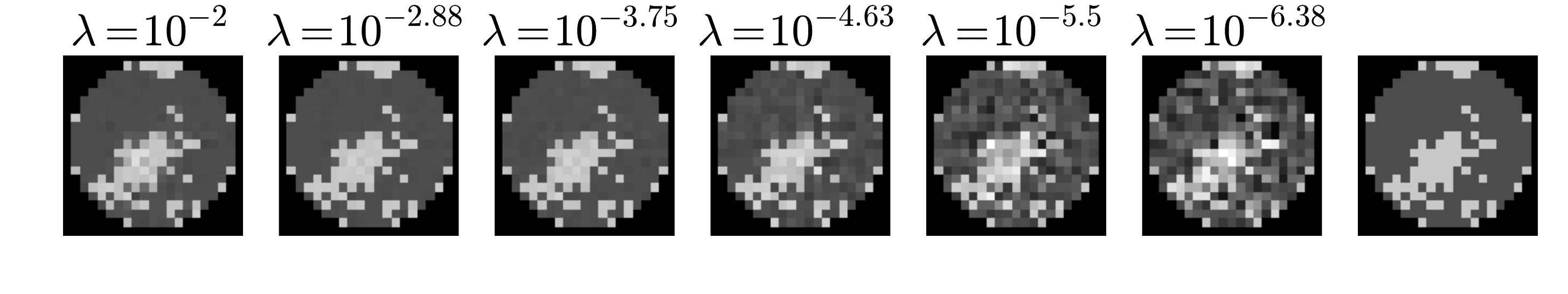}
 \caption{The noise level used for validation is illustrated qualitatively in these reconstructed  images with five values of regularization parameter $\lambda$, each reconstructed using 10 iterations of IRLS. 
 These images are from separate noise realizations and 
 correspond to the five regularization parameter strengths used in the evaluation below. The image on the far right
 is the numerical phantom used.
 The display window is [0.17, 0.25] cm$^{-1}$. \label{fig:recons}}
 \end{figure}

We investigate up to 8 iterations of IRLS and values of $\lambda$ ranging from 10$^{-2}$ to 10$^{-6.3}$. 
For the majority of the $\lambda$ settings we investigate here, 8 iterations is sufficient to reconstruct an image which is visibly indistinguishable from the converged solution. 
The phantom chosen is a digital breast phantom. For the small scale validation, a 32 $\times$ 32 pixel image is used, along with 64 detector pixels and 12 projection views. This 
configuration has sparse projection-view sampling and requires TV minimization for accurate reconstruction of noiseless data. The acquisition geometry used is circular fan-beam, with 
a magnification factor of 1.92, and the full circular field-of-view is inscribed in the reconstructed image. 
The range of $\lambda$ values investigated is chosen through visual inspection of noisy images so that both under- and over-regularized images are investigated 
(see Figure \ref{fig:recons}). Finally, each iteration of the IRLS algorithm is solved using the method of conjugate gradients. 

\new{As an initial validation of our approximation, we inspect variance and covariance images for the proposed method, as well as sample variance and covariance derived from noise realizations. In
additions to this visual inspection,} several comparison metrics are used to evaluate the approximation of image covariance. First, a simple root-mean-squared error (RMSE) between the elements of the approximated and sample covariance 
matrices is computed. Likewise, RMSE between the matrix diagonals (variance maps) is calculated. In each case, we normalize the RMSE by the mean image pixel variance derived from the noise realizations. This
allows for a performance comparison across regularization parameter values. Next, we compare the two covariance matrices using a metric motivated by Pearson's R, in order to 
investigate the linearity of the approximation with respect to the sample covariance. We define this metric as 
\begin{equation}
R^2 = \left( \frac{\mathrm{Cov}\left(K_1,K_2\right)}{\sigma_{K_1}\sigma_{K_2}}\right)^2,
\end{equation}
where the term in the numerator is the sample covariance between all elements of $K_1$ and all elements of $K_2$, and the terms in the denominator are the sample standard deviation of 
all elements within $K_1$ and all elements within $K_2$, respectively. This $R^2$ metric will be equal to unity when a single linear function relates all elements of $K_1$ to elements of $K_2$. 
Similarly, we define an equivalent metric where we consider only pixel variances and neglect the off-diagonal elements of $K_1$ and $K_2$. 

The above metric summarizes how well two matrices agree to within an arbitrary shift and scaling. In other words, it characterizes the goodness-of-fit of the model $K_1 = aK_2 + b$ for scalars $a$ and $b$. 
 However, it is also informative to quantify the absolute predictive value of our approximation. 
For this, we construct a metric which is a variation on the previous regression metric $R^2$, in which the goodness-of-fit between the matrices is evaluated for the model $K_1 = K_2$. We define this metric as
\begin{equation}
\tilde{R}^2 = 1- \frac{\frac{1}{N}\sum_i\left( K_1^i - K_2^i\right)^2}{\mathrm{Var}\left(K_2\right)},
\end{equation}
where the summation is over all $N$ elements of the matrices, and as before the denominator term is the sample variance among the elements of $K_2$. In terms of linear regression, the term $K_2$ represents the 
samples obtained, while $K_1$ represents the modeled values. Therefore, we insert the sample covariance matrix as $K_2$ and the approximated covariance matrix as $K_1$. 

Finally, we evaluate our approximation with a metric proposed by Ref. \cite{forstner_metric_2003}. The primary appeal of this metric is that it is invariant to matrix inversion, so that the applicability of 
our approximation to Hotelling observer metrics, where the covariance is inverted, can be assessed. This metric is defined as 
\begin{equation}
d = \sqrt{\sum_i \mathrm{ln}^2 s_i},
\end{equation}
where $s_i$ represents the $i$th singular value of the matrix $K_2 K_1^{-1}$.

\subsection{Results of Validation on a Small System}

\begin{figure}[ht]
  \centering
 \includegraphics[width=0.65\columnwidth]{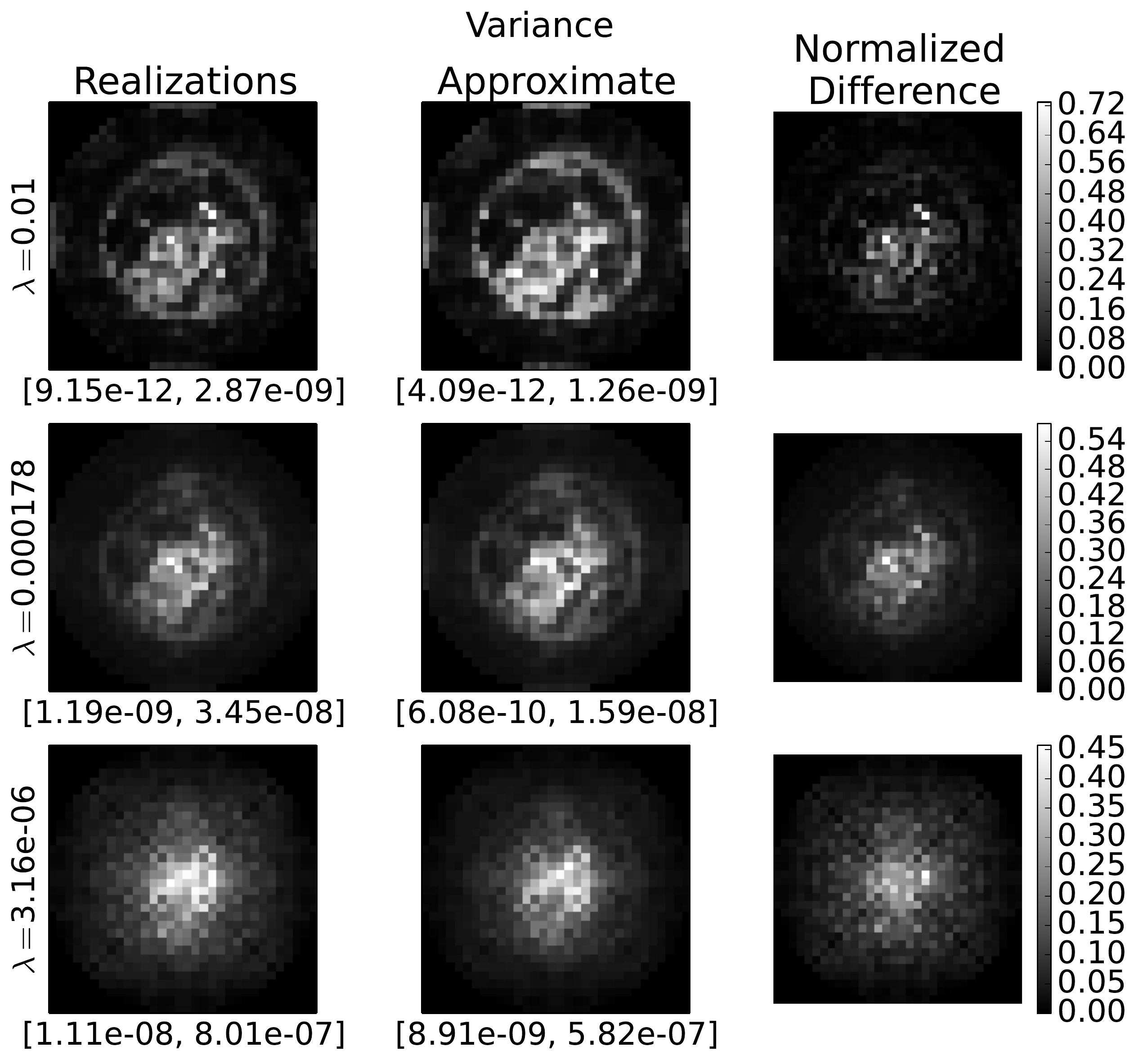}
 \caption{Variance images from three different regularization parameter strengths. The display window of each image is given in the figure. The display windows are different so that the structure of the 
 variance map estimate can be assessed. The right column provides the difference image between our approximation and the sample covariance from realizations, normalized to the maximum sample variance. \label{fig:var1}}
 \end{figure}

\new{We begin by presenting variance and covariance images from our proposed method, along with sample variance and covariance images derived from 2,500 noise realizations.}
Figure \ref{fig:var1} shows variance maps obtained from realizations (left column) versus those obtained with our 
proposed approximation (center column) after 6 IRLS iterations. Each row corresponds to a different setting of the regularization parameter $\lambda$. The display windows for each image are also provided in the figure. The 
right column shows the difference between the approximation and the result from realizations, normalized by the maximum image variance. Clearly, while the proposed approximation consistently underestimates 
the absolute image pixel variance, the structure of the variance map is well approximated. \new{A comparison based on RMSE is sensitive both to overall structural differences in the covariance matrix, as well as to 
offset and scaling of the covariance, which depending on the application may or may not be of interest. Isolation of these two types of comparison is the motivation for the $R^2$ and $\tilde{R}^2$ metrics previously discussed.} 
 
\begin{figure}[ht]
 \centering
 \includegraphics[width=0.65\columnwidth]{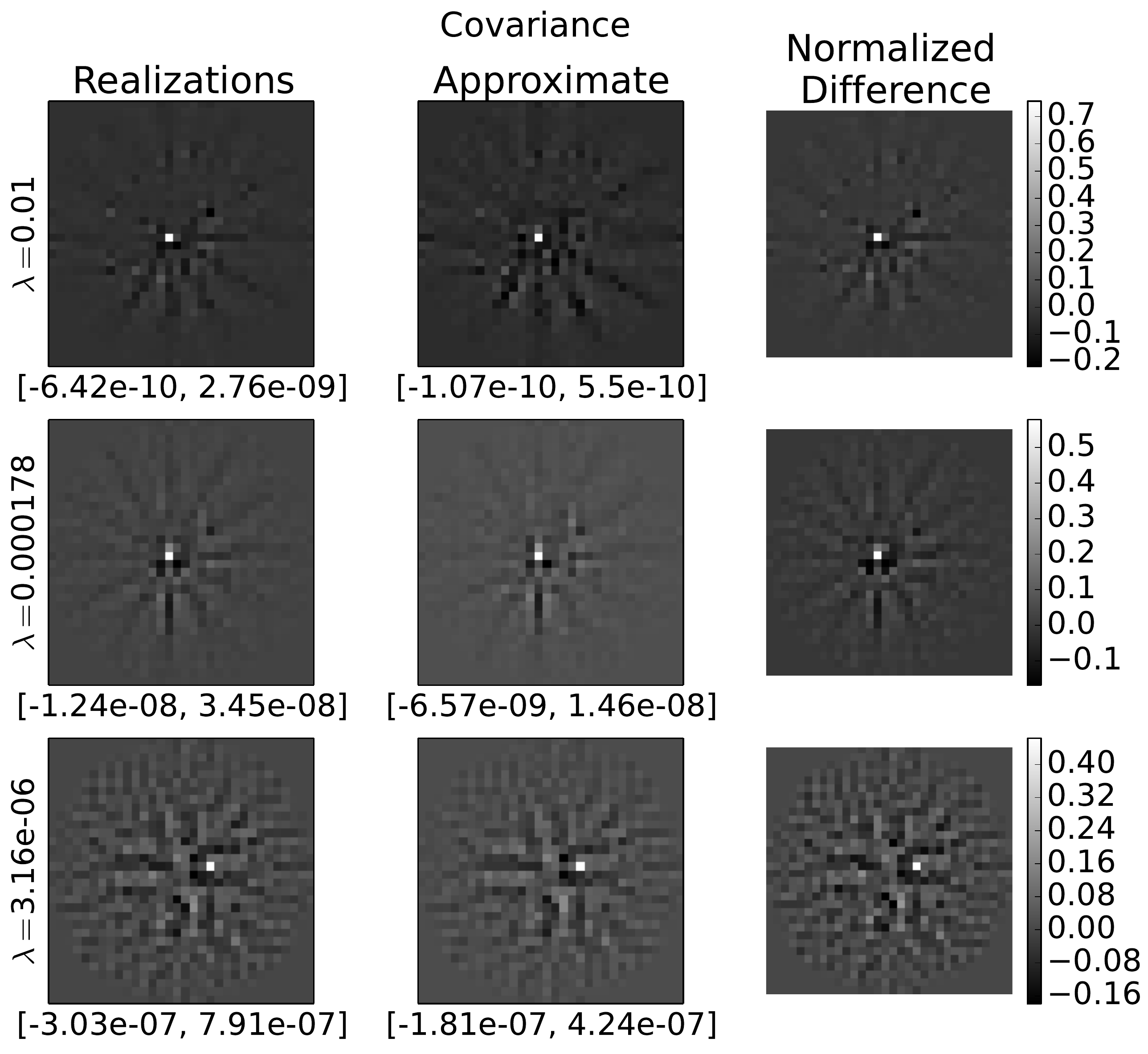}
 \caption{Individual rows from the covariance matrices for three regularization parameter values. These rows had the worst RMSE between the two matrices of any rows, so they are the 
 worst case approximations for each regularization strength. The color scale for the difference image is in \% of the maximum covariance (i.e. percentage of the variance of the pixel corresponding to the 
 row being visualized). \label{fig:cov1}}
 \end{figure}

 Similarly, Figure \ref{fig:cov1} shows results for image covariance with a single pixel for the proposed approximation as well as for the sample covariance obtained from realizations. The pixel whose covariance is 
 shown has highest RMSE for a given parameter combination. Again, as in the case of the variances, the approximation underestimates image pixel covariance but preserves the covariance structure. In terms of 
 Hotelling observer assessment, this suggests that \new{if the proposed method were used to estimate the Hotelling template, the resulting template would have an overall structure close to that of the true Hotelling template.
 However, the slight inaccuracy in the covariance's scale would lead to a positive offset in the estimated HO SNR. Ultimately, the importance of absolutely quantifying the covariance magnitude is dependent on the 
 application of interest. In subsequent sections, we focus primarily on questions of object-dependence and stationarity, which can be addressed by only considering the structure of the image covariance.}

 \begin{figure}[ht!]
 \centering
 \includegraphics[width=0.57\columnwidth]{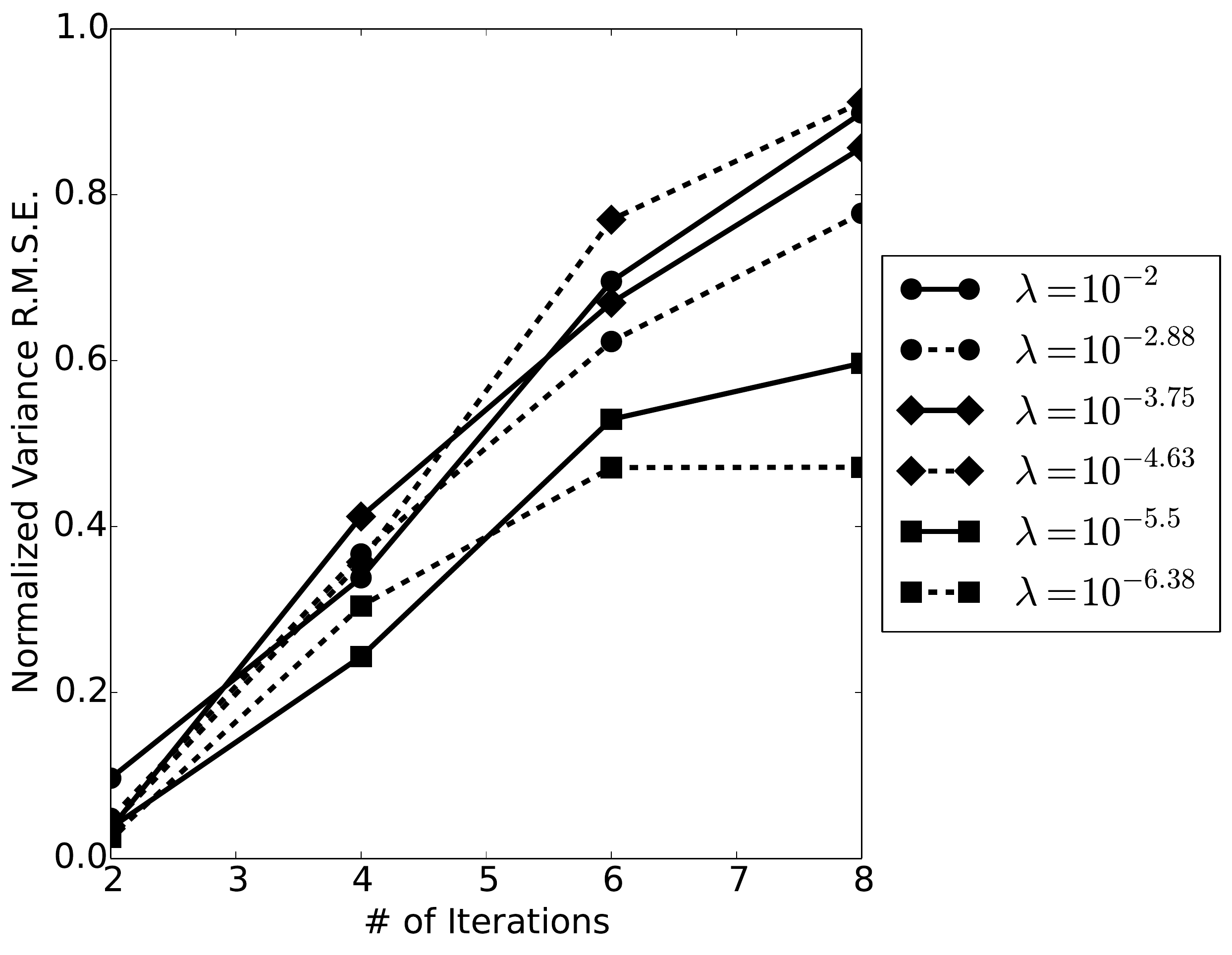} \includegraphics[width=0.42\columnwidth]{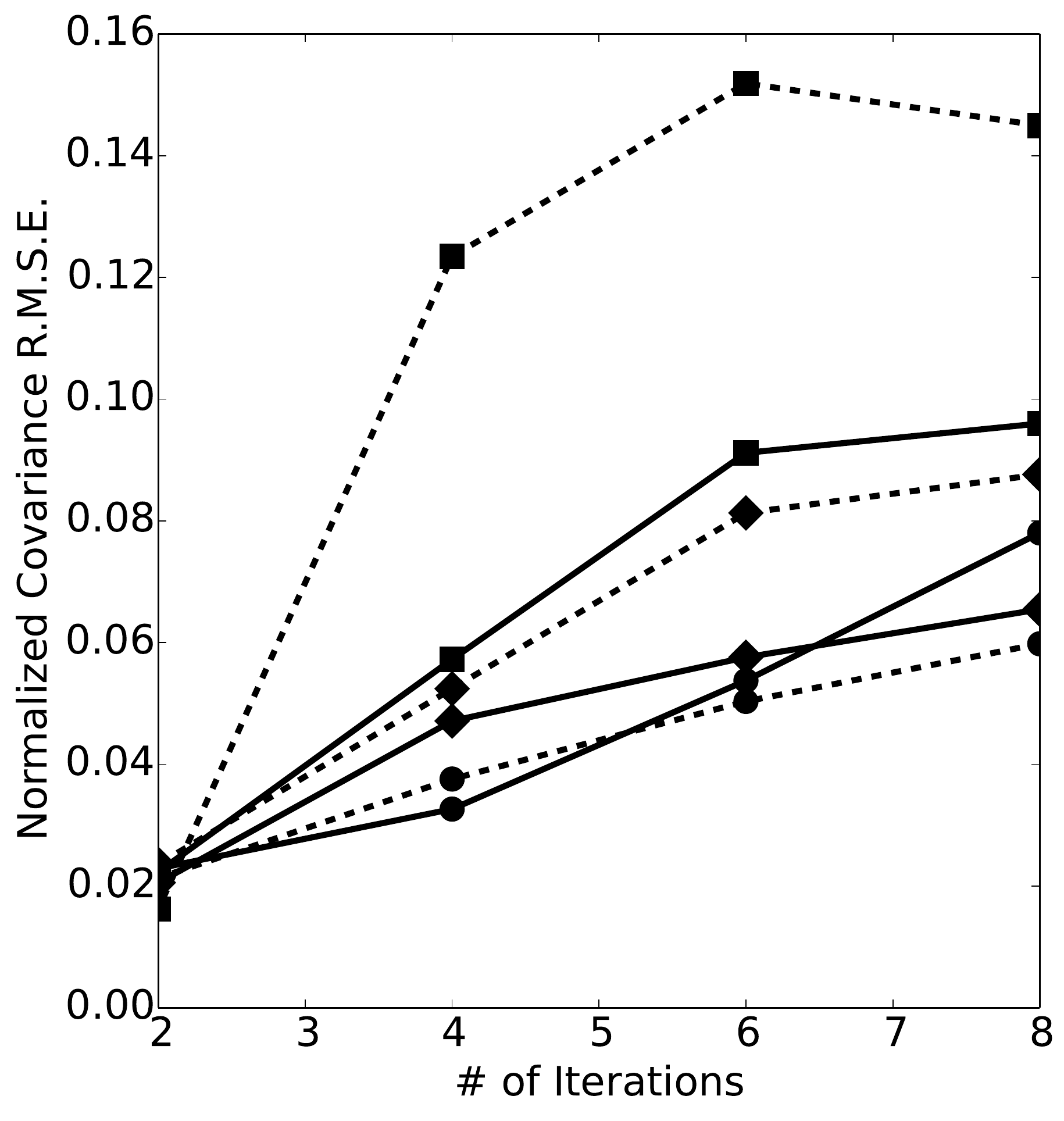} 
 \caption{The normalized RMSE between the approximated covariance matrix and the covariance matrix derived from realizations
 for the 32x32 image is shown for a range of regularization parameters including only the variance terms (left) and including all of the covariance terms (right). In this and subsequent figures, error bars arising 
 from the statistical uncertainty of the sample covariance estimates are too small to be seen.
\label{fig:parm_sweep1}}
\end{figure}

\new{In order to quantitatively evaluate our method, we now apply the various metrics previously described in Section \ref{validation_method}.}
Figure \ref{fig:parm_sweep1} shows the comparison of our approximation of the image covariance $K_\mathbf{f}$ with a sample covariance matrix derived from 2,500 independent noise realizations
in terms of RMSE. The RMSE values have been normalized to the mean sample variance since the overall noise level varies greatly across the range of $\lambda$ setting used. Error bars arising from the 
statistical uncertainty of the sample covariance are too small to be visualized on this plot. The general 
trend that error increases with successive iterations of the IRLS algorithm is evident, however these results do not convey an obvious dependence of variance accuracy on the regularization parameter. 
Covariance RMSE appears to have a general trend toward increasing with decreased regularization strength. However, apart from the result that the approximation is more accurate at early iterations, 
RMSE does not provide a great deal of insight.

  \begin{figure}[ht!] 
  \centering
  \includegraphics[width=0.57\columnwidth]{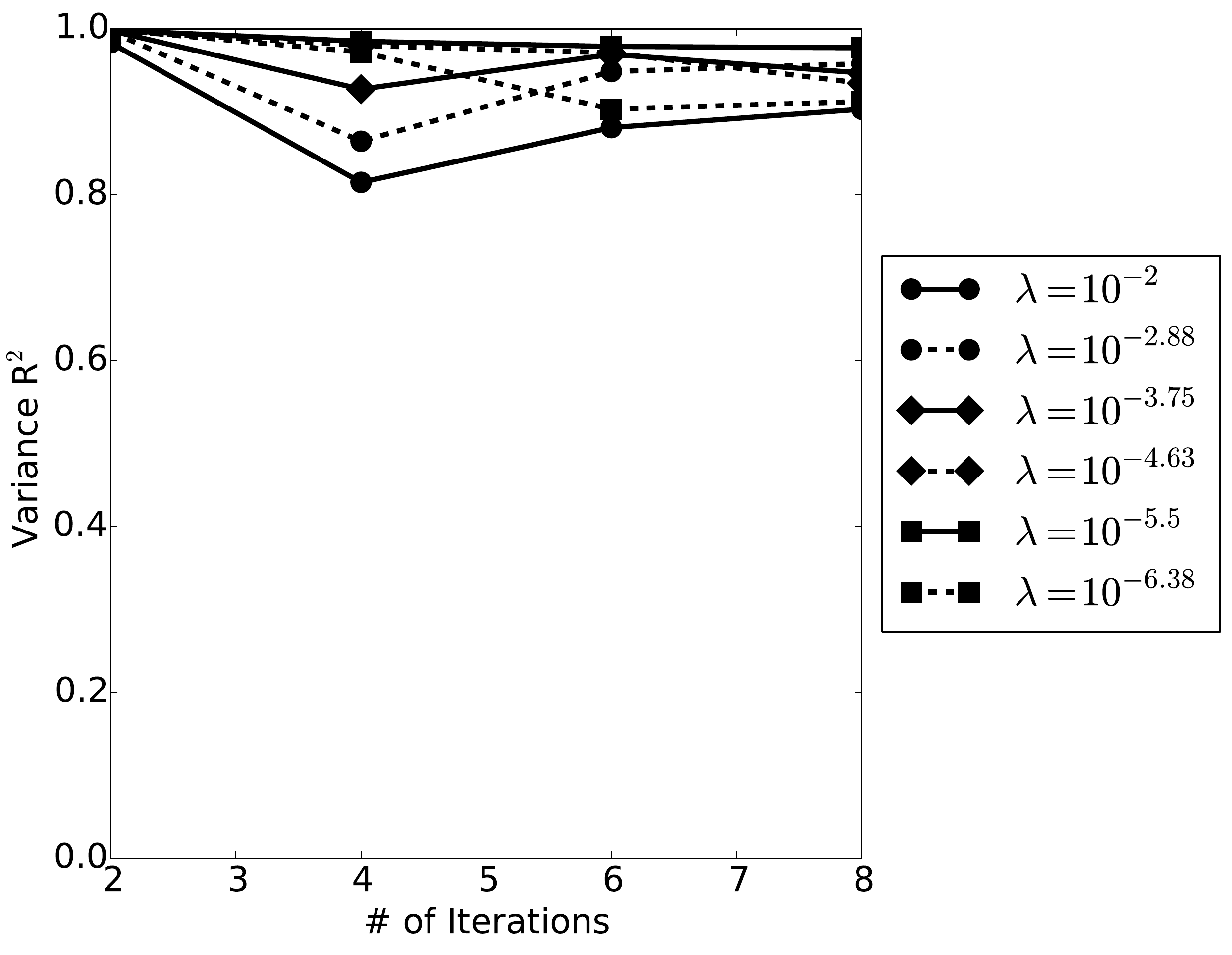} \includegraphics[width=0.41\columnwidth]{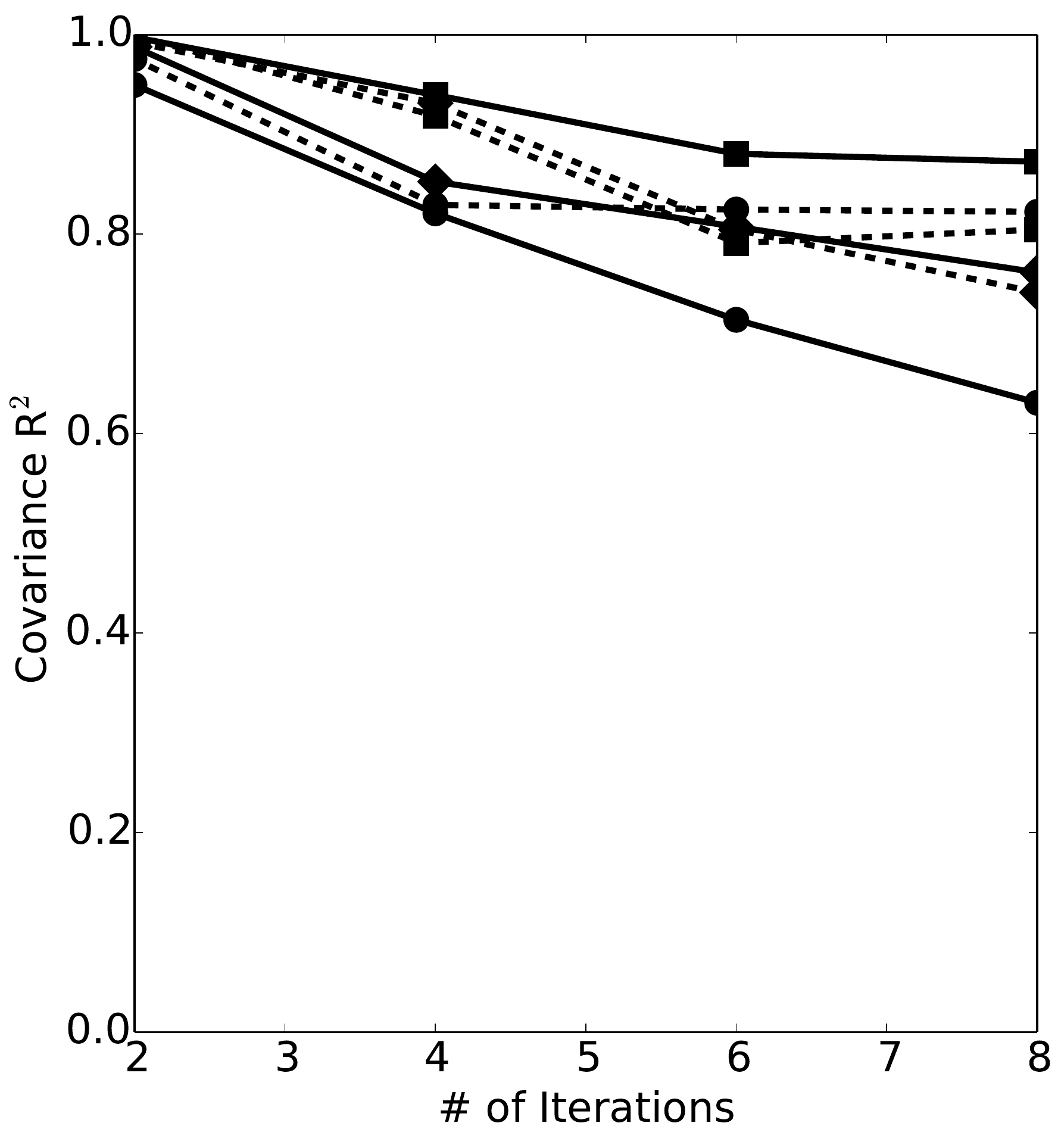} 
  \caption{Coefficients of determination $R^2$ for the variances and covariances between the approximation method and noise realizations.\label{fig:parm_sweep2}}
  \end{figure}

For a more interpretable metric, Figure \ref{fig:parm_sweep2} shows the coefficient of determination $R^2$ for each parameter setting investigated. The absolute scale of this metric is meaningful in that 
a value of 1 indicates a perfect linear agreement between the approximated and sample variance or covariance. \new{In this way, we can isolate structural errors in the variance and covariance approximations from the error in 
scale shown in Figures \ref{fig:var1} and \ref{fig:cov1}.} The linearity assumption appears relatively robust for variances and covariances across 
a wide range of parameter values. Higher $\lambda$ values, corresponding to stronger regularization, degrade the approximation slightly in terms of linearity for both variances and covariances. As with RMSE, the impact of
iteration number also worsens the approximation gradually, however this effect seems stronger for covariances than for variances. Overall the linearity of our approximated covariance with respect to the actual covariance appears
well validated, implying that variance maps and covariances can be accurately approximated in terms of their overall structure.

\begin{figure}[ht!] 
  \centering
  \includegraphics[width=0.57\columnwidth]{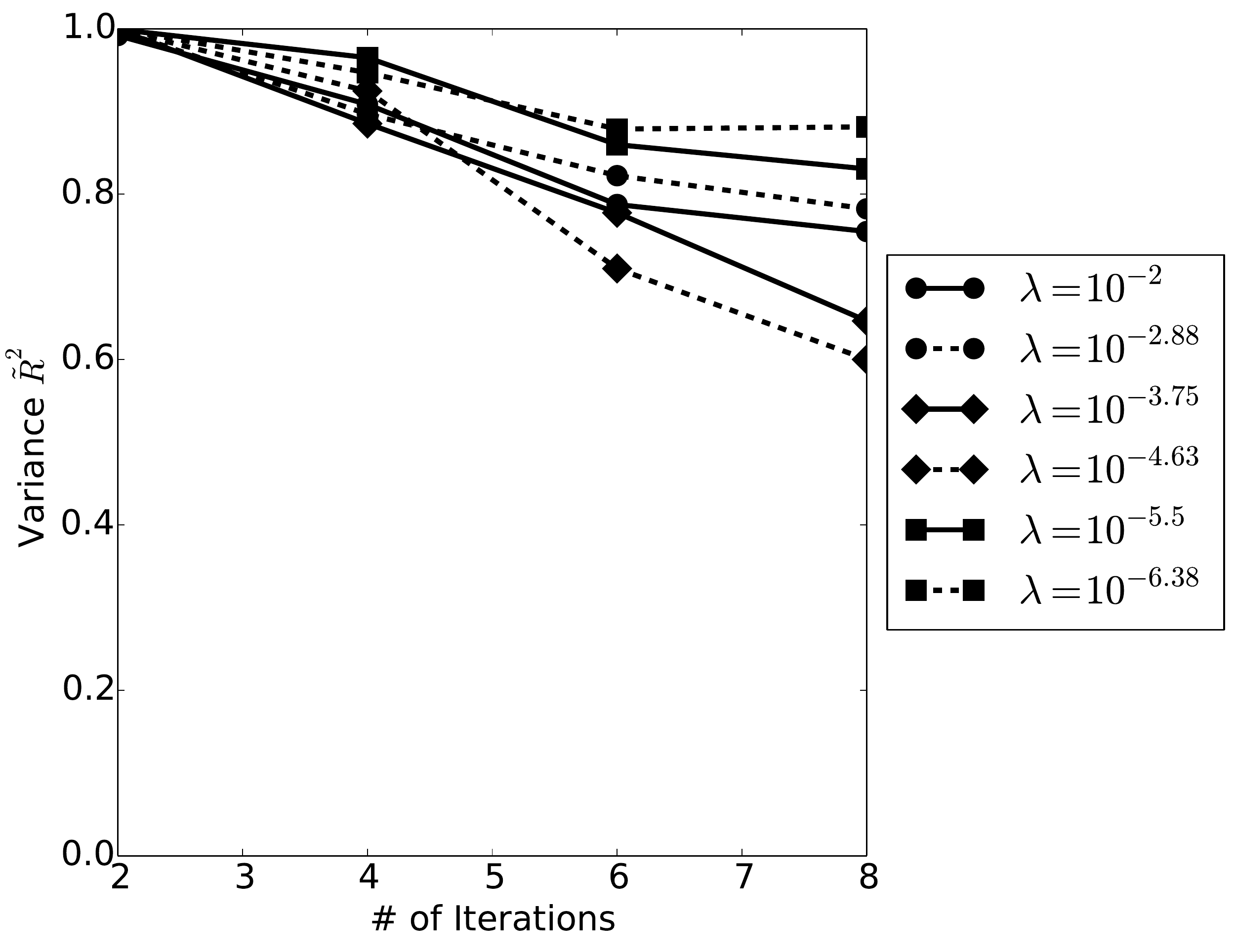} \includegraphics[width=0.41\columnwidth]{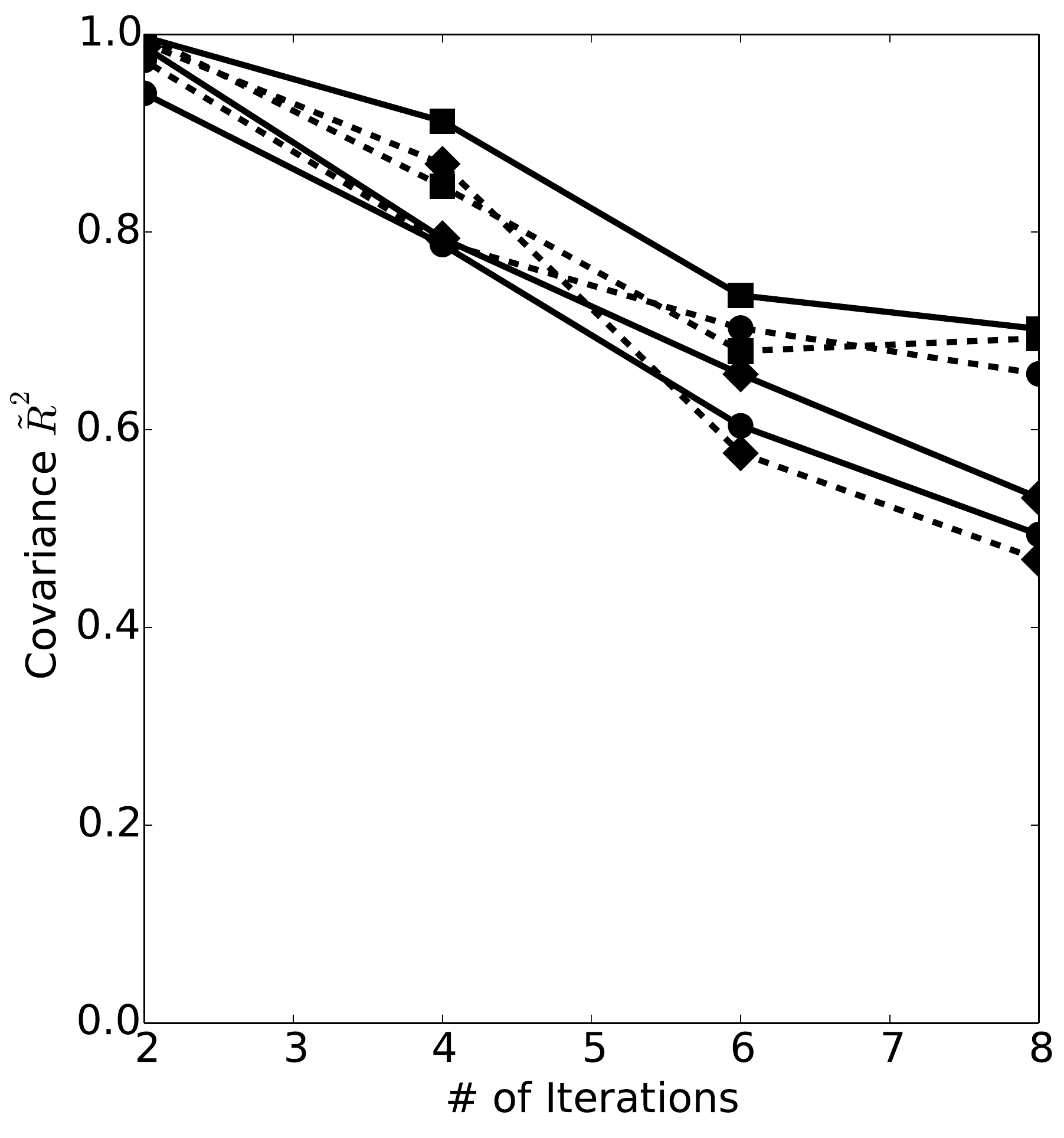} 
  \caption{The regression metric $\tilde{R}^2$ with a model of equality between the approximation and the sample covariance is illustrated here as a function of regularization parameter and iteration. 
  Results are shown for the variances alone (Left) as well as for the full covariance matrix (Right).\label{fig:parm_sweep3}}
  \end{figure}

In order to assess the absolute quantitative accuracy of our covariance approximation, Figure \ref{fig:parm_sweep3} illustrates the dependence of the modified coefficient $\tilde{R}^2$ on iteration number and $\lambda$ value. 
As expected, these values are slightly lower than the $R^2$ values, since the model of equality between the approximated and sample covariance matrices is stricter than a linear model. However, the drop in 
the metric is modest, particularly since six iterations of IRLS was frequently sufficient to approximate the converged TV solution to within visually discernible difference. 

\begin{figure}
  \centering
  \includegraphics[width=0.6\columnwidth]{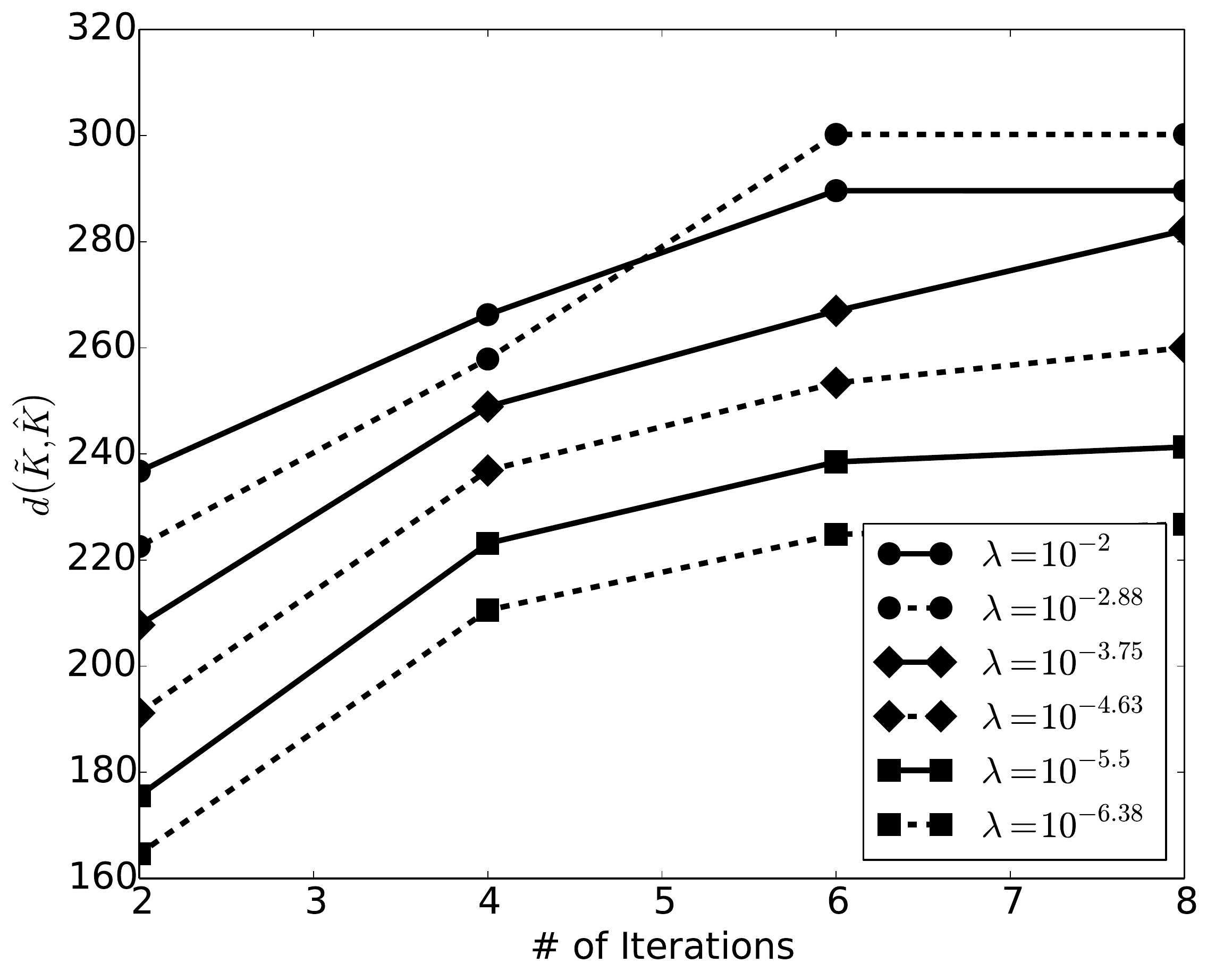}
  \caption{The covariance distance metric, which is independent of matrix inversion, for a range of regularization parameters (left) and noise levels (right).\label{fig:parm_sweep4}}
 \end{figure}

  Finally, Figure \ref{fig:parm_sweep4} shows the results of assessment using the distance metric which is invariant to matrix inversion. For applications involving inversion of the covariance matrix, such as use of the
  Hotelling observer, these results are particularly relevant. More than in any other metric, a clear dependence on regularization parameter emerges, with lighter regularization (smaller $\lambda$) consistently 
  improving the covariance approximation. Likewise, a monotonic increase in the distance metric reveals the impact of iteration, however it is interesting to note that for several values of $\lambda$, the 
  distance metric seems to plateau 
  at higher iterations. This could imply that continuing to run the IRLS algorithm to tighter convergence would eventually have a diminishing impact on the approximation of model observer metrics. 
  
\section{Examples and Approach for Larger Systems}
\label{sec:large_system}


Next, we turn our attention to some specific applications of the the proposed methodology for covariance approximation. Specifically, each question we address in this section relates 
to issues underlying the assessment of image quality when using IIR. In each of the following examples, we 
have increased the image size to 128 $\times$ 128 pixels, and proportionately increased the data detector sampling to 256 detector pixels. We set the number of 
projection views to 50, as we find that this is just within the realm of sparsity where the TV term is necessary for accurate reconstruction of noise-free data. Otherwise, all 
previous experimental conditions are held constant. The 
computation was performed on a system with enough RAM to store and directly invert matrices of dimension $N_\textrm{pixels} \times N_\textrm{pixels}$, \new{so the matrix inverses 
of Eqn. \ref{eqn:jac} were precomputed, as opposed to being solved with conjugate-gradients or other linear solvers.}

\begin{figure}[ht!]
\centering
\includegraphics[width=0.3\columnwidth ]{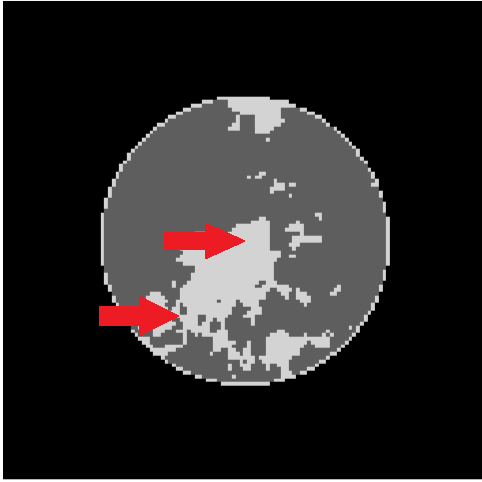} \includegraphics[width=0.3\columnwidth]{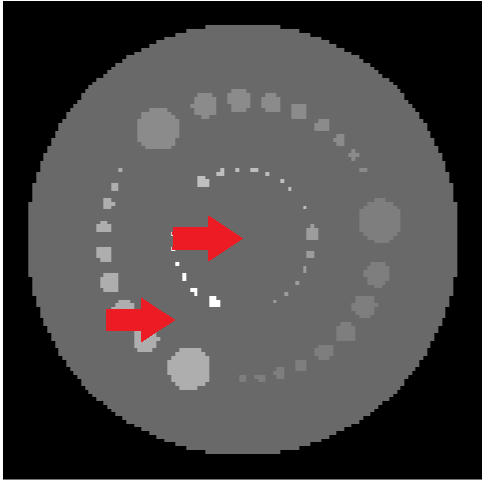} \\
\vspace{0.1cm}
\includegraphics[width=0.3\columnwidth]{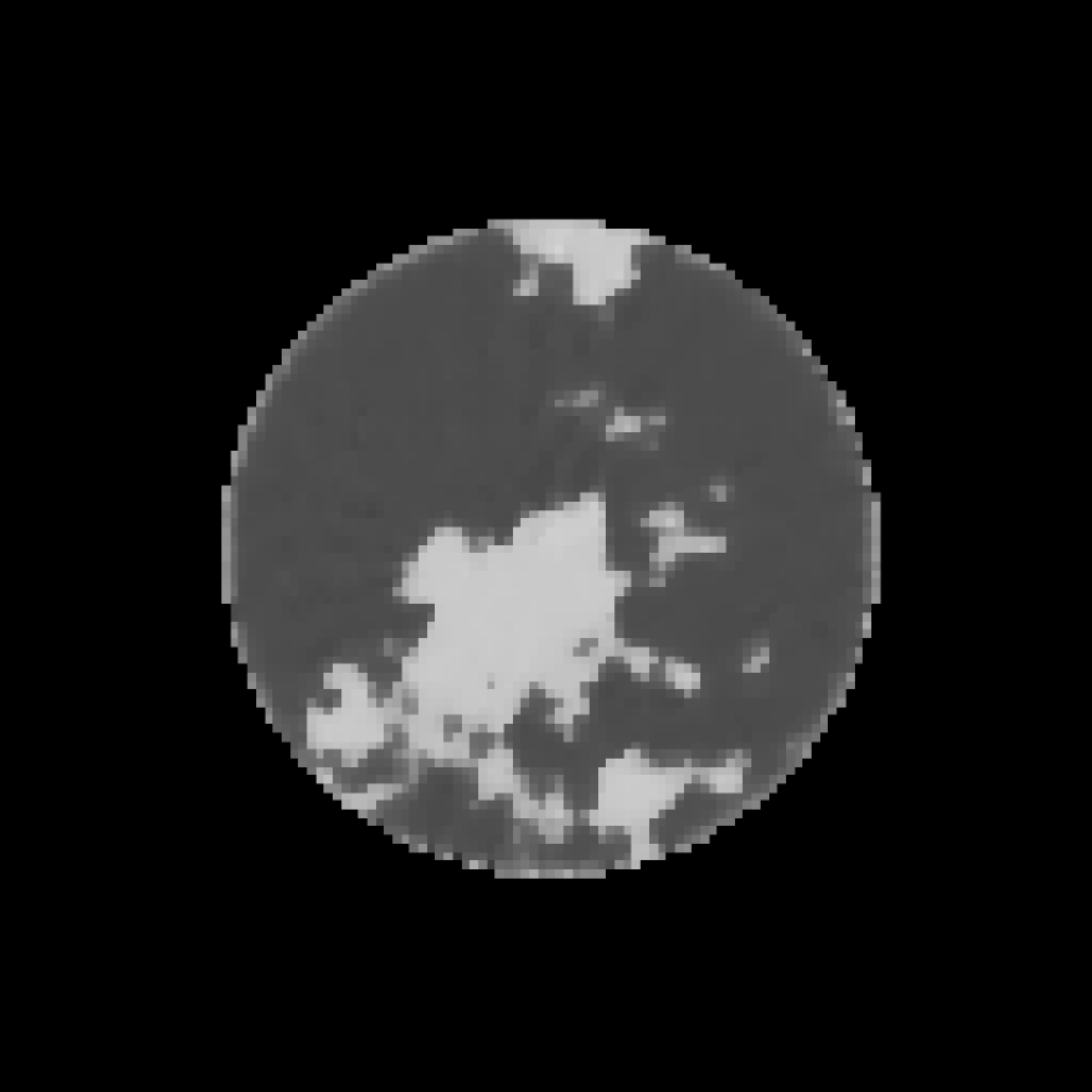} \includegraphics[width=0.3\columnwidth]{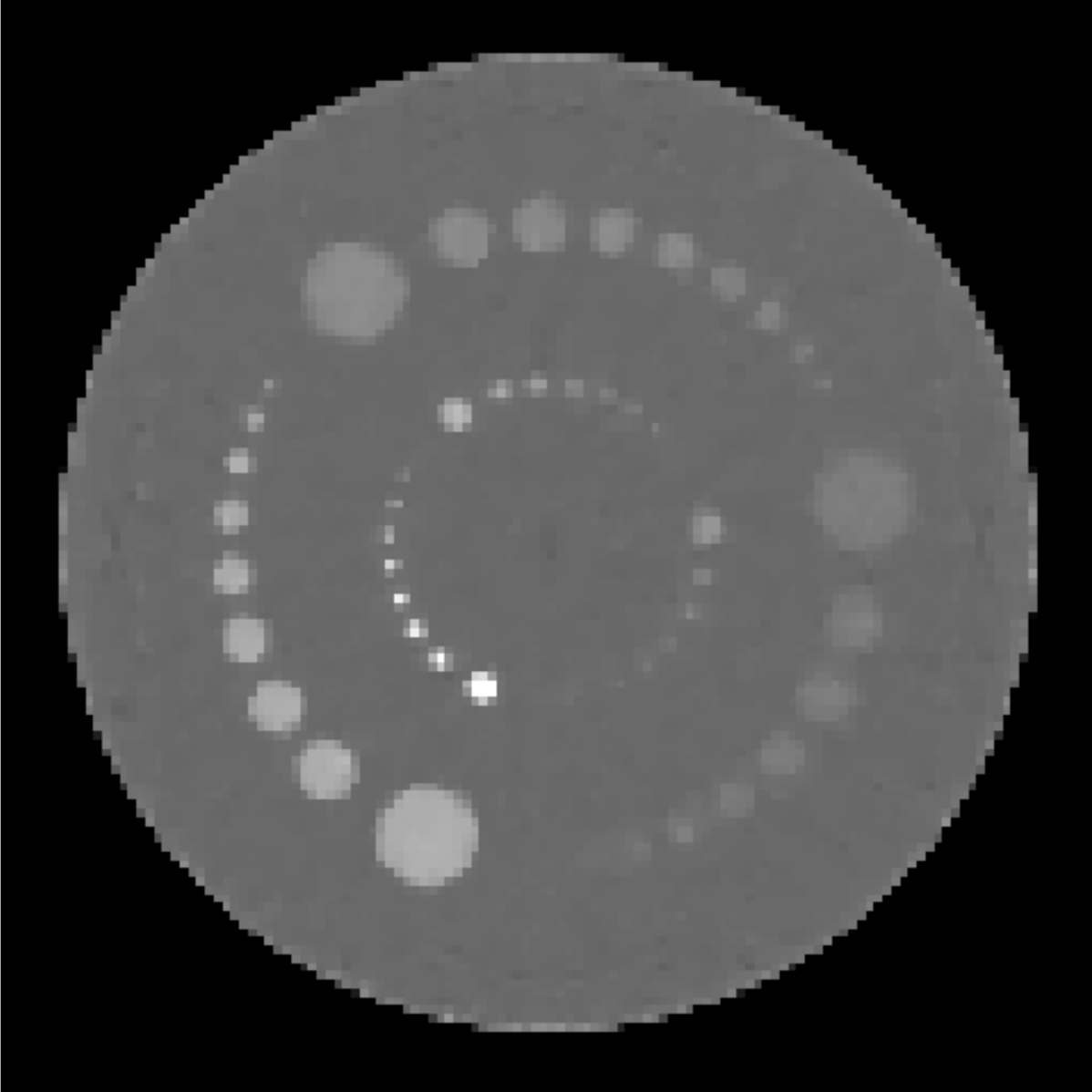} \\
\vspace{0.1cm}
\includegraphics[width=0.3\columnwidth]{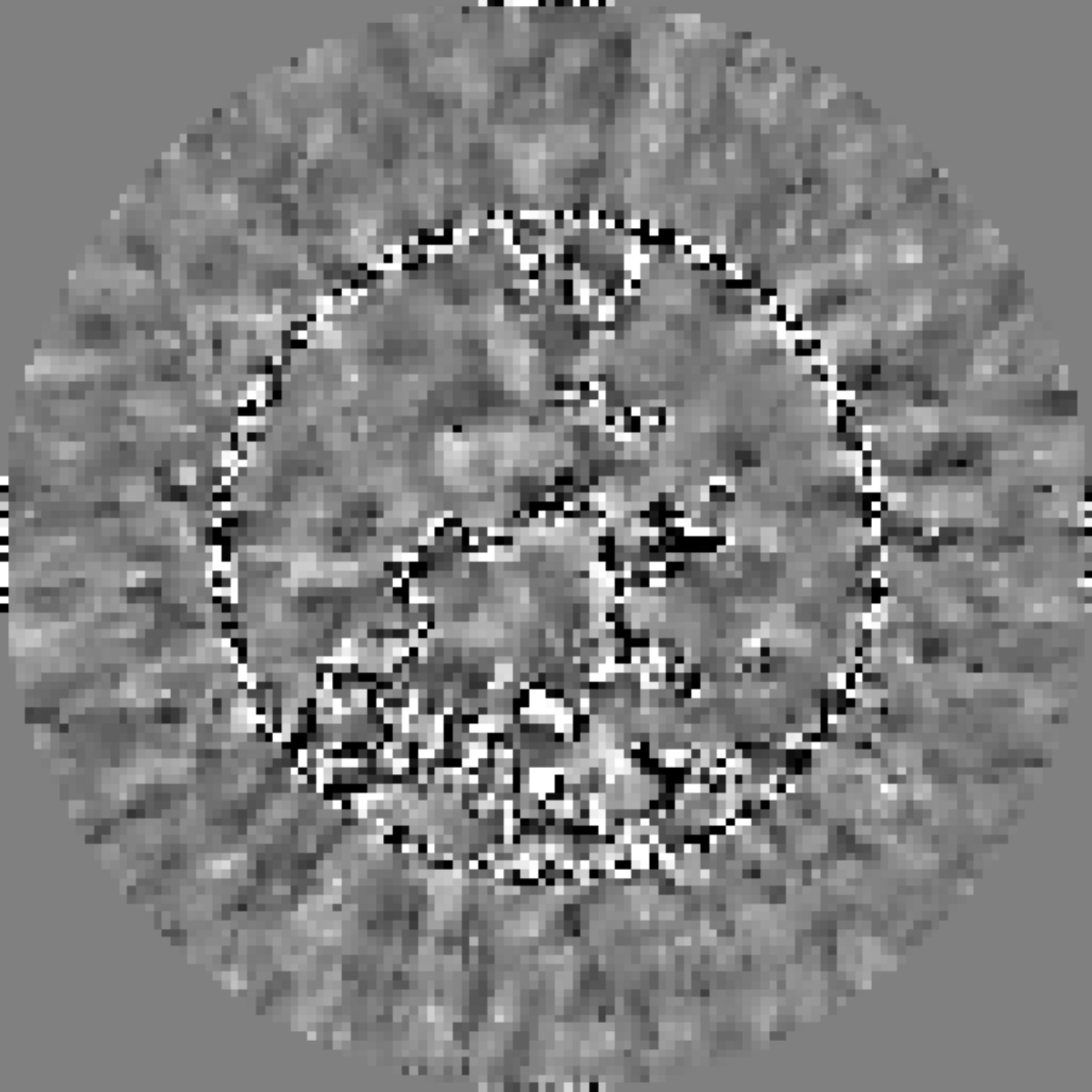} \includegraphics[width=0.3\columnwidth]{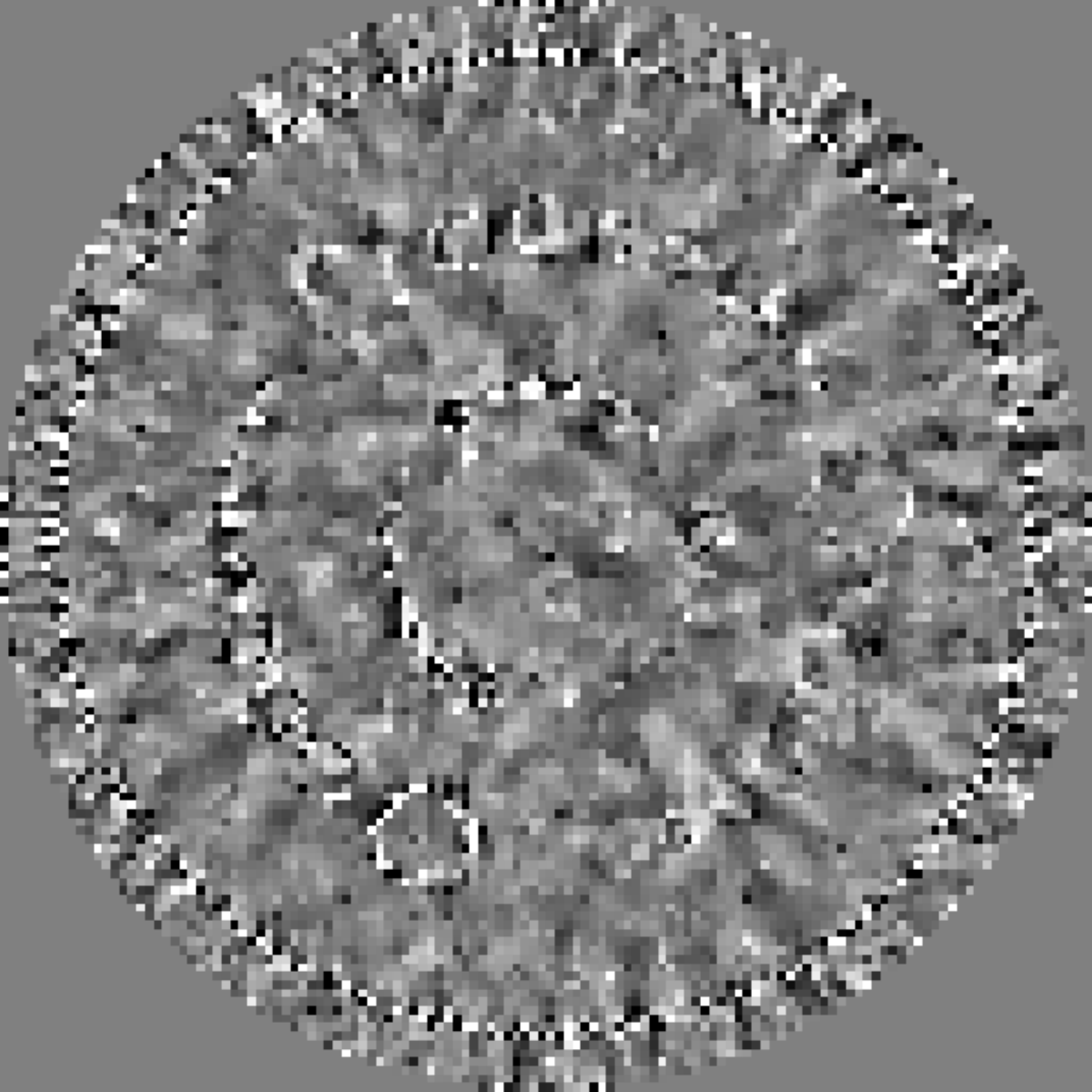}
\caption{\textbf{Top:} The two numerical phantoms used in this study, a numerical breast phantom (left), and a disk phantom (right). The display windows are
[0.17, 0.25] cm$^{-1}$ and [0.17, 0.35] cm$^{-1}$, respectively. Arrows indicate the locations where local noise properties are studied in subsequent sections.
\textbf{Middle:} Reconstructed noise realizations generated using the same phantoms. The display windows are identical to the images in the top row. Although heavy 
regularization is applied, some noise is still visible in the uniform regions of the phantoms.
\textbf{Bottom:} Difference images between two noisy IRLS reconstructions of each phantom are shown in order to visualize the noise structure. The display windows are 
[-0.001, 0.001] cm$^{-1}$ and [-0.003, 0.003] cm$^{-1}$ for the left and right images, respectively.
\label{fig:big_recons}}
\end{figure}

For the remainder of our investigation, we consider two numerical phantoms: a numerical breast phantom as before, but with finer pixelization than in Section \ref{validation_method}, and a continuously-defined numerical disk phantom. 
The GMI sparsity is nearly identical for the two phantoms, 7.5\% and 7.4\% nonzero elements for the breast phantom and (discretized) disk phantom images, respectively.
Further, we consider only one set of reconstruction parameters, fixing $\lambda=0.1$ and 6 iterations of IRLS. This is because in this section we primarily wish to demonstrate the application
of the proposed approximation when the TV penalty plays a predominant role in the reconstruction. The numerical phantoms used and examples of 
noise realizations are shown in Figure \ref{fig:big_recons} (a discretized version of the continuous numerical disk phantom is shown). The red arrows indicate locations where local 
noise properties are investigated in subsequent sections.
While the heavy regularization evident in the reconstructions greatly lowers the noise magnitude, it also introduces patchy pixel correlations
characteristic of TV penalties. It is the impact of this characteristic noise structure that we hope to assess here. 

\subsection{Accurate Preservation of Noise Structure}

\begin{figure}[ht]
\centering
\includegraphics[width=0.3\columnwidth]{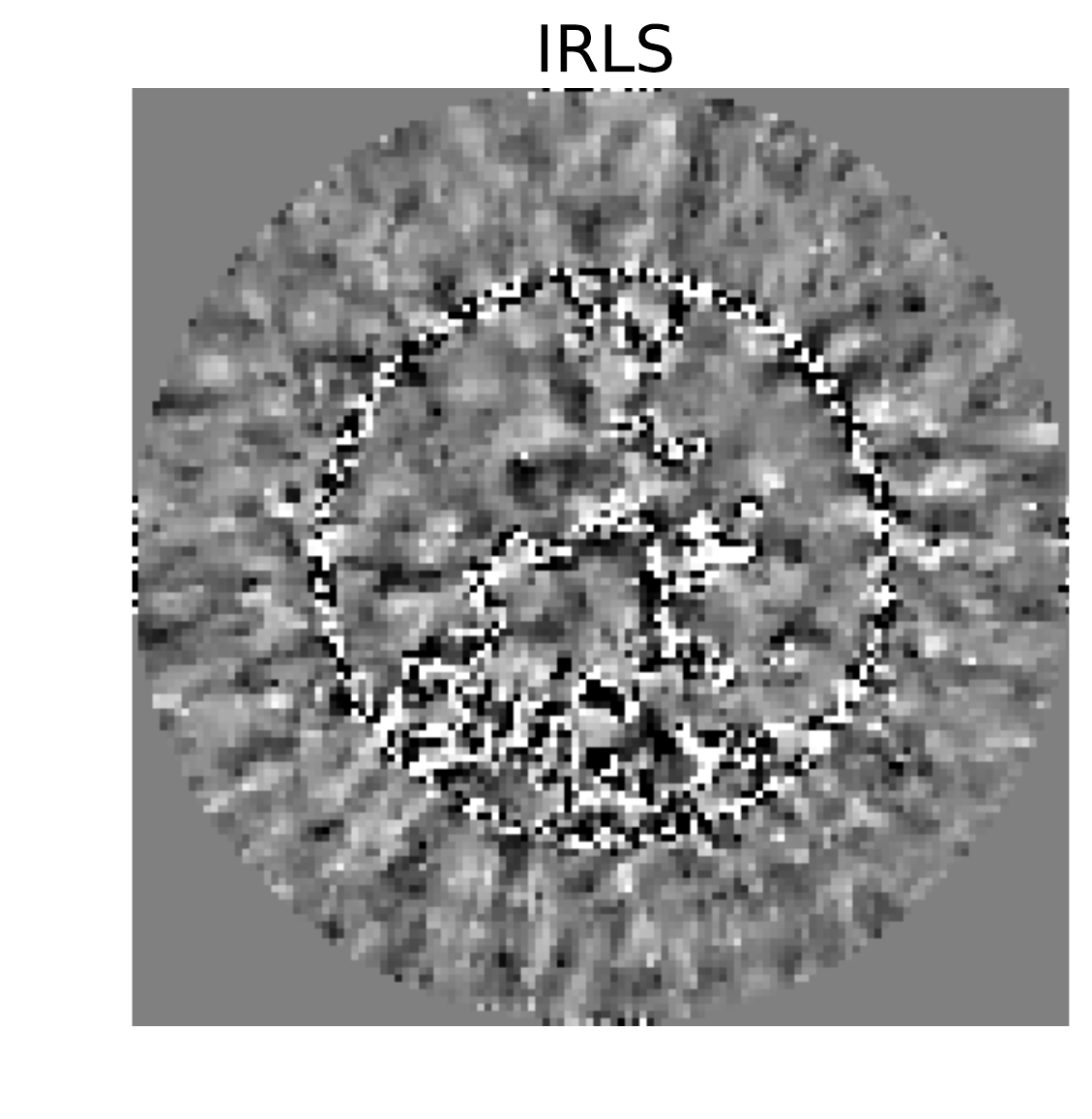} \includegraphics[width=0.3\columnwidth]{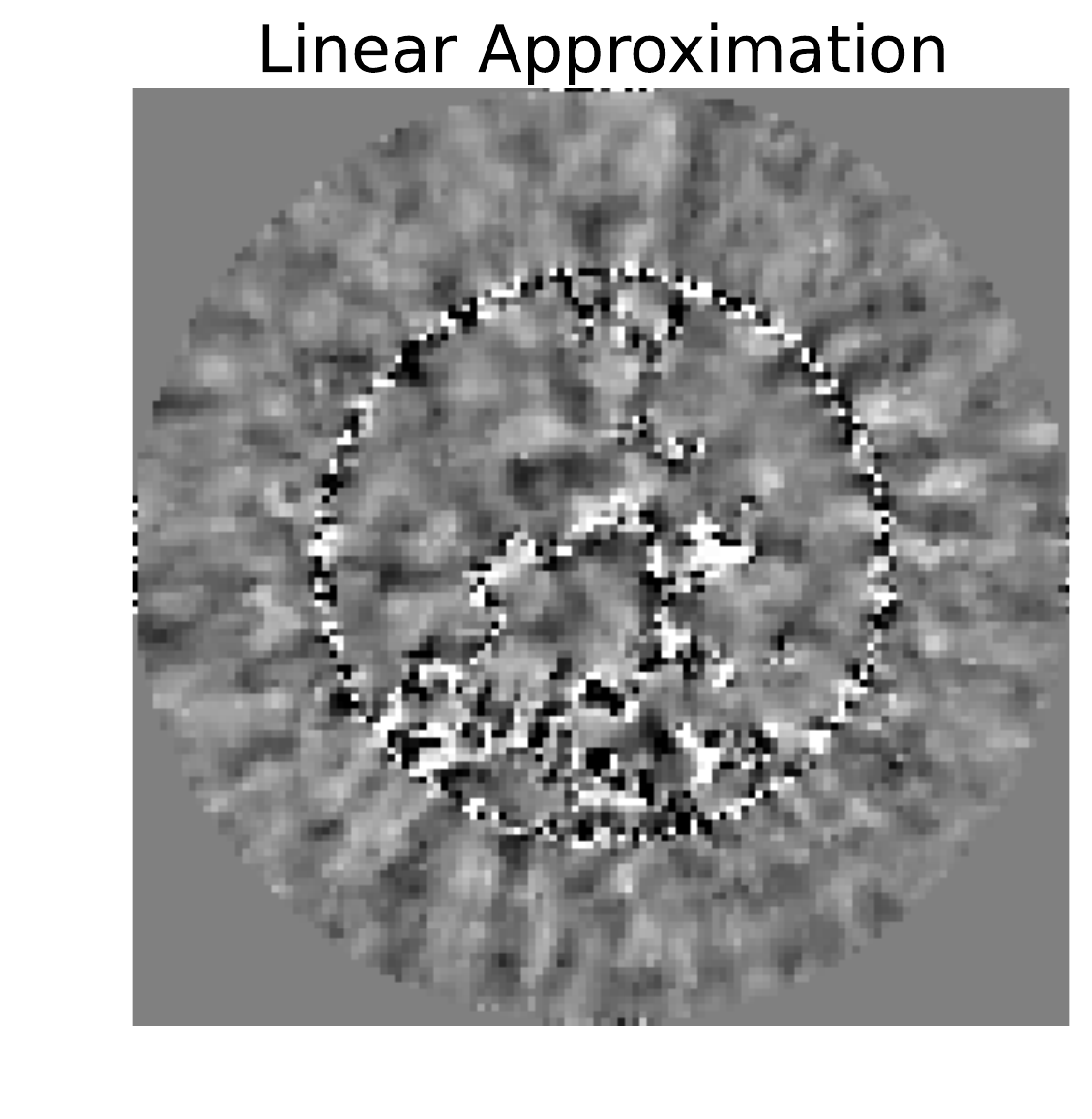}
\caption{Shown here are difference images between two independent noise realizations. On the left, the two images have been reconstructed using 6 iterations of the IRLS algorithm. 
On the right, the same data realizations were propagated through the linearization of Eqn. \ref{eqn:jac}. The patchy noise structure which is evident in the IRLS reconstructions 
is well preserved in the linearization approach, meaning that the characteristic noise texture of TV minimization can potentially be well approximated with only 2nd-order image statistics.
As in the small system examples, note that the noise magnitude appears to be the primary source of error in the linear approximation. \label{fig:patches}}  
\end{figure}

Since most anecdotal discussion of noise structure in TV-based reconstruction describes a ``patchy'' appearance, we would like any approximations which make up our 
noise model to maintain this appearance. It is not immediately obvious that the 2nd-order image statistics adequately preserve the characteristic noise structure of TV reconstruction.
Therefore, as a simple subjective validation of this, we perform two reconstructions of simulated independent breast phantom data realizations and investigate the resulting difference image. 
Likewise, we propagate the same data realizations through the linearization of the reconstruction defined in Eqn. \ref{eqn:linearization}, using the expression for $J_n$ in Eqn. \ref{eqn:jac}, and compute their difference. The result is shown in 
Figure \ref{fig:patches}, with a tight display window centered at 0. Clearly, although the overall noise magnitude is somewhat larger in the IRLS reconstructions, the linear model still 
captures the essential features of the image noise texture, specifically the lumpy appearance in the uniform phantom regions. 

\subsection{Object Dependence of Noise}

Having validated that our approximation method is reasonably accurate for small images and a range of parameters, and having demonstrated that the approximation qualitatively preserves 
TV-based noise texture, we next turn our attention to the issue of object-dependent noise. This issue is of central importance to image quality assessment, since the actual object of interest is 
rarely available for system evaluation, and often stylized phantoms are used for assessing the performance of IIR. The assumption of approximate object-independence is desirable because 
it enables system assessment for a generic phantom in hopes that the evaluation performed will generalize to a wide array of actual patient data. \new{However, this assumption is never completely 
valid for practical applications of IIR. Therefore our purpose here is to demonstrate the application of our covariance approximation method to the quantification of object-dependence of image noise.} 

\begin{figure}[ht]
\centering
\includegraphics[width=0.3\columnwidth ]{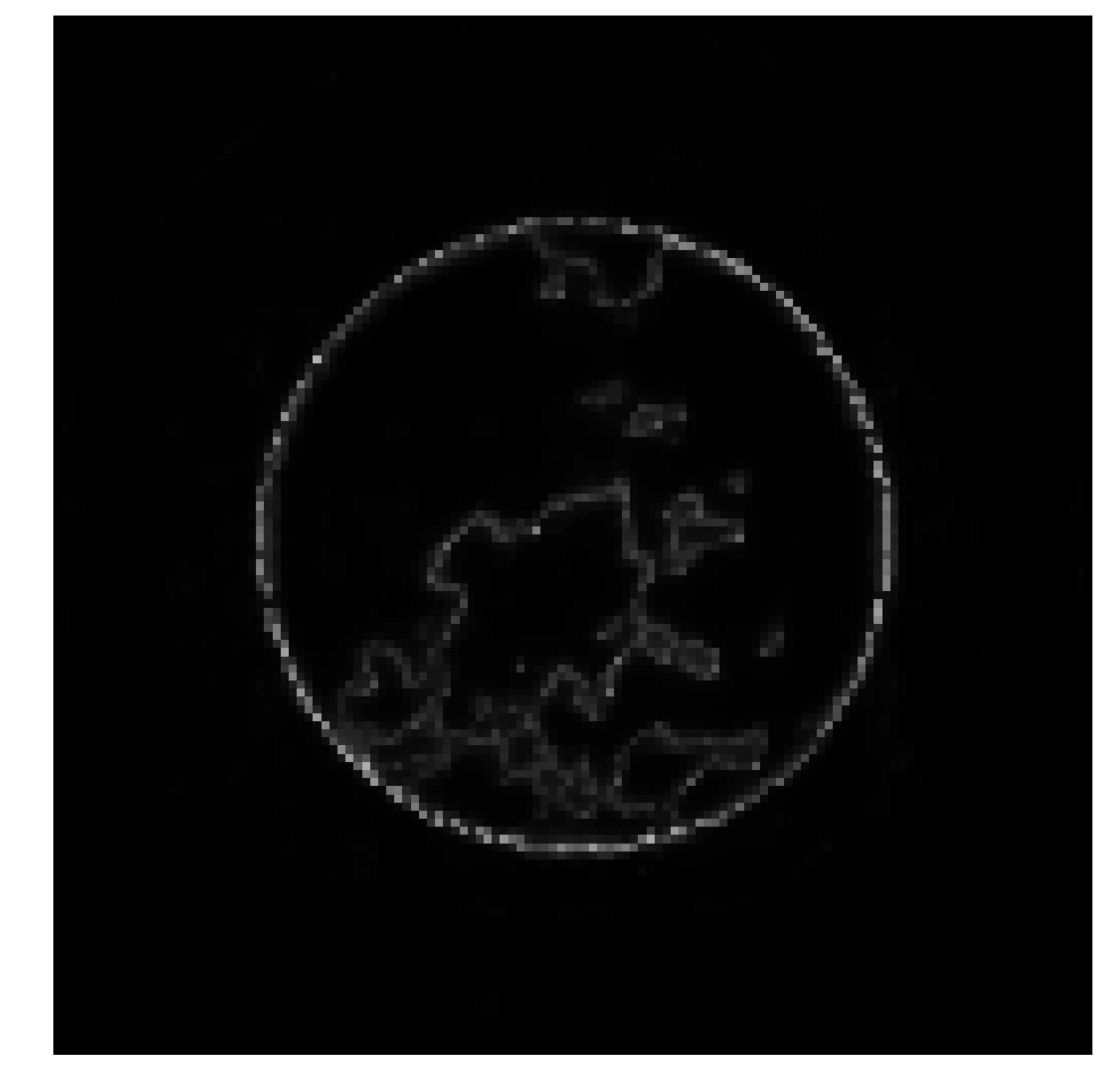} \includegraphics[width=0.3\columnwidth]{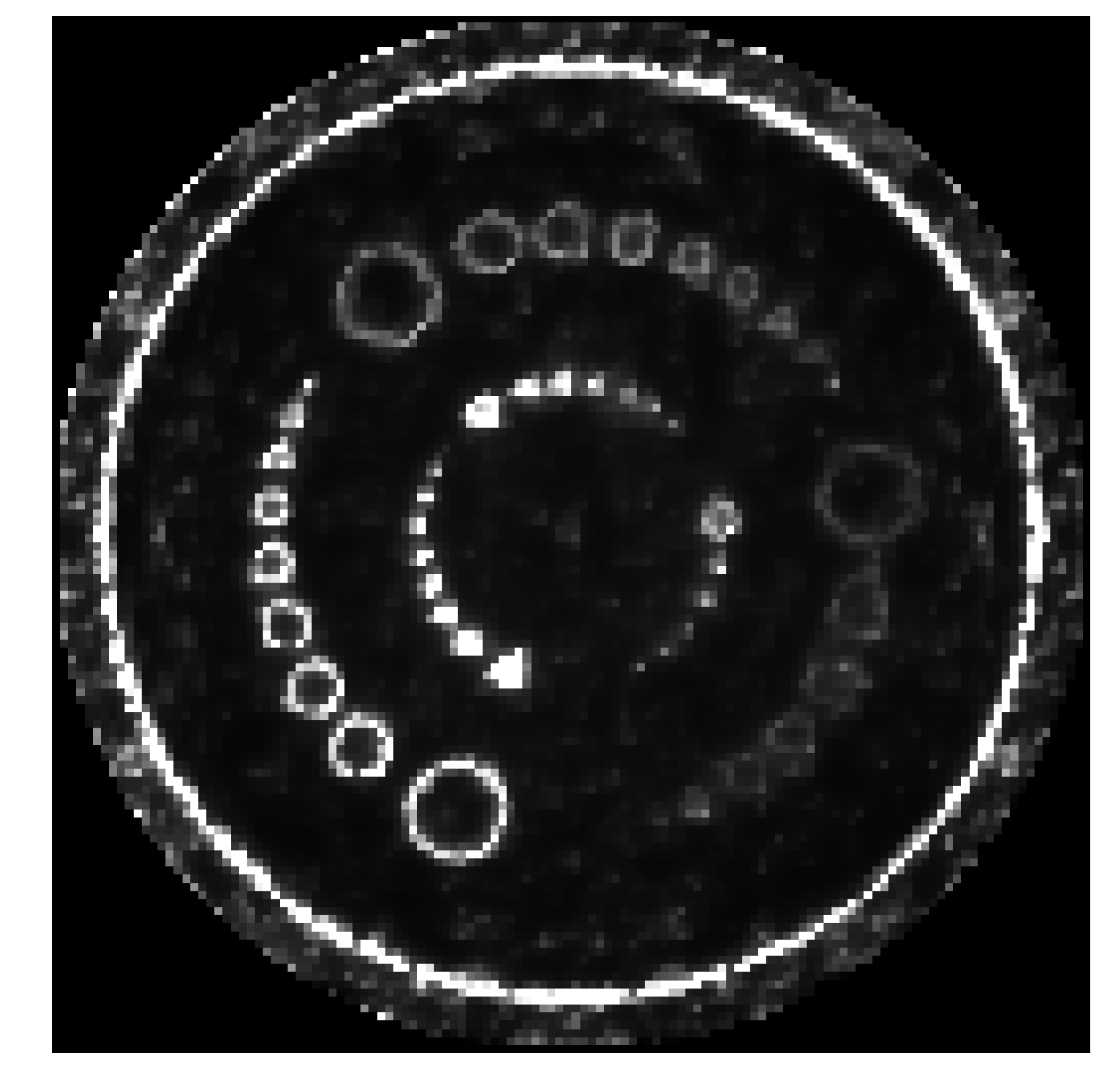}
\caption{Variance maps for each phantom determined by approximation method. The display window in each case is [0, 2E-6] cm$^{-2}$.  Note the strong noise dependene on local edge information, as previously reported by others
 \cite{koehler_noise_2009, rose_noise_2015}.\label{fig:varmaps}}
\end{figure}

First, we investigate the dependence of image variance on the object being imaged. Figure \ref{fig:varmaps} shows variance maps for the two numerical phantoms used. Clearly, image pixel 
variance is strongly object-dependent, with pixels located near edges displaying higher noise levels that pixels in uniform regions. This is to be expected since the TV penalty is most active in 
regions of the image where pixel constancy is likely to be enforced. This effect was first observed by Koehler and Proksa \cite{koehler_noise_2009} 
and also confirmed with realization studies by Rose \emph{et al.} \cite{rose_noise_2015}. Logically, one would expect this effect could play an even greater role in actual data, since the objects being imaged are not piece-wise constant, and
regularization is not typically so strong as to eliminate visualization of the physiological variability within a given organ or tissue type.

 While pixel variance is obviously important in terms of image quality, 
of equal or greater concern is pixel covariance. This is particularly true when task-based metrics are employed, as covariances play in important role in many radiological tasks. Detection tasks, for instance, 
are often modeled using stylized phantoms with small or low-contrast cylindrical or spherical objects placed in a large, uniform background. For IIR, there is no guarantee that the covariance, and hence task performance metrics,
obtained with these phantoms is in any way related to task-performance in a realistic object. Further, even for realistic phantoms it may be important to consider a range of backgrounds, signal locations, and tasks 
in order to construct meaningful image quality metrics when the image covariance is object-dependent. 

\begin{figure}[ht]
\includegraphics[width=\columnwidth]{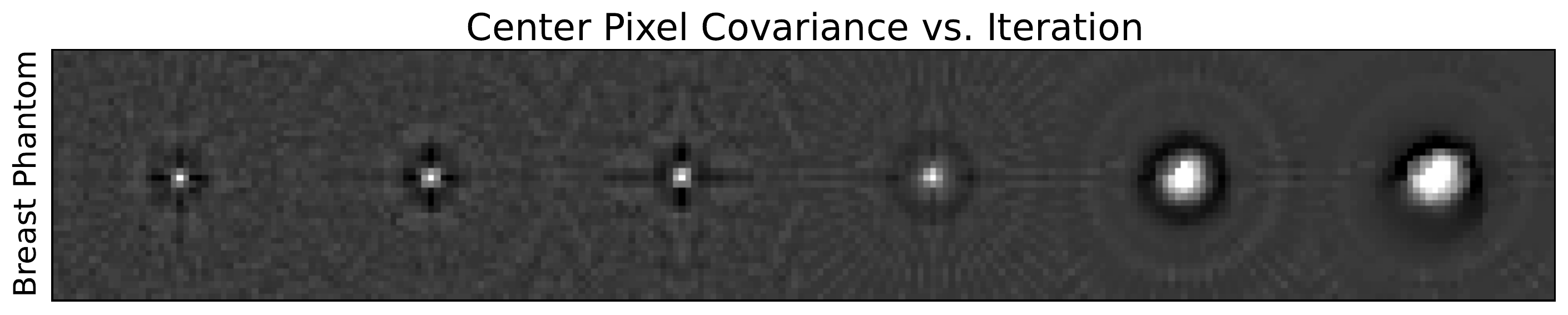} \\
\includegraphics[width=\columnwidth]{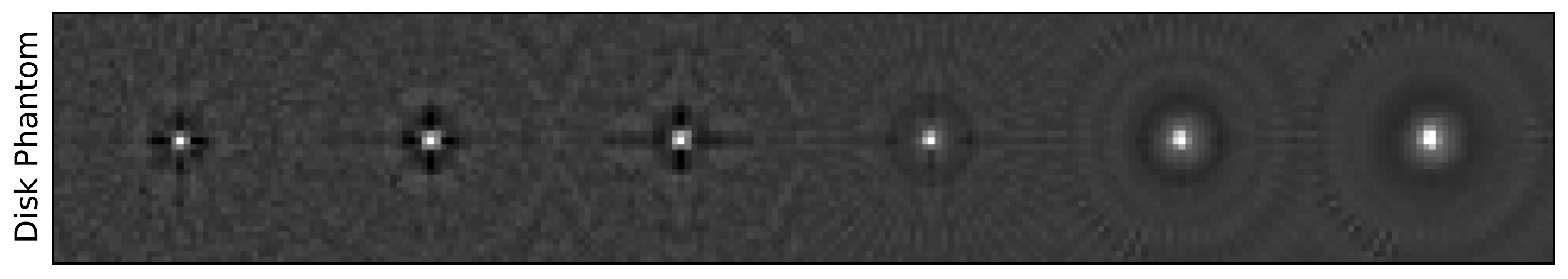}
\caption{Evolution of the correlation structure with increasing iteration number from left to right. The center pixel is in the center of the full image (breast phantom on the top, disk phantom on the bottom).
The display window of each image is set so that white corresponds to 50\% of the maximum covariance and the gray level of 0 is kept constant.
This enables visualization of the structural evolution of the correlation independent of the decrease in overall noise level with each iteration. 
\label{fig:evolution1}}
\end{figure}

\begin{figure}[ht]
\includegraphics[width=\columnwidth ]{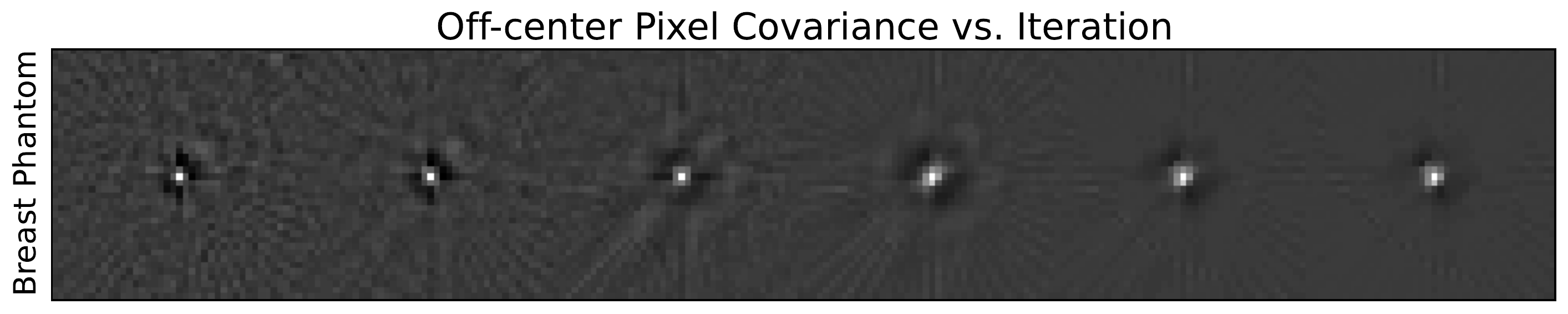} \\
\includegraphics[width=\columnwidth]{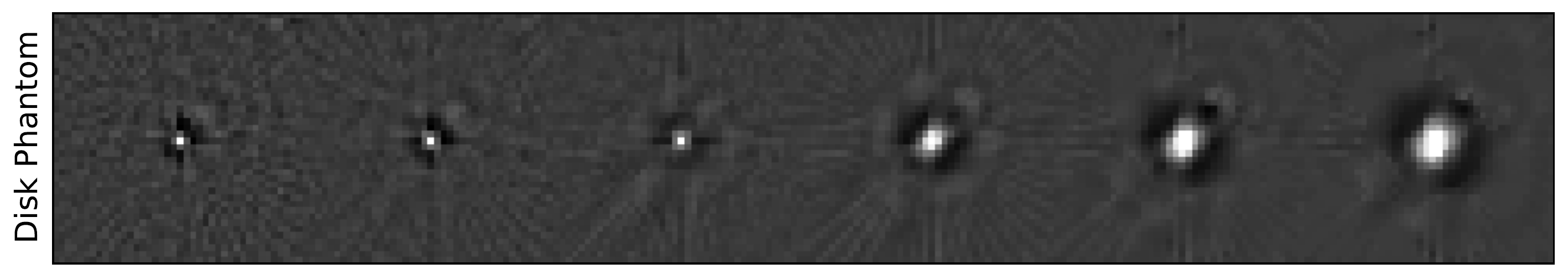}
\caption{Same as Figure \ref{fig:evolution1}, but for a pixel which lies on a tissue boundary in the breast phantom. This illustrates the object-dependence of the noise structures in images reconstructed with 
TV minimization. In general, any IIR approach potentially produces object-dependent noise. Additionally, when compared to Figure \ref{fig:evolution1}, the non-stationarity of the image noise becomes 
apparent. \label{fig:evolution2}}
\end{figure}

In order to illustrate this phenomenon of object-dependent covariance, Figures \ref{fig:evolution1} and \ref{fig:evolution2} show the covariance structure in the numerical breast phantom and disk phantom as it 
evolves through successive iterations of IRLS. By covariance structure, we mean the covariance of each image pixel with a single fixed pixel. This can be interpreted as a single row of the image covariance matrix. 
The window and level settings are dynamic between iterations so that the overall structure of the covariance can be visualized (successive iterations decrease the overall noise level). Figure \ref{fig:evolution1} 
shows the covariance with a pixel that is in a central, uniform region of each phantom. Meanwhile, Figure \ref{fig:evolution2} corresponds to a pixel located near tissue boundaries in the breast phantom, and 
in the corresponding location of the disk phantom, which is locally uniform. These locations are indicated with red arrows in Figure \ref{fig:big_recons}.

For early iterations of IRLS, the reconstructed image noise structure remains largely object-independent. As the reconstruction progresses, however, the correlation pattern in the image begins to 
diverge for the two phantoms. This is to be expected for the pixel located near a boundary in the breast phantom (Figure \ref{fig:evolution2}), since local gradient magnitude information will affect the amount of regularization. 
However, the result illustrated in Figure \ref{fig:evolution1} is somewhat surprising, where the divergence between noise correlation structures indicates that object structure far from a region of interest can have a 
significant impact on local noise properties. \new{In general, this sort of non-local object dependence presents a particular challenge for task-based evaluation of IIR algorithms using physical phantoms, since it suggests that 
the realism of a phantom far from a signal of interest has the potential to affect the resulting covariance-dependent image quality metrics.}

\subsection{TV-based Regularization and Local Stationarity}

\begin{figure}[ht]
\centering
\includegraphics[width=0.3\columnwidth]{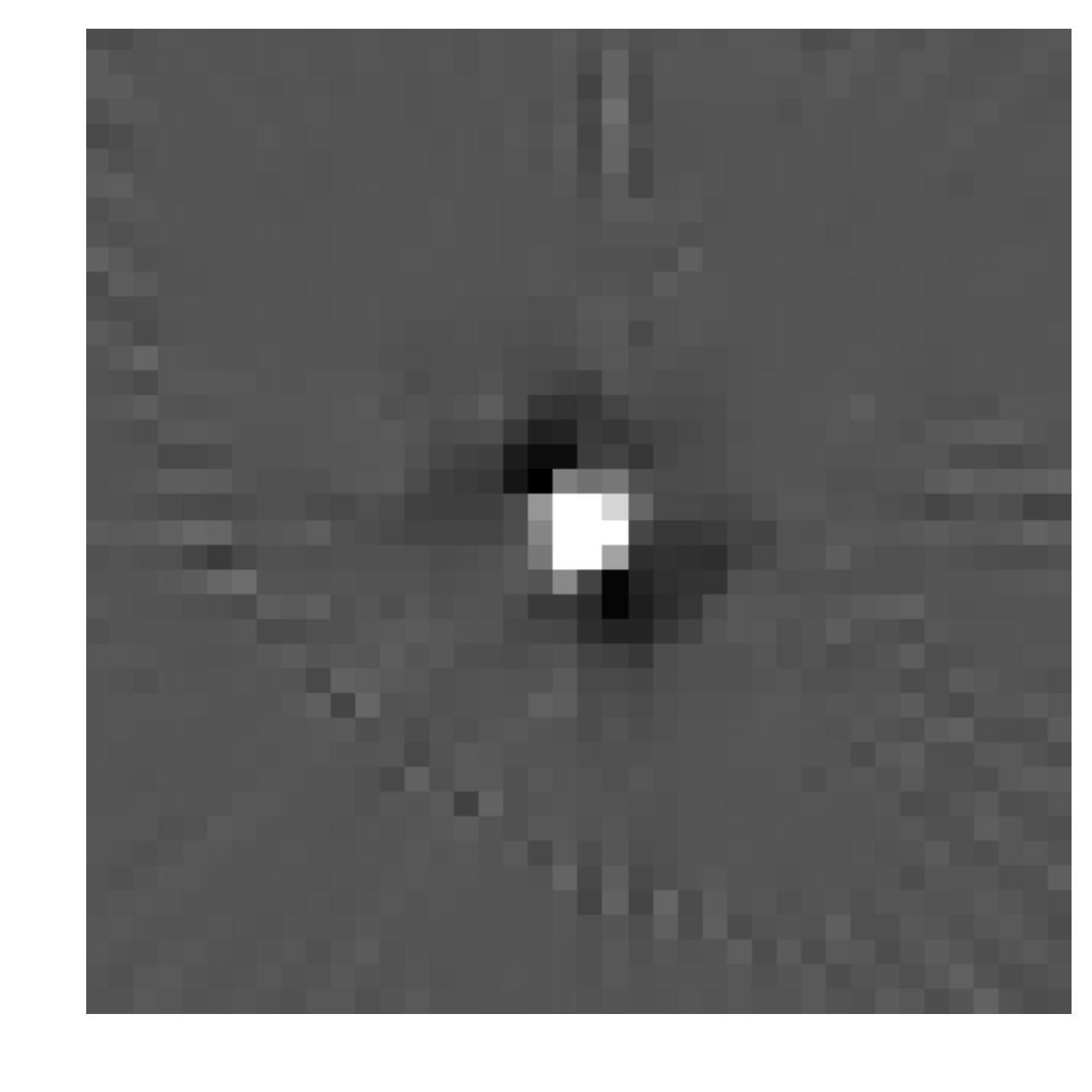} \includegraphics[width=0.3\columnwidth]{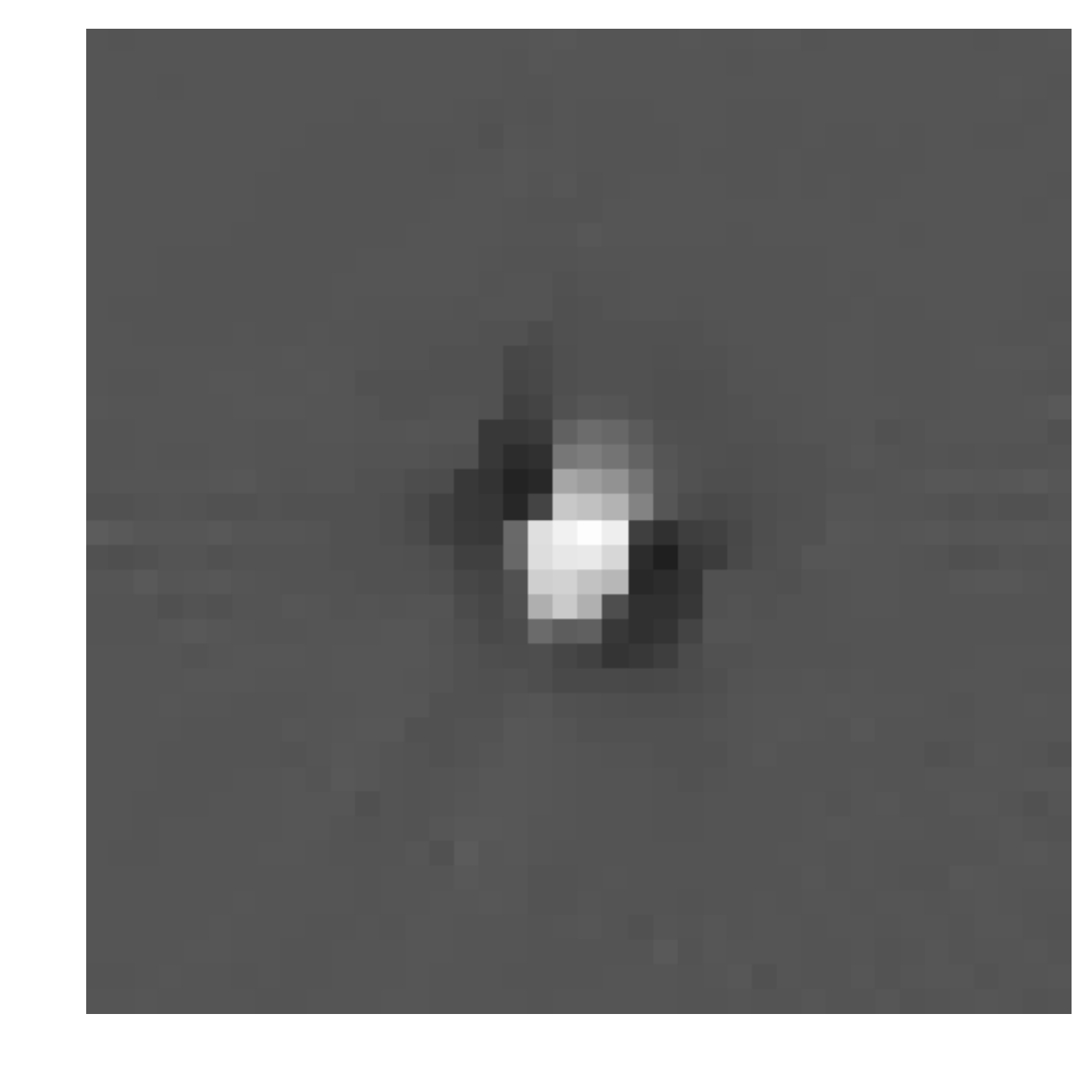}
\caption{Each image shows the correlation structure in a region of the breast phantom reconstruction near a tissue boundary. The left image is centered on the boundary pixel in Figure \ref{fig:stat_test}. The 
right image is centered on a point slightly to the right of the left image's center. The distance between the two 
points is only 4 pixels, but non-stationarity is clearly evident through the change in the correlation structure. The display window is [-2$\times 10^{-8}$, 4$\times 10^{-8}$] cm$^{-2}$ for both figures. \label{fig:stat_test2}}
\end{figure}

For the comparisons in the preceding section, we were careful to compare image covariance between objects in identical physical locations. A related question  is whether or not the covariance 
structure is invariant to small changes in location within a given object. This property, known as local stationarity, is an assumption which underlies any image quality metric involving the noise power-spectrum (NPS).
As previously stated, stationarity is often invoked in order to enable discussion of conventional image quality metrics, such as detective quantum efficiency (DQE), the NPS, or any model observer
which is defined in the Fourier domain. In Figure \ref{fig:stat_test2}, we demonstrate \new{the application of our covariance approximation to qualitatively investigate local stationarity.}
The left side of Figure \ref{fig:stat_test2} 
shows the covariance with the peripheral pixel highlighted in the breast phantom in Figure \ref{fig:big_recons}. The right side of the figure shows the covariance structure when the covarying pixel is shifted 
to the right by 4 pixels (about 5mm). \new{For this object, local stationarity is then valid to the extent that the left and right sides of the figure are the same.}

While visualization of the image covariance can be informative, a more direct means of probing the validity of the stationarity assumption is to investigate image noise in the discrete Fourier transform (DFT) domain. 
An equivalent means of stating the assumption of stationarity is that a DFT diagonalizes the image covariance matrix. This property is a primary motivation for assuming stationarity, 
since diagonalization of the covariance matrix allows for efficient computation of a range of image quality metrics, including Hotelling observer performance. In our notation, stationarity implies that
\begin{equation}
FK_\mathbf{f}F^\dagger = D,
\end{equation}
where $F$ denotes the DFT, $\dagger$ is the Hermitian adjoint, and $D$ denotes a diagonal matrix. Given our present approximation for $K_\mathbf{f}$ and the fact that 
$F^\dagger=\frac{1}{N_\textrm{pixels}}F^{-1}$, this assumption can be straight-forwardly investigated. Consider a vector with dimensionality equal to the reconstructed 
image and with a single non-zero element at pixel $i$. We will denote this vector by $e_i$. It is then straightforward to compute 
\begin{equation}
\label{eqn:stat_test}
\tilde{e}_i = FK_\mathbf{f}F^{\dagger} e_i
\end{equation}
and investigate the structure of this new vector $\tilde{e}_i$. \new{The magnitude and extent of non-zero elements in $\tilde{e}_i$ is then an indication of the extent to which image noise is stationary.}

\begin{figure}[ht]
\centering
\includegraphics[width=0.3\columnwidth]{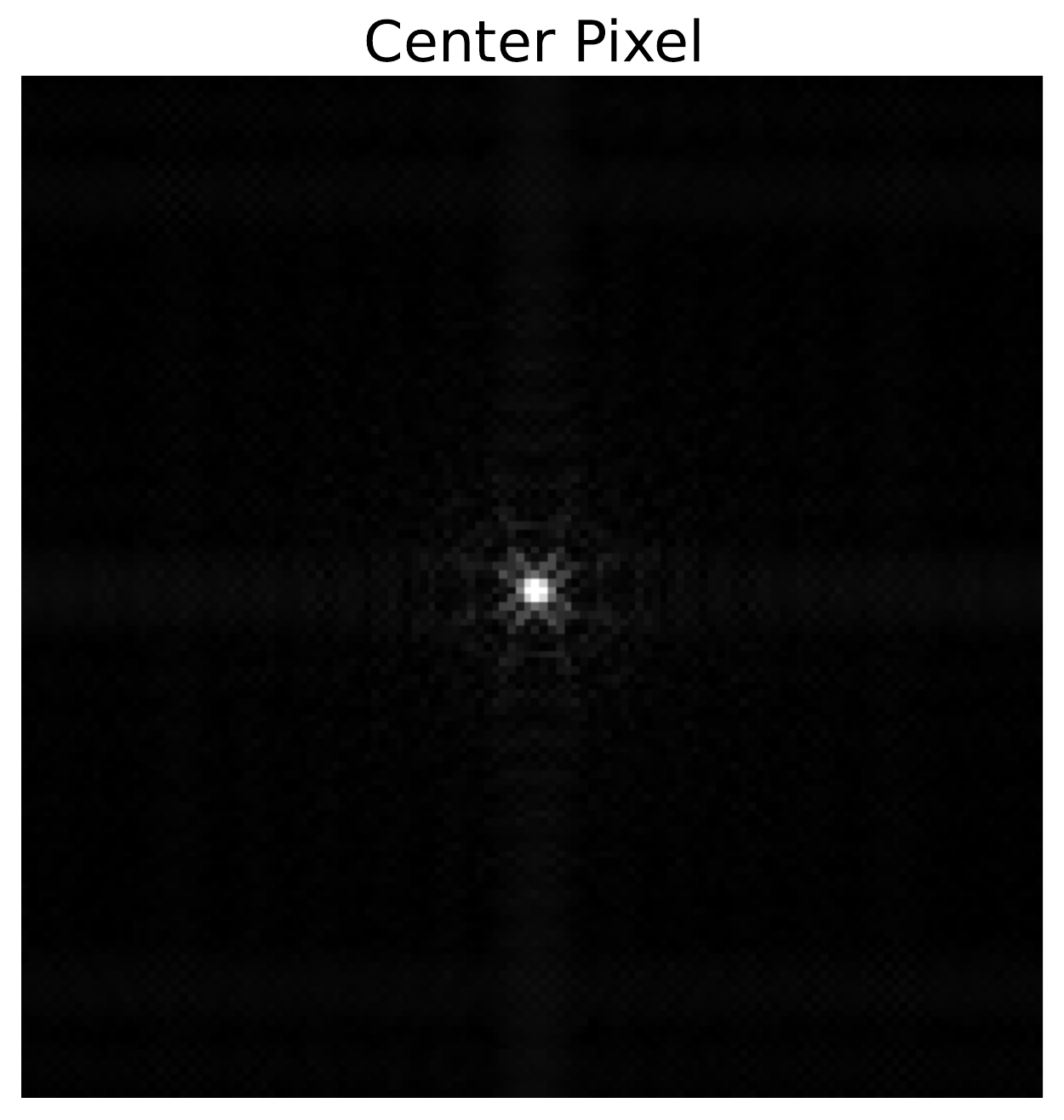} \includegraphics[width=0.3\columnwidth]{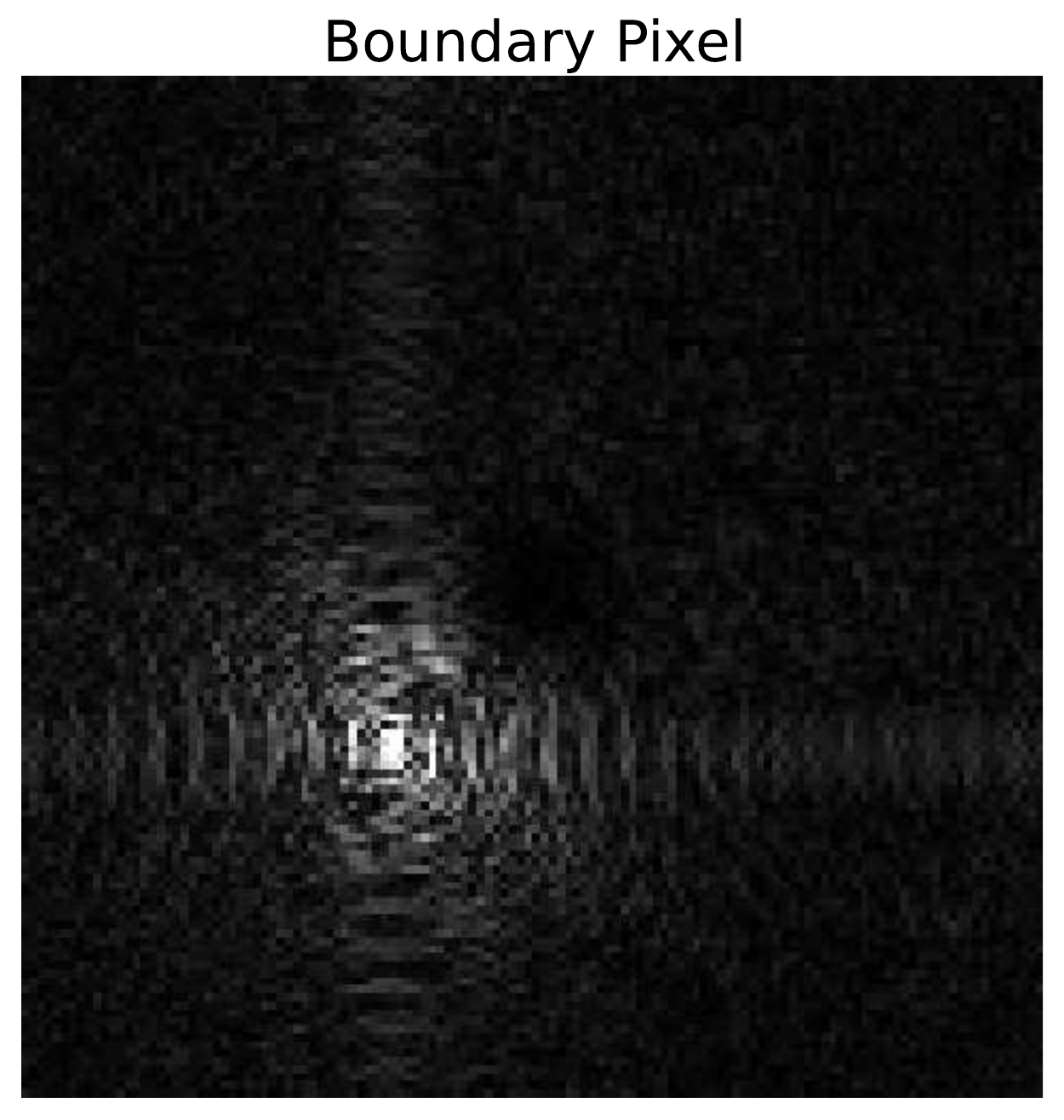}
\caption{ Shown are the resulting vectors $\tilde{e}_i$ from Eqn. \ref{eqn:stat_test}, reshaped into 2-dimensional images. The existence of many pixels which are non-zero is a direct indication that 
the DFT does not diagonalize the image covariance matrix, hence invalidating the assumption of stationarity. The display window is set so that black corresponds to 0, while white corresponds to 25\% of the 
maximum image value.\label{fig:stat_test}}
\end{figure}

Figure \ref{fig:stat_test} shows the results of performing this test at the two pixel locations marked in Figure \ref{fig:big_recons} in the breast phantom. It is important to note, however, 
that locations in the image no longer correspond to physical locations since we are now operating in the DFT domain. 
\new{Interestingly, 
when $i$ is chosen to correspond to the location of the central pixel, there are few elements of $\tilde{e}_i$ with values significantly different from zero. However, moving beyond the central pixels of the DFT-domain 
image produces vectors $\tilde{e}_i$ with many more non-zero components, as seen on the right of Figure \ref{fig:stat_test}.}

\begin{figure}[ht]
\centering
\includegraphics[width=0.8\columnwidth]{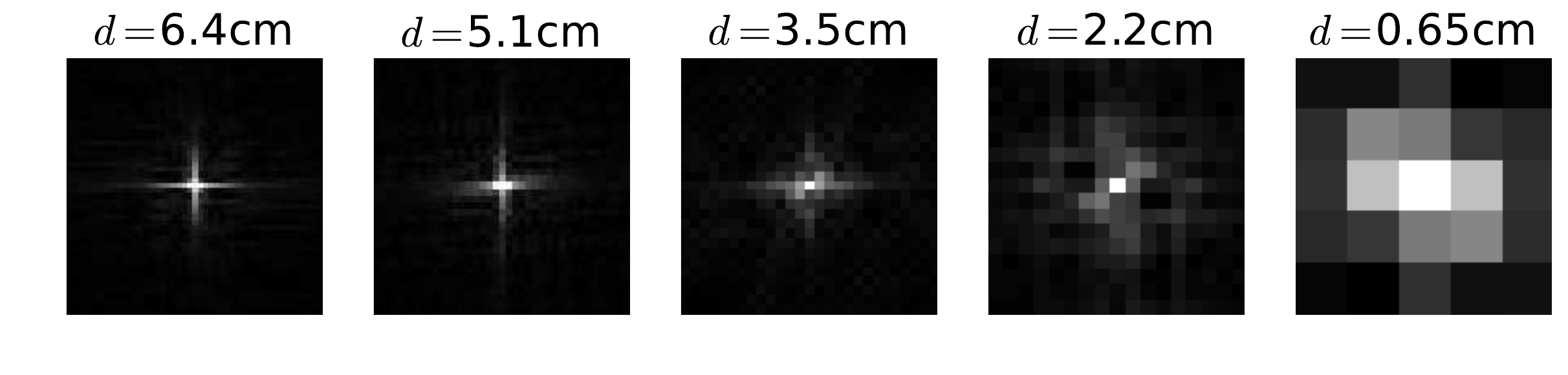} 
\caption{ These results, similar to those in Figure \ref{fig:stat_test}, demonstrate that local stationarity is similarly not satisfied, despite restriction of the image ROI to small sizes. The headings of the figures 
denote the width of the square ROI used. The window 
for each image is set as in Figure \ref{fig:stat_test}. \label{fig:local_stat}}
\end{figure}

While this example is informative, typically \emph{local} stationarity is assumed, rather than global stationarity, and it is \new{a quantitative measure} of global stationarity which is demonstrated in Figure \ref{fig:stat_test}. 
Recall, however, that 
for this example, the covariance matrix was directly stored in computer memory. This enables us to access only those components of $K_\mathbf{f}$ which correspond to a local region of interest (ROI) in 
the image. In order to investigate the local stationarity assumption, the same procedure described above for global stationarity was implemented using only the covariance of pixels within square image ROIs 
centered about the off-center pixel highlighted in Figure \ref{fig:big_recons}. The resulting vectors $\tilde{e}_i$ are shown as image ROIs in Figure \ref{fig:local_stat}. Each ROI image is labeled with the 
spatial extent of the square ROI used. While the structure of the non-zero components changes somewhat as the ROI size decreases, there is no ROI size for which \new{there is only a single non-zero component. Since the DFT 
does not necessarily diagonalize the covariance matrix in this case, this example
suggests that the use of any image quality metrics which rely on local stationarity should be justified on a case-by-case basis, especially when edge-preserving penalties are used in CT IIR.} 

Lastly, while we do not present results for images larger than $128 \times 128$ pixels in the present work, it is worth briefly discussing the feasibility of this extension. The first difficulty is in obtaining 
an estimate of $\bar{\mathbf{w}}_n$ for each iteration of IRLS. This is because pixel variance estimates are required for every pixel and every iteration. As stated 
in Section \ref{wbar}, these variance estimates do not need to be precise, and a variety of approaches exist for efficiently constructing approximate variance maps. 
The subsequent difficulty is in performing the linear solve in algorithms \ref{alg1} and \ref{alg2}. Virtually any first-order method can solve these systems efficiently enough 
to view single rows of the covariance matrix in isolation. Further, in our experience methods based on conjugate gradients can likely enable computation of Hotelling observer performance for 
a single reconstruction implementation, with covariance matrix 
inner products being computed on a single CPU of our system in roughly 10 seconds for a 512 $\times$ 512 image. Therefore, for the current implementation on our system, given estimates of $\bar{f}_n$ and 
$\bar{w}_n$, full calculation of HO SNR would likely 
take on the order of several weeks, which is feasible, if somewhat time-consuming. However, careful optimization of parameters with the Hotelling observer and construction of 
full and precise image variance maps would require a more efficient means of estimating $\bar{w_n}$ for each iteration. A more efficient statistical model for the mean and variance of these weights than that 
presented in Section \ref{wbar} could address this issue.

\section{Conclusion}
In this work, we have presented a method for the approximation of CT image covariance when the edge-preserving TV penalty is used. The method relies on the ability to apply an IRLS algorithm 
to the solution of the TV-penalized objective, so that noise can be propagated through a series of quadratic subproblems. The resulting approximation is non-stochastic and does not rely on the collection of many 
noisy images. The method was validated by comparison to sample covariance matrices of small (32 $\times$ 32 pixel) 
images obtained through many independent noise realizations. The method appears robust for a wide range of reconstruction parameter settings, and enables several pertinent issues to be addressed with regard to 
image quality in CT. 

\new{In conclusion, we have applied the proposed covariance approximation in order to construct variance maps and visualize image pixel correlations, as well as to address questions of object-dependence and 
stationarity. These last two issues are particularly relevant, as our findings highlight the need for realistic task-specific simulation and phantom development when evaluating images obtained with IIR, since noise is 
likely to be non-stationary and highly object-dependent. 
Future work will directly address the computational strategies discussed in this work which make the proposed method feasible for larger images. Similarly, the application of our methodology to 
full task-based image quality assessment will be a future direction of this research. }

\label{sec:conclusion}


%% file: Chapter7-Conclusion/Conclusion.tex
\chapter{Summary and Conclusions}
\label{ch:conclusion}
In this thesis, we have developed and applied a framework for objective assessment of image quality in x-ray CT 
based upon the Hotelling observer (HO). We have demonstrated that CT system optimization can be performed 
efficiently, and that the resulting system and reconstruction design is objectively optimal with respect to the HO.
Further, we have provided evidence that typical reconstruction methods using FBP lead either to close agreement 
between the HO and humans, or at least that optimal HO parameters are also likely to be optimal for humans (Chapter \ref{ch:humans}). 

In Chapter \ref{ch:sampling}, we provided a mathematical intuition to justify restriction of the HO to an ROI-HO and 
validated this intuition in simulation of parallel-beam FBP reconstruction. Specifically, we related the issue of long-range 
pixel correlation and image pixel size to the concept of aliasing, showing that because pixels outside of an ROI are dependent 
on the interior of the ROI, they do not substantially contribute to HO metric evaluation. Further, this hypothesis was 
validated by showing that HO efficiency begins to plateau when the ROI boundary used passes a given distance from a small signal.
This transition into diminishing HO SNR improvement is gradual, but seems to depend primarily on system geometry rather than image 
pixel size, thus implying that the ROI-HO method is a practical and robust means of performing parameter sweeps to optimize a system. 

Next, in Chapter \ref{ch:roi} we validated and evaluated the ROI-HO by comparing the results of optimizing a single parameter to results 
obtained with other established approaches to HO metrics. Namely, we showed that the ROI-HO gave nearly identical results to the 
channelized HO (CHO) with efficient channels, when the assumptions upon which the CHO is predicated were satisfied. Additionally, we
performed a comparison with common Fourier-based approximations to the HO and demonstrated the potential inconsistencies 
presented by the Fourier approach. Lastly, we demonstrated the difficulty of performing system optimization with HO metrics 
computed from sample statistics by illustrating the large number of realizations necessary to compute these metrics with minimal 
statistical variability. 

Chapter \ref{ch:breastct} provided an example of a full system optimization using our proposed method. The system chosen 
was simulated based on an existing prototype breast CT system. This enabled us to compare our conclusions regarding 
parameter selection with conclusions obtained through alternative means by the researches who developed the system.
Our results were consistent with previous findings in terms of optimal parameter values, and the absolute performance predicted 
in our simulations was close to demonstrated human performance. The ROI-HO metrics consistently corresponded to superior 
performance to humans, which is to be expected since the HO is ideal in our studies. 

Lastly, in Chapter \ref{ch:tv} we laid the groundwork for the extension of HO methods to images obtained through total-variation- (TV) based
image reconstruction. In particular, we demonstrated a means of approximating the image covariance for images obtained with 
unconstrained TV minimization. The approximation was validated by comparing to sample covariances for small images. We then applied 
the approximation to answering several pertinent questions regarding noise behavior in TV-based reconstruction. 

In conclusion, this thesis develops and demonstrates a method for CT algorithm and system design based on the HO which is 
efficient to compute and straightforward to implement, provided basic reconstruction software is available. The work developed in the 
first part of the thesis is immediately applicable to aiding in protocol and system design for a wide array of clinical applications, and therefore 
has the potential to facilitate an improvement in clinical utility as well as to speed the development and implementation of novel systems. 
Finally, the work regarding noise in optimization-based reconstruction is an important initial development which could ultimately lead to 
objective task-based assessment of TV-based reconstruction algorithms. The clinical implications of improved assessment for optimization-based
reconstruction are substantial, since a task-based framework could help to ensure that algorithm-enabled dose-reduction does not come at the 
cost of decreased diagnostic utility. 

%% file: thesis.bbl
\begin{thebibliography}{100}

\bibitem{wunderlich_new_2013}
A.~Wunderlich and F.~Noo, ``New {Theoretical} {Results} on {Channelized}
  {Hotelling} {Observer} {Performance} {Estimation} {With} {Known} {Difference}
  of {Class} {Means},'' {\em IEEE Transactions on Nuclear Science}, vol.~60,
  pp.~182--193, Feb. 2013.

\bibitem{kwan_evaluation_2006}
A.~L.~C. Kwan, J.~M. Boone, K.~Yang, and S.-Y. Huang, ``Evaluation of the
  spatial resolution characteristics of a cone-beam breast {CT} scanner,'' {\em
  Medical Physics}, vol.~34, pp.~275--281, Dec. 2006.

\bibitem{koehler_noise_2009}
T.~Koehler and R.~Proksa, ``Noise properties of maximum likelihood
  reconstruction with edge-preserving regularization in transmission
  tomography,'' {\em Proceedings of the Fully 3D Meeting}, pp.~263--266, 2009.

\bibitem{rose_noise_2015}
S.~Rose, E.~Y. Sidky, X.~Pan, and M.~S. Andersen, ``Noise properties of {CT}
  images reconstructed by use of a constrained total-variation,
  data-discrepancy minimization,'' {\em Med. Phys.}, vol.~42, 2015.

\bibitem{barrett_radiological_1996}
H.~H. Barrett and W.~Swindell, {\em Radiological imaging: the theory of image
  formation, detection, and processing}.
\newblock Academic Press, 1996.

\bibitem{kak_principles_1988}
A.~C. Kak and M.~Slaney, {\em Principles of {Computerized} {Tomographic}
  {Imaging}}.
\newblock IEEE Press, 1988.

\bibitem{sidky_constrained_2011}
E.~Y. Sidky, Y.~Duchin, X.~Pan, and C.~Ullberg, ``A constrained,
  total-variation minimization algorithm for low-intensity x-ray {CT},'' {\em
  Medical physics}, vol.~38, p.~S117, 2011.

\bibitem{tian_low-dose_2011}
Z.~Tian, X.~Jia, K.~Yuan, T.~Pan, and S.~B. Jiang, ``Low-dose {CT}
  reconstruction via edge-preserving total variation regularization,'' {\em
  Physics in medicine and biology}, vol.~56, no.~18, p.~5949, 2011.

\bibitem{bian_evaluation_2010}
J.~Bian, J.~H. Siewerdsen, X.~Han, E.~Y. Sidky, J.~L. Prince, C.~A. Pelizzari,
  and X.~Pan, ``Evaluation of sparse-view reconstruction from
  flat-panel-detector cone-beam {CT},'' {\em Physics in medicine and biology},
  vol.~55, no.~22, p.~6575, 2010.

\bibitem{han_algorithm-enabled_2011}
X.~Han, J.~Bian, D.~Eaker, T.~Kline, E.~Sidky, E.~Ritman, and X.~Pan,
  ``Algorithm-{Enabled} {Low}-{Dose} {Micro}-{CT} {Imaging},'' {\em IEEE
  Transactions on Medical Imaging}, vol.~30, no.~3, pp.~606--620, 2011.

\bibitem{sidky_enhanced_2009}
E.~Y. Sidky, X.~Pan, I.~S. Reiser, R.~M. Nishikawa, R.~H. Moore, and D.~B.
  Kopans, ``Enhanced imaging of microcalcifications in digital breast
  tomosynthesis through improved image-reconstruction algorithms,'' {\em
  Medical physics}, vol.~36, p.~4920, 2009.

\bibitem{barrett_foundations_2004}
H.~H. Barrett and K.~J. Myers, {\em Foundations of {Image} {Science}}.
\newblock Hoboken, New Jersey: John Wiley \& Sons, Inc., 2004.

\bibitem{trees_detection_1968}
H.~L.~V. Trees, {\em Detection, estimation, and modulation theory ({Part}
  {I})}.
\newblock Academic Press, 1968.

\bibitem{wagner_assortment_1972}
R.~F. Wagner and K.~E. Weaver, ``An assortment of image quality indexes for
  radiographic film-screen combinations—can they be resolved?,'' in {\em
  Application of {Optical} {Instrumentation} in {Medicine}}, pp.~83--94, 1972.

\bibitem{pineda_analysis_2008}
A.~R. Pineda, J.~H. Siewerdsen, and D.~J. Tward, ``Analysis of image noise in
  3d cone-beam {CT}: {Spatial} and {Fourier} domain approaches under conditions
  of varying stationarity,'' in {\em Medical {Imaging}}, pp.~69131Q--69131Q,
  2008.

\bibitem{abbey_observer_1996}
C.~K. Abbey, H.~H. Barrett, and D.~W. Wilson, ``Observer signal-to-noise ratios
  for the {ML}-{EM} algorithm,'' in {\em Medical {Imaging} 1996}, pp.~47--58,
  International Society for Optics and Photonics, 1996.

\bibitem{abbey_linear_1995}
C.~K. Abbey and H.~H. Barrett, ``Linear iterative reconstruction algorithms:
  study of observer performance,'' {\em Information Processing in Medical Im},
  pp.~65--76, 1995.

\bibitem{soares_noise_1995}
E.~J. Soares, H.~H. Barrett, and C.~K. Abbey, ``Noise characterization and
  objective image-quality assessment of {SPECT} imaging,'' in {\em Information
  {Processing} in {Medical} {Imaging}}, vol.~3, p.~353, Kluwer Academic Pub,
  1995.

\bibitem{barrett_model_1993}
H.~H. Barrett, J.~Yao, J.~P. Rolland, and K.~J. Myers, ``Model observers for
  assessment of image quality,'' {\em Proceedings of the National Academy of
  Sciences}, vol.~90, no.~21, pp.~9758--9765, 1993.

\bibitem{fiete_psychophysical_1987}
R.~Fiete, H.~Barrett, E.~Cargill, K.~Myers, and W.~Smith, ``Psychophysical
  validation of the {Hotelling} trace criterion as a metric for system
  performance,'' in {\em Proceedings {SPIE} {Medical} {Imaging}}, vol.~767,
  pp.~298--305, 1987.

\bibitem{tsui_comparison_1978}
B.~Tsui, C.~Metz, F.~Atkins, S.~Starr, and R.~Beck, ``A comparison of optimum
  detector spatial resolution in nuclear imaging based on statistical theory
  and on observer performance,'' {\em Physics in medicine and biology},
  vol.~23, no.~4, p.~654, 1978.

\bibitem{myers_visual_1985}
K.~J. Myers, {\em Visual {Perception} in {Correlated} {Noise} (models).}
\newblock PhD thesis, 1985.

\bibitem{burgess_human_2001}
A.~E. Burgess, F.~L. Jacobson, and P.~F. Judy, ``Human observer detection
  experiments with mammograms and power-law noise,'' {\em Medical physics},
  vol.~28, p.~419, 2001.

\bibitem{myers_effect_1985}
K.~J. Myers, H.~H. Barrett, M.~C. Borgstrom, D.~D. Patton, and G.~W. Seeley,
  ``Effect of noise correlation on detectability of disk signals in medical
  imaging,'' {\em JOSA A}, vol.~2, no.~10, pp.~1752--1759, 1985.

\bibitem{fiete_hotelling_1987}
R.~D. Fiete, H.~H. Barrett, W.~E. Smith, and K.~J. Myers, ``Hotelling trace
  criterion and its correlation with human-observer performance,'' {\em JOSA
  A}, vol.~4, no.~5, pp.~945--953, 1987.

\bibitem{rolland_effect_1992}
J.~P. Rolland and H.~H. Barrett, ``Effect of random background inhomogeneity on
  observer detection performance,'' {\em JOSA A}, vol.~9, no.~5, pp.~649--658,
  1992.

\bibitem{jenkins_spectral_1968}
G.~M. Jenkins and D.~G. Watts, ``Spectral analysis,'' 1968.

\bibitem{dainty_image_1974}
J.~C. Dainty and R.~Shaw, {\em Image {Science}: {Principles}, {Analysis} and
  {Evaluation} of {Photographic}-type {Imaging} {Processes}}, vol.~29.
\newblock London: Academic, 1974.

\bibitem{blackman_measurement_1959}
R.~B. Blackman, J.~W. Tukey, J.~W. Tukey, and J.~W. Tukey, {\em The measurement
  of power spectra: from the point of view of communications engineering},
  vol.~190.
\newblock Dover New York, 1959.

\bibitem{rossmann_measurement_1964}
K.~Rossmann, ``Measurement of the modulation transfer function of radiographic
  systems containing fluorescent screens,'' {\em Physics in Medicine and
  Biology}, vol.~9, no.~4, p.~551, 1964.

\bibitem{rossmann_spatial_1968}
K.~Rossmann, ``The spatial frequency spectrum: a means for studying the quality
  of radiographic imaging systems,'' {\em Radiology}, vol.~90, no.~1,
  pp.~1--13, 1968.

\bibitem{baek_local_2011}
J.~Baek and N.~J. Pelc, ``Local and global 3d noise power spectrum in cone-beam
  {CT} system with {FDK} reconstruction,'' {\em Medical Physics}, vol.~38,
  pp.~2122--2131, Apr. 2011.

\bibitem{pineda_beyond_2012}
A.~R. Pineda, D.~J. Tward, A.~Gonzalez, and J.~H. Siewerdsen, ``Beyond noise
  power in 3d computed tomography: {The} local {NPS} and off-diagonal elements
  of the {Fourier} domain covariance matrix,'' {\em Medical Physics}, vol.~39,
  pp.~3240--3252, May 2012.

\bibitem{defrise_performance_1994}
M.~Defrise, A.~Geissbuhler, and D.~W. Townsend, ``A performance study of 3d
  reconstruction algorithms for positron emission tomography,'' {\em Physics in
  Medicine and Biology}, vol.~39, p.~305, Mar. 1994.

\bibitem{yao_predicting_1992}
J.~Yao and H.~H. Barrett, ``Predicting human performance by a channelized
  {Hotelling} observer model,'' in {\em San {Diego}'92}, pp.~161--168,
  International Society for Optics and Photonics, 1992.

\bibitem{hanley_meaning_1982}
J.~A. Hanley and B.~J. McNeil, ``The meaning and use of the area under a
  receiver operating characteristic ({ROC}) curve,'' {\em Radiology}, vol.~143,
  pp.~29--36, Apr. 1982.

\bibitem{park_singular_2009}
S.~Park, J.~M. Witten, and K.~J. Myers, ``Singular {Vectors} of a {Linear}
  {Imaging} {System} as {Efficient} {Channels} for the {Bayesian} {Ideal}
  {Observer},'' {\em IEEE Trans. Med. Imag.}, vol.~28, pp.~657--68, May 2009.

\bibitem{beutel_handbook_2000}
J.~Beutel, H.~L. Kundel, and R.~L.~V. Metter, {\em Handbook of {Medcial}
  {Imaging}}, vol.~1 - Physics and Psychophysics.
\newblock SPIE Press, 2000.

\bibitem{witten_partial_2010}
J.~Witten, S.~Park, and K.~Myers, ``Partial {Least} {Squares}: {A} {Method} to
  {Estimate} {Efficient} {Channels} for the {Ideal} {Observers},'' {\em IEEE
  Transactions on Medical Imaging}, vol.~29, pp.~1050--1058, Apr. 2010.

\bibitem{nocedal_numerical_2006}
J.~Nocedal and S.~Wright, {\em Numerical {Optimization}, 2nd ed.}
\newblock Springer, 2006.

\bibitem{laroque_evaluation_2007}
S.~J. LaRoque, E.~Y. Sidky, D.~C. Edwards, and X.~Pan, ``Evaluation of the
  channelized {Hotelling} observer for signal detection in 2d tomographic
  imaging,'' in {\em Medical {Imaging}:{\textbackslash} {Image} {Perception},
  {Observer} {Performance}, and {Technology} {Assessment}} (Y.~Jiang and
  B.~Sahiner, eds.), vol.~6515 of {\em Proc. {SPIE}}, p.~651514, 2007.

\bibitem{sidky_accurate_2008}
E.~Y. Sidky, S.~J. LaRoque, and X.~Pan, ``Accurate computation of the
  {Hotelling} template for {SKE}/{BKE} detection tasks,'' in {\em Medical
  {Imaging}:{\textbackslash} {Image} {Perception}, {Observer} {Performance},
  and {Technology} {Assessment}} (B.~Sahiner and D.~J. Manning, eds.),
  vol.~6917 of {\em Proc. {SPIE}}, pp.~69170W--69170W--5, 2008.

\bibitem{sidky_-depth_2008}
E.~Sidky and X.~Pan, ``In-depth analysis of cone-beam {CT} image reconstruction
  by ideal observer performance on a detection task,'' in {\em {IEEE} {Nuclear}
  {Science} {Symposium} {Conference} {Record}, 2008. {NSS} '08},
  pp.~5161--5165, 2008.

\bibitem{wunderlich_image_2008}
A.~Wunderlich and F.~Noo, ``Image covariance and lesion detectability in direct
  fan-beam x-ray computed tomography,'' {\em Physics in Medicine and Biology},
  vol.~53, pp.~2471--2493, May 2008.

\bibitem{abbey_human-_2001}
C.~K. Abbey and H.~H. Barrett, ``Human- and model-observer performance in
  ramp-spectrum noise: effects of regularization and object variability,'' {\em
  Journal of the Optical Society of America A}, vol.~18, pp.~473--488, Mar.
  2001.

\bibitem{papoulis_probability_2002}
A.~Papoulis and S.~U. Pillai, {\em Probability, random variables and stochastic
  processes with errata sheet}.
\newblock New York: McGraw-Hill Education, 4~ed., 2002.

\bibitem{gong_microcalcification_2004}
X.~Gong, A.~A. Vedula, and S.~J. Glick, ``Microcalcification detection using
  cone-beam {CT} mammography with a flat-panel imager,'' {\em Physics in
  Medicine and Biology}, vol.~49, pp.~2183--2195, June 2004.

\bibitem{kijewski_noise_1987}
M.~F. Kijewski and P.~F. Judy, ``The noise power spectrum of {CT} images,''
  {\em Physics in Medicine and Biology}, vol.~32, no.~5, p.~565, 1987.

\bibitem{sanchez_investigation_2013}
A.~A. Sanchez, E.~Y. Sidky, and X.~Pan, ``Investigation of {Template}
  {Structure} for a {Cone}-{Beam} {CT} {Signal} {Detection} {Task},'' in {\em
  Proceedings of the {Fully} 3D {Meeting}}, vol.~11, pp.~444--447, 2013.

\bibitem{wagner_unified_1985}
R.~F. Wagner and D.~G. Brown, ``Unified {SNR} analysis of medical imaging
  systems,'' {\em Physics in Medicine and Biology}, vol.~30, no.~6, p.~489,
  1985.

\bibitem{burgess_efficiency_1981}
A.~E. Burgess, R.~F. Wagner, R.~J. Jennings, and H.~B. Barlow, ``Efficiency of
  human visual signal discrimination.,'' {\em Science}, 1981.

\bibitem{metz_roc_1986}
C.~E. Metz, ``{ROC} methodology in radiologic imaging,'' {\em Investigative
  radiology}, vol.~21, no.~9, pp.~720--733, 1986.

\bibitem{gallas_validating_2003}
B.~D. Gallas and H.~H. Barrett, ``Validating the use of channels to estimate
  the ideal linear observer,'' {\em JOSA A}, vol.~20, no.~9, pp.~1725--1738,
  2003.

\bibitem{chan_classifier_1999}
H.-P. Chan, B.~Sahiner, R.~F. Wagner, and N.~Petrick, ``Classifier design for
  computer-aided diagnosis: {Effects} of finite sample size on the mean
  performance of classical and neural network classifiers,'' {\em Medical
  Physics}, vol.~26, no.~12, p.~2654, 1999.

\bibitem{boone_accurate_1997}
J.~M. Boone and J.~A. Seibert, ``An accurate method for computer-generating
  tungsten anode x-ray spectra from 30 to 140 {kV},'' {\em Medical physics},
  vol.~24, no.~11, pp.~1661--1670, 1997.

\bibitem{boone_molybdenum_1997}
J.~M. Boone, T.~R. Fewell, and R.~J. Jennings, ``Molybdenum, rhodium, and
  tungsten anode spectral models using interpolating polynomials with
  application to mammography,'' {\em Medical Physics}, vol.~24, no.~12,
  pp.~1863--1874, 1997.

\bibitem{boone_spectral_1998}
J.~M. Boone, ``Spectral modeling and compilation of quantum fluence in
  radiography and mammography,'' vol.~3336, pp.~592--601, 1998.

\bibitem{johns_x-ray_1987}
P.~C. Johns and M.~J. Yaffe, ``X-ray characterisation of normal and neoplastic
  breast tissues,'' {\em Physics in medicine and biology}, vol.~32, no.~6,
  p.~675, 1987.

\bibitem{boone_technique_2005}
J.~M. Boone, A.~L. Kwan, J.~A. Seibert, N.~Shah, K.~K. Lindfors, and T.~R.
  Nelson, ``Technique factors and their relationship to radiation dose in
  pendant geometry breast {CT},'' {\em Medical Physics}, vol.~32, p.~3767,
  2005.

\bibitem{wagner_multivariate_1993}
R.~F. Wagner, D.~G. Brown, J.-P. Guedon, K.~J. Myers, and K.~A. Wear,
  ``Multivariate gaussian pattern classification: {Effects} of finite sample
  size and the addition of correlated or noisy features on summary measures of
  goodness,'' in {\em Information {Processing} in {Medical} {Imaging}} (H.~H.
  Barrett and A.~F. Gmitro, eds.), no.~687 in Lecture {Notes} in {Computer}
  {Science}, pp.~507--524, Springer Berlin Heidelberg, Jan. 1993.

\bibitem{fukunaga_effects_1989}
K.~Fukunaga and R.~Hayes, ``Effects of sample size in classifier design,'' {\em
  IEEE Transactions on Pattern Analysis and Machine Intelligence}, vol.~11,
  pp.~873--885, Aug. 1989.

\bibitem{boone_dedicated_2001}
J.~M. Boone, T.~R. Nelson, K.~K. Lindfors, and J.~A. Seibert, ``Dedicated
  {Breast} {CT}: {Radiation} {Dose} and {Image} {Quality} {Evaluation}1,'' {\em
  Radiology}, vol.~221, no.~3, pp.~657--667, 2001.

\bibitem{chen_cone-beam_2002}
B.~Chen and R.~Ning, ``Cone-beam volume {CT} breast imaging: {Feasibility}
  study,'' {\em Medical Physics}, vol.~29, p.~755, 2002.

\bibitem{yang_dedicated_2007}
W.~T. Yang, S.~Carkaci, L.~Chen, C.-J. Lai, A.~Sahin, G.~J. Whitman, and C.~C.
  Shaw, ``Dedicated cone-beam breast {CT}: {Feasibility} study with surgical
  mastectomy specimens,'' {\em AJR. American journal of roentgenology},
  vol.~189, no.~6, p.~1312, 2007.

\bibitem{mckinley_development_2012}
R.~L. McKinley, M.~P. Tornai, L.~A. Tuttle, D.~Steed, and C.~M. Kuzmiak,
  ``Development and initial demonstration of a low-dose dedicated fully 3d
  breast {CT} system,'' in {\em Breast {Imaging}}, pp.~442--449, Springer,
  2012.

\bibitem{gong_computer_2006}
X.~Gong, S.~J. Glick, B.~Liu, A.~A. Vedula, and S.~Thacker, ``A computer
  simulation study comparing lesion detection accuracy with digital
  mammography, breast tomosynthesis, and cone-beam {CT} breast imaging,'' {\em
  Medical physics}, vol.~33, p.~1041, 2006.

\bibitem{glick_breast_2007}
S.~J. Glick, ``Breast {CT},'' {\em Annu. Rev. Biomed. Eng.}, vol.~9,
  pp.~501--526, 2007.

\bibitem{pisano_diagnostic_2005}
E.~D. Pisano, C.~Gatsonis, E.~Hendrick, M.~Yaffe, J.~K. Baum, S.~Acharyya,
  E.~F. Conant, L.~L. Fajardo, L.~Bassett, and C.~D'Orsi, ``Diagnostic
  performance of digital versus film mammography for breast-cancer screening,''
  {\em New England Journal of Medicine}, vol.~353, no.~17, pp.~1773--1783,
  2005.

\bibitem{lindfors_dedicated_2008}
K.~K. Lindfors, J.~M. Boone, T.~R. Nelson, K.~Yang, A.~L.~C. Kwan, and D.~F.
  Miller, ``Dedicated {Breast} {CT}: {Initial} {Clinical} {Experience}1,'' {\em
  Radiology}, vol.~246, pp.~725--733, Mar. 2008.

\bibitem{rose_human_1974}
A.~Rose, ``Human {Vision},'' in {\em Vision}, pp.~29--53, Springer, 1974.

\bibitem{shen_suei02:_2013}
Y.~Shen, Y.~Zhong, C.~Lai, T.~Wang, and C.~Shaw, ``{SU}‐{E}‐{I}‐02:
  {Effects} of {Projection} {View} {Sampling} {On} {CT} {Numbers} and {Noise}
  {Level} in {Cone} {Beam} {Breast} {CT},'' {\em Medical Physics}, vol.~40,
  pp.~125--125, June 2013.

\bibitem{yang_computer_2007}
K.~Yang, A.~L.~C. Kwan, and J.~M. Boone, ``Computer modeling of the spatial
  resolution properties of a dedicated breast {CT} system,'' {\em Medical
  Physics}, vol.~34, pp.~2059--2069, May 2007.

\bibitem{yang_noise_2008}
K.~Yang, A.~L.~C. Kwan, S.-Y. Huang, N.~J. Packard, and J.~M. Boone, ``Noise
  power properties of a cone-beam {CT} system for breast cancer detection,''
  {\em Medical Physics}, vol.~35, no.~12, p.~5317, 2008.

\bibitem{lai_visibility_2007}
C.-J. Lai, C.~C. Shaw, L.~Chen, M.~C. Altunbas, X.~Liu, T.~Han, T.~Wang, W.~T.
  Yang, G.~J. Whitman, and S.-J. Tu, ``Visibility of microcalcification in cone
  beam breast {CT}: {Effects} of x-ray tube voltage and radiation dose,'' {\em
  Medical Physics}, vol.~34, no.~7, p.~2995, 2007.

\bibitem{metz_basic_1978}
C.~E. Metz, ``Basic principles of {ROC} analysis,'' in {\em Seminars in nuclear
  medicine}, vol.~8, pp.~283--298, 1978.

\bibitem{glick_evaluating_2007}
S.~J. Glick, S.~Thacker, X.~Gong, and B.~Liu, ``Evaluating the impact of x-ray
  spectral shape on image quality in flat-panel {CT} breast imaging,'' {\em
  Medical Physics}, vol.~34, no.~1, p.~5, 2007.

\bibitem{packard_effect_2012}
N.~J. Packard, C.~K. Abbey, K.~Yang, and J.~M. Boone, ``Effect of slice
  thickness on detectability in breast {CT} using a prewhitened matched filter
  and simulated mass lesions,'' {\em Medical Physics}, vol.~39, no.~4, p.~1818,
  2012.

\bibitem{chen_association_2013}
L.~Chen, C.~K. Abbey, and J.~M. Boone, ``Association between power law
  coefficients of the anatomical noise power spectrum and lesion detectability
  in breast imaging modalities,'' {\em Physics in Medicine and Biology},
  vol.~58, p.~1663, Mar. 2013.

\bibitem{wagner_multiplex_1981}
R.~F. Wagner, D.~G. Brown, and C.~E. Metz, ``On {The} {Multiplex} {Advantage}
  {Of} {Coded} {Source}/{Aperture} {Photon} {Imaging},'' vol.~0314, pp.~72--76,
  1981.

\bibitem{hanson_rayleigh_1991}
K.~Hanson and K.~J. Myers, ``Rayleigh task performance as a method to evaluate
  image reconstruction algorithms,'' in {\em Maximum {Entropy} and {Bayesian}
  {Methods}}, The {Fundamental} {Theories} of {Physics}, pp.~303--312,
  Springer, 1991.

\bibitem{myers_rayleigh_1998}
K.~J. Myers, R.~F. Wagner, and K.~M. Hanson, ``Rayleigh task performance in
  tomographic reconstructions: comparison of human and machine performance,''
  {\em Image Processing}, pp.~628--637, 1998.

\bibitem{sidky_accurate_2006}
E.~Y. Sidky, C.-M. Kao, and X.~Pan, ``Accurate image reconstruction from
  few-views and limited-angle data in divergent-beam {CT},'' {\em Journal of
  X-ray Science and Technology}, vol.~14, no.~2, pp.~119--139, 2006.

\bibitem{sidky_image_2008}
E.~Y. Sidky and X.~Pan, ``Image reconstruction in circular cone-beam computed
  tomography by constrained, total-variation minimization,'' {\em Physics in
  medicine and biology}, vol.~53, no.~17, p.~4777, 2008.

\bibitem{defrise_algorithm_2011}
M.~Defrise, C.~Vanhove, and X.~Liu, ``An algorithm for total variation
  regularization in high-dimensional linear problems,'' {\em Inverse Problems},
  vol.~27, no.~6, p.~065002, 2011.

\bibitem{jensen_implementation_2012}
T.~L. Jensen, J.~H. Jørgensen, P.~C. Hansen, and S.~H. Jensen,
  ``Implementation of an optimal first-order method for strongly convex total
  variation regularization,'' {\em BIT Numerical Mathematics}, vol.~52, no.~2,
  pp.~329--356, 2012.

\bibitem{jorgensen_quantifying_2013}
J.~S. Jorgensen, E.~Y. Sidky, and X.~Pan, ``Quantifying admissible
  undersampling for sparsity-exploiting iterative image reconstruction in x-ray
  {CT},'' {\em Medical Imaging, IEEE Transactions on}, vol.~32, no.~2,
  pp.~460--473, 2013.

\bibitem{ramani_splitting-based_2012}
S.~Ramani and J.~A. Fessler, ``A splitting-based iterative algorithm for
  accelerated statistical {X}-ray {CT} reconstruction,'' {\em Medical Imaging,
  IEEE Transactions on}, vol.~31, no.~3, pp.~677--688, 2012.

\bibitem{chen_prior_2008}
G.-H. Chen, J.~Tang, and S.~Leng, ``Prior image constrained compressed sensing
  ({PICCS}): a method to accurately reconstruct dynamic {CT} images from highly
  undersampled projection data sets,'' {\em Medical physics}, vol.~35, no.~2,
  pp.~660--663, 2008.

\bibitem{sidky_image_2010}
E.~Y. Sidky, M.~A. Anastasio, and X.~Pan, ``Image reconstruction exploiting
  object sparsity in boundary-enhanced {X}-ray phase-contrast tomography,''
  {\em Optics express}, vol.~18, no.~10, pp.~10404--10422, 2010.

\bibitem{ritschl_improved_2011}
L.~Ritschl, F.~Bergner, C.~Fleischmann, and M.~Kachelrie{\textbackslash}s~s,
  ``Improved total variation-based {CT} image reconstruction applied to
  clinical data,'' {\em Physics in medicine and biology}, vol.~56, no.~6,
  p.~1545, 2011.

\bibitem{han_optimization-based_2012}
X.~Han, J.~Bian, E.~L. Ritman, E.~Y. Sidky, and X.~Pan, ``Optimization-based
  reconstruction of sparse images from few-view projections,'' {\em Physics in
  medicine and biology}, vol.~57, no.~16, p.~5245, 2012.

\bibitem{song_sparseness_2007}
J.~Song, Q.~H. Liu, G.~A. Johnson, and C.~T. Badea, ``Sparseness prior based
  iterative image reconstruction for retrospectively gated cardiac
  micro-{CT},'' {\em Medical Physics}, vol.~34, pp.~4476--4483, Nov. 2007.

\bibitem{chen_temporal_2009}
G.-H. Chen, J.~Tang, and J.~Hsieh, ``Temporal resolution improvement using
  {PICCS} in {MDCT} cardiac imaging,'' {\em Medical physics}, vol.~36, no.~6,
  pp.~2130--2135, 2009.

\bibitem{bergner_investigation_2010}
F.~Bergner, T.~Berkus, M.~Oelhafen, P.~Kunz, T.~Pan, R.~Grimmer, L.~Ritschl,
  and M.~Kachelrie{\textbackslash}s~s, ``An investigation of 4d cone-beam {CT}
  algorithms for slowly rotating scanners,'' {\em Medical physics}, vol.~37,
  no.~9, pp.~5044--5053, 2010.

\bibitem{ritschl_iterative_2012}
L.~Ritschl, S.~Sawall, M.~Knaup, A.~Hess, and M.~Kachelrie{\textbackslash}s~s,
  ``Iterative 4d cardiac micro-{CT} image reconstruction using an adaptive
  spatio-temporal sparsity prior,'' {\em Physics in medicine and biology},
  vol.~57, no.~6, p.~1517, 2012.

\bibitem{kuntz_constrained_2013}
J.~Kuntz, B.~Flach, R.~Kueres, W.~Semmler, M.~Kachelrie{\textbackslash}s~s, and
  S.~Bartling, ``Constrained reconstructions for 4d intervention guidance,''
  {\em Physics in medicine and biology}, vol.~58, no.~10, p.~3283, 2013.

\bibitem{rudin_nonlinear_1992}
L.~I. Rudin, S.~Osher, and E.~Fatemi, ``Nonlinear total variation based noise
  removal algorithms,'' {\em Physica D: Nonlinear Phenomena}, vol.~60,
  pp.~259--268, Nov. 1992.

\bibitem{delaney_globally_1998}
A.~Delaney and Y.~Bresler, ``Globally convergent edge-preserving regularized
  reconstruction: an application to limited-angle tomography,'' {\em IEEE
  Transactions on Image Processing}, vol.~7, pp.~204--221, Feb. 1998.

\bibitem{vogel_iterative_1996}
C.~Vogel and M.~Oman, ``Iterative {Methods} for {Total} {Variation}
  {Denoising},'' {\em SIAM Journal on Scientific Computing}, vol.~17,
  pp.~227--238, Jan. 1996.

\bibitem{panin_total_1999}
V.~Panin, G.~Zeng, and G.~Gullberg, ``Total variation regulated {EM} algorithm
  [{SPECT} reconstruction],'' {\em IEEE Transactions on Nuclear Science},
  vol.~46, pp.~2202--2210, Dec. 1999.

\bibitem{persson_total_2001}
M.~Persson, D.~Bone, and H.~Elmqvist, ``Total variation norm for
  three-dimensional iterative reconstruction in limited view angle
  tomography,'' {\em Physics in Medicine and Biology}, vol.~46, p.~853, Mar.
  2001.

\bibitem{lawson_contributions_1961}
C.~L. Lawson, {\em Contributions to the theory of linear least maximum
  approximation}.
\newblock PhD thesis, University of California, Los Angeles–Mathematics.,
  1961.

\bibitem{beaton_fitting_1974}
A.~E. Beaton and J.~W. Tukey, ``The fitting of power series, meaning
  polynomials, illustrated on band-spectroscopic data,'' {\em Technometrics},
  vol.~16, no.~2, pp.~147--185, 1974.

\bibitem{wohlberg_iteratively_2007}
B.~Wohlberg and P.~Rodriguez, ``An {Iteratively} {Reweighted} {Norm}
  {Algorithm} for {Minimization} of {Total} {Variation} {Functionals},'' {\em
  IEEE Signal Processing Letters}, vol.~14, pp.~948--951, Dec. 2007.

\bibitem{chartrand_iteratively_2008}
R.~Chartrand and W.~Yin, ``Iteratively reweighted algorithms for compressive
  sensing,'' in {\em Acoustics, speech and signal processing, 2008. {ICASSP}
  2008. {IEEE} international conference on}, pp.~3869--3872, IEEE, 2008.

\bibitem{sidky_constrained_2014}
E.~Y. Sidky, R.~Chartrand, J.~M. Boone, and {Xiaochuan Pan}, ``Constrained
  {TpV} {Minimization} for {Enhanced} {Exploitation} of {Gradient} {Sparsity}:
  {Application} to {CT} {Image} {Reconstruction},'' {\em IEEE Journal of
  Translational Engineering in Health and Medicine}, vol.~2, pp.~1--18, 2014.

\bibitem{leipsic_adaptive_2010}
J.~Leipsic, T.~M. LaBounty, B.~Heilbron, J.~K. Min, G.~B.~J. Mancini, F.~Y.
  Lin, C.~Taylor, A.~Dunning, and J.~P. Earls, ``Adaptive {Statistical}
  {Iterative} {Reconstruction}: {Assessment} of {Image} {Noise} and {Image}
  {Quality} in {Coronary} {CT} {Angiography},'' {\em American Journal of
  Roentgenology}, vol.~195, pp.~649--654, Sept. 2010.

\bibitem{vorona_reducing_2011}
G.~A. Vorona, R.~C. Ceschin, B.~L. Clayton, T.~Sutcavage, S.~S. Tadros, and
  A.~Panigrahy, ``Reducing abdominal {CT} radiation dose with the adaptive
  statistical iterative reconstruction technique in children: a feasibility
  study,'' {\em Pediatric Radiology}, vol.~41, pp.~1174--1182, May 2011.

\bibitem{ren_comparison_2012}
Q.~Ren, S.~K. Dewan, M.~Li, J.~Li, D.~Mao, Z.~Wang, and Y.~Hua, ``Comparison of
  adaptive statistical iterative and filtered back projection reconstruction
  techniques in brain {CT},'' {\em European Journal of Radiology}, vol.~81,
  pp.~2597--2601, Oct. 2012.

\bibitem{barrett_noise_1994}
H.~H. Barrett, D.~W. Wilson, and B.~M. Tsui, ``Noise properties of the {EM}
  algorithm. {I}. {Theory},'' {\em Physics in medicine and biology}, vol.~39,
  no.~5, p.~833, 1994.

\bibitem{wilson_noise_1994}
D.~W. Wilson, B.~M. Tsui, and H.~H. Barrett, ``Noise properties of the {EM}
  algorithm. {II}. {Monte} {Carlo} simulations,'' {\em Physics in medicine and
  biology}, vol.~39, no.~5, p.~847, 1994.

\bibitem{soares_noise_2000}
E.~J. Soares, C.~L. Byrne, and S.~J. Glick, ``Noise characterization of
  block-iterative reconstruction algorithms. {I}. {Theory},'' {\em Medical
  Imaging, IEEE Transactions on}, vol.~19, no.~4, pp.~261--270, 2000.

\bibitem{soares_noise_2005}
E.~Soares, S.~Glick, and J.~Hoppin, ``Noise characterization of block-iterative
  reconstruction algorithms: {II}. {Monte} {Carlo} simulations,'' {\em IEEE
  Transactions on Medical Imaging}, vol.~24, pp.~112--121, Jan. 2005.

\bibitem{wang_noise_1997}
W.~Wang and G.~Gindi, ``Noise analysis of {MAP} - {EM} algorithms for emission
  tomography,'' {\em Physics in Medicine and Biology}, vol.~42, p.~2215, Nov.
  1997.

\bibitem{qi_unified_2003}
J.~Qi, ``A unified noise analysis for iterative image estimation,'' {\em
  Physics in Medicine and Biology}, vol.~48, p.~3505, Nov. 2003.

\bibitem{fessler_mean_1996}
J.~A. Fessler, ``Mean and variance of implicitly defined biased estimators
  (such as penalized maximum likelihood): {Applications} to tomography,'' {\em
  Image Processing, IEEE Transactions on}, vol.~5, no.~3, pp.~493--506, 1996.

\bibitem{qi_fast_1999}
J.~Qi and R.~M. Leahy, ``Fast computation of the covariance of {MAP}
  reconstructions of {PET} images,'' in {\em Proc. {SPIE}}, vol.~3661,
  pp.~344--355, 1999.

\bibitem{qi_resolution_2000}
J.~Qi and R.~M. Leahy, ``Resolution and noise properties of {MAP}
  reconstruction for fully 3-{D} {PET},'' {\em Medical Imaging, IEEE
  Transactions on}, vol.~19, no.~5, pp.~493--506, 2000.

\bibitem{zhang-oconnor_fast_2007}
Y.~Zhang-O'Connor and J.~Fessler, ``Fast {Predictions} of {Variance} {Images}
  for {Fan}-{Beam} {Transmission} {Tomography} {With} {Quadratic}
  {Regularization},'' {\em IEEE Transactions on Medical Imaging}, vol.~26,
  pp.~335--346, Mar. 2007.

\bibitem{li_noise_2011}
Y.~Li, ``Noise propagation for iterative penalized-likelihood image
  reconstruction based on {Fisher} information,'' {\em Physics in Medicine and
  Biology}, vol.~56, p.~1083, Feb. 2011.

\bibitem{chun_noise_2013}
S.~Y. Chun and J.~Fessler, ``Noise {Properties} of {Motion}-{Compensated}
  {Tomographic} {Image} {Reconstruction} {Methods},'' {\em IEEE Transactions on
  Medical Imaging}, vol.~32, pp.~141--152, Feb. 2013.

\bibitem{dutta_quantitative_2013}
J.~Dutta, S.~Ahn, and Q.~Li, ``Quantitative statistical methods for image
  quality assessment,'' {\em Theranostics}, vol.~3, no.~10, p.~741, 2013.

\bibitem{ahn_analysis_2008}
S.~Ahn and R.~Leahy, ``Analysis of {Resolution} and {Noise} {Properties} of
  {Nonquadratically} {Regularized} {Image} {Reconstruction} {Methods} for
  {PET},'' {\em IEEE Transactions on Medical Imaging}, vol.~27, pp.~413--424,
  Mar. 2008.

\bibitem{sidky_image_2007}
E.~Y. Sidky, R.~Chartrand, and X.~Pan, ``Image reconstruction from few views by
  non-convex optimization,'' in {\em Nuclear {Science} {Symposium} {Conference}
  {Record}, 2007. {NSS}'07. {IEEE}}, vol.~5, pp.~3526--3530, IEEE, 2007.

\bibitem{rodriguez_efficient_2009}
P.~Rodriguez and B.~Wohlberg, ``Efficient {Minimization} {Method} for a
  {Generalized} {Total} {Variation} {Functional},'' {\em IEEE Transactions on
  Image Processing}, vol.~18, pp.~322--332, Feb. 2009.

\bibitem{rodriguez_comparison_2012}
P.~Rodriguez and B.~Wohlberg, ``A comparison of the computational performance
  of {Iteratively} {Reweighted} {Least} {Squares} and alternating minimization
  algorithms for ℓ 1 inverse problems,'' in {\em Image {Processing} ({ICIP}),
  2012 19th {IEEE} {International} {Conference} on}, pp.~3069--3072, IEEE,
  2012.

\bibitem{forstner_metric_2003}
W.~Förstner and B.~Moonen, ``A metric for covariance matrices,'' in {\em
  Geodesy-{The} {Challenge} of the 3rd {Millennium}}, pp.~299--309, Springer,
  2003.

\end{thebibliography}
